\documentclass[12pt]{book}
\usepackage{amsmath, amssymb}

\newcommand{\F}{\noindent}
\newcommand{\qqq}{\qquad\qquad}
\newcommand{\qq}{\qquad}
\newcommand{\q}{\quad}
\newcommand{\SP}{\smallskip}
\newcommand{\MP}{\medskip}
\newcommand{\BP}{\bigskip}

\newcommand{\beq}{\begin{eqnarray}}
\newcommand{\ene}{\end{eqnarray}}
\newcommand{\beqn}{\begin{eqnarray*}}
\newcommand{\enen}{\end{eqnarray*}}

\newcommand{\N}{{\mbox{\bf N}}}
\newcommand{\HH}{{\cal H}}
\newcommand{\UU}{{\cal U}}
\newcommand{\OO}{{\cal O}}
\newcommand{\DD}{{\cal D}}
\newcommand{\C}{{\mbox{\bf C}}}
\newcommand{\RR}{{\cal  R}}
\newcommand{\la}{{\lambda}}

\newcommand{\tT}{{\widetilde T}}
\newcommand{\tH}{{\widetilde H}}

\newcommand{\SSS}{{\cal{S}}}
\newcommand{\TT}{{\cal T}}
\newcommand{\ff}{{\widetilde f}}
\newcommand{\ep}{{\epsilon}}
\newcommand{\FF}{{\cal F}}
\newcommand{\de}{{\delta}}
\newcommand{\deo}{{\delta_0}}
\newcommand{\del}{{\delta_1}}
\newcommand{\al}{{\alpha}}
\newcommand{\be}{{\beta}}
\newcommand{\sig}{{\sigma}}
\newcommand{\lan}{{\langle}}
\newcommand{\ran}{{\rangle}}
\newcommand{\parti}{{\partial}}
\newcommand{\wzeta}{{\widehat \zeta}}
\newcommand{\tf}{{\tilde f}}
\newcommand{\deL}{{\de'}}
\newcommand{\sigL}{{\sig'}}
\newcommand{\La}{{\Lambda}}
\newcommand{\ellj}{{\ell(j)}}
\newcommand{\wtP}{{\widetilde P}}
\newcommand{\whP}{{\widehat P}}
\newcommand{\Mdm}{{M_d^m}}

\newcommand{\AAA}{{\bf A}}
\newcommand{\BBB}{{\bf B}}
\newcommand{\PP}{{\bf P}}
\newcommand{\aaa}{{\mbox{\bf a}}}

\newcommand{\qqqq}{{\mbox{\bf q}}}
\newcommand{\kk}{{\mbox{\bf k}}}
\newcommand{\xx}{{\mbox{\bf x}}}

\newcommand{\eq}[1]{(\ref{#1})}

\newcommand{\nom}{\nonumber}

\begin{document}

\newtheorem{df}{Definition}[chapter]

\newtheorem{thm}[df]{Theorem}

\newtheorem{ass}[df]{Assumption}

\newtheorem{lem}[df]{Lemma}

\newtheorem{pro}[df]{Proposition}

\newtheorem{axm}[df]{Axiom}

\

\vskip16pt
\renewcommand{\thefootnote}{\fnsymbol{footnote}}

\newcommand{\namelistlabel}[1]{\mbox{#1}\hfil}
\newenvironment{namelist}[1]{%
\begin{list}{}
{\let\makelabel\namelistlabel
\settowidth{\labelwidth}{#1}
\setlength{\leftmargin}{1.1\labelwidth}}
}{%
\end{list}}

\title{\bf{QUANTUM MECHANICS}\footnote{\copyright 1998--2003 by Hitoshi Kitada, All Rights Reserved}}
\author{Hitoshi Kitada
{\footnote{Graduate School of Mathematical Sciences,
University of Tokyo,
Komaba, Meguro, Tokyo 153-8914, Japan,
e-mail: kitada@kims.ms.u-tokyo.ac.jp, 
web page: http://kims.ms.u-tokyo.ac.jp/}}}
\date{December 28, 2003}
\maketitle

\vskip8pt

\renewcommand{\thefootnote}{\arabic{footnote}}


%
%
%
%
%
%
%
%
%
%
%
%
%

\renewcommand{\thepage}{%
\roman{page}}

\setcounter{page}{2}

\chapter*{Preface}

I consider in this book a formulation of Quantum Mechanics, which is often abbreviated as QM. Usually QM is formulated based on the notion of time and space, both of which are thought {\it a priori} given quantities or notions. However, when we try to define the notion of velocity or momentum, we encounter a difficulty as we will see in chapter \ref{chap:1}. The problem is that if the notion of time is given {\it a priori}, the velocity is definitely determined when given a position, which contradicts the uncertainty principle of Heisenberg.

We then set the basis of QM on the notion of position and momentum operators as in chapter \ref{chap:2}. Time of a local system then is defined approximately as a ratio $|x|/|v|$ between the space coordinate $x$ and the velocity $v$, where $|x|$, etc. denotes the absolute value or length of a vector $x$. In this formulation of QM, we can keep the uncertainty principle, and time is a quantity that does not have precise values unlike the usually supposed notion of time has.

The feature of local time is that it is a time proper to each local system, which is defined as a finite set of quantum mechanical particles. We now have an infinite number of local times that are unique and proper to each local system.

Based on the notion of local time, the motion inside a local system is described by the usual Schr\"odinger equation. We investigate such motion in a given local system in part \ref{localmotion}. This is a usual quantum mechanics.

After some excursion of the investigation of local motion, we consider in part \ref{observation-part} the relative relation or motion between plural local systems. We regard each local system's center of mass as a classical particle. Then as the relative coordinate inside a local system is independent of its center of mass, we can set an arbitrary rule on the relation among those centers of mass of local systems. We adopt the principles of general relativity as the rules that govern the relations of plural local systems. By the reason that the center of mass and the inner coordinate are independent, we can combine quantum mechanics and general relativity consistently.

We give an approximate Hamiltonian that explains partially the usual relativistic quantum mechanical phenomena in chapter \ref{chap:9}.

In the final part \ref{Conclusions}, we consider some contradictory aspect of mathematics in chapter \ref{inconsistency-chap}. Although this does not give directly that mathematics is inconsistent, this will give an introduction to the next chapter \ref{StationaryUniverse}, where starting with the contradictory nature of the semantics of set theory in the sense that if we consider all sentences of set theory, they are contradictory, we regard that the Universe that is described by ourselves is of contradictory nature, and can be described as a superposition of all possible, infinite number of waves. As this is the state of the Universe, the Universe is described as a stationary state describing a superposition of all waves. We then give a formulation of the Universe and local systems inside it, in the form of a theory described by Axiom 1 to Axiom 5 in chapter \ref{StationaryUniverse}. In the final chapter \ref{LocalMotion}, we will prove that there is at least one Universe wave function $\phi$ in which all local systems have local motions and thus local times. This concludes our formulation of Quantum Mechanics.

\ 
\BP

\begin{flushright}
Hitoshi Kitada\\
\ \\
Dec. 15, 2003,
Tokyo
\end{flushright}

\ 
\newpage

\chapter*{Notation}

We here explain some notations which will be used in the text. $C^k(R^m)$ $(k=0,1,2,\cdots,\infty)$ is a space of $k$ times continuously differentiable functions $f(x)$ of $x\in R^m$. $C_0^k(R^m)$ is a subspace of $C^k(R^m)$ whose element $f\in C^k(R^m)$ has compact support in $R^m$. In particular, $C_0^\infty(R^m)$ is a space of infinitely differentiable functions on $R^m$ with compact support. We also use the notation $C_0^\infty(G)$ for a region $G$ in $R^m$ to denote the space of functions with continuous derivatives up to order $k$ with the support contained in $G$. $\SSS=\SSS(R^m)$ denotes a space of rapidly decreasing functions $f$ on $R^m$. Namely $f\in\SSS$ means that $f$ is an infinitely differentiable function satisfying
\beq
\sup_{x\in R^m}\left| |x|^k\partial_x^\alpha f(x)\right|<\infty
\ene
for all integers $k=0,1,2,\cdots$ and multi-indices $\alpha=(\alpha_1,\cdots,\alpha_m)$, where each $\alpha_j$ is a non-negative integer and
\beq
\partial_x=(\partial/\partial x_1,\cdots\partial/\partial x_m),\q \partial_x^\alpha
=(\partial/\partial x_1)^{\alpha_1}\cdots (\partial/\partial x_m)^{\alpha_m}.
\ene
The most important notion is the Hilbert space $L^2(R^m)$ with $m=1,2,\cdots$. It is a space of functions $f(x)$ on $R^m$ satisfying
\beq
\Vert f \Vert = (f,f)^{1/2}<\infty,\label{L2norm}
\ene
where the inner product is given by
\beq
(f,g)=\int_{R^m} f(x)\overline{g(x)}dx.
\ene
Concretely it is obtained by a completion of $\SSS$ or of $C_0^\infty(R^m)$ with respect to the norm \eq{L2norm}.
Along with this Hilbert space we use weighted $L^2$ space: $L^2_s=L^2_s(R^m)$ $(s\in R^1)$, which is a completion of $\SSS$ with respect to the norm
\beq
\Vert f\Vert_s=\Vert f\Vert_{L^2_s}=\left( \int_{R^m}  |f(x)|^2 \langle x\rangle^{2s}dx\right)^{1/2},
\ene
where
$\langle x\rangle=(1+|x|^2)^{1/2}$. $L^2_s(R^m)$ is also a Hilbert space with the inner product:
\beq
(f,g)_s=(f,g)_{L^2_s}=\int f(x) \overline{g(x)} \langle x\rangle^{2s}dx.
\ene
$\FF$ denotes Fourier transformation from $\SSS$ onto itself:
\beq
\FF f(\xi)=(2\pi)^{-m/2}\int e^{-i\xi x}f(x)dx.
\ene
$\FF$ is extended to a unitary operator from $L^2(R^m)$ onto itself:
\beq
\Vert \FF f\Vert=\Vert f\Vert.
\ene
We define Sobolev space $H^s=H^s(R^m)$ of order $s\in R^1$ as the Fourier image of $L^2_s(R^m)$. Thus it is a completion of $\SSS$ with respect to the norm
\beq
\Vert f\Vert_{H^s}=\left( \int_{R^m}|\langle D_x\rangle^s f(x)|^2 dx\right)^{1/2}.
\ene
Here
\beq
\langle D_x\rangle^s=\FF^{-1}\langle \xi\rangle^s \FF,
\ene
where $\langle \xi\rangle^s$ denotes a multiplication operator by $\langle \xi\rangle^s$ in $L^2(R^m_\xi)$. Namely its domain ${\cal D}(\langle \xi\rangle^s)$ is  a set of functions $f$ satisfying
\beq
\langle \xi\rangle^s f(\xi) \in L^2(R^m_\xi)
\ene
and its value applied to $f\in {\cal D}(\langle \xi\rangle^s)$ is
$
\langle \xi\rangle^s f(\xi)\in L^2(R^m).
$
$H^s(R^m)$ is also a Hilbert space with the inner product:
\beq
(f,g)_{H^s}=\int_{R^m} \langle D_x\rangle^s f(x) \overline{\langle D_x\rangle^s g(x)}dx.
\ene
We further use weighted Sobolev space $H^s_\delta(R^m)$ $(\delta\in R^1)$. This is a completion of $\SSS$ with respect to the norm
\beq
\Vert f\Vert_{H^s_\delta}=\left(\int_{R^m} |\langle D_x\rangle^s f(x)|^2 \langle x\rangle^{2\delta}dx\right)^{1/2},
\ene
and is a Hilbert space with the inner product
\beq
(f,g)_{H^s_\delta}=\int_{R^m} \langle D_x\rangle^s f(x) \overline{\langle D_x\rangle^s g(x)} \langle x\rangle^{2\delta}dx.
\ene
$S^{m-1}$ denotes the unit sphere of $R^m$ with surface element $d\omega$. $L^2(S^{m-1})$ is a space of the functions $\varphi(\omega)$ satisfying
\beq
\Vert \varphi\Vert_{L^2(S^{m-1})}=\left(\int_{S^{m-1}}|\varphi(\omega)|^2 d\omega\right)^{1/2}<\infty.
\ene
$L^2(S^{m-1})$ becomes a Hilbert space with the inner product
\beq
(\varphi,\psi)_{L^2(S^{m-1})}=\int_{S^{m-1}}\varphi(\omega)\overline{\psi(\omega)}d\omega.
\ene
We remark that $L^2_s(R^m)$ and $L^2_{-s}(R^m)$ are dual spaces each other with respect to the inner product of $L^2(R^m)$. Similarly $H^s(R^m)$ and $H^s_\delta(R^m)$ are dual spaces of $H^{-s}(R^m)$ and $H^{-s}_{-\delta}(R^m)$, respectively. $L^2(R^m)$ is a dual space of itself. We denote by $B(\HH_1,\HH_2)$ the Banach space of bounded operators from a Hilbert space $\HH_1$ into another Hilbert space $\HH_2$.
The notation
\beq
A:=B\q \mbox{or}\q B=:A
\ene
means that $A$ is defined by $B$.

For a self-adjoint operator $H$ in a Hilbert space $\HH$, we denote the corresponding resolution of the identity by $E_H(\lambda)$ $(\la\in R^1)$ that satisfies
\beq
&&E_H(\lambda)E_H(\mu)=E_H(\min(\lambda,\mu)),\nonumber\\
&& \mbox{s-}\lim_{\lambda\to -\infty}E_H(\lambda)=0,\q \mbox{s-}\lim_{\lambda\to\infty}E_H(\lambda)=I,\nonumber\\
&&
E_H(\lambda+0)=E_H(\lambda)\nonumber
\ene
where
$E_H(\lambda+0)=\mbox{s-}\lim_{\mu \downarrow \lambda}E_H(\mu)$. Such a family $\{E_H(\lambda)\}_{\lambda\in R^1}$ is uniquely determined by the relation
$$
H=\int_{-\infty}^\infty \lambda dE_H(\lambda).
$$
An operator valued measure $E_H(B)$ is defined by the relation $E_H((a,b])=E_H(b)-E_H(a)$ from the resolution $\{E_H(\la)\}$ of the identity, and is extended to general Borel sets $B$ as a countably additive measure using the properties above of $E_H(\lambda)$. 

Let
$$
P(\lambda)=E_H(\lambda)-E_H(\lambda-0)
$$
for $\lambda\in R^1$. $P(\lambda)\ne 0$ if and only if $\lambda$ is an eigenvalue of $H$.
The eigenspace or pure point spectral subspace $\HH_{p}(H)$ for a selfadjoint operator $H$ in a Hilbert space $\HH$ is defined by
$$
\HH_{p}(H)=\mbox{the closed linear hull of }\{ f\ | \ Hf=\lambda f\ \mbox{for some}\ \lambda\in R^1\}.
$$
The orthogonal projection $P_H$ onto $\HH_{p}(H)$ is called eigenprojection for $H$. The continuous spectral subspace $\HH_c(H)$ for $H$ is defined by
$$
\HH_c(H)=\{ f\ |\ E_H(\lambda)f\ \mbox{is strongly continuous with respect to}\ \lambda\in R^1\}.
$$
Then it is seen that $\HH_c(H)=\HH_{p}(H)^\perp$.
The absolutely continuous subspace $\HH_{ac}(H)$ for $H$ is defined by
\beq
&&\HH_{ac}(H)=\{ f\ |\ \mbox{The measure }(E_H(\Delta)f,f)=\Vert E_H(\Delta)f\Vert^2\ \mbox{is}\nonumber\\
&&\hskip80pt
\mbox{absolutely continuous with respect to Lebesgue measure}\}.\nonumber
\ene
The singular continuous subspace $\HH_{sc}(H)$ is then defined by
$$
\HH_{sc}(H)=\HH_c(H)\ominus\HH_{ac}(H).
$$
Thus
$$
\HH_c(H)=\HH_{ac}(H)\oplus\HH_{sc}(H).
$$
The part $H_{p}$, $H_{c}$, $H_{ac}$, $H_{sc}$ of $H$ in $\HH_{p}(H)$,
$\HH_c(H)$, $\HH_{ac}(H)$, $\HH_{sc}(H)$ are spectrally discontinuous, 
spectrally continuous, spectrally absolutely continuous and spectrally singular continuous, respectively. The spectra of these operators, $\sigma(H_{p}), \sigma(H_c), \sigma(H_{ac}), \sigma(H_{sc})$ are called point spectrum, continuous spectrum, absolutely continuous spectrum, and singular continuous spectrum of $H$, respectively.

\pagebreak

\ 

\pagebreak

\tableofcontents

\ 

\newpage

\ 

\newpage

\ 

\newpage

\setcounter{page}{1}

\renewcommand{\thepage}{%
\arabic{page}}

\part{Local Systems}\label{LocalSystem}

\pagestyle{myheadings}
\markboth{CHAPTER 1.  USUAL TIME CONTRADICTS THE UNCERTAINTY PRINCIPLE}{}

\chapter{Quantum Mechanical Time Contradicts
 The Uncertainty Principle}\label{chap:1}

In classical Newtonian mechanics, one can define mean velocity $v$ by
 $v=x/t$ of a scattered particle that starts from the origin at time $t=0$ and
 arrives at position $x$ at time $t$, if we assume that the
 coordinates of space and time are given in an {\it a priori}
 sense. This definition of velocity and hence that of momentum
 do not produce any problems, which assures that in classical regime
 there is no problem in the notion of space-time. Also in classical
 relativistic view, this would be valid insofar as we discuss the
 motion of a particle in the coordinates of the observer's.

Let us consider quantum mechanical case where the space-time
 coordinates are given {\it a priori}. Then the mean velocity of
 a scattered particle that starts from a point around the origin at time $0$
 and arrives at a point around $x$ at time $t$ should be defined
 as $v=x/t$. The longer the time length $t$ is, the more exact
 this value will be, if the errors of the positions
 at time $0$ and $t$ are the same extent, say $\delta>0$, for all $t$.
 This is a definition of the velocity, so this must hold in exact sense
 if the definition works at all. Thus 
\beq
\mbox{we have a precise value of (mean) momentum $p=mv$
 at a large time $t$}
\label{0}
\ene
with $m$ being the mass of the particle. Note that the mean momentum
 approaches the momentum at time $t$ when $t\to\infty$ as the
 interaction of the particle with other particles vanishes
 as $t\to\infty$ because the particle we are considering is a scattered one so that it escapes to the infinity as $t\to\infty$.

However in quantum mechanics, the uncertainty principle prohibits
 the position and momentum from taking exact values simultaneously.
 For illustration we consider a normalized state $\psi$ such that
 $\Vert \psi\Vert=1$ in one dimensional case. Then the expectation
 values of the position and momentum operators $Q=x$ and
 $P=\displaystyle{\frac{\hbar}{i}\frac{d}{d x}}$ on the state $\psi$ are given by
$$
q=(Q\psi,\psi),\quad p=(P\psi,\psi)
$$
respectively, and these operators satisfy commutation relation:
$$
[P,Q]=PQ-QP=\frac{\hbar}{i}.
$$
Further
their variances are
$$
\Delta q=\Vert (Q-q)\psi\Vert,\quad
\Delta p=\Vert (P-p)\psi\Vert.
$$
Thus their product satisfies the inequality
\beq
\Delta q\cdot\Delta p &=& \Vert(Q-q)\psi\Vert\Vert(P-p)\psi\Vert
\ge|((Q-q)\psi,(P-p)\psi)|\nonumber\\
&=&|(Q\psi,P\psi)-qp|\ge|\mbox{Im}((Q\psi,P\psi)-qp)|\nonumber\\
&=&|\mbox{Im}(Q\psi,P\psi)|
=\left|\frac{1}{2}((PQ-QP)\psi,\psi)\right|\nonumber\\
&=&\left|\frac{1}{2}\frac{\hbar}{i}\right|
=\frac{\hbar}{2}.\nonumber
\ene
Namely
\beq
\Delta q\cdot\Delta p\ge \frac{\hbar}{2}.\label{1}
\ene
This uncertainty principle means that there is a least value
 $\hbar/2(>0)$ for the product of the variances of position and
 momentum so that the independence between position and momentum
 is assured in an absolute sense that there is no way to let
 position and momentum correlate exactly as in classical views. 

Applying \eq{1} to the above case of the particle that starts
 from the origin at time $t=0$ and arrives at $x$ at time $t$,
 we have at time $t$
\beq
\Delta p>\frac{\hbar}{2\delta}\label{2}
\ene
because we have assumed the error $\Delta q$ of the coordinate
 $x$ of the particle at time $t$ is less than $\delta>0$.
 But the argument \eq{0} above tells that $\Delta p\to 0$ when 
 $t\to\infty$, contradicting \eq{2}.

This observation shows that, if given a pair of {\it a priori}
 space and time coordinates, quantum mechanics becomes contradictory.

A possible solution to this apparent contradiction of the notion of space-time in quantum mechanics would be to regard the independent
 quantities, space and momentum operators, as the fundamental
 quantities of quantum mechanics, and disregard the notion of time from the framework of quantum mechanics. As time $t$ can be introduced
 as a ratio $x/v$ on the basis of the notion of space
 and momentum in this view (see Definition \ref{Time}),
 time is a redundant notion that
 should not be given a role independent of space and momentum.

It might be thought that in this view we lose the relation $v=x/t$
 that is necessary for the notion of time to be valid, if space and
 momentum operators are independent as we have seen.
 However there can be found a relation like $x/t=v$ as an approximate
 relation that holds to the extent that the relation does not
 contradict the uncertainty principle (see Theorem \ref{Enss} below).

As a consequence of the abandonment of the {\it a priori} given time in quantum mechanics, the quantum jump (or wave function collapse) that is assumed to occur in usual quantum mechanics whenever the particles are observed
 becomes unnecessary in our formulation.

This misconception of quantum jump comes from our unconscious inheritance of the classical view
 of time to quantum mechanics that the motion is governed by time, and hence the system must evolve along with this given time coordinate. This unconscious assumption urges us to think we observe a definite eigenstate that has sharp values of the quantity that we observe
 and jumps or collapses must occur when we make observation at one moment in time coordinate. However, what one is able to observe actually is not the eigenstates. No stable eigenstates can be observed as eigenstates, as will be seen in chapter \ref{chap:4}, \eq{unobservable}. Even if we can observe eigenstates, they are necessarily destroyed and become unstable scattered states. We thus observe just the scattering
 states or processes. We define time as the evolution of these scattering states. Then no eigenstates need appear in the formulation of quantum mechanics.  Jumps and eigenstates are ghosts arising from our customary thought that we are accustomed to based on the
 passed classical notion of time that has been assumed given {\it a priori}. In more exact words,
 the usual quantum mechanical theory is an overdetermined
 system that involves too many independent variables: space,
 momentum, and time. In that framework time is inevitably not free from
 the classical image that velocity is defined by
 $v=x/t$, thus yielding a contradiction discussed above. What is responsible for this misunderstanding is our lack of recognition that the revolution by quantum mechanics of our common sense is too far to be caught by our conventional understanding of the world.

\pagestyle{headings}

\chapter{Position and Momentum}\label{chap:2}

We consider $N$ ($N\ge 1$) particles, which are moving in the Euclidean space $R^3$. We label them as $1,2,\cdots,N$.

Let $X_j=(X_{j1},X_{j2},X_{j3})$ and $P_j=(P_{j1},P_{j2},P_{j3})$ $(j=1,2,\cdots,N)$ denote the position and momentum operators of the $j$-th particle. Namely $X_{jk}$ $(k=1,2,3)$ is a multiplication operator in $L^2(R^{3N})$ defined by
$$
(X_{jk}f)(x)=x_{jk}f(x),\quad (x=(x_{11},x_{12},x_{13},x_{21},\cdots,x_{N1},x_{N2},x_{N3}) \in R^{3N})
$$
and $P_{j}$ is the differential operator
$$
(P_{j}f)(x)=\hbar D_{x_{j}}f(x)=\frac{\hbar}{i}\frac{\partial f}{\partial {x_j}}(x):=\frac{\hbar}{i}\left(\frac{\partial f}{\partial x_{j1}}(x),\frac{\partial f}{\partial x_{j2}}(x),\frac{\partial f}{\partial x_{j3}}(x)\right).
$$
Here $\hbar=\frac{h}{2\pi}$, where $h$ is the Planck's constant.
Their domains are
\beq
&&D(X_{jk})=\{ f | f \in L^2(R^{3N}), x_{jk}f(x) \in L^2(R^{3N})\},\nonumber\\
&&D(P_{jk})=\{ f | f \in L^2(R^{3N}), \frac{\partial f}{\partial x_{jk}}(x)\in L^2(R^{3N})\},\nonumber
\ene
where the differentiation is understood in the distribution sense.

$X_{jk}$ and $P_{j'k'}$ satisfy the canonical commutation relation. We write $[A,B]=AB-BA$ for two operators $A$ and $B$ in $L^2(R^{3N})$.
\beq
&&[X_{jk},X_{j'k'}]=0,\nonumber\\
&&[P_{jk},P_{j'k'}]=0,\label{canonical}\\
&&[X_{jk},P_{j'k'}]=i\hbar\delta_{jj'}\delta_{kk'},\nonumber
\ene
where $\delta_{j\ell}$ is Kronecker's delta.

Let $m_j>0$ be the mass of the $j$-th particle.
The Hamiltonian of the system is defined by
\beq
H=\sum_{j=1}^N\frac{1}{2m_j}P_j^2 + \sum_{1\le i<j\le N}V_{ij}(X_i-X_j),
\ene
where $P_j^2=\sum_{k=1}^3P_{jk}^2$ and $V_{ij}(x)$ $(x\in R^3)$ is a real-valued pair potential which describes the interaction between the particles $i$ and $j$. Since this interaction depends only on the relative position $x_i-x_j\in R^3$ of the particles, we can remove the center of mass from the Hamiltonian. Namely, denoting the old variables $x_i$ by $X_i$ with some abuse of notation, we introduce new variables $x_i$ as follows.
We first define the center of mass of the $N$-particle system by
\beq
X_C=\frac{m_1X_1+\cdots+m_NX_N}{m_1+\cdots+m_N},
\ene
and then define $x_i$ as Jacobi coordinates:
\beq
x_i=X_{i+1}-
\frac{m_1X_1+\cdots+m_iX_i}{m_1+\cdots+m_i}, \q
i=1,2,\cdots,n=N-1.\label{(1.4)}
\ene
Accordingly, we define the momentum operators $P_C=(P_{C1},P_{C2},P_{C3})$ and $p_i=(p_{i1},p_{i2},p_{i3})$:
$$
P_{C} =\frac{\hbar}{i}\frac{\partial}{\partial X_{C}},\q
p_{i}=\frac{\hbar}{i}\frac{\partial}{\partial x_{i}}.
$$
It is clear that these satisfy the canonical commutation relation.
Using these new $X_C,P_C,x_i,p_i$,
we can rewrite $H$ as
\beq
H={\tilde H}
+H_{C},\label{(1.5)}
\ene
where
\beq
&&\tilde H=\sum_{i=1}^{n}\frac{1}{2\mu_i}p_i^2+\sum_{i<j}V_{ij}(x_{ij}),\nonumber\\
&&H_{C}=\frac{1}{2\sum_{j=1}^Nm_j}P_C^2,\nonumber
\ene
with $x_{ij}$ being the expression of $X_i-X_j$ in the new coordinates,
and $\mu_i>0$ is the reduced mass defined by the relation:
$$
\frac{1}{\mu_i} =\frac{1}{m_{i+1}}+\frac{1}{m_1+\cdots+m_i}.
$$
The new coordinates give a decomposition $L^2(R^{3N})=L^2(R^3)\otimes L^2(R^{3n})$ and in this decomposition, $H$ is written as
$$
H=H_{C}\otimes I+I\otimes {\tilde H}.
$$
$H_{C}$ is the well-known Laplacian, and what we are concerned with is the relative motion of the $N$-particles. Thus we have only to consider $\tilde H$ in the Hilbert space $\HH=L^2(R^{3n})$. We rewrite this $\tilde H$ as $H$:
\beq
H=H_0+V=\sum_{i=1}^{n}\frac{1}{2\mu_i}p_i^2+\sum_{i<j}V_{ij}(x_{ij})
=-\sum_{i=1}^{n}\frac{\hbar^2}{2\mu_i}\Delta_{x_i}+\sum_{i<j}V_{ij}(x_{ij}),
\label{(1.6)}
\ene
where
$$
\Delta_{x_i}=\sum_{k=1}^3\frac{\partial^2}{\partial x_{ik}^2}.
$$
This means that we consider the Hamiltonian $H$ in \eq{(1.5)} restricted to the subspace
\beq
(m_1+\cdots+m_N)X_C=m_1X_1+\cdots+m_NX_N=0\label{(1.7)}
\ene
of $R^{3N}$.
We equip this subspace with the inner product:
\beq
\langle x,y\rangle=\sum_{i=1}^n \mu_i x_i\cdot y_i,\label{(1.8)}
\ene
where $\cdot$ denotes the Euclidean scalar product. With respect to this inner product, the changes of variables between Jacobi coordinates in \eq{(1.4)} are realized by orthogonal transformations on the space $R^{3n}$ defined by \eq{(1.7)}, while $\mu_i$ and $x_i$ depend on the order of the constitution of Jacobi coordinates in \eq{(1.4)}. If we use this inner product, $H_0$ can be written as:
\beq
H_0=\frac{1}{2}\langle v,v\rangle,
\ene
where $v=(v_1,\cdots,v_n)=(\mu_1^{-1}p_1,\cdots,\mu_n^{-1}p_n)$ is the velocity operator.

It is known that $H$ is a selfadjoint operator in $\HH$ under suitable decay assumptions on the pair potentials $V_{ij}(x)$ as $|x|\to\infty$. We consider such a situation in the followings, and precise conditions on $V_{ij}(x)$ will be given when necessary.
\BP

We summarize the assumptions we made in this chapter as the following two Axioms \ref{axiom2} and \ref{axiom3}.

\begin{axm}\label{axiom2}
Let $n\ge 1$ and $F_{n+1}$ be a finite subset of 
${\N}=\{1,2,\cdots\}$ with $\sharp(F_{n+1})=n+1$. Then for any 
$j\in F_{n+1}$, there are selfadjoint operators 
$X_j =(X_{j 1},X_{j2},X_{j3})$ and $P_j =(P_{j1},P_{j2},P_{j3})$ in
a tensor product $\HH^{n+1}=\HH\otimes\cdots\otimes\HH$ of $(n+1)$ times of a separable Hilbert space $\HH$, and constants $m_j>0$ such that
$$
[X_{j\ell},X_{k m}]=0,\q [P_{j\ell},P_{k m}]=0,
\q [X_{j\ell},P_{k m}]=i\delta_{jk}\delta_{\ell m},
$$
$$
\sum_{j\in F_{n+1}} m_j X_j=0,\q \sum_{j\in F_{n+1}} P_j=0.
$$
\end{axm}

The Stone-von Neumann theorem and Axiom \ref{axiom2} specify the space
 dimension (see  \cite{[A-M]}, p.452) as 3 dimension. Namely $\HH^n$ is represented as and can be identified with $L^2(R^{3n})$ in the following.

\begin{axm}\label{axiom3}
Let $n\ge0$ and $F_N$ $(N=n+1)$ be a finite subset
 of ${\N}=\{1,2,\cdots\}$ with $\sharp(F_N)=N$. Let 
$\{ F_N^\ell\}_{\ell=0}^\infty$ be the countable totality of such $F_N$. Then
 the local Hamiltonian $H_{n\ell}$ $(\ell\ge0)$ is of the form
$$
H_{n\ell}=H_{n\ell0}+V_{n\ell},\q V_{n\ell}=\sum_{{\scriptstyle \alpha=(i,j)}
\atop{\scriptstyle 1\le i<j<\infty,\ i,j\in F_N^\ell}} V_\alpha(x_\alpha)
$$
on $C_0^\infty(R^{3n})$, where $x_\alpha=x_i-x_j$ $(\alpha=(i,j))$ with $x_i$ being
 the position vector of the $i$-th particle, and 
$V_\alpha(x_\alpha)$ is a real-valued measurable function of 
$x_\alpha\in R^3$ which is $H_{n\ell0}$-bounded with 
$H_{n\ell0}$-bound of $V_{n\ell}$ less than 1.
$H_{n\ell0}=H_{(N-1)\ell0}$ is the free Hamiltonian of the 
$N$-particle system, which has the form
$$
-\sum_{\ell=1}^n\sum_{k=1}^3 \frac{\hbar^2}{2\mu_\ell}
\frac{\partial^2}{\partial x_{\ell k}^2}\quad \mbox{with } \mu_\ell>0 \mbox{ being reduced mass}.
$$
We remark that the subscript $\ell$ in $H_{n\ell}$ distinguishes different systems with the same number $N=n+1$ of particles.
\end{axm}

This Axiom implies that $H_{n\ell}=H_{(N-1)\ell}$ is uniquely
 extended to a selfadjoint operator bounded from below in 
$\HH^n=\HH^{N-1} =L^2(R^{3(N-1)})$ by the Kato-Rellich theorem.
We write $\HH_{n\ell}=\HH^n$ to indicate that the space $\HH^n$ is associated with the Hamiltonian $H_{n\ell}$, and use the notation $(H_{n\ell},\HH_{n\ell})$ to make explict this relation. We will call this pair $(H_{n\ell},\HH_{n\ell})$ a local system (see Definition \ref{local}).

We do not include vector potentials in the Hamiltonian 
$H_{n\ell}$ of Axiom \ref{axiom3}, for we take the position that what 
is elementary is the electronic charge, and the magnetic 
forces are the consequence of the motions of charges.

\chapter{Time}\label{chap:3}

\section{Definition of local time}

In the previous chapter, we introduced position and momentum operators and defined a Hamiltonian of an $N$-particle system. All analyses of quantum-mechanical theory are done under the basis of these notions.

The reader might have noticed we do not introduce Schr\"odinger equation at all. In the usual theory of quantum mechanics, Schr\"odinger equation is one of the basic assumptions of the theory, without which no analysis of motion of quantum-mechanical particles could be done.

The usual theory of quantum mechanics assumes the {\it a priori} existence of time when it introduces Schr\"odinger equation. And the motion of particles is analyzed by the use of the equation along that {\it a priori} given time.

We reverse the order. We first define time of the system under consideration on the basis of the position and momentum operators. Then we introduce the Schr\"odinger equation by using that notion of time, which is proper to each system of quantum-mechanical particles. Thus our basic notions of the theory are just position and momentum operators that satisfy the canonical commutation relations.

In this sense, we discard the usual notion of space-time, which is assumed as a fundamental basis of any physical theory. Instead we adopt position and momentum as the fundamental basis of quantum theory. Our degree of freedom in describing nature is thus 6 in place of 4 of space-time that the usual theory assumes. This increase of freedom would make us possible to see nature's properties more precisely than the usual physical theory would.

We leave such precise analysis to the future, and return to the usual description of nature by Schr\"odinger equation. To do so, we first introduce clock and time of an $N$-particle system whose Hamiltonian is given by $H$ in \eq{(1.6)}.

Since $H$ in \eq{(1.6)} is selfadjoint under suitable assumptions on the decay rate of pair potentials $V_{ij}(x)$, we can construct the unitary operator
\beq
\exp(-itH/\hbar)\label{clock}
\ene
for all real numbers $t\in R^1$. We remark that $H$ is defined by \eq{(1.6)}, and hence $\exp(-itH/\hbar)$ is constructed on the basis of the mere notion of position and momentum operators.

\begin{df}\label{Time}
We call the unitary group $\exp(-itH/\hbar)$ in \eq{clock} the {\rm (local or proper) clock} of the system that we are considering, and $t$ in the exponent of $\exp(-itH/\hbar)$ the {\rm (local)} {\rm time} of the system whose Hamiltonian $H$ is given by \eq{(1.6)}.
\end{df}

\section{Justification of local time as a notion of time}\label{section3.2}

To see that this definition of time coincides with our intuition, we introduce some notion used in many body scattering theory.

Let $b=\{ C_1,\cdots,C_k\}$ be a decomposition of the set $\{1,2,\cdots,N\}$ into $k$ disjoint subsets $C_1,\cdots,C_k$ of $\{1,2,\cdots,N\}$. If we denote the number of the elements of a set $S$ by $\sharp(S)$ or $|S|$, we can write $k=\sharp(b)=|b|$. Such a $b$ is called a cluster decomposition of $\{1,2,\cdots,N\}$.

A clustered Jacobi coordinate $x=(x_b,x^b)$ associated with a cluster decomposition $b=\{ C_1,\cdots,C_k\}$ is obtained by first choosing a Jacobi coordinate
$$
x^{(C_\ell)}=(x_1^{(C_\ell)},\cdots,x_{\sharp(C_\ell)-1}^{(C_\ell)})\in R^{3(\sharp(C_\ell)-1)},\q (\ell=1,\cdots,k)
$$
for the $\sharp(C_\ell)$ particles in the cluster $C_\ell$, and then by choosing an intercluster Jacobi coordinate
$$
x_b=(x_1,\cdots,x_{k-1})\in R^{3(k-1)}
$$
for the centers of mass of the $k$ clusters $C_\ell$. Then $x^b=(x^{(C_1)},\cdots,x^{(C_k)})\in R^{3(N-k)}$ and $x=(x_b,x^b)\in R^{3(N-1)}=R^{3n}$, and the corresponding canonically conjugate momentum operator is
\beq
&&p=(p_b,p^b),\q p_b=(p_1,\cdots,p_{k-1}),\q p^b=(p^{(C_1)},\cdots,p^{(C_k)})\nonumber\\
&&p_i=\frac{\hbar}{i}\frac{\partial}{\partial x_i},\q
p^{(C_\ell)}=(p_1^{(C_\ell)},\cdots,p_{\sharp(C_\ell)-1}^{(C_\ell)}),\q
 p_i^{(C_\ell)}=\frac{\hbar}{i}\frac{\partial}{\partial x_i^{(C_\ell)}}\nonumber
\ene
Accordingly $\HH=L^2(R^{3n})$ is decomposed:
$$
\HH=\HH_b\otimes\HH^b,\q \HH_b=L^2(R_{x_b}^{3(k-1)}),\q \HH^b=L^2(R_{x^b}^{3(N-k)}).
$$
In this coordinates system, $H_0$ in \eq{(1.6)} is decomposed:
\beq
&&H_0=T_b+H_0^b,\nonumber\\
&&T_b=-\sum_{\ell=1}^{k-1}\frac{\hbar^2}{2M_\ell}\Delta_{x_\ell},\\
&&H_0^b=-\sum_{\ell=1}^k\sum_{i=1}^{\sharp(C_\ell)-1}\frac{\hbar^2}{2\mu_i^{(C_\ell)}}\Delta_{x_i^{(C_\ell)}},\nonumber
\ene
where $\Delta_{x_\ell}$ and $\Delta_{x_i^{(C_\ell)}}$ are 3-dimensional Laplacians and $M_\ell$ and $\mu_i^{(C_\ell)}$ are the reduced masses.
If we introduce the inner product in the space $R^{3n}$ as in \eq{(1.8)}:
\beq
&&\langle x,y\rangle = \langle (x_b,x^b),(y_b,y^b)\rangle = 
\langle x_b,y_b\rangle+\langle x^b,y^b\rangle \nonumber\\
&&= \sum_{\ell=1}^{k-1}M_\ell x_\ell\cdot y_\ell+\sum_{\ell=1}^k\sum_{i=1}^{\sharp(C_\ell)-1}\mu_i^{(C_\ell)} x_i^{(C_\ell)}\cdot y_i^{(C_\ell)},\nonumber
\ene
and velocity operator
$$
v=(v_b,v^b)=M^{-1}p=(m_b^{-1}p_b,(\mu^b)^{-1}p^b),
$$
where $M=\left(\begin{array}{cc} m_b& 0\\ 0& \mu^b\end{array}\right)$ is the $3n$-dimensional diagonal mass matrix whose diagonals are given by $M_1,\cdots,M_{k-1},\mu_1^{(C_1)},\cdots,\mu_{\sharp(C_k)-1}^{(C_k)}$,
then $H_0$ is written as
$$
H_0=\frac{1}{2}\langle v,v\rangle=T_b+H_0^b=\frac{1}{2}\langle v_b,v_b\rangle+\frac{1}{2}\langle v^b,v^b\rangle.
$$

We next decompose the sum of pair potentials in \eq{(1.6)}:
$$
\sum_{i<j}V_{ij}(x_{ij})=V_b+I_b,
$$
where
\beq
&&V_b=\sum_{C_\ell\in b}V_{C_\ell},\nonumber\\
&&V_{C_\ell}=\sum_{\{i,j\}\subset C_\ell} V_{ij}(x_{ij}),\nonumber\\
&&I_b=\sum_{\forall C_\ell\in b: \{i,j\}\notin C_\ell}V_{ij}(x_{ij}).\nonumber
\ene
By definition, $V_{C_\ell}$ depends only on the variable $x^{(C_\ell)}$ inside the cluster $C_\ell$. Similarly, $V_b$ depends only on the variable $x^b=(x^{(C_1)},\cdots,x^{(C_k)})\in R^{3(N-\sharp(b))}$, while $I_b$ depends on all components of the variable $x$.

Then $H$ in \eq{(1.6)} is decomposed:
\beq
&&H=H_b + I_b=T_b\otimes I+ I\otimes H^b + I_b,\nonumber\\
&&H_b=H-I_b=T_b\otimes I+ I\otimes H^b,\label{(2.3)} \\ 
&&H^b=H_0^b+V_b.\nonumber
\ene

We denote by $P_b$ the orthogonal projection onto the pure point spectral subspace (or eigenspace) $\HH_{p}^b=\HH_{p}(H^b)$ for $H^b$ of $\HH^b$. We use the same notation $P_b$ for the obvious extention $I\otimes P_b$ to the total space $\HH$. For $\sharp(b)=N$, we set $P_b=I$, and for $\sharp(b)=1$, we write $P_b=P_H=P$. Let $M=1,2,\cdots$ and $P_b^M$ denote an $M$-dimensional partial projection of $P_b$ such that s-$\lim_{M\to\infty}P_b^M=P_b$. We define for an $\ell$-dimensional multi-index $M=(M_1,\cdots,M_\ell)$ ($M_j\ge 1$) and $\ell=1,\cdots,n=N-1$
\beq
{\widehat P}_\ell^M=\left(I-\sum_{\sharp(b_\ell)=\ell}P_{b_\ell}^{M_\ell}\right)\cdots\left(I-\sum_{\sharp(b_2)=2}P_{b_2}^{M_2}\right)(I-P^{M_1}).\label{(2.4)}
\ene
(Note that for $\sharp(b)=1$, $b=\{C\}$ with $C=\{1,2,\cdots,N\}$. Thus $P^{M_1}$ is an $M_1$-dimensional partial projection into the eigenspace of $H$.)
We further define for a $\sharp(b)$-dimensional multi-index $M_b=(M_1,\cdots,M_{\sharp(b)-1},M_{\sharp(b)})=({\widehat M}_b,M_{\sharp(b)})$
\beq
{\widetilde P}_b^{M_b}=P_b^{M_{\sharp(b)}}{\widehat P}_{\sharp(b)-1}^{{\widehat M}_b},\q
2\le \sharp(b) \le N.
\ene
Then it is clear that
\beq
\sum_{2\le \sharp(b)\le N}{\widetilde P}_b^{M_b}={\widehat P}_1^{M_1}=I-P^{M_1},
\label{(2.6)}
\ene
provided that the component $M_j$ of $M_b$ depends only on the number $j$ but not on $b$. In the following we use such $M_b$'s only.

Related with those notions, we denote by $\HH_c=\HH_c(H)$ the orthogonal complement $\HH_{p}(H)^\perp$ of the eigenspace $\HH_{p}=\HH_{p}(H)$ for the total Hamiltonian $H$. Namely $\HH_c(H)$ is the continuous spectral subspace for $H$. We note that $\HH_c(H)=(I-P_a)\HH$ for a unique $a$ with $|a|=1$, and that for $f\in\HH$, $(I-P^{M_1})f\to (I-P_a)f\in \HH_c(H)$ as $M_1\to\infty$. We use freely the notations of functional calculus for selfadjoint operators, e.g. $E_H(B)$ is the spectral measure for $H$ as defined in the section of notation.

Let $v_b$, as above, denote the velocity operator between the clusters in $b$. It is expressed as $v_b=m_b^{-1}p_b$ for some $3(\sharp(b)-1)$-dimensional diagonal mass matrix $m_b$. To see the meaning of our time in Definition \ref{Time}, we prepare the following

\begin{thm}[\cite{[En]}]\label{Enss}
Let $N=n+1\ge 2$ and let $H$ be the Hamiltonian $H$ in \eq{(1.6)} or \eq{(2.3)} for an $N$-body quantum-mechanical system. Assume that $|X^b|P_b^M$ is a bounded operator for any integer $M\ge 1$. Let suitable conditions on the smoothness and the decay rate of the pair potentials $V_{ij}(x_{ij})$ be satisfied: E.g., assume
$$
|V_{ij}(x)|+|x\cdot(\nabla_x V_{ij})(x)|\to 0\ \mbox{\rm (as}\ |x|\to \infty).
$$
Let $f\in \HH$. Then there exist a sequence $t_m\to\pm\infty$ (as $m\to\pm\infty$) and a sequence $M_b^m$ of multi-indices whose components all tend to $\infty$ as $m\to\pm\infty$ such that for all cluster decompositions $b$ with $2\le \sharp(b)\le N$, for all $\varphi\in C_0^\infty(R_{x_b}^{3(\sharp(b)-1)})$, $R>0$, and $\alpha=\{i,j\}$ that is not included in any $C_\ell\in b$,
\beq
&&\Vert \chi_{\{x| |x_\alpha|<R\}}{\widetilde P}_b^{M_b^m}e^{-it_mH/\hbar}f\Vert \to 0\label{(2.7)}\\
&&\Vert (\varphi(X_b/t_m)-\varphi(v_b)){\widetilde P}_b^{M_b^m}e^{-it_mH/\hbar}f\Vert \to 0\label{(2.8)}
\ene
as $m\to\pm\infty$. Here $\chi_S$ is the characteristic function of a set $S$.
\end{thm}

\noindent
{\it Proof\ \footnote[1]{Proof here follows that of \cite{[En]}.}:}
Since mass factors and Planck constant in the definition of the Hamiltonian $H$ are unessential, we may assume $\hbar=1$ and
$$
H=H_0+V,\q H_0=\frac{1}{2}D^2=-\frac{1}{2}\Delta, \q V=\sum_{i<j}V_{ij}(x_{ij}).
$$
where
$$
D=\frac{1}{i}\frac{\partial}{\partial x},\q x\in R^{3n}.
$$

Let $f\in \HH$ satisfy $(1+|X|)^2f\in \HH$ and $f=E_H(B)f$ for some bounded open set $B$ of $R^1$. Note that such $f$'s are dense in $\HH$. We compute, noting \eq{(2.6)} and writing ${\widetilde P}_b^{M_b}={\widetilde P}_b$ and $\ff=(I-P^{M_1})f$
\beq
&&\sum_{2\le\sharp(b)\le N}e^{itH}\left(\frac{X}{t}-D\right)^2{\widetilde P}_be^{-itH}f\label{(2.9)}\\
&&=e^{itH}\left(\frac{X}{t}-D\right)^2e^{-itH}\ff\nonumber\\
&&=e^{itH}\left(\frac{X^2}{t^2}-\frac{2A}{t}+D^2\right)e^{-itH}\ff\nonumber\\
&&=\frac{1}{t^2}\left(e^{itH}X^2e^{-itH}\ff-X^2\ff\right)-\frac{2}{t}e^{itH}Ae^{-itH}\ff+2e^{itH}H_0e^{-itH}\ff+\frac{X^2}{t^2}\ff.\nonumber
\ene
Here $A=\frac{1}{2}(X\cdot D+D\cdot X)$. The first term on the RHS is equal to
$$
\frac{1}{t^2}\int_0^t e^{isH}i[H_0,X^2]e^{-isH}\ff ds.
$$
By the relation $i[H_0,X^2]=2A$, \eq{(2.9)} is equal to
\beq
&&\frac{2}{t^2}\left(\int_0^t e^{isH} Ae^{-isH}\ff ds -te^{itH}Ae^{-itH}\ff\right)+2e^{itH}H_0e^{-itH}\ff+\frac{X^2}{t^2}\ff.\nonumber
\ene
The formula in the first parentheses equals
\beq
&&\int_0^t e^{isH} Ae^{-isH}\ff ds -te^{itH}Ae^{-itH}\ff\nonumber\\
&&=\int_0^t\frac{d}{d\tau}\left(\int_0^\tau e^{isH} Ae^{-isH}\ff ds -\tau e^{i\tau H}A e^{-i\tau H}\ff\right)d\tau\nonumber\\
&&=-\int_0^t s e^{isH}i[H,A]e^{-isH}\ff ds.\nonumber
\ene
Here for any $b$ with $2\le |b|=\sharp(b)\le N$
$$
i[H,A]=i[T_b,A]+i[I_b,A]+i[H^b,A]=2T_b+i[I_b,A]+i[H^b,A^b],
$$
where
$$
A^b=\frac{1}{2}(X^b\cdot D^b+D^b\cdot X^b).
$$
Thus we have
\beq
&&\sum_{2\le\sharp(b)\le N}e^{itH}\left(\frac{X}{t}-D\right)^2{\widetilde P}_be^{-itH}f\label{appro0}\\
&&=-\frac{4}{t^2}\sum_{2\le|b|\le N}\int_0^t se^{isH}T_b{\widetilde P}_b e^{-isH}\ff ds+2e^{itH}H_0e^{-itH}\ff\nonumber\\
&&-\frac{2}{t^2}\sum_{2\le|b|\le N}\int_0^t se^{isH}i[I_b,A]{\widetilde P}_be^{-isH}\ff ds\nonumber\\
&&-\frac{2}{t^2}\sum_{2\le|b|\le N}\int_0^t se^{isH}i[H^b,A^b]{\widetilde P}_b e^{-isH}\ff ds+\frac{X^2}{t^2}\ff.\nonumber
\ene

\begin{lem}\label{RAGE}
 Let $B(s)$ be a continuous family of uniformly bounded operators in $\HH$ with respect to $s$ in uniform operator topology. Let $B\subset R^1$ be a bounded open set satisfying $E_H(B)\HH\subset \HH_c(H)$ and let $2\le |b| \le N$. Then there is a constant $\epsilon_M>0$ that goes to $0$ when $M_j$'s in the multi-indices $M_b$'s tend to $\infty$ such that as $T\to\infty$
\beq
\left\Vert\frac{1}{T}\int_0^T B(s) F(|x_\alpha|<R){\widetilde P}_b^{M_b}e^{-isH} E_H(B)ds\right\Vert \sim_{\epsilon_M} 0
\ene
for any pair $\alpha=\{i,j\}$ with $\alpha\notin C_\ell$ for all $C_\ell\in b$. Here $\sim_{\epsilon_M}$ means that the norm of the difference of the both sides is smaller than $\epsilon_M$ as $T\to\infty$, and
$F(|x_\alpha|<R)$ denotes a smooth positive cut off function which is $1$ on the set $S=\{x | |x_\alpha|<R\}\subset R^{3n}$ and is $0$ outside some neighborhood of $S$.
\end{lem}

By this lemma, the third term on the RHS of \eq{appro0} vanishes as $t\to\infty$ within the small error $\epsilon_M>0$ determined by the values of $M_j$ in the multi-indices $M_b=(M_1,\cdots,M_\ell)$. The last term on the RHS of \eq{appro0} also vanishes by $X^2f\in\HH$.
Thus as $t\to\infty$ we have asymptotically
\beq
&&\sum_{2\le\sharp(b)\le N}e^{itH}\left(\frac{X}{t}-D\right)^2{\widetilde P}_be^{-itH}f\\
&&\q \sim_{\epsilon_M} -\frac{4}{t^2}\sum_{2\le|b|\le N}\int_0^t se^{isH}T_b{\widetilde P}_b e^{-isH}\ff ds+2e^{itH}H_0e^{-itH}\ff\nonumber\\
&&\q\q\q-\frac{2}{t^2}\sum_{2\le|b|\le N}\int_0^t se^{isH}i[H^b,A^b]{\widetilde P}_b e^{-isH}\ff ds.\nonumber
\ene
Taking the inner product of the last term with $\ff=(I-P^{M_1})f$ and noting by Lemma \ref{RAGE} that as $t\to\infty$
\beq
\sum_{2\le|b|\le N}\sum_{2\le |d|\le N, d\ne b}\frac{2}{t^2}\int_0^t s(\ff,e^{isH}({\widetilde P}_d)^*i[H^b,A^b]{\widetilde P}_b e^{-isH}\ff) ds\sim_{\epsilon_M}0,
\ene
we have by \eq{(2.6)} as $t\to\infty$
\beq
&&\frac{2}{t^2}\sum_{2\le|b|\le N}\int_0^t s(\ff,e^{isH}i[H^b,A^b]{\widetilde P}_b e^{-isH}\ff) ds\nonumber\\
&&\sim_{\epsilon_M} \frac{2}{t^2}\sum_{2\le|b|\le N}\int_0^t s(\ff,e^{isH}({\widetilde P}_b)^*i[H^b,A^b]{\widetilde P}_b e^{-isH}\ff) ds\nonumber\\
&&\sim \frac{1}{t}\sum_{2\le|b|\le N}\int_0^t (\ff,e^{isH}({\widetilde P}_b)^*i[H^b,A^b]{\widetilde P}_b e^{-isH}\ff) ds,\nonumber
\ene
provided that the limit as $t\to\infty$ of the RHS exists, which we will prove below. Here $\sim$ means $\sim_0$.
Letting $t(s)=s-mS$ for $mS\le s<(m+1)S$ for any fixed $S>0$, we have by Lemma \ref{RAGE} and some commutator arguments as $t\to\infty$
\beq
&&\sum_{2\le|b|\le N}\frac{1}{t}\int_0^t (\ff,e^{isH}({\widetilde P}_b)^*i[H^b,A^b]{\widetilde P}_b e^{-isH}\ff) ds\nonumber\\
&&\sim_{\epsilon_M}\sum_{2\le|b|\le N}
\frac{1}{t}\int_0^t (\ff,e^{i(s-t(s))H}({\widetilde P}_b)^*e^{it(s)H_b}i[H^b,A^b]{\widetilde P}_b e^{-isH}\ff) ds.\nonumber
\ene
This can further be reduced and is asymptotically equal to as $t\to\infty$ with an error $\epsilon_M>0$
\beq
\sum_{2\le|b|\le N}\frac{1}{t}\int_0^t (\ff,e^{i(s-t(s))H}({\widetilde P}_b)^*e^{it(s)H_b}i[H^b,A^b]e^{-it(s)H_b}{\widetilde P}_b e^{-i(s-t(s))H}\ff) ds.\nonumber
\ene
Noting $s-t(s)=mS$ for $mS\le s <(m+1)S$, we rewrite this for $t=nS$
\beq
&&\frac{1}{nS}\sum_{m=0}^{n-1}\int_0^S (\ff,e^{imSH}({\widetilde P}_b)^* e^{isH_b}i[H^b,A^b]e^{-isH_b}{\widetilde P}_b e^{-imSH}\ff) ds\\
&&=\frac{1}{n}\sum_{m=0}^{n-1}\frac{1}{S}\int_0^S
\frac{d}{ds}(\ff,e^{imSH}({\widetilde P}_b)^* e^{isH_b}A^b e^{-isH_b}{\widetilde P}_b e^{-imSH}\ff) ds\nonumber\\
&&=\frac{1}{n}\sum_{m=0}^{n-1}\frac{1}{S}[(\ff,e^{imSH}({\widetilde P}_b)^* e^{iSH_b}A^be^{-iSH_b}{\widetilde P}_b e^{-imSH}\ff)
-(\ff,e^{imSH}({\widetilde P}_b)^* A^b{\widetilde P}_b e^{-imSH}\ff)].\nonumber
\ene
Writing ${\widetilde P}_b=\sum_{j=1}^L P_{b,E_j}{\widehat P}_{|b|-1}$ with $P_{b,E_j}$ being the one dimensional eigenprojection of $H^b$ with eigenvalue $E_j$, we see that the RHS is bounded by
\beq
&&\sum_{j=1}^L\frac{1}{n}\sum_{m=0}^{n-1}\frac{1}{S}\bigl|(\ff,e^{imSH}({\widetilde P}_b)^* e^{iS(H^b-E_j)}A^b P_{b,E_j}{\widehat P}_{|b|-1} e^{-imSH}\ff)\nonumber\\
&&\q\q\q\q\q-(\ff,e^{imSH}({\widetilde P}_b)^* A^b P_{b,E_j}{\widehat P}_{|b|-1} e^{-imSH}\ff)\bigr|.\nonumber
\ene
This is arbitrarily small when $S>0$ is fixed sufficiently large, by our assumption $\Vert |X^b|P_{b,E_j}\Vert < \infty$.

Summarizing, we have proved when $t\to\infty$
\beq
&&\left(\ff,\sum_{2\le\sharp(b)\le N}e^{itH}\left(\frac{X}{t}-D\right)^2{\widetilde P}_be^{-itH}f\right)\label{appro1}\\
&&\sim_{4\epsilon_M} -2\left(\ff,\sum_{2\le|b|\le N}\left[\frac{2}{t^2}\int_0^t se^{isH}T_b{\widetilde P}_b e^{-isH}\ff ds-e^{itH}T_b{\widetilde P}_be^{-itH}\ff\right]\right)\nonumber\\
&&\q+2\left(\ff,\sum_{2\le|b|\le N}e^{itH}H_0^b{\widetilde P}_be^{-itH}\ff\right),\nonumber
\ene
where we have used \eq{(2.6)} and $H_0=T_b+H_0^b$. The uniform boundedness in $t$ of the operator
$$
(1+|X|^2)^{-1}(H-i)^{-1}e^{itH}\left(\frac{X}{t}-D\right)^2,
$$
and that the projections $P_{b}^{M_{|b|}}$ are of finite dimension yield for large $R>1$ with an arbitrarily small error $\delta_R>0$
\beq
&&\left(\ff,\sum_{2\le\sharp(b)\le N}e^{itH}\left(\frac{X}{t}-D\right)^2{\widetilde P}_be^{-itH}f\right)\label{appro}\\
&&\approx \left(\ff,\sum_{2\le\sharp(b)\le N}e^{itH}\left(\frac{X}{t}-D\right)^2F(|x^b|<R){\widetilde P}_be^{-itH}\ff\right).\nonumber
\ene
Here
\beq
\left(\frac{X}{t}-D\right)^2&=&\left(\frac{X_b}{t}-D_b\right)^2+\left(\frac{(X^b)^2}{t^2}-\frac{2A^b}{t}+2H_0^b\right).\nonumber
\ene
Thus the RHS of \eq{appro} is asymptotically equal to as $t\to\infty$
\beq
&&\left(\ff,\sum_{2\le\sharp(b)\le N}e^{itH}\left(\frac{X_b}{t}-D_b\right)^2F(|x^b|<R){\widetilde P}_be^{-itH}\ff\right)\\
&&+2\left(\ff,\sum_{2\le\sharp(b)\le N}e^{itH}H_0^bF(|x^b|<R){\widetilde P}_be^{-itH}\ff\right).\nonumber
\ene
Comparing this with \eq{appro1} and removing $F(|x^b|<R)$ with a small error $\delta_R>0$, we have as $t\to\infty$
\beq
&&\left(\ff,\sum_{2\le\sharp(b)\le N}e^{itH}\left(\frac{X_b}{t}-D_b\right)^2{\widetilde P}_be^{-itH}f\right)\label{appro3}\\
&&\sim_{4\epsilon_M+2\delta_R}-2\left(\ff,\sum_{2\le|b|\le N}\left[\frac{2}{t^2}\int_0^t se^{isH}T_b{\widetilde P}_b e^{-isH}\ff ds-e^{itH}T_b{\widetilde P}_be^{-itH}\ff\right]\right).\nonumber
\ene
The both sides do not depend on the cut off $F(|x^b|<R)$, thus we can replace
$4\epsilon_M+2\delta_R$ by $4\epsilon_M$.
\MP

We set
\beq
&&F(t)=\sum_{2\le|b|\le N}\left[\frac{2}{t^2}\int_0^t se^{isH}T_b{\widetilde P}_b e^{-isH} ds-e^{itH}T_b{\widetilde P}_be^{-itH}\right],\nonumber\\
&&G(t)=\sum_{2\le|b|\le N}\left[\frac{2}{t^2}\int_0^t se^{isH}({\widetilde P}_b)^*T_b{\widetilde P}_b e^{-isH} ds-e^{itH}({\widetilde P}_b)^*T_b{\widetilde P}_be^{-itH}\right].\nonumber
\ene
By Lemma \ref{RAGE}, we have when $T\to\infty$ for a large fixed $A>1$
\beq
\sum_{2\le|b|\le N}\sum_{2\le|d|\le N, d\ne b}\frac{2}{A}\int_T^{T+A}\left[\frac{2}{t^2}\int_0^t se^{isH}({\widetilde P}_d)^*T_b{\widetilde P}_b e^{-isH}\ff ds-e^{itH}({\widetilde P}_d)^*T_b{\widetilde P}_be^{-itH}\ff\right]dt\sim_{\epsilon_M}0.\nonumber
\ene
Thus by \eq{(2.6)}, the time mean of the RHS of \eq{appro3} is equal to
\beq
-\frac{2}{A}\int_T^{T+A}\left(\ff,F(t)\ff\right)dt
\sim_{\epsilon_M}-\frac{2}{A}\int_T^{T+A}(\ff,G(t)\ff)dt\nonumber
\ene
asymptotically as $T\to\infty$.
Since the function $H(t)=\frac{2}{t^2}\int_0^t s(\ff,e^{isH}({\widetilde P}_b)^*T_b{\widetilde P}_b e^{-isH}\ff)ds$ is real valued, continuously differentiable, uniformly bounded, and its derivative with respect to $t$ goes to $0$ as $t\to\infty$, we can find a sequence $T_k\to\infty$ as $k\to\infty$ for each fixed $A>1$ such that the RHS of the above formula goes to $0$ as $T=T_k\to\infty$ (see Lemma 8.15 in \cite{[En2]}):
\beq
&&-\lim_{k\to\infty}\frac{2}{A}\int_{T_k}^{T_k+A}(\ff,G(t)\ff)dt\nonumber\\
&&
=-\lim_{k\to\infty}\frac{2}{A}\int_{T_k}^{T_k+A}
\left[\frac{2}{t^2}\int_0^t s(\ff,e^{isH}({\widetilde P}_b)^*T_b{\widetilde P}_b e^{-isH}\ff) ds-(\ff,e^{itH}({\widetilde P}_b)^*T_b{\widetilde P}_be^{-itH}\ff)\right]dt
\nonumber\\
&&
=\lim_{k\to\infty}\frac{1}{A}\int_{T_k}^{T_k+A}t\frac{dH}{dt}(t)dt\nom=0.\nonumber
\ene
These and \eq{appro3} give when $T_k\to\infty$
\beq
\frac{1}{A}\int_{T_k}^{T_k+A}\sum_{2\le\sharp(b)\le N}\left(\ff,e^{itH}\left(\frac{X_b}{t}-D_b\right)^2{\widetilde P}_be^{-itH}f\right)
dt\sim_{5\epsilon_M}0.\nonumber
\ene
By \eq{(2.6)} and Lemma \ref{RAGE}, the LHS is equal to
\beq
&&\frac{1}{A}\int_{T_k}^{T_k+A} \sum_{2\le\sharp(d)\le N}\sum_{2\le\sharp(b)\le N}\left(f,e^{itH}({\widetilde P}_d)^*\left(\frac{X_b}{t}-D_b\right)^2{\widetilde P}_be^{-itH}f\right)dt\nonumber\\
&&\sim_{\epsilon_M}\frac{1}{A}\int_{T_k}^{T_k+A}\sum_{2\le\sharp(b)\le N}\left\Vert\left(\frac{X_b}{t}-D_b\right){\widetilde P}_be^{-itH}f\right\Vert^2 dt
\ene
 asymptotically as $T_k\to\infty$.
Thus we have proved that for given components $M_j$'s of multi-indices $M_b$'s and any fixed large $A>1$
\beq
\frac{1}{A}\int_{T_k}^{T_k+A}\sum_{2\le\sharp(b)\le N}\left\Vert\left(\frac{X_b}{t}-D_b\right){\widetilde P}_be^{-itH}f\right\Vert^2 dt\sim_{6\epsilon_M}0
\ene
as $T_k\to\infty$. By Lemma \ref{RAGE}, we further have as $T_k\to\infty$
\beq
\frac{1}{A}\int_{T_k}^{T_k+A}\sum_{2\le\sharp(b)\le N}\left[\left\Vert\left(\frac{X_b}{t}-D_b\right){\widetilde P}_be^{-itH}f\right\Vert^2
+\sum_{\forall C_\ell\in b: \alpha\notin C_\ell}\left\Vert F(|x_\alpha|<R){\widetilde P}_be^{-itH}f\right\Vert\right] dt
\sim_{7\epsilon_M}0, \nonumber
\ene
where we first fix $A>1$ large enough so that the second term is less than $\ep_M$ and then we let $T_k\to\infty$ (see the proof of Lemma \ref{RAGE} below).
We can thus take a sequence $\{t_m\}$ tending to $\infty$ and  sequences $M_j^m$ that also tend to $\infty$ so that
\beq
&&\sum_{2\le\sharp(b)\le N}\left\Vert\left(\frac{X_b}{t_m}-D_b\right){\widetilde P}_b^{M_b^m}e^{-it_mH}f\right\Vert^2 \to 0\label{(2.21)}
\ene
and
for all $R>0$, and $\alpha=\{i,j\}$ that is not in any $C_\ell\in b$
\beq
\Vert F(|x_\alpha|<R){\widetilde P}_b^{M_b^m} e^{-it_mH} f\Vert \to 0,
\ene
when $m\to\infty$.
\eq{(2.7)} and \eq{(2.8)} follow from these by density argument.
The case $t_m\to -\infty$ is treated similarly.
\BP

There remains to prove Lemma \ref{RAGE}.

\BP

\noindent
{\it Proof of Lemma \ref{RAGE}:} We prove a more general version of Lemma \ref{RAGE}: Under the assumption of the lemma, we have as $T\to\infty$
\beq
\left\Vert\frac{1}{T}\int_0^T B(s) F(|x_\alpha|<R)F(|x^b|<R){\widehat P}_{|b|-1}^{{\widehat M}_b} e^{-isH} E_H(B)ds\right\Vert \sim_{\epsilon_M}0\label{newl}
\ene
for any $\alpha=\{i,j\}$ such that $\alpha\notin C_\ell$ for all $C_\ell\in b$.

\BP

We prove \eq{newl} by induction on $k=|b|$.

\begin{lem}\label{RAGE-basic} \eq{newl} for $|b|=2$ holds.
\end{lem}

\noindent
{\it Proof:} Since $\Vert F(|x|>S)F(|x_\alpha|<R)F(|x^b|<R)\Vert\to 0$ as $S\to\infty$ when $|b|=2$, $R<\infty$ and $\alpha\notin C_\ell$ for any $C_\ell\in b$ $(\ell=1,2)$, we have only to show
\beq
\lim_{T\to\infty}\left\Vert\frac{1}{T}\int_0^T B(s)F(|x|<R)E_H(B)e^{-isH}ds\right\Vert=0.
\ene
The operator $F(|x|<R)E_H(B)$ is a compact operator. Thus it suffices to prove the lemma with $F(|x|<R)E_H(B)$ replaced by a one dimensional operator $Kf=(f,\phi)\psi$, where $\phi\in\HH_c(H)$. Then
\beq
\left\Vert\frac{1}{T}\int_0^T B(s)Ke^{-isH}ds\right\Vert^2&=&\left\Vert\frac{1}{T}\int_0^T e^{isH}K^*B(s)^*ds\right\Vert^2\\
&=&\sup_{\Vert f\Vert=1}\left\Vert\frac{1}{T}\int_0^T e^{isH}K^*B(s)^*fds\right\Vert^2\nonumber\\
&=&\sup_{\Vert f\Vert=1}\frac{1}{T^2}\int_0^T\int_0^T (B(s)^* f,\psi)(\psi,B(t)^* f)(e^{-i(t-s)H}\phi,\phi)dt ds\nonumber\\
&\le&C \frac{1}{T^2}\int_0^T\int_0^T |(e^{-i(t-s)H}\phi,\phi)|dt ds\nonumber\\
&\le& C \frac{1}{T}\int_{-T}^T |(e^{-itH}\phi,\phi)|dt\nonumber
\ene
for some constant $C>0$.
By Schwarz inequality, the RHS is bounded by
\beq
\sqrt{2}C\left(\frac{1}{T}\int_{-T}^T |(e^{-itH}\phi,\phi)|^2 dt\right)^{\frac{1}{2}}.
\ene
Noting that the measure $\mu(\lambda)=(E_H(\lambda)\phi,\phi)$ is continuous by $\phi\in \HH_c(H)$, we calculate the formula inside the parentheses:
\beq
&&\frac{1}{T}\int_{-T}^T\int_{R^1}\int_{R^1}e^{-i(\lambda-\lambda')t}d\mu(\lambda)d\mu(\lambda')dt\\
&&=2\int_{R^1}\int_{R^1}\frac{\sin\{(\lambda-\lambda')T\}}{(\lambda-\lambda')T}d\mu(\lambda)d\mu(\lambda').\nonumber
\ene
Dividing the integration region $R_{(\lambda,\lambda')}^2$ into $|\lambda-\lambda'|\le \epsilon$ and the other, we obtain a bound:
\beq
2\int_{|\lambda-\lambda'|\le \epsilon}d\mu(\lambda)d\mu(\lambda')+\frac{2}{\epsilon T}.\nonumber
\ene
The first term can be small arbitrarily if $\epsilon>0$ is small enough, since the measure $\mu(\lambda)$ is continuous. Then letting $T\to\infty$ we can let the second term go to $0$. $\Box$

\BP

Now assume \eq{newl} for $|b|<k$ $(3\le k\le N)$. Let $b=\{C_1,\cdots,C_{|b|}\}$ with $|b|=k$ and assume $\alpha=\{i,j\}$ connects the clusters $C_1$ and $C_2$ of $b$. We denote the new cluster decomposition by $d=\{C_1\cup C_2,C_3,\cdots,C_k\}$. Then $|d|=k-1$, and $K_1=F(|x_\alpha|<R)F(|x^b|<R)$ bounds the variable $x^d$. We decompose ${\widehat P}_{k-1}^{{\widehat M}_b}$ (see \eq{(2.4)}) in \eq{newl} as
\beq
{\widehat P}_{k-1}^{{\widehat M}_b}\label{bound}&=&(I-P_d^{M_{k-1}}){\widehat P}_{k-2}^{{\widehat M}_d}
-\sum_{b_{k-1}\ne d} P_{b_{k-1}}^{M_{k-1}}{\widehat P}_{k-2}^{{\widehat M}_d},\label{decomp}
\ene
where ${\widehat M}_d=(M_1,\cdots,M_{k-2})$. Each $P_{b_{k-1}}^{M_{k-1}}$ on the second term bounds the variable $x^{b_{k-1}}$ with $|b_{k-1}|=k-1$. Since $b_{k-1}\ne d$ and $F(|x_\alpha|<R)F(|x^b|<R)$ bounds the variable $x^d$, $F(|x_\alpha|<R)F(|x^b|<R)P_{b_{k-1}}^{M_{k-1}}$ connects at least one pair of different two clusters in $b_{k-1}$. Thus the terms in the second summand on the RHS of \eq{decomp} are treated by the induction hypothesis. Thus we have to show when $T\to\infty$
\beq
\left\Vert\frac{1}{T}\int_0^T B(s) F(|x_\alpha|<R)F(|x^b|<R)
(I-P_d^{M_{k-1}}){\widehat P}_{k-2}^{{\widehat M}_d}e^{-isH}E_H(B)ds\right\Vert\sim_{\epsilon_M} 0.\label{b1}
\ene
Let $S>0$ be arbitrary but fixed and let as before $t(s)=s-mS$ for $mS\le s <(m+1)S$. The norm of \eq{b1} is bounded by
\beq
&&\left\Vert\frac{1}{T}\int_0^T B(s) K_1
(I-P_d^{M_{k-1}})
e^{-it(s)H_d}e^{it(s)H}
{\widehat P}_{k-2}^{{\widehat M}_d}e^{-isH}E_H(B)ds\right\Vert\label{int0}\\
&&+
\left\Vert\frac{1}{T}\int_0^T B(s) K_2
(I-e^{-it(s)H_d}e^{it(s)H})
{\widehat P}_{k-2}^{{\widehat M}_d}e^{-isH}E_H(B)ds\right\Vert,\nonumber
\ene
where $K_1=F(|x_\alpha|<R)F(|x^b|<R)$ and $K_2=K_1 (I-P_d^{M_{k-1}})$.

Since $H-H_d=I_d$,
\beq
I-e^{-it(s)H_d}e^{it(s)H}=\int_0^{t(s)}e^{-i\tau H_d}i(H_d-H)e^{i\tau H}d\tau \q (0\le t(s)<S) \label{int}
\ene
is a sum of the terms, each of which bounds at least one variable $x_\beta$ with $\beta=\{k,m\}$ connecting two different clusters of $d$. Noting that
$F(|x^d|<CR)\ge K_1=F(|x_\alpha|<R)F(|x^b|<R)$ holds for some constant $C>0$,
we can treat the second term of \eq{int0} by induction hypothesis.

The first term in \eq{int0} is rewritten as
\beq
\left\Vert\frac{1}{T}\int_0^T B(s) K_1
(I-P_d^{M_{k-1}})
e^{-it(s)H_d}
{\widehat P}_{k-2}^{{\widehat M}_d}e^{-i(s-t(s))H}E_H(B)ds\right\Vert\label{main-contri}
\ene
with some remainder terms. These remainder terms come from the commutator of $e^{it(s)H}$ and ${\widehat P}_{k-2}^{{\widehat M}_d}$, and can be treated by induction hypothesis. Since $s-t(s)=mS$, \eq{main-contri} is rewritten for $t=nS$
\beq
\left\Vert\frac{1}{nS}\sum_{m=0}^{n-1}\int_0^S B(s+mS) K_1
(I-P_d^{M_{k-1}})
e^{-isH_d}
{\widehat P}_{k-2}^{{\widehat M}_d}E_H(B)ds e^{-imSH}\right\Vert.\label{last}
\ene
Since $K_1$ bounds $x^d$, the difference
\beq
K_1\{(I-P_d^{M_{k-1}})-(I-P_d)\}=K_1(P_d-P_d^{M_{k-1}})
\ene
tends to $0$ as $M_{k-1}\to\infty$ in operator norm. Thus we can replace $K_1(I-P_d^{M_{k-1}})$ in \eq{last} by $K_1(I-P_d)$ with an error $\ep_M$. This step yields the error $\epsilon_M$ in the lemma. To estimate \eq{last}, letting $S>0$ large but fixed, we first get an energy cut off for $H^d$ from $E_H(B)$ by some commutator arguments, whose commutators are treated by induction hypothesis. Then we can apply \eq{newl} for $|b|=2$ to \eq{last} with $H$ replaced by $H^d$. The proof of \eq{newl} and Theorem \ref{Enss} is complete. $\Box$

\BP

We now turn to our purpose of stating Theorem \ref{Enss}. I.e., let us see how our definition \ref{Time} of time coincides with the usual notion of time. \eq{(2.8)} of Theorem \ref{Enss} or more intuitive form \eq{(2.21)}: for all $b$ with $2\le |b|\le N$
\beq
\left\Vert\left(\frac{x_b}{t_m}-v_b\right){\widetilde P}_b^{M_b^m}e^{-it_mH/\hbar}f\right\Vert \to 0 \q (m\to\infty)\label{2.8-1}
\ene
means that the ratio of the position vector $x_b$ and the velocity vector $v_b$ are proportional to time we have defined in Definition \ref{Time}. Namely we have  at least schematically
\beq
\frac{|x_b|}{|v_b|}\sim t\label{time}
\ene
as $t$ tends to $\pm\infty$ along some sequence $t=t_m$. 
In this sense, time $t$ of the system we are considering is determined independently of the cluster decomposition $b$ and of the particles inside the system. Thus $t$ has a usual sense of time as a {\it common} parameter of motion of the system in accordance with the notion of `common time' in Newton's sense: ``relative, apparent, and common time, is some sensible and external (whether accurate or unequable) measure of duration by the means of motion, $\cdots$"
(I. Newton \cite{[New]} p.6).

Once we have defined time in this way that coincides with our intuition, the motion of the particles is described in the sense of \eq{time} by the evolution $\exp(-itH/\hbar)f$ for an initial value wave function $f$ at time $t=0$. In this sense, we can say that the motion of the particles inside the system obeys the identical equality
\beq
\left(\frac{\hbar}{i}\frac{d}{dt}+H\right) \exp(-itH/\hbar)f=0.\label{Schroedinger}
\ene
This is the usual Schr\"odinger equation. Thus in the context where time is already defined as in Definition \ref{Time}, we can say that the motion is governed by Schr\"odinger equation as usual quantum mechanics assumes as an axiom.

\section{Uncertainty of time}

We have however to remind that this view to nature is {\it only} an approximation. Time $t$ of the system having Hamiltonian $H$ in \eq{(1.6)} satisfies the asymptotic relation \eq{(2.8)} or \eq{2.8-1}, but this is an asymptotic relation and does not give any exact relation like $x_b=tv_b$ which is assumed to hold in classical context. In quantum mechanical context, as we have seen in chapter \ref{chap:1} there is known the uncertainty principle that yields that the values of position and momentum are not determined in precise sense simultaneously, which has been known as a discrepancy between classical and quantum mechanics. 

In our context, this uncertainty is understood as that of time in the following sense. Position and momentum are fundamentally independent quantities that satisfy canonical commutation relation \eq{canonical} given in chapter \ref{chap:2}. This canonical commutation relation gives the uncertainty relation (see chapter \ref{chap:1}) by usual commutation arguments about their variances from the expectation values:
$$
\Delta q\cdot\Delta p\ge \frac{\hbar}{2}.
$$
This uncertainty means that position and momentum are independent in the sense that there is no way to let position and momentum correlate exactly as in classical views. In the usual formulation where time is given {\it a priori} as a fundamental quantity, it is anticipated as in chapter \ref{chap:1} that the position $x$ and mean velocity $v$ of a particle are related by the relation $x=tv$ if the particle starts from the origin at time $t=0$. In this ordinary formulation the uncertainty principle is a cause of a question why position and momentum do not relate exactly to each other. However, in our formulation where we do not introduce {\it a priori} time $t$ as a fundamental quantity but derive it from position and momentum operators as in Definition \ref{Time},
we need not make any anticipation about their relation other than the ones that follow from \eq{(2.8)} or its schematic expression \eq{time}.
 The uncertainty principle works so as to prohibit the relation $x_b=tv_b$ from being an exact relation and \eq{(2.8)} gives the most possible extent so that the relation $x_b=tv_b$ holds {\it without contradicting} the uncertainty principle (for a more precise estimate than \eq{(2.8)},
 see e.g., \cite{[De]}).
Thus our local time is defined from the outset to have the uncertainty that is allowed by the relation \eq{(2.8)}.


\chapter{Local Systems}\label{chap:4}

So far we have considered an $N$-particle system and postulated the existence of position and momentum operators for that system. On the basis of these notions we defined time of that $N$-particle system. Thus space-momentum and time constitute the notions proper to each $N$-particle system. We call such a system {\it local system}.

\begin{df}\label{local}
A pair $(H_{n\ell},\HH_{n\ell})$ of a Hilbert space $\HH_{n\ell}=\HH^n=L^2(R^{3(N-1)})$ $(N=n+1)$ and an $N$-body Hamiltonian $H_{n\ell}$ is called a local system.
\end{df}

Then local time $t$ of a local system $(H_{n\ell},\HH_{n\ell})$ is defined as in Definition \ref{Time}. We sometimes denote this time as $t_{(H_{n\ell},\HH_{n\ell})}$ indicating the local system under consideration. Here we recall that the label $\ell$ distinguishes different local systems with the same number $N=n+1$ of particles, as remarked in Axiom \ref{axiom3} of chapter \ref{chap:2}.

\BP

There are two regions of research related with local systems. One is the investigation of the properties of the motion inside each local system. Another is the relation between local systems.

The former constitutes the same region as the usual quantum mechanics. One point that is different from the usual view is that the motion is only possible when the initial state $f$ is a scattering state, i.e. when it belongs to the continuous spectral subspace $\HH_c(H)$ of the Hamiltonian $H$, because the usual bound state is unobservable: Let $f$ be a bound state of $H$ with eigenvalue $E$:
\beq
Hf=Ef.
\ene
Then its evolution is given by
\beq
\exp(-itH/\hbar)f(x)=\exp(-itE/\hbar)f(x).
\ene
Therefore, such a state does never change in the sense that the probability density of the existence in the configuration space given by
\beq
|\exp(-itH/\hbar)f(x)|^2=|\exp(-itE/\hbar)f(x)|^2=|f(x)|^2\label{unobservable}
\ene
is a constant of motion. Thus such a state cannot be observed except for the case that the state $f$ changes to another state $g$, by some disturbance from the outside, that involves a part belonging to the continuous spectral subspace of $H$\ \footnote[2]{I quote the following passage from p. 667 of Paul Busch and Pekka J. Lahti \cite{[BL]}: ``In fact, defining (preparing) a physical system in a pure state implies that it is isolated from its environment. Therefore, strictly speaking, it cannot be observed, since an observation entails an interaction which amounts to suspending the system's isolation."}. In this case what can change in accordance with time is the scattering part of $g$ only. Thus the investigation of quantum-mechanical motion is that of scattering states. The bound states that are considered in usual textbooks should be understood as a certain kind of resonances, which are close to bound states. They are never pure bound states. Thus our purpose in the next part II is to study the evolution $\exp(-itH/\hbar)f$ of scattering states $f$ of an $N$-body Hamiltonian $H$ in a fixed Hilbert space $\HH=L^2(R^{3n})$, where $n=N-1\ge 1$.

After some excursion of studying the motion inside a local system, we turn to the investigation of the relation between plural number of local systems in part III. This subject is related with observation. In the explanation of observation, we regard the center of mass of each local system a classical particle behaving under the general principle of relativity and the principle of equivalence. Namely, we postulate that, when one observes other local system than oneself's, the observer observes the motion of the inside of the other local system as following the general theory of relativity, and that the inside of an observed local system is divided in accordance with the purpose of each particular observation. Those subsystems resulting from the division are regarded as classical particles identified with their centers of mass. This postulate would be justified if we see that the observation necessarily identifies the objects of the observation and as a consequence the observer decomposes the observed local system into some number of sublocal systems. Then the observer's investigation would be the relative motion of those sublocal systems. Our postulate here is, therefore, the identification of the center of mass of each subsystem with a classical particle that behaves following classical general relativity. This is our basic assumption on observation.

The problem that the reader might propose at this stage would be why the quantum mechanics which is assumed to have Euclidean space-time structure in the previous parts can be consistent with the postulates of general relativity. This crucial point will be answered at the beginning of part III. The main point is that the space-momentum or space-time structures of different local systems are mutually independent, and hence we can postulate any laws to the relative motion between local systems. We choose the general theory of relativity as our laws that govern the relative motion of local systems, because the theory is known to give sufficiently precise predictions matching the observations or experiments.

We remark that it is possible to interpret special theory of relativity as assuming this kind of setting.  Namely, in special theory of relativity, the space-time {\it inside} a ``stationary system\footnote[3]{We here use this wording to mean an inertial frame of reference, following p. 38 of A. Einstein \cite{[Ein]}.} " can be interpreted as Euclidean, while the relation between two stationary systems moving with {\it non-zero} relative speed follows Lorentz transformation and hence in this case the two systems have Lorentz metrics. When we consider inside a stationary system, the Lorentz transformation inside itself can be interpreted as Galilei transformation, because the relative speed with respect to itself is zero. Thus, we arrive at an interpretation of special theory of relativity that the space-time is Euclidean inside a stationary system, while Minkowskian space-time {\it only} appears when one considers observation of a stationary system moving with non-zero velocity with respect to observer's system. The main purpose of the special theory of relativity was to explain the phenomena that occur when making observation of a moving systems from another stationary system. Thus it is understandable that the possibility of adopting Euclidean geometry inside a stationary system has been overlooked for a long time. If one would want to get a unified theory including the zero relative velocity, it is quite natural to regard the Galilei transformation as a special case of Lorentz transformation with zero relative velocity. Therefore the consistency of the Euclidean space-time inside a local system and the outer curved space-time is already inherent in the formulation of special theory of relativity.

The next problem after we show the consistency of the quantum-mechanical Euclidean structure and the general relativistic Riemannian structure of space-time, is the explanation of the results of actual observations or experiments. This we will do to the extent we reach at the present stage of our theory in the remainder of part III. The main ingredients in these explanations are the results we obtain in part II, where normal quantum-mechanical motions are analyzed. Explanation of observation will be done in the same way as in special theory of relativity by assuming that the actually observed values are the ones that are obtained by modifying the usual results of quantum-mechanical calculation by the general relativistic change of coordinates from the observed system to the observer's system.

\newpage



\noindent
{\bf Exercise}
\BP

We assume we are given a 3-dimensional coordinate $x=(x_1,x_2,x_3)$ of 3-dimensional Euclidean space $R^3$ and a corresponding 3-dimensional momentum operator $p=(p_1,p_2,p_3)$ conjugate to $x$, i.e. we define
$$
p_j=\frac{\hbar}{i}\frac{\partial}{\partial x_j},\quad j=1,2,3,
$$
where $\hbar=h/2\pi$, $h$ being Planck constant. We note in the momentum representation $x_j$ works as a differential operator:
$$
{\cal F}x_j{\cal F}^{-1}=i\hbar\frac{\partial}{\partial p_j}.
$$
Here ${\cal F}$ is the Fourier transformation defined by
$$
{\cal F}g(p)=(2\pi\hbar)^{-3/2}\int_{R^3} \exp(-ip\cdot x/\hbar)g(x)dx,
$$
where $p\cdot x=\sum_{j=1}^3 p_j x_j$ is the inner product of the vectors $x$ and $p$.
We assume we are given a time parameter $t$ that takes real values. We then define 3-dimensional time and energy operators $T=(t_1,t_2,t_3)$ and $E=(e_1,e_2,e_3)$ for $t\ne 0$ by
\begin{align*}
&t_j=tp_j|p|^{-1},\\
&e_j=\frac{1}{4t}(|p|x_j+x_j|p|),
\end{align*}
where
$$
|p|=\left(\sum_{j=1}^3 p_j^2\right)^{1/2}
$$
is the positive square root of a nonnegative operator $\sum_{j=1}^3 p_j^2=-\hbar^2\Delta_x$ with $\Delta_x$ being Laplacian with respect to $x$.
We note that these operators $t_j$ and $e_j$ initially defined on the space ${\cal D}={\cal F}^{-1}C_0^\infty(R^3-\{0\})$ can be extended to selfadjoint operators in $L^2(R^3)$, where $C_0^\infty(R^3-\{0\})$ is the space of $C^\infty$-functions with their supports compact in $R^3-\{0\}$ and $L^2(R^3)$ is the space of square integrable functions on $R^3$. Clearly they have dimensions of time and energy, respectively, and for $\pm t>0$, $\pm T=\pm(t_1,t_2,t_3)$ has the same direction as momentum $p$ and satisfies
$$
\pm|T|=\pm\left(\sum_{j=1}^3 t_j^2\right)^{1/2}=t.
$$
For $\pm t>0$, $\pm E=\pm(e_1,e_2,e_3)$ has almost the same direction as position $x$, and when $x$ and $p$ denote position and momentum of a scattering particle with mass $m$ whose evolution is governed by a Hamiltonian $H$, we have
$$
|E|\exp(-itH)g=\left(\sum_{j=1}^3 e_j^2\right)^{1/2}\exp(-itH)g\sim \frac{|p|^2}{2m}\exp(-itH)g
$$
in $L^2(R^3)$ asymptotically as $t\to\pm \infty$ for a scattering state $g\in L^2(R^3)$, since we have $x_j/t\sim p_j/m$ on $\exp(-itH)g$ as $t\to\pm\infty$ along some sequence $t_k\to\pm\infty$. Thus these operators can be regarded as a 3-dimensional version of quantum mechanical time and energy.
\BP

Now show that the following uncertainty relation holds between $T$ and $E$.
\BP

\F
{\bf Theorem.} Let $f\in {\cal D}\subset L^2(R^3)$ with its $L^2$-norm $\Vert f\Vert=\langle f,f\rangle^{1/2}=1$, where
$$
\langle f,g\rangle =\int_{R^3}f(x)\overline{g}(x)dx
$$
is the inner product of $L^2(R^3)$ with $\overline{g}(x)$ being the complex conjugate of $g(x)$. Let
\begin{align*}
&{\tilde t}_j = \langle t_j f,f\rangle,\quad {\tilde e}_j=\langle e_jf,f\rangle,\\
&{\tilde T}=({\tilde t}_1,{\tilde t}_2,{\tilde t}_3),\quad {\tilde E}=
({\tilde e}_1,{\tilde e}_2,{\tilde e}_3)
\end{align*}
be the expectation values of these operators on the state $f$. 
Let the variances of $T$ and $E$ be defined by
\begin{align*}
&\Delta T=\Vert (T-{\tilde T})f\Vert=\left(\sum_{j=1}^3\Vert(t_j-{\tilde t}_j)f\Vert^2\right)^{1/2},\\
&\Delta E=\Vert(E-{\tilde E})f\Vert=\left(\sum_{j=1}^3\Vert(e_j-{\tilde e}_j)f\Vert^2\right)^{1/2}.
\end{align*}
Then we have the uncertainty relation:
$$
\Delta T\Delta E\ge \frac{\hbar}{2}.
$$
\BP

\part{Motion Inside a Local System}\label{localmotion}

In this part, we will see how the result of Enss \cite{[En]} stated in Theorem \ref{Enss} gives a solution of the fundamental problem of scattering theory, which is thought to give a basis of any physical observation in the sense stated in the last paragraph of chapter \ref{chap:1}.

Before Enss found the so-called Enss' time dependent method in \cite{[E1]}, scattering theory had been investigated mainly by stationary method which utilizes the Laplace transform of the propagator $\exp(-itH/\hbar)$. Enss gave a simple method to prove the fundamental property called ``asymptotic completeness" of wave operators in \cite{[E1]} by directly utilizing the propagator $\exp(-itH/\hbar)$. Since then the proof of asymptotic completeness can take an elegant form compared with the stationary method by the predecessors (e.g., \cite{[Ikebe]}, \cite{KK}, and the references therein). Though we have to remark that in the case of many-body scattering, some stronger estimates are necessary to establish the asymptotic completeness, and in this book we leave some of those estimates to the references \cite{[De]}, \cite{SS}, etc.

Theorem \ref{Enss} of Enss played an important role in the interpretation of our formulation of local time in chapter \ref{chap:3}. It would be impressive to see that his method also plays an important role in the main part of scattering theory.

We will return to the stationary theory of the Universe in part \ref{Conclusions}, where the original idea of stationary method by the predecessors will be revived as an important view to nature. These will show that both stationary and time-dependent methods are fundamental in our understanding of nature. We will see in chapter \ref{StationaryUniverse} that time-dependent method is based on an artificial introduction of local time after the introduction of the stationary universe. This could be anticipated from our introduction of the notion of local time in chapter \ref{chap:3}.

\chapter{Free Hamiltonian}\label{chap:5}

\section{Spectral representation of free Hamiltonian}\label{spectral}

In this chapter we assume $\hbar=1$ and consider the free Hamiltonian:
\beq
H_0=-\frac{1}{2}\Delta=-\frac{1}{2}\sum_{j=1}^m \frac{\partial^2}{\partial x_j^2}=\frac{1}{2}D_x^2=\frac{1}{2}\sum_{j=1}^m D_{x_j}^2,\label{free-Hamiltonian}
\ene
defined for functions on $R^m$ with general dimension $m=1,2,\cdots$, where
$$
D_x=(D_{x_1},\cdots,D_{x_m}),\q D_{x_j}=\frac{1}{i}\frac{\partial}{\partial x_j}.
$$
When mass factors and Planck constant are necessary, it is easy to include them as in chapter \ref{chap:2} or \ref{chap:3}.

We initially define $H_0$ for functions $f$ from the space $\cal S$ of rapidly decreasing functions:
\beq
H_0f(x)=-\frac{1}{2}\sum_{j=1}^m \frac{\partial^2 f}{\partial x_j^2}(x)
\ene
and extend it to a selfadjoint operator in $\HH=L^2(R^m)$ with domain ${\cal D}(H_0)$ being the Sobolev space of order 2: $H^2(R^m)=\{f\in L^2(R^m)|\ \int_{R^m}|(1+D_x^2)f(x)|^2dx<\infty\}$. We denote the selfadjoint extension also by $H_0$.

In the following, we denote by $(f,g)$ the inner product of $\HH=L^2(R^m)$ and by $\Vert f\Vert$ the norm of $\HH$.

Let $\FF$ be the Fourier transformation: For $f\in\cal S$
\beq
\FF f(\xi)=(2\pi)^{-\frac{m}{2}}\int_{R^m} e^{-i\xi x}f(x) dx,
\ene
where $\xi x=\sum_{j=1}^m \xi_j x_j$ is the Euclidean scalar product of $R^m$. Then we extend $\FF$ as a unitary transformation from $\HH$ onto $\HH$:
\beq
\Vert \FF f\Vert =\Vert f\Vert.
\ene
If we denote by $|\xi|^2/2$ the multiplication operator by the function $|\xi|^2/2$ defined in $L^2(R^m_\xi)$, it is easy to see
\beq
H_0={\FF}^{-1} (|\xi|^2/2)\FF\label{Hamiltonian}.
\ene
{}From this follows that there is no eigenvalue of $H_0$: for any $f\in {\cal D}(H_0)$
\beq
\exists \lambda\in R^1: H_0f=\lambda f \Rightarrow f=0.\nom
\ene
Thus $\HH_{p}(H_0)=\{0\}$, and $\HH_c(H_0)={\HH_{p}(H_0)}^\perp=\HH=L^2(R^m)$.

Letting $\chi_S$ denote the characteristic function of a set $S$, we note that 
the function $(\chi_{(-\infty,\lambda]}(|\xi|^2/2)g,h)$ $(g,h\in\HH)$ is of bounded variation with respect to $\lambda\in R^1$, and it vanishes for $\la\le 0$.
Thus we have the relation for any $g,h\in \SSS$
\beq
((|\xi|^2/2)g,h)=\int_{0}^\infty \lambda d_\lambda(\chi_{(-\infty,\lambda]}(|\xi|^2/2)g,h).\label{resol}
\ene
This relation can be extended to $h\in \HH$ and $g$ satisfying
\beq
\int_{0}^\infty \lambda^2 d\Vert \chi_{(-\infty,\lambda]}(|\xi|^2/2)g\Vert^2< \infty,
\ene
which is equivalent to
\beq
g\in \FF H^2(R^m).\label{domain}
\ene
We define for $f\in\HH$
\beq
E_0(\lambda)f=\FF^{-1}\chi_{(-\infty,\lambda]}(|\xi|^2/2)\FF f.\label{resH0}
\ene
Then by Plancherel formula, we have from \eq{Hamiltonian}, \eq{resol} and \eq{domain} that
\beq
(H_0f,g)=\int_{0}^\infty \lambda d(E_0(\lambda)f,g)\nom
\ene
for $f\in {\cal D}(H_0)=H^2(R^m)$ and $g\in\HH$.

It is easy to see by definition that $E_0(\lambda)$ satisfies\label{spectral-resolution}
\beq
&&E_0(\lambda)E_0(\mu)=E_0(\min(\lambda,\mu)),\label{res1}\\
&& \mbox{s-}\lim_{\lambda\to -\infty}E_0(\lambda)=0,\q \mbox{s-}\lim_{\lambda\to\infty}E_0(\lambda)=I,\label{res2}\\
&&
E_0(\lambda+0)=E_0(\lambda)\label{res3}
\ene
where
$E_0(\lambda+0)=\mbox{s-}\lim_{\mu \downarrow \lambda}E_0(\mu)$.

A family of operators that satisfies conditions \eq{res1}-\eq{res3} is called a resolution of the identity. It is known (see, e.g., Chapter XI of \cite{Yosida}) that there is a one-to-one correspondence between the resolutions of the identity $\{E(\lambda)\}$ and selfadjoint operators $H$ defined in a Hilbert space $\HH$ by the relation
\beq
(Hf,g)=\int_{-\infty}^\infty \lambda d(E(\lambda)f,g),\q f\in {\cal D}(H),\q g\in \HH.\label{spectral-exp}
\ene
Thus $\{E_0(\lambda)\}$ defined by \eq{resH0} is the resolution of the identity corresponding to the selfadjoint operator $H_0$. We remark that \eq{spectral-exp} yields the relation for any continuous function $F(\la)$
\beq
(F(H)f,g)=\int_{-\infty}^\infty F(\la) d(E(\lambda)f,g),\q f\in {\cal D}(F(H)),\q g\in \HH.\label{functional-calc}
\ene

Since for $\la>0$
\beq
(\chi_{(-\infty,\lambda]}(|\xi|^2/2)g,h)&=&\int_{|\xi|^2/2\le \lambda}g(\xi)\overline{h(\xi)}d\xi\nom\\
&=&\int_{0}^\lambda\int_{S^{m-1}}g(\sqrt{2\mu}\omega)\overline{h(\sqrt{2\mu}\omega)}d\omega (2\mu)^{(m-2)/2}d\mu,\nonumber
\ene
we have for $g,h\in \SSS$ and $\la>0$
\beq
\frac{d}{d\lambda}(\chi_{(-\infty,\lambda]}(|\xi|^2/2)g,h)=(2\lambda)^{(m-2)/2}
\int_{S^{m-1}}g(\sqrt{2\lambda}\omega)\overline{h(\sqrt{2\lambda}\omega)}
d\omega,\nom
\ene
where $d\omega$ is the surface element of the $(m-1)$-dimensional unit sphere $S^{m-1}$.
Thus we have for $f,g\in\SSS$ and $\la>0$
\beq
\frac{d}{d\lambda}(E_0(\lambda)f,g)&=&\frac{d}{d\lambda}(\chi_{(-\infty,\lambda]}(|\xi|^2/2)\FF f, \FF g)\label{spec}\\
&=&(2\lambda)^{(m-2)/2}\int_{S^{m-1}}(\FF f)(\sqrt{2\lambda}\omega)\overline{(\FF g)(\sqrt{2\lambda}\omega)}d\omega .\nonumber
\ene
We set for $\la>0$
\beq
\FF(\lambda)f(\omega)=(2\lambda)^{(m-2)/4}(\FF f)(\sqrt{2\lambda}\omega).\label{Fourier-la}
\ene
We remark that by \eq{Hamiltonian} we have
\beq
\FF(\la)H_0 f(\omega)=\la \FF(\la)f(\omega),
\ene
thus the adjoint $\FF(\la)^*$ that is defined below is an {\it eigenoperator} of $H_0$ in the sense that it satisfies
\beq
H_0\FF(\la)^* = \la \FF(\la)^*.
\ene
\eq{spec} is now written as
\beq
\frac{d}{d\lambda}(E_0(\lambda)f,g)=(\FF(\lambda) f,\FF(\la) g)_{L^2(S^{m-1})},\label{specdensity}
\ene
where $(\varphi,\psi)_{L^2(S^{m-1})}$ is the inner product of the Hilbert space $L^2(S^{m-1})$. Here if we let $g(\rho)=(\FF f)(\rho\ \cdot)\in L^2(S^{m-1})$ for $\rho>0$, we have for $\la>0$
\beq
\FF(\lambda)f&=& (\sqrt{2\lambda})^{(m-2)/2}(\FF f)(\sqrt{2\lambda}\ \cdot)\label{Fourier}\\
&=&(2\lambda)^{-1/4}\ (\sqrt{2\lambda})^{(m-1)/2}g(\sqrt{2\lambda})\nonumber\\
&=&(2\lambda)^{-1/4} (2\pi)^{-1/2}\int_{-\infty}^\infty e^{i\sqrt{2\lambda}r}\FF_\rho(\phi(\rho)\rho^{(m-1)/2}g)(r) dr\nonumber,
\ene
where $\phi \in C_0^\infty((0,\infty))$ such that supp $\phi\subset(0,\infty)$ and $\phi(\sqrt{2\la})=1$, and ${\FF_\rho g}(r)$ is the Fourier transform of $g(\rho)$ with respect to $\rho\in R^1$. Then $\Vert \FF(\lambda)f\Vert_{L^2(S^{m-1})}$ is bounded by for $s>1/2$
\beq
&&(2\pi)^{-1/2}(2\lambda)^{-1/4}\left(\int_{-\infty}^\infty\langle r\rangle^{-2s}dr\right)^{1/2}
\left(\int_{-\infty}^\infty\langle r\rangle^{2s}\Vert \FF_\rho(\phi(\rho)\rho^{(m-1)/2}g)(r)\Vert_{L^2(S^{m-1})}^2 dr \right)^{1/2}\nonumber\\
&&\le C_s\lambda^{-1/4}\Vert \phi(\rho)\rho^{(m-1)/2}g\Vert_{H^s((0,\infty),L^2(S^{m-1}))},\label{F(la)f}
\ene
where $\langle r\rangle=(1+|r|^2)^{1/2}$, and with $\langle D_\rho\rangle^s=\FF_\rho^{-1}\langle r\rangle^s\FF_\rho$
\beq
&&H^s((0,\infty),L^2(S^{m-1}))=\ 
\mbox{the completion of}\ 
C_0^\infty((0,\infty),L^2(S^{m-1}))\\
&&\mbox{with respect to the norm}\ 
\Vert h\Vert=\left(\int_0^\infty \Vert \langle D_\rho \rangle^s h(\rho)\Vert^2_{L^2(S^{m-1})}d\rho\right)^{1/2}.\nom
\ene
By a calculation, the RHS of \eq{F(la)f} is bounded by
\beq
&&C_{s\phi}\lambda^{-1/4}\left(\int_0^\infty\Vert \langle D_\rho\rangle^s g(\rho)\Vert_{L^2(S^{m-1})}^2 \rho^{m-1}d\rho\right)^{1/2}\nonumber\\
&&\le C_{s\phi} \lambda^{-1/4}\Vert f\Vert_{L^2_s},\nonumber
\ene
where $L^2_s=L_s^2(R^m)$ is the Hilbert space with inner product
\beq
(f,g)_{L^2_s}=\int_{R^m}\langle x\rangle^{2s}f(x)\overline{g(x)}dx.
\ene
Thus we have proved the estimate: For any $\delta>0$ and $s>1/2$, there is a constant $C_{s\delta}>0$ such that for $\la>\delta$
\beq
\Vert \FF(\lambda)f\Vert_{L^2(S^{m-1})}\le C_{s\delta}\lambda^{-1/4}\Vert f\Vert_{L^2_s}.\label{Fest}
\ene
Further, using the expression \eq{Fourier} and estimating the difference $\FF(\la)f-\FF(\mu)f$ similarly to the above, we see that $\FF(\la)$ is continuous in $\la>\delta$:
\beq
\Vert \FF(\la)f-\FF(\mu)f\Vert_{L^2(S^{m-1})}\le C_{s\delta} \epsilon(\la,\mu)\Vert f\Vert_{L^2_s},\label{continuity}
\ene
where with $0<\theta<s-1/2$
\beq
\epsilon(\la,\mu)=\left(\int_{-\infty}^\infty
|e^{i\sqrt{2\la}r}-e^{i\sqrt{2\mu}r}|^2\langle r\rangle^{-2s}dr\right)^{1/2}\le 
C_{\theta\delta}|\la-\mu|^\theta
\to 0\label{diff}
\ene
as $\mu\to\la$ with $\la,\mu>\delta$.

On the other hand, we have for $\varphi\in L^2(S^{m-1})$ and $g\in L^2_s$
\beq
(\FF(\lambda)^*\varphi,g)_{L^2(R^m)}&=&(\varphi,\FF(\lambda)g)_{L^2(S^{m-1})}\label{adjoint}\\
&=&\int_{R^m}(2\lambda)^{(m-2)/4}(2\pi)^{-m/2}\int_{S^{m-1}}e^{i\sqrt{2\lambda}\omega y}\varphi(\omega)d\omega \overline{g(y)}dy,\nonumber
\ene
which and \eq{Fest} yield
\beq
\FF(\lambda)^*\varphi(x)=(2\pi)^{-m/2}(2\lambda)^{(m-2)/4}\int_{S^{m-1}}e^{i\sqrt{2\la}x\omega}\varphi(\omega)d\omega\label{Fourier-adjoint}
\ene
and
for any $s>1/2$ and $\la>\delta(>0)$
\beq
\Vert \FF(\lambda)^* \varphi\Vert_{L^2_{-s}(R^m)}\le C_{s\delta}\la^{-1/4}\Vert\varphi\Vert_{L^2(S^{m-1})}.\nom
\ene
By \eq{continuity} and \eq{adjoint}, we further have for $\la,\mu>\delta$
\beq
\Vert \FF(\la)^*\varphi-\FF(\mu)^*\varphi\Vert_{L^2_{-s}}\le C_{s\delta} \epsilon(\la,\mu)\Vert \varphi \Vert_{L^2(S^{m-1})}.\nom
\ene
Combining these estimates with \eq{specdensity} and \eq{continuity}, we have for $\la, \mu>\delta$ and $s>1/2$
\beq
\left\Vert \frac{dE_0}{d\la}(\la)\right\Vert_{L^2_s\to L^2_{-s}}\le C_{s\delta}\la^{-1/2},\nom
\ene
and
\beq
\left\Vert \frac{dE_0}{d\la}(\la)-\frac{dE_0}{d\la}(\mu)\right\Vert_{L^2_s\to L^2_{-s}}\le C_{s\delta}\epsilon(\la,\mu),\nom
\ene
where $\epsilon(\la,\mu)\to 0$ as $\mu\to\la$.

We now recall the well-known relation for Poisson integral:
\beq
\frac{1}{\pi}\int_a^b\frac{\ep}{(\la-\mu)^2+\ep^2}h(\la)d\la\to \left\{
\begin{array}{ll}
0& \q (\mu<a\ \mbox{or}\ \mu>b)\\
h(\mu)& \q (a<\mu<b)
\end{array}
\right.\nom
\ene
as $\ep\downarrow 0$ for a continuous function $h(\la)$. We apply this relation with setting
\beq
h(\la)=\left(\frac{dE_0}{d\la}(\la)f,g\right), \q (f,g\in L^2_s,\q s>1/2).\nom
\ene
Then we have for $0<a<\mu<b<\infty$
\beq
h(\mu)&=&\left(\frac{dE_0}{d\la}(\mu)f,g\right)\\
&=&\lim_{\ep\downarrow 0}
\frac{1}{\pi}\int_a^b\frac{\ep}{(\la-\mu)^2+\ep^2}\left(\frac{dE_0}{d\la}(\la)f,g\right)d\la\nonumber\\
&=&\lim_{\ep\downarrow 0}
\frac{1}{2\pi i}\int_a^b\left(\frac{1}{\la-\mu-i\ep}-\frac{1}{\la-\mu+i\ep}\right)\left(\frac{dE_0}{d\la}(\la)f,g\right)d\la.\nonumber
\ene
By \eq{functional-calc}, the RHS is equal to
\beq
&&\lim_{\ep\downarrow 0}
\frac{1}{2\pi i}
\int_{R^1}
\left(\frac{1}{\la-\mu-i\ep}-\frac{1}{\la-\mu+i\ep}\right)d(E_0(\la)E_0(B)f,g)\\
&&=
\lim_{\ep\downarrow 0}
\frac{1}{2\pi i}
((R_0(z)-R_0(\bar{z}))E_0(B)f,g),\nonumber
\ene
where $B=(a,b)$, $z=\mu+i\ep$, and
\beq
R_0(z)=(H_0-z)^{-1}
\ene
is a resolvent of $H_0$. Thus we have an expression of $\displaystyle{\frac{dE_0}{d\la}(\mu)}$:
\beq
\left(\frac{dE_0}{d\la}(\mu)f,g\right)=\lim_{\ep\downarrow 0}
\frac{1}{2\pi i}
((R_0(z)-R_0(\bar{z}))E_0(B)f,g)\nom
\ene
for $\mu\in B=(a,b)$ $(0<a<b<\infty)$ and $f,g\in L^2_s$ $(s>1/2)$.

We consider termwise boundary values of $R_0(z)=R_0(\mu\pm i\ep)$ as $\ep\downarrow 0$. To do so, we write for $f,g \in \SSS$ using Fourier transform
\beq
(R_0(\mu+i\ep)f,g)&=&\int_{R^m}\frac{1}{\xi^2/2-(\mu+i\ep)}{\hat f}(\xi)\overline{{\hat g}(\xi)}d\xi \label{resolv}\\
&=& \int_0^\infty 
(2\la)^{(m-2)/2}
\frac{1}{(\la-\mu)-i\ep} 
\int_{S^{m-1}}{\hat f}
(\sqrt{2\la}\omega)
\overline{{\hat g}
(\sqrt{2\la}\omega)}d\omega d\la,\nonumber
\ene
where $\hat f$ denotes the Fourier transform of $f$. Setting
\beq
h(\la)=(2\la)^{(m-2)/2}\int_{S^{m-1}}
{\hat f}(\sqrt{2\la}\omega)\overline{{\hat g}(\sqrt{2\la}\omega)}d\omega=(\FF(\la)f,\FF(\la)g)_{L^2(S^{m-1})},\label{h}
\ene
we rewrite the RHS of \eq{resolv} as
\beq
\int_{-\mu}^\infty\frac{\la}{\la^2+\ep^2}h(\la+\mu)d\la+\pi i\frac{1}{\pi}\int_0^\infty\frac{\ep}{(\la-\mu)^2+\ep^2}h(\la)d\la.\label{resolvent}
\ene
The second integral on the RHS is a Poisson integral, thus the limit as $\ep\downarrow 0$ exists and satisfies the estimate by \eq{Fest}:
\beq
\left|\lim_{\ep\downarrow 0}\frac{1}{\pi}\int_0^\infty\frac{\ep}{(\la-\mu)^2+\ep^2}h(\la)d\la\right|
\le C_{s\mu}\Vert f\Vert_{L^2_{s}} \Vert g\Vert_{L^2_s}\label{Poiss}
\ene
for $s>1/2$, where $C_{s\mu}>0$ is bounded locally uniformly with respect to $\mu>0$.
The first term on the RHS of \eq{resolvent} is written for $\delta>0$ as
\beq
&&\int_{-\mu}^\infty\frac{\la}{\la^2+\ep^2}h(\la+\mu)d\la\label{dec}\\
&&=\left(\int_{-\mu}^{-\delta}+\int_\delta^\infty\right)\frac{\la}{\la^2+\ep^2}h(\la+\mu)d\la+\int_{-\delta}^\delta\frac{\la}{\la^2+\ep^2}h(\la+\mu)d\la.\nonumber
\ene
The limit as $\ep\downarrow0$ of the first term is equal to
\beq
\left(\int_{-\mu}^{-\delta}+\int_\delta^\infty\right)\frac{1}{\la}h(\la+\mu)d\la
=\int_{G_\delta}\frac{1}{|\xi|^2/2-\mu}{\hat f}(\xi)\overline{{\hat g}(\xi)}d\xi,\nom
\ene
where $G_\delta$ is the sum of the regions $|\xi|^2/2\le \mu-\delta$ and $|\xi|^2/2\ge \mu+\delta$.
This is bounded by
\beq
\frac{1}{\delta}\Vert f\Vert_{L^2}\Vert g\Vert_{L^2}.\label{first}
\ene
The second term on the RHS of \eq{dec} is equal to
\beq
\int_0^\delta\frac{\la}{\la^2+\ep^2}(h(\mu+\la)-h(\mu-\la))d\la=
\int_0^\delta\frac{\la^{1+\theta}}{\la^2+\ep^2}\ 
\frac{h(\mu+\la)-h(\mu-\la)}{\la^\theta}d\la,\nom
\ene
where $0<\theta<s-1/2$.
Thus by applying \eq{continuity} to the definition \eq{h} of $h(\la)$, we see that the limit as $\ep\downarrow 0$ exists and is bounded by
\beq
\sup_{0<\la<\delta}\frac{|h(\mu+\la)-h(\mu-\la)|}{\la^\theta}\int_0^\delta \la^{\theta-1}d\la
\le C_{s\theta\mu}\Vert f\Vert_{L^2_{s}} \Vert g\Vert_{L^2_{s}}\nom
\ene
for $s>1/2$ and $s-1/2>\theta>0$, where $C_{s\theta\mu}>0$ is bounded when $\mu$ moves in a compact subset of $(0,\infty)$.
Thus the second term on the RHS of \eq{dec} is bounded by
\beq
C_{s\mu}\Vert f\Vert_{L^2_s}\Vert g\Vert_{L^2_s}\nom
\ene
for $s>1/2$. Combining this with \eq{first} and \eq{Poiss}, we obtain the existence of the boundary value $R_0(\mu+i0)f=\lim_{\ep\downarrow 0}R_0(\mu+i\ep)f$ in $L^2_{-s}(R^m)$ and the estimate
\beq
\Vert R_0(\mu+i0)\Vert_{L^2_s\to L^2_{-s}}\le C\nom
\ene
locally uniformly in $\mu>0$, where $s>1/2$. Similarly, the same estimate holds also for $R_0(\mu-i0)$. Reexamining the arguments above, we also see that $R_0(\mu\pm i0)$ is continuous with respect to $\mu>0$ in operator norm from $L^2_s(R^m)$ into $L^2_{-s}(R^m)$ for $s>1/2$.

\newcommand{\wxi}{{\widehat \xi}}

Summarizing, we have proved:

\begin{thm}\label{Spectral-H0}
The resolution $\{E_0(\la)\}$ of the identity for $H_0$ is expressed for $f,g\in L^2_s(R^m)$ $(s>1/2)$ and $\la>0$ as:
\beq
\frac{d}{d\la}(E_0(\la)f,g)&=&\frac{1}{2\pi i}((R_0(\la+i0)-R_0(\la-i0))f,g)\label{spectral-H0}\\
&=&(\FF(\la)f,\FF(\la)g)_{L^2(S^{m-1})}.\nonumber
\ene
Here $\{R_0(\la\pm i0)\}_{\la>0}$ and $\{\FF(\la)\}_{\la>0}$ are continuous families of bounded operators from $L^2_s(R^{m})$ $(s>1/2)$ into $L^2_{-s}(R^m)$ and
 $L^2(S^{m-1})$, respectively, defined by for $f\in L^2_s(R^m)$
\beq
&&R_0(\la \pm i0)f=\lim_{\ep\downarrow 0}R_0(\la\pm i\ep)f,\\
&&\FF(\la)f(\omega)=(2\la)^{(m-2)/4}(\FF f)(\sqrt{2\la}\omega).
\ene
In particular, $R_0(\la\pm i0)$ and $\FF(\la)$ are H\"older continuous of order $\theta$ $(0<\theta<s-1/2)$ locally uniformly with respect to $\la>0$, in the uniform operator topology of $B(L^2_s(R^m),L^2_{-s}(R^m))$ and $B(L^2_s(R^m),L^2(S^{m-1}))$, respectively.
\end{thm}

\section{Spatial asymptotics of the free resolvent}

Since $H_0$ is a selfadjoint operator in $\HH=L^2(R^m)$, $H_0$ generates a unitary group
\beq
U_0(t)=\exp(-itH_0)=e^{-itH_0}\q (t\in R^1)
\ene
such that
\beq
U_0(t)U_0(s)=U_0(t+s)\q (t,s\in R^1).
\ene
If we move to the momentum space ${\widehat \HH}=L^2(R^m_\xi)$ by Fourier transformation, we have for $g\in {\widehat \SSS}=\FF \SSS$ by \eq{Hamiltonian}
\beq
{\widehat U}_0(t)g(\xi):=(\FF U_0(t) \FF^{-1}g)(\xi)=e^{-it|\xi|^2/2}g(\xi).\label{FourierU}
\ene
By Lebesgue's dominated convergence theorem, this implies that ${\widehat U}_0(t)g$ is continuous in $t\in R^1$ as an $L^2(R^m_\xi)$-valued function of $t\in R^1$ for general $g\in \HH$.

For $f\in \SSS$
\beq
e^{-itH_0}f(x)&=&U_0(t)f(x)=(\FF^{-1} {\widehat U}_0(t) \FF f)(x)\label{evolution}\\
&=&(2\pi)^{-m/2}\int_{R^m}e^{ix\xi}e^{-it|\xi|^2/2} (\FF f)(\xi) d\xi\nonumber\\
&=&(2\pi)^{-m}\int_{R^m}e^{ix\xi}e^{-it|\xi|^2/2} \int_{R^m}e^{-i\xi y}f(y)dy d\xi\nonumber
\ene
Let $\chi(\xi)\in {\widehat \SSS}$ such that $0\le \chi(\xi)\le 1$ and $\chi(0)=1$, and set for $\ep>0$
\beq
\chi_\ep(\xi)=\chi(\ep\xi).\nom
\ene
Then $\chi_\ep\in{\widehat \SSS}$ for each $\ep>0$ and $\chi_\ep(\xi)\to 1$ as $\ep\downarrow 0$ for each $\xi\in R^m_\xi$. Thus we can rewrite \eq{evolution} using Fubini's theorem as
\beq
e^{-itH_0}f(x)=(2\pi)^{-m}\lim_{\ep\downarrow 0}\int_{R^m}\int_{R^m}e^{i(x\xi-t|\xi|^2/2-\xi y)}\chi_\ep(\xi)f(y)dy d\xi.\label{Os0}
\ene
Obviously this limit does not depend on the choice of $\chi\in {\widehat \SSS}$. We call this type of the limits of integrals oscillatory integrals, and write
\beq
e^{-itH_0}f(x)=(2\pi)^{-m}\mbox{Os-}\int\int_{R^{2m}}e^{i(x\xi-t|\xi|^2/2-\xi y)}f(y)dy d\xi.\label{Os}
\ene
For the simplicity of notation, we introduce a variable
\beq
{\wxi}=((2\pi)^{-1}\xi_1,\cdots,(2\pi)^{-1}\xi_m).\label{wxi}
\ene
Then \eq{Os} can be written as
\beq
e^{-itH_0}f(x)=\mbox{Os-}\int\int_{R^{2m}}e^{i(x\xi-t|\xi|^2/2-\xi y)}f(y)dy d\wxi.\label{Os1}
\ene
We often drop the integration region $R^{2m}$, when it is obvious from the context, and write
\beq
e^{-itH_0}f(x)=\mbox{Os-}\int \int e^{i(x\xi-t|\xi|^2/2-\xi y)}f(y)dy d\wxi.\label{Os2}
\ene

We introduce some notations. We call $\alpha=(\alpha_1,\cdots,\alpha_m)$ with the components $\alpha_j$ being nonnegative integers a multi-index. Then we define
\beq
&&D_x^\alpha=D_{x_1}^{\alpha_1}\cdots D_{x_m}^{\alpha_m},\q
x^\alpha=x_1^{\alpha_1}\cdots x_m^{\alpha_m},\nonumber\\
&&|\alpha|=\alpha_1+\cdots+\alpha_m,\\
&&\langle D_x\rangle=(1+D_x^2)^{1/2}=(1-\Delta_x)^{1/2}.\nonumber
\ene
Noting the relation
\beq
D_y^\alpha (e^{-i\xi y})=(-1)^{|\alpha|}\xi^\alpha (e^{-i\xi y}),\label{y-partial}
\ene
and integrating by parts inside the integral \eq{Os0} with respect to $y$, we obtain for $f\in\SSS$
\beq
e^{-itH_0}f(x)&=&(2\pi)^{-m}\lim_{\ep\downarrow 0}\int_{R^m}\int_{R^m}e^{i(x\xi-t|\xi|^2/2-\xi y)}\chi_\ep(\xi)f(y)dy d\xi\nom\\
&=&\lim_{\ep\downarrow 0}\int\int e^{i(x\xi-t|\xi|^2/2-\xi y)}\chi_\ep(\xi)\langle \xi\rangle^{-2m}(\langle D_y\rangle^{2m}f)(y)dy d\wxi
\nonumber\\
&=&\int\int e^{i(x\xi-t|\xi|^2/2-\xi y)}
\langle \xi\rangle^{-2m}(\langle D_y\rangle^{2m}f)(y)dy d\wxi.
\nonumber
\ene
If $f$ does not belong to $\SSS$ but just satisfies, e.g. the conditions
\beq
\sup_{y\in R^m}|D_y^\alpha f(y)|<\infty\nom
\ene
for all multi-indices $\alpha$, we then 
define $e^{-itH_0}f$ by
\beq
e^{-itH_0}f(x)&=&(2\pi)^{-m}\lim_{\ep\downarrow 0}\int_{R^m}\int_{R^m}e^{i(x\xi-t|\xi|^2/2-\xi y)}\chi_\ep(\xi)\chi_\ep(y)f(y)dy d\xi,\nom
\ene
and integrate by parts with respect to $y$ and $\xi$ using the relations \eq{y-partial} and
\beq
D_\xi^\alpha (e^{-i\xi y})=(-1)^{|\alpha|}y^\alpha (e^{-i\xi y}).\label{xi-partial}
\ene
Then we obtain
\beq
e^{-itH_0}f(x)=\lim_{\ep\downarrow 0} \int\int e^{-i\xi y}\langle D_\xi\rangle^{2m}(e^{i(x\xi-t|\xi|^2/2)}\langle \xi\rangle^{-4m}\chi_\ep(\xi))
\langle y\rangle^{-2m} \langle D_y\rangle^{4m}(\chi_\ep(y) f(y))dy d\wxi.\nonumber
\ene
Noting that $\chi_\ep(\xi)$ satisfies
\beq
|D_\xi^\alpha (\chi_\ep(\xi))|=|\ep^{|\alpha|}(D_\xi^\alpha\chi)(\ep\xi)|\le 
C_{\alpha} \ep^{|\alpha|}\nom
\ene
for any multi-index $\alpha$, we see that the equality
\beq
e^{-itH_0}f(x)=\int\int e^{-i\xi y}\langle D_\xi\rangle^{2m}(e^{i(x\xi-t|\xi|^2/2)}\langle \xi\rangle^{-4m})
\langle y\rangle^{-2m}(\langle D_y\rangle^{4m}f)(y)dy d\wxi\label{partial-integrated-form}
\ene
holds. With some smoothness assumptions on the integrands, we always have in this way an expression of oscillatory integrals that does not contain the damping factors like $\chi_\ep(\xi)$ or $\chi_\ep(y)$. We will use these techniques in the next section in considering the behavior of $e^{-itH_0}$.

By \eq{functional-calc}, we have the relation for $\mu\in R^1$ and $\ep\ne 0$
\beq
(R_0(\mu\pm i\ep)f,g)=\int_0^\infty (\la-\mu\mp i\ep)^{-1} d(E_0(\la)f,g)
\ene
at least for $f,g\in \SSS$. Noting the relation for $\ep>0$
\beq
(\la-\mu\mp i\ep)^{-1}=i\int_0^{\pm\infty}e^{it(\mu\pm i\ep-\la)}dt,
\ene
and using Fubini's theorem, we can write
\beq
(R_0(\mu\pm i\ep)f,g)&=&i\int_0^{\pm\infty}\int_0^\infty e^{it(\mu\pm i\ep-\la)}
 d(E_0(\la)f,g)dt\nom\\
&=&i\int_0^{\pm\infty} (e^{it(\mu\pm i\ep-H_0)}f,g)dt.\nonumber
\ene
Since $\Vert e^{it(\mu\pm i\ep-H_0)}f\Vert\le e^{-\ep |t|}\Vert f\Vert$ in respective signs, we have from this for $f\in \HH$
\beq
R_0(\mu\pm i\ep)f=i\int_0^{\pm\infty} e^{it(\mu\pm i\ep-H_0)}f dt,\q (\ep>0, \mu\in R^1).\label{resolvent-t}
\ene
Since $e^{itH_0}f$ $(f\in\HH)$ is continuous in $\HH$ with respect to $t\in R^1$ by the remark after \eq{FourierU}, the integral can be understood as a Riemann integral.

Using the second line of \eq{evolution}, we can rewrite \eq{resolvent-t} for $f\in \SSS$:
\beq
R_0(\mu\pm i\ep)f(x)=(2\pi)^{-m/2}i\int_0^{\pm\infty}\int e^{i(x\xi-t(\xi^2-2\mu)/2)}e^{-\ep|t|}{\hat f}(\xi)d\xi dt.\nom
\ene
where ${\hat f}=\FF f$. In what follows, assuming $\mu>0$, we apply stationary phase method to this integral and derive an asymptotic expansion as $r=|x|\to\infty$.

To do so, we assume ${\hat f}\in C_0^\infty(R^m_\xi-\{0\})$ with supp ${\hat f}\subset \{\xi | \ (0<)a\le |\xi|\le b(<\infty)\}$, and in the integral make a change of variables
\beq
x=r\omega, \q r=|x|, \q t=rs, \qq (\omega\in S^{m-1}).\nom
\ene
For the sake of simplicity, we consider the $+$ case only. The $-$ case can be treated similarly. Then we obtain
\beq
-i(2\pi)^{m/2}(R_0(\mu+ i\ep)f)(r\omega)&=&I(r\omega)=I_{\mu\ep}(r\omega)\label{resolv-space}\\
&:=& r\int_0^\infty\int_{R^m}e^{ir(\omega\xi-s(\xi^2-2\mu)/2)}e^{-\ep rs}
{\hat f}(\xi)d\xi ds.\nonumber
\ene
We set
\beq
\phi=\phi(\mu,\omega;s,\xi)=\omega\xi-s(\xi^2-2\mu)/2.\nom
\ene
Then
\beq
\partial_\xi \phi=\omega-s\xi,\q \partial_s\phi=-\xi^2/2+\mu,\nom
\ene
where $\partial_\xi=(\partial/\partial \xi_1,\cdots,\partial/\partial \xi_m)$, etc. The solution of $\partial_\xi\phi=0$ and $\partial_s\phi=0$ are given by
\beq
\xi=\xi_c:=\sqrt{2\mu}\omega,\q s=s_c:=\frac{1}{\sqrt{2\mu}}.\nom
\ene
We first divide the integral $I(r\omega)$ as a sum of the integral near $s=0$ and the one away from $s=0$. Let $\varphi(s)\in C_0^\infty(R^1_s)$ with supp $\varphi\subset \{s|\ |s|<\frac{1}{2}\min(s_c,\frac{1}{2b})\}$ and $\varphi(s)=1$ for $|s|\le \frac{1}{4}\min(s_c,\frac{1}{2b})$, and consider
\beq
I_0(r\omega)=r\int_0^\infty\int e^{ir\phi}e^{-\ep rs}{\hat f}(\xi)\varphi(s)d\xi ds.\nom
\ene
Noting that on supp $\varphi$ and supp ${\hat f}$
\beq
|\partial_\xi\phi|=|\omega-s\xi|\ge 1-\frac{b}{2b}=\frac{1}{2}>0\nom
\ene
and using the relation
\beq
r^{-\ell}(|\partial_\xi\phi|^{-2}i^{-1}\partial_\xi\phi\cdot \partial_\xi)^\ell e^{ir\phi}=e^{ir\phi},\nom
\ene
we integrate by parts with respect to $\xi$ inside the integral $I_0(r\omega)$. Then we obtain
\beq
|I_0(r\omega)|\le C_\ell r^{1-\ell}\nom
\ene
for all $\ell=1,2,\cdots$ with the constant $C_\ell>0$ depending on $\ell$ but not on $\ep>0$. Thus in the limit as $r\to\infty$, we have only to consider the integral
\beq
I_\infty(r\omega)=r\int_0^\infty\int e^{ir\phi}e^{-\ep rs}{\hat f}(\xi)(1-\varphi)(s)d\xi ds.\nom
\ene

We next take a function $\chi_\delta\in C_0^\infty(R^{m+1})$ and $\chi_{\delta r}\in C_0^\infty(R^{m+1})$ for $\delta>0$, $r>0$ and $1>\theta>0$ such that
\beq
\chi_\delta(s,\xi)=\left\{
\begin{array}{ll}
1& \q(|(s,\xi)|\le \delta)\\
0& \q(|(s,\xi)|\ge 2\delta)
\end{array}
\right.\nom
\ene
\beq
\chi_{\delta r}(s,\xi)=
\chi_\delta(r^\theta((s,\xi)-(s_c,\xi_c)))
\nom
\ene
and divide the integral $I_\infty(r\omega)$ as
\beq
I_\infty(r\omega)=I_1(r\omega)+I_2(r\omega),\nom
\ene
where
\beq
&&I_1(r\omega)=r\int_0^\infty\int e^{ir\phi}e^{-\ep rs}{\hat f}(\xi)\chi_{\delta r}(s,\xi)(1-\varphi)(s)d\xi ds,\label{I1}\\
&&I_2(r\omega)=r\int_0^\infty\int e^{ir\phi}e^{-\ep rs}{\hat f}(\xi)(1-\chi_{\delta r})(s,\xi)(1-\varphi)(s)d\xi ds.
\ene
On supp $(1-\chi_{\delta r})(s,\xi)$, we have by an argument using the definition of $(s_c,\xi_c)$
\beq
|\partial_\xi\phi|+|\partial_s\phi|\ge \rho r^{-2\theta}\nom
\ene
for some $\rho=\rho_\delta>0$. Thus if we define a differential operator $P$ whose transposed operator is
\beq
{}^tP=i^{-1}(|\partial_\xi\phi|^2+|\partial_s\phi|^2)^{-1}(\partial_{(s,\xi)}\phi\cdot \partial_{(s,\xi)}),\nom
\ene
then we have
\beq
r^{-\ell}({}^tP)^\ell e^{ir\phi}=e^{ir\phi}.\nom
\ene
Using this relation, we integrate by parts in the integral $I_2(r\omega)$ and obtain
\beq
|I_2(r\omega)|\le C_\ell r^{1-\ell(1-2\theta)}\nom
\ene
for any $\ell=1,2,\cdots$ and $0<\theta<1/2$ uniformly in $\ep>0$. Here to assure the integrability with respect to $s$ in $I_2(r\omega)$, we have used the estimates
\beq
(|\partial_\xi\phi|+|\partial_s\phi|)^{-1}\le |\partial_\xi\phi|^{-1}=|\omega-s\xi|^{-1}\le C|s|^{-1}\nonumber
\ene
which holds for large $s>1$ on supp ${\hat f}(1-\varphi)\subset \{(s,\xi)|\ (0<)a\le |\xi|\le b(<\infty),\  s\ge \frac{1}{4}\min(s_c,\frac{1}{2b})\}$. When $s>0$ is small we can use the bound $r^{2\theta}$ as we have obtained above. Thus we can neglect $I_2(r\omega)$ in the limit $r\to\infty$.

We now evaluate $I_1(r\omega)$. We note that the integration region of the integral $I_1(r\omega)$ is included in the compact set supp $\chi_{\delta r}(s,\xi)$ of $R^{m+1}$, thus the limit as $\ep\downarrow 0$ exists and we can drop the factor $e^{-\ep rs}$ from $I_1(r\omega)$:
\beq
I_1(r\omega)=r\int_0^\infty\int e^{ir\phi}{\hat f}(\xi)\chi_{\delta r}(s,\xi)(1-\varphi)(s)d\xi ds.\label{In1}
\ene
To estimate this, we make a Taylor expansion of $\phi=\phi(s,\xi)=\phi(\mu,\omega;s,\xi)$ around $(s_c,\xi_c)$:
\beq
\phi(s,\xi)=\phi(s_c,\xi_c)+\partial_{(s,\xi)}\phi(s_c,\xi_c)
\left(\begin{array}{c}{\tilde s}\\ {\tilde \xi}\end{array}\right)
+\frac{1}{2}
\left\langle {J}(s,\xi)
\left(\begin{array}{c}{\tilde s}\\ {\tilde \xi}\end{array}\right),
\left(\begin{array}{c}{\tilde s}\\ {\tilde \xi}\end{array}\right)
\right\rangle,\label{Taylor}
\ene
where ${\tilde s}=s-s_c, {\tilde \xi}=\xi-\xi_c$ and $\langle X,Y\rangle=\sum_{j=1}^{m+1} X_j Y_j$ is a scalar product of $X,Y\in R^{m+1}$. By the definition of $(s_c,\xi_c)$, the second term vanishes.
The Hessian matrix ${J}(s,\xi)$ is given by
\beq
J(s,\xi)
=
\left(
\begin{array}{cc}
\partial_s^2\phi(s,\xi)& \partial_s\partial_\xi\phi(s,\xi)\\
\partial_\xi\partial_s\phi(s,\xi)& \partial_\xi^2\phi(s,\xi)
\end{array}
\right)
=
\left(
\begin{array}{cc}
0& -\xi\\
-{}^t\xi& -s I_m
\end{array}
\right)
\ene
with $I_m$ being the unit matrix of order $m$. Since $s_c=\frac{1}{\sqrt{2\mu}}>0$ and $\xi_c=\sqrt{2\mu}\omega\ne 0$, the matrix ${J}(s,\xi)$ is non-singular on supp $\chi_{\delta r}$ when $r>1$ is large enough. Further since ${J}(s,\xi)$ is real symmetric, we can take an orthogonal matrix $P=P(s,\xi)$ such that
${}^tPJ(s,\xi)P$ is a non-singular diagonal matrix:
\beq
A=A(s,\xi):={}^tPJ(s,\xi)P
=
\left(
\begin{array}{ccccc}
\frac{-s+\sqrt{s^2+4\xi^2}}{2}& 0& 0& \cdots& 0\\
0&\frac{-s-\sqrt{s^2+4\xi^2}}{2}& 0& \cdots& 0\\
0& 0& -s& \cdots& 0\\
\cdots& \cdots& \cdots& \cdots& \cdots\\
0& 0& 0& \cdots& -s
\end{array}
\right).
\ene
Since $A$ is diagonal, there is a diagonal matrix $Q=Q(s,\xi)$ such that
\beq
{}^tQAQ=
\left(
\begin{array}{ccccc}
1& 0& 0& \cdots& 0\\
0&        -1& 0& \cdots& 0\\
0& 0& -1& \cdots& 0\\
\cdots& \cdots& \cdots& \cdots& \cdots\\
0& 0& 0& \cdots& -1
\end{array}
\right)=:{\cal E}.
\ene
Then since
\beq
{}^tQ{}^tPJ(s,\xi)PQ={\cal E},\nom
\ene
we have
\beq
|\det (PQ)|=|\det J(s,\xi)|^{-1/2}.\nom
\ene
In partcular if we put
$$
P_c=P(s_c,\xi_c),\quad Q_c=Q(s_c,\xi_c),
$$
we have
\beq
{}^tQ_c{}^tP_cJ(s_c,\xi_c)P_cQ_c={\cal E},\quad |\det (P_cQ_c)|=|\det J(s_c,\xi_c)|^{-1/2}.\nom
\ene
We set
\beq
(s,\xi)(r,\omega)=(s_c,\xi_c)(r,\omega)+\frac{1}{\sqrt{r}}P_cQ_cy,\q y\in R^{m+1}.
\label{change}
\ene
If we set
\beq
{\widetilde{\cal E}}={}^tQ_c{}^tP_cJ(s,\xi)P_cQ_c,\nom
\ene
we have
\beq
|{\widetilde{\cal E}}-{\cal E}|\le C|J(s,\xi)-J(s_c,\xi_c)|\le C(|s-s_c|+|\xi-\xi_c|)\le Cr^{-\theta}\label{wideE}
\ene
on supp $\chi_{\delta r}$, and we have
\beq
\phi(s,\xi)=\phi(s_c,\xi_c)+r^{-1}\frac{1}{2}\langle {\widetilde{\cal E}}y,y\rangle.\nom
\ene
Making a change of variables \eq{change} and inserting this relation into the definition \eq{In1} of $I_1(r\omega)$, we obtain
\beq
&&I_1(r\omega)
=r^{(1-m)/2} e^{ir\phi(s_c,\xi_c)}|\det J(s_c,\xi_c)|^{-1/2} \int_{R^{m+1}}
e^{i\frac{1}{2}\left\langle \widetilde{\cal E}y,y\right\rangle} u(r,y)dy,\nonumber
\ene
where
$$
u(r,y)={\hat f}(\xi(r,y))\chi_{\delta r}((s,\xi)(r,y))(1-\varphi)(s(r,y)).
$$
Noting \eq{change}, we have for any multi-index $\alpha$
\beq
|\partial_y^\alpha u(r,y)|\le C_\alpha \frac{1}{r^{|\alpha|(1/2-\theta)}},\label{estimate}
\ene
where $C_\alpha>0$ is a constant independent of $r,y$.
We now consider
\beq
J(r\omega)=\int_{R^{m+1}}e^{\frac{i}{2}\langle \widetilde{\cal E}y,y\rangle}u(r,y)dy,\nom\\
K(r\omega)=\int_{R^{m+1}}e^{\frac{i}{2}\langle {\cal E}y,y\rangle}u(r,y)dy.\nom
\ene
The difference between the two is
\beq
|J(r\omega)-K(r\omega)|&\le&\int_{R^{m+1}}|e^{\frac{i}{2}\langle \widetilde{\cal E}y,y\rangle}-e^{\frac{i}{2}\langle {\cal E}y,y\rangle}||u(r,y)|dy\nom\\
&\le&C_\kappa\int_{|y|\le Cr^{1/2-\theta}}|\langle(\widetilde{\cal E}-{\cal E})y,y\rangle|^\kappa dy\label{sa}
\ene
for an arbitrary $1\ge \kappa>0$, where we have used that on supp $\chi_{\delta r}$, $|(s,\xi)-(s_c,\xi_c)|=|P_cQ_cy/\sqrt{r}|\le Cr^{-\theta}$. Using \eq{wideE}, we have that
\eq{sa} is bounded by
\beq
&\le& C_\kappa r^{(1/2-\theta)(m+1)}r^{-\kappa\theta}r^{2(1/2-\theta)\kappa}\nom\\
&=& r^{(1/2-\theta)(m+1)-\kappa(3\theta-1)}.\label{bound1}
\ene
Thus if $1/3<\theta<1/2$ is close to 1/2, we can take $\kappa$ as follows:
\beq
1\ge \kappa>\frac{1/2-\theta}{3\theta-1}(m+1)(>0),\nom
\ene
and we have the negative power on the RHS of \eq{bound1}, thus
\beq
|J(r\omega)-K(r\omega)|\to 0\quad\mbox{  as  }r\to\infty.\nom
\ene
Thus we have only to consider $K(r\omega)$ instead of $J(r\omega)$ when considering the asymptotic behavior of $I_1(r\omega)$.

Taking the Fourier transform of the two factors in the integrand of $K(r\omega)$, we have by Plancherel formula
\beq
K(r\omega)&=&|\det {\cal E}|^{-1/2}e^{\pi i \mbox{\scriptsize{sgn}}({\cal E})/4}\int_{R^{m+1}}e^{-\frac{i}{2}\langle {\cal E}^{-1}\eta,\eta\rangle}{\hat u}(r,\eta)d\eta\label{J}\\
&=&e^{-(m-1)\pi i/4}\int_{R^{m+1}}e^{-\frac{i}{2}\langle {\cal E}\eta,\eta\rangle}{\hat u}(r,\eta)d\eta,\nom
\ene
where sgn$({\cal E})=1-m$ is the signature of ${\cal E}$ and ${\hat u}(r,\eta)$ is the Fourier transform of $u(r,y)$ with respect to $y$. Taking the Taylor expansion of the exponential function in the integrand, we have
\beq
\left|e^{-\frac{i}{2}\langle {\cal E}\eta,\eta\rangle}-\sum_{j=0}^{\nu-1}\frac{1}{j!}(-i\langle {\cal E}\eta,\eta\rangle/2)^j\right|\le \frac{1}{\nu !}|\langle {\cal E}\eta,\eta\rangle/2|^\nu.
\ene
Inserting this into \eq{J}, we obtain
\beq
\left|K(r\omega)-(2\pi)^{(m+1)/2}e^{-(m-1)\pi i/4}\sum_{|\alpha|<2\nu}c_\alpha D_y^\alpha u(r,0)\right|
&\le& C_\nu \sum_{|\beta|=2\nu}\left|\int \eta^\beta {\hat u}(r,\eta)d\eta\right| \label{Jexpan}\\
&\le & C'_\nu \sum_{2\nu\le|\beta|\le 2\nu+m+2}\int |D_y^\beta {u}(r,y)|dy,\nom
\ene
where
\beq
c_\alpha=\frac{1}{\alpha !}\left.\partial_\eta^{\alpha} (e^{-\frac{i}{2}\langle {\cal E}\eta,\eta\rangle})\right|_{\eta=0}
\ene
vanishes for odd $|\alpha|$.
Since the support of $u(r,y)$ with respect to $y$ is included in the ball of radius $cr^{1/2-\theta}$ with center $0$ in $R^{m+1}$, we obtain the bound of the RHS of \eq{Jexpan} from \eq{estimate}:
\beq
C r^{((m+1)-2\nu)(1/2-\theta)}.
\ene
Taking $\nu>(m+1)/2$, we have an expansion formula for $I_1(r\omega)$. In particular, as the first approximation we get
 for large $r>1$
\beq
\left|I_1(r\omega)-(2\pi)^{(m+1)/2}r^{(1-m)/2}e^{-(m-1)\pi i/4}e^{ir\phi(s_c,\xi_c)}|\det J(s_c,\xi_c)|^{-1/2}{\hat f}(\xi_c)\right|
= o(r^{-(m-1)/2}).\nonumber
\ene
Noting that
\beq
\phi(s_c,\xi_c)=\sqrt{2\mu},\q |\det J(s_c,\xi_c)|=(2\mu)^{-(m-3)/2},\nom
\ene
we obtain returning to \eq{resolv-space}
\beq
&&R_0(\mu+ i0)f(r\omega)\nom\\
&&=\sqrt{2\pi} e^{-(m-3)\pi i/4}(2\mu)^{(m-3)/4}e^{i\sqrt{2\mu}r}r^{-(m-1)/2}(\FF f)(\sqrt{2\mu}\omega)+ o(r^{-(m-1)/2})\nonumber
\ene
as $r\to\infty$ for ${\hat f}\in C_0^\infty(R^m-\{0\})$.
Similarly we can show
\beq
&&R_0(\mu- i0)f(r\omega)\nom\\
&&=\sqrt{2\pi} e^{(m-3)\pi i/4}(2\mu)^{(m-3)/4}e^{-i\sqrt{2\mu}r}r^{-(m-1)/2}(\FF f)(-\sqrt{2\mu}\omega)+ o(r^{-(m-1)/2})\nonumber
\ene
as $r\to\infty$.

Reversing the order of expression we obtain a relation between the Fourier transform $\FF(\mu)$ defined by \eq{Fourier-la} and the spatial asymptotics of the free resolvent $R_0(\mu\pm i0)$:

\begin{thm}\label{Fourier-H0}
For $\FF f\in C_0^\infty(R^m-\{0\})$ and $\mu>0$, $\omega\in S^{m-1}$, one has
\beq
\FF(\mu)f(\pm\omega)=(2\pi)^{-1/2}e^{\pm(m-3)\pi i/4}(2\mu)^{1/4}\lim_{r\to\infty}r^{(m-1)/2}e^{\mp i\sqrt{2\mu}r}(R_0(\mu\pm i0)f)(r\omega).
\ene
\end{thm}

\section{Propagation estimates for the free evolution}\label{propagationest}

In this section we consider some estimates of $e^{-itH_0}$ which are stronger version of Theorem \ref{Enss} in the free Hamiltonian case. For this purpose we introduce the notion of pseudo-differential operator (we call this a $\psi$do in the following). $P$ is called a $\psi$do with symbol $p(x,\xi)$, if it is written
for $f\in\SSS$
\beq
Pf(x)=\mbox{Os-}\int_{R^m}\int_{R^m} e^{i(x-y)\xi}p(x,\xi)f(y)dyd\wxi,\label{pseudo}
\ene
where $\wxi$ is the variable defined by \eq{wxi}. We call $p(x,\xi)$ the symbol of the $\psi$do $P$ and write $p(x,\xi)=\sigma(P)(x,\xi)$. For \eq{pseudo} to be well-defined as an oscillatory integral, we need to assume some smoothness conditions on the symbol $p(x,\xi)$ of $P$. E.g. let us assume that $p(x,\xi)$ is $C^\infty$ with respect to $(x,\xi)$ and satisfies the estimates
\beq
\sup_{(x,\xi)\in R^{2m}}|\partial_x^\alpha\partial_\xi^\beta p(x,\xi)|<\infty
\ene
for all multi-indices $\alpha$ and $\beta$. Let as before $\chi\in\SSS(R^m)$ such that $\chi(0)=1$, and set $\chi_\ep(\xi)=\chi(\ep\xi)$, and define \eq{pseudo} by
\beq
Pf(x)=\lim_{\ep\downarrow 0}\int\int e^{i(x-y)\xi}p(x,\xi)\chi_\ep(\xi)f(y)dy d\wxi.\label{pseudo2}
\ene
Then using the relation
\beq
(1+D_y^2) e^{-iy\xi}=(1+|\xi|^2)e^{-iy\xi},\nonumber
\ene
we integrate by parts inside the integral \eq{pseudo2}:
\beq
Pf(x)&=&\lim_{\ep\downarrow 0}\int\int
e^{i(x-y)\xi}p(x,\xi)\chi_\ep(\xi)(1+|\xi|^2)^{-m}(1+D_y^2)^{m} f(y) dyd\wxi\nom\\
&=& 
\int \int
e^{i(x-y)\xi}p(x,\xi)(1+|\xi|^2)^{-m}(1+D_y^2)^{m} f(y) dyd\wxi.\nonumber
\ene
Thus $Pf$ is well-defined for $f\in\SSS$ as an oscillatory integral independently of the choice of the damping factor $\chi_\ep$. We write $Pf$ as $Pf=p(X,D_x)f$ to indicate the symbol $p(x,\xi)$ used in the definition \eq{pseudo}.

As other forms of the definition of $\psi$do, we can adopt
\beq
Qf(x)=q(D_x,X')f(x)=\mbox{Os-}\int\int e^{i(x-y)\xi}q(\xi,y)f(y)dy d\wxi,\nom
\ene
or
\beq
Pf(x)=p(X,D_x,X')f(x)=\mbox{Os-}\int\int e^{i(x-y)\xi}p(x,\xi,y)f(y)dy d\wxi,\nom
\ene
where the symbols $q(\xi,y)$ and $p(x,\xi,y)$ satisfy
\beq
&&\sup_{(\xi,y)\in R^{2m}}|\partial_\xi^\alpha\partial_y^\beta q(\xi,y)|<\infty,\nom\\
&&\sup_{(x,\xi,y)\in R^{3m}}|\partial_x^\alpha\partial_\xi^\beta\partial_y^\gamma p(x,\xi,y)|<\infty\nom
\ene
for all multi-indices $\alpha,\beta,\gamma$.
The relation between these expressions is given by

\newcommand{\weta}{{\widehat \eta}}

\begin{pro}\label{pseudo-pro}
Let $p(x,\xi,y)$ be as above. Then $Pf=p(X,D_x,X')f$ is written as
\beq
Pf(x)&=&p_L(X,D_x)f(x)=\mbox{\rm Os-}\int\int e^{i(x-y)\xi}p_L(x,\xi)f(y)dyd\wxi\\
&=&p_R(D_x,X')f(x)=\mbox{\rm Os-}\int\int e^{i(x-y)\xi}p_R(\xi,y)f(y)dyd\wxi,
\ene
where $p_L(x,\xi)$ and $p_R(\xi,y)$ are defined by
\beq
&&p_L(x,\xi)=\mbox{\rm Os-}\int\int e^{-iy\eta}p(x,\xi+\eta,x+y) dy d\weta,\\
&&p_R(\xi,y)=\mbox{\rm Os-}\int\int e^{iz\eta}p(y+z,\xi+\eta,z) dz d\weta.
\ene
\end{pro}
Proof is easily done by using Fourier transformation, and is left to the reader.
\BP

It is easy to see that for $P=p(X,D_x,X')$, the adjoint operator $P^*$ is given by
\beq
P^*f(x)=\mbox{Os-}\int\int e^{i(x-y)\xi}\overline{p(y,\xi,x)}f(y)dy d\wxi.\nom
\ene
Thus if we consider the operator $P^*P$, we have the integral operator
\beq
P^*Pf(x)=\mbox{Os-}\int\int e^{i(x-y)\xi} r(x,\xi,y)f(y)dy d\wxi,\nom
\ene
where
\beq
r(x,\xi,y)=\mbox{Os-}\int\int e^{-iz\eta}\overline{p(x+z,\xi+\eta,x)}p(x+z,\xi,y)dz d\weta.\label{prod}
\ene
If we define the semi-norms for the symbol $p(x,\xi,y)$ by
\beq
|p|_\ell=\sup_{|\alpha|+|\beta|+|\gamma|\le \ell}\sup_{x,\xi,y\in R^m}
|\partial_x^\alpha\partial_\xi^\beta\partial_y^\gamma p(x,\xi,y)|\nom
\ene
for $\ell=0,1,2,\cdots$, then we have for $r$ above
\beq
|r|_\ell\le C_\ell^2|p|_{\ell'}^2\label{2-prod}
\ene
for some constant $C_\ell>0$, where
\beq
\ell'=\ell+2m_0,\q m_0=2[m/2+1].\nom
\ene
Here for a real number $s$, $[s]$ denotes the greatest integer that does not exceed $s$. This is seen by integrating by parts in \eq{prod} by using the relations
\beq
(1+D_z^2)e^{-iz\eta}=(1+|\eta|^2)e^{-iz\eta},\q
(1+D_\eta^2)e^{-iz\eta}=(1+|z|^2)e^{-iz\eta},\label{diff-rel}
\ene
and noting that
\beq
\int_{R^m}\int_{R^m} (1+|\eta|^2)^{-[m/2+1]}(1+|z|^2)^{-[m/2+1]} dz d\eta<\infty.\label{integrability}
\ene

For the product of $\nu+1$ $(\nu\ge 1)$ $\psi$do's
\beq
P_jf(x)=\mbox{Os-}\int\int e^{i(x-y)\xi}p_j(x,\xi,y)f(y)dyd\wxi, \q(j=1,2,\cdots,\nu+1),\nom
\ene
we rewrite $P_jf$:
\beq
P_jf(x)=\mbox{Os-}\int\int e^{-iy\xi}p_j(x,\xi,x+y)f(x+y)dyd\wxi\nom
\ene
and calculate the product $Q_{\nu+1}=P_1\cdots P_{\nu+1}$. Then we have
\beq
Q_{\nu+1}f(x)=\mbox{Os-}\int\int e^{i(x-x')\xi}q_{\nu+1}(x,\xi,x')f(x')dx' d\xi.\nom
\ene
Here
\beq
&&q_{\nu+1}(x,\xi,x')\label{q_nu}\\
&&=\mbox{Os-}\overbrace{\int\cdots\int}^{2\nu}
e^{-i\sum_{j=1}^\nu y^j\eta^j}
\prod_{j=1}^\nu p_j(x+{\bar{y}}^{j-1},\xi+\eta^j,x+{\bar y}^j)p_{\nu+1}(x+{\bar y}^\nu,\xi,x')d\mbox{\boldmath $y$}^\nu d{\widehat{\mbox{\boldmath $\eta$}}}^\nu,
\nonumber
\ene
where
\beq
&&{\bar y}^0=0,\q {\bar y}^j=y^1+\cdots+y^j\ (j=1,2,\cdots,\nu),\nom\\
&&d\mbox{\boldmath $y$}^\nu=dy^1\cdots dy^\nu,\q 
d{\widehat{\mbox{\boldmath $\eta$}}}^\nu=d\weta^1\cdots d\weta^\nu.\nom
\ene
Using \eq{diff-rel}, we integrate by parts in \eq{q_nu}:
\beq
&&q_{\nu+1}(x,\xi,x')\label{q_nu-2}\\
&&=\mbox{Os-}\overbrace{\int\cdots\int}^{2\nu}
e^{-i\sum_{j=1}^\nu y^j\eta^j}\prod_{\ell=1}^\nu(1+|y^\ell|^{m_0})^{-1}(1+D_{\eta^\ell}^{m_0})\nonumber\\
&&\times\prod_{j=1}^\nu p_j(x+{\bar{y}}^{j-1},\xi+\eta^j,x+{\bar y}^j)p_{\nu+1}(x+{\bar y}^\nu,\xi,x')d\mbox{\boldmath $y$}^\nu d{\widehat{\mbox{\boldmath $\eta$}}}^\nu.\nonumber
\ene
We then make a change of variables
\beq
z^j=y^1+\cdots+y^j \q(j=1,2,\cdots,\nu),\nom
\ene
which is equivalent to
\beq
y^j=z^j-z^{j-1},\q z^0=0.\nom
\ene
Noting
\beq
\sum_{j=1}^\nu y^j\eta^j=\sum_{k=1}^\nu z^k(\eta^k-\eta^{k+1}),\q \eta^{k+1}=0,\nom
\ene
we again integrate by parts in \eq{q_nu-2}:
\beq
&&q_{\nu+1}(x,\xi,x')\label{q_nu-3}\\
&&=\mbox{Os-}\overbrace{\int\cdots\int}^{2\nu}
e^{-i\sum_{k=1}^\nu z^k(\eta^k-\eta^{k+1})}
\prod_{k=1}^\nu (1+|\eta^k-\eta^{k+1}|^{m_0})^{-1}
(1+D_{z_k}^{m_0})\nonumber\\
&&\times\prod_{\ell=1}^\nu(1+|z^\ell-z^{\ell-1}|^{m_0})^{-1}(1+D_{\eta^\ell}^{m_0})\nonumber\\
&&\times\prod_{j=1}^\nu p_j(x+{z}^{j-1},\xi+\eta^j,x+{z}^j)p_{\nu+1}(x+{z}^\nu,\xi,x')d\mbox{\boldmath $z$}^\nu d{\widehat{\mbox{\boldmath $\eta$}}}^\nu.\nonumber
\ene
By \eq{integrability}, we now obtain the estimate:
\beq
|q_{\nu+1}(x,\xi,x')|\le C_0^{\nu+1}\prod_{j=1}^{\nu+1}|p_j|_{3m_0}\nom
\ene
for some constant $C_0>0$ independent of $\nu$. Differentiating the both sides of \eq{q_nu} and estimating similarly, we have with $\ell_j$ being 3-dimensional mulit-indices
\beq
|q_{\nu+1}(x,\xi,x')|_\ell\le C_0^{\nu+1}\sum_{|\ell_1+\cdots+\ell_{\nu+1}|\le \ell}\prod_{j=1}^{\nu+1}|p_j|_{3m_0+|\ell_j|}.\label{nu-prod}
\ene
Note that the constant $C_\ell$ in \eq{2-prod} depends on $\ell$, while in the present estimate \eq{nu-prod} it is replaced by the sum
$\sum_{|\ell_1+\cdots+\ell_{\nu+1}|\le \ell}$, which enables us to take the  constant $C_0$ independent of $\ell$.

Using this estimate we prove

\begin{thm}\label{L^2-bound}
Let $p(x,\xi,y)$ satisfy $|p|_\ell<\infty$ for all $\ell\le 3m_0$. Then
$P=p(X,D_x,X')$ defines a bounded operator from $\HH=L^2(R^m)$ into itself satisfying
\beq
\Vert P\Vert\le C_0|p|_{3m_0}\label{L2bound}
\ene
for the constant $C_0>0$ in \eq{nu-prod}.
\end{thm}
{\it Proof \footnote[4]{Proof here follows that of \cite{Kumano-go} p.224}:} Let $\chi\in C_0^\infty(B)$, $\chi(0)=1$ and $0\le \chi(x)\le 1$, where $B=\{(x,\xi,x')|\ \max(|x|,|\xi|,|x'|)\le 1\}$, and set for $0<\epsilon<1$
\beq
p_\ep(x,\xi,x')=\chi(\ep x,\ep\xi,\ep x')p(x,\xi,x').\nom
\ene
Set
\beq
K(x,x')=\int_{R^m}e^{i(x-x')\xi}p_\ep(x,\xi,x')d\wxi.\nom
\ene
Then
\beq
P_\ep f(x)=p_\ep(X,D_x,X')f(x)=\int_{R^m} K(x,x')f(x')dx'\nom
\ene
satisfies the estimate for $f\in \SSS$
\beq
\Vert P_\ep f\Vert_{\HH}\le V_\ep^2 |p|_0 \Vert f\Vert_{\HH},\label{ball-est}
\ene
where $V_\ep>0$ is the volume of the ball $|\xi|<\ep^{-1}$ in $R^m_\xi$.

Now set for $\nu=2^\ell$ $(\ell=0,1,2,\cdots)$
\beq
Q_{\ep\nu}=\overbrace{(P_\ep^*P_\ep)\cdots(P_\ep^*P_\ep)}^{\nu}.\nom
\ene
Then by $\sigma(P_\ep^*)(x,\xi,x')=\overline{p_\ep(x',\xi,x)}$ and by \eq{q_nu}, the support of the symbol $\sigma(Q_{\ep\nu})(x,\xi,x')$ of $Q_{\ep\nu}$ is included in $B_\ep=\{(x,\xi,x')|\ \ep(x,\xi,x')\in B\}$. Thus for this $Q_{\ep\nu}$, \eq{ball-est} also holds:
\beq
\Vert Q_{\ep\nu} f\Vert\le V_\ep^2 |\sigma(Q_{\ep\nu})|_0 \Vert f\Vert\q (f\in \SSS).\label{Q-est}
\ene
By $\Vert Q_{\ep\nu}f\Vert^2=(Q_{\ep\nu}^*Q_{\ep\nu}f,f)\le \Vert Q_{\ep\nu}^*Q_{\ep\nu}\Vert\Vert f\Vert^2$, 
we have 
$\Vert Q_{\ep\nu}\Vert^2\le \Vert Q_{\ep\nu}^* Q_{\ep\nu}\Vert =\Vert Q_{\ep(2\nu)}\Vert$. Thus repeating this argument, we have
\beq
\Vert P_\ep\Vert^2\le \Vert P_\ep^*P_\ep\Vert=\Vert Q_{\ep1}\Vert\le 
\Vert Q_{\ep2}\Vert^{1/2}\le \cdots\le
\Vert Q_{\ep(2^\ell)}\Vert^{1/2^{\ell}}\nom
\ene
 for any $\ell=0,1,2,\cdots$. Inserting \eq{Q-est} into this and applying \eq{nu-prod}, we obtain
\beq
\Vert P_\ep\Vert\le (V_\ep^2 |\sigma(Q_{\ep 2^\ell})|_0)^{1/2^{\ell+1}}\le V_\ep^{1/2^{\ell}}C_0|p_\ep|_{3m_0}.\nom
\ene
Letting $\ell\to\infty$ we get
\beq
\Vert P_\ep\Vert\le C_0|p_\ep|_{3m_0}.\nom
\ene
Since, as $\ep\to 0$, $P_\ep f\to Pf$ in $\HH$ for $f\in\SSS$ and $|p_\ep|_{3m_0}\to|p|_{3m_0}$, we obtain the theorem. $\Box$

\BP

With these preparations, we investigate the so-called propagation estimates for $e^{-itH_0}$. Propagation estimates are the estimates with respect to time $t$ of the operator norm of
\beq
P_1 e^{-itH_0} P_2,\nom
\ene
where $P_j$ are the $\psi$do's of the following types:
\beq
&&P_s=\langle x\rangle^{-s}, \q P_sq(D_x)\q (s\ge 0)\label{weight}\\
&&P_{+}f(x)=\mbox{Os-}\int\int e^{i(x-y)\xi}p_+(\xi,y)f(y)dyd\wxi,\label{ps+}\\
&&P_{-}f(x)=\mbox{Os-}\int\int e^{i(x-y)\xi}p_-(x,\xi)f(y)dyd\wxi.\label{ps-}
\ene
Here $q$ and $p_\pm$ are the symbols satisfying for all $\ell=0,1,2,\cdots$
\beq
|q|_\ell+|p_\pm|_\ell<\infty\nom
\ene
and for some $\theta\in (-1,1)$ and $\rho>0$ with $\theta+\rho<1$ and $\theta-\rho>-1$
\beq
&&|\partial_x^\alpha\partial_\xi^\beta p_+(\xi,x)|\le C_{\ell\alpha\beta}\langle x\rangle^{-\ell}\q (\cos(x,\xi):=\frac{x\xi}{|x||\xi|}<\theta+\rho)\label{p+}\\
&&|\partial_x^\alpha\partial_\xi^\beta p_-(x,\xi)|\le C_{\ell\alpha\beta}\langle x\rangle^{-\ell}\q (\cos(x,\xi)>\theta-\rho).\label{p-}
\ene
For a technical reason to avoid the singularities at $x=0$ and $\xi=0$, we also assume for some $\sigma>0$
\beq
&&q(\xi)=0\q(|\xi|<\sigma)\label{energy-cut-off}\\
&&p_+(\xi,x)=0,\q p_-(x,\xi)=0 \q(|x|<\sigma\ \mbox{or}\ |\xi|<\sigma).\label{energy-cut-off-2}
\ene
Of course by Proposition \ref{pseudo-pro}, we can consider other forms of $\psi$do's by rewriting them as the $\psi$do's of the form of \eq{ps+} or \eq{ps-}.

What we prove are the followings:

\begin{thm}\label{Propa-1}
Let $P_s$ be as above. Then we have for any $s\ge 0$
\beq
\Vert P_s q(D_x) e^{-itH_0} P_s \Vert \le C_s\langle t\rangle^{-s}\q (t\in R^1),\label{weight-est}
\ene
where the constant $C_s>0$ is independent of $t\in R^1$.
\end{thm}

\begin{thm}\label{Propa-2}
Let $P_s$ and $P_\pm$ be as above. Then we have for any $s\ge 0$ and $s\ge \delta\ge 0$
\beq
&&\Vert P_s e^{-itH_0}P_+\langle x\rangle^\delta\Vert\le C_{s\delta} \langle t\rangle^{-s+\delta}\q (t\ge 0),\label{psp+}\\
&&\Vert \langle x\rangle^\delta P_- e^{-itH_0}P_s\Vert\le C_{s\delta} \langle t\rangle^{-s+\delta}\q (t\ge 0),\label{p-ps}
\ene
where the constant $C_{s\delta}>0$ is independent of $t$.
\end{thm}

\begin{thm}\label{Propa-3}
Let $P_\pm$ be as above. Then we have for any $s\ge 0$ and $\delta\ge 0$
\beq
\Vert \langle x\rangle^\delta P_- e^{-itH_0} P_+\langle x\rangle^\delta\Vert\le C_{s\delta} \langle t\rangle^{-s}\q (t\ge 0),\label{p-p+est}\\
\Vert \langle x\rangle^\delta P_+ e^{-itH_0} P_-\langle x\rangle^\delta\Vert\le C_{s\delta} \langle t\rangle^{-s}\q (t\le 0),\label{p+p-est}
\ene
where the constant $C_{s\delta}>0$ is independent of $t$.
\end{thm}

If we take a Laplace transform of \eq{weight-est} when $s>1$, we obtain for $\ep>0$
\beq
\left\Vert i\int_0^\infty P_sq(D_x)e^{it(\la+i\ep)}e^{-itH_0}P_s dt \right\Vert\le C_s,\nom
\ene
where $C_s>0$ is independent of $\ep>0$. Since the integral is equal to $P_sq(D_x)R_0(\la+i\ep)P_s=\langle x\rangle^{-s}q(D_x)(H_0-(\la+i\ep))^{-1}\langle x\rangle^{-s}$, this implies a partial result of Theorem \ref{Spectral-H0} for $s>1$.

Also taking the Laplace transform of \eq{p-p+est} and \eq{p+p-est}, we obtain for $\delta\ge 0$
\beq
\Vert P_\mp R_0(\la\pm i\ep)P_\pm\Vert_{L^2_{-\delta}\to L^2_{\delta}}\le C_{\delta},
\ene
where $C_{\delta}>0$ is independent of $\ep>0$. 

The reason that these theorems are called propagation estimates is as follows: In the case of Theorem \ref{Propa-1}, $P_s$ on the RHS of $P_s e^{-itH_0}P_sf$ when applied to a function $f\in\HH=L^2(R^m)$ restricts the initial state $f$ in a localized region around 0 to the order $\langle x\rangle^{-s}$. $P_s$ on the LHS restricts the propagated state $e^{-itH_0}P_sf$ also in a localized region around $0$ to the same extent. According to the propagator $e^{-itH_0}$, as is indicated by Theorem \ref{Enss}, the wave function $f$ restricted to a region $G$ in $R^m_x$ should propagate along a line parallel to the ``idealized" velocity $v$ with $|v|\ge \sigma(>0)$: If $f$ is localized in a region $G$ of $R^m$, physically the region $G$ should move to the region $G+tv=\{x+tv|\ x\in G\}$ after time $t$. Thus the localization factor $P_s$ on the LHS should effect so that $\Vert P_se^{-itH_0}P_sf\Vert$ decays as $t\to\infty$. Theorem \ref{Propa-1} tells that the rate of this decay is $t^{-s}$.

In the case of \eq{psp+} of Theorem \ref{Propa-2}, $P_+$ on the RHS of $P_s e^{-itH_0}P_+$ restricts the initial function to the phase space region where $\cos(x,v)\ge \theta+\rho$ by \eq{p+}. Then the state $e^{-itH_0}P_+f$ propagates toward the direction $v\ne 0$ which is almost parallel to $x$. Thus the location of the state should be separated from $0$ when $t\to\infty$. Theorem \ref{Propa-2} gives the rate of this separation.

In the case of Theorem \ref{Propa-3}, similarly to \eq{psp+} of Theorem \ref{Propa-2}, the state $e^{-itH_0}P_+f$ propagates toward the region in the phase space where $x$ and $v$ are almost parallel to each other. Since $P_-$ on the LHS restricts the state to the region where $x$ and $v$ are anti-parallel, the state $P_-e^{-itH_0}P_+f$ should decay. Theorem \ref{Propa-3} gives the rate of this decay.

We now prove these theorems.

\BP

\F
{\it Proof of Theorem \ref{Propa-1}:} It suffices to prove \eq{weight-est} for even integers $s\ge 0$, because we have only to interpolate them to get the desired estimates. The case $s=0$ is obvious. Let $s>0$ be an even integer. 
We recall \eq{Os2}:
\beq
e^{-itH_0}f(x)=\mbox{Os-}\int \int e^{i(x\xi-t|\xi|^2/2-\xi y)}f(y)dy d\wxi,\label{Os2'}
\ene
and the arguments after it, where it was shown that \eq{Os2'} is rewritten as \eq{partial-integrated-form} in which integration by parts with respect to $\xi$ and $y$ can be performed any times.
Thus using the relation
\beq
(1-t\xi D_\xi)e^{-it\xi^2/2}=(1+|t\xi|^2)e^{-it\xi^2/2}=:h(t,\xi)e^{-it\xi^2/2},\nom
\ene
we integrate by parts in \eq{Os2'} to get
\beq
q(D_x) e^{-itH_0}f(x)&=&
\mbox{Os-}\int \int e^{-it|\xi|^2/2}\left((1+t\xi D_\xi)h(t,\xi)^{-1}\right)^s[q(\xi)e^{-i\xi y}e^{ix\xi}]f(y)dy d\wxi\nonumber\\
&=&\mbox{Os-}\int \int e^{i(x\xi-t|\xi|^2/2)}\sum_{j=1}^J Q_j(x)h_j(t,\xi)P_j(y)e^{-i\xi y}f(y)dy d\wxi.\nonumber
\ene
Here $J$ is an integer, $Q_j(x)$ and $P_j(y)$ are polynomials of $x$ and $y$ of order up to $s$, and $h_j(t,\xi)$ satisfies
\beq
|h_j(t,\xi)|_\ell\le C_\ell\langle t\rangle^{-s}\q (\ell=0,1,2,\cdots)\nom
\ene
by the inequality
\beq
(1+|t\xi|^2)^{-s/2}\le C\langle t\rangle^{-s},\nom
\ene
which holds on supp $q$ by \eq{energy-cut-off}. From these, we obtain
\beq
P_s q(D_x) e^{-itH_0} P_sf(x)=\sum_{j=1}^J\mbox{Os-}\int \int e^{i(x\xi-t|\xi|^2/2-\xi y)}u_j(x)h_j(t,\xi)s_j(y)f(y)dy d\wxi,\label{psps}
\ene
where $u_j(x)$ and $s_j(y)$ satisfy
\beq
|u_j|_\ell+|s_j|_\ell<\infty\nom
\ene
for all $\ell=0,1,2,\cdots$. Noting that \eq{psps} can be written as
\beq
P_s q(D_x) e^{-itH_0} P_sf=\sum_{j=1}^J u_j(X)h_j(t,D_x)e^{-itH_0}s_j(X')f,\nom
\ene
we use Theorem \ref{L^2-bound} to conclude the proof. $\Box$
\BP

\F
{\it Proof of Theorem \ref{Propa-2}:} We prove \eq{psp+} only. Another is proved similarly.
We have only to prove the estimate for an even integer $s\ge 0$ as above. We write
\beq
e^{-itH_0}P_+\langle x\rangle^\delta f(x)=\mbox{Os-}\int\int e^{ix\xi}e^{-i(t\xi^2/2+\xi y)}p_+(\xi,y)\langle y\rangle^\delta f(y)dyd\wxi,\nom
\ene
and integrate by parts with respect to $\xi$ using the relation
\beq
(1+|t\xi+y|^2)^{-1}(1-(t\xi+y)D_\xi)e^{-i(t\xi^2/2+\xi y)}=e^{-i(t\xi^2/2+\xi y)}.\nom
\ene
Noting
\beq
&&(1+|t\xi+y|)^{-1}(1+|y|)^{-1}\le C(1+|t\xi|)^{-1},\nom\\
&&|t\xi+y|^{-1}\le C(|t\xi|+|y|)^{-1}\q (\cos(y,\xi)\ge \theta+\rho)\nom
\ene
and applying \eq{p+} and \eq{energy-cut-off-2} to the result of the integration by parts,
we obtain
\beq
e^{-itH_0}P_+\langle x\rangle^\delta f(x)=\sum_{k=1}^KP_k(x)e^{-itH_0}s_{k}(t;D_x,X')\langle y\rangle^\delta f(y),\label{poly-p+}
\ene
where $K$ is some integer, $P_k(x)$ is a polynomial of $x$ of order at most $s$, and each symbol $s_{k}(t;\xi,y)$ satisfies
\beq
|s_{k}(t;\xi,y)|_\ell\le C_\ell\langle t\rangle^{-s+\delta}\nom
\ene
for all $\ell=0,1,2,\cdots$. Thus $P_s$ on the LHS of \eq{psp+} damps the growth of $P_k(x)$, and we obtain the desired estimate by using Theorem \ref{L^2-bound}. $\Box$
\BP

\F
{\it Proof of Theorem \ref{Propa-3}:} We consider the case $t\ge 0$ only. The other case is proved similarly. We divide $P_-$ and $P_+$ as
\beq
P_-=P_{--}+P_{-+},\q P_+=P_{+-}+P_{++},\nom
\ene
where the respective symbols satisfy
\beq
&&\mbox{supp}\ p_{--}(x,\xi)\subset \{(x,\xi)|\ \cos(x,\xi)<\theta-\rho/3, |x|\ge \sigma,|\xi|\ge\sigma \}\nom\\
&&\mbox{supp}\ p_{-+}(x,\xi)\subset \{(x,\xi)|\ \cos(x,\xi)>\theta-2\rho/3,  |x|\ge \sigma,|\xi|\ge\sigma  \}\nom\\
&&\mbox{supp}\ p_{+-}(\xi,y)\subset \{(\xi,y)|\ \cos(y,\xi)<\theta+2\rho/3,  |y|\ge \sigma,|\xi|\ge\sigma \}\nom\\
&&\mbox{supp}\ p_{++}(\xi,y)\subset \{(\xi,y)|\ \cos(y,\xi)>\theta+\rho/3,  |y|\ge \sigma,|\xi|\ge\sigma\}\nom
\ene
and
\beq
|\langle x\rangle^k p_{-+}|_\ell <\infty,\q |\langle y\rangle^k p_{+-}|_\ell <\infty\label{cross}
\ene
for any $k,\ell=0,1,2,\cdots$. Then we have to estimate the four terms:
\beq
&&\langle x\rangle^\delta P_{--}e^{-itH_0}P_{+-}\langle x\rangle^\delta,\q
\langle x\rangle^\delta P_{--}e^{-itH_0}P_{++}\langle x\rangle^\delta,\nom\\
&&\langle x\rangle^\delta P_{-+}e^{-itH_0}P_{+-}\langle x\rangle^\delta,\q
\langle x\rangle^\delta P_{-+}e^{-itH_0}P_{++}\langle x\rangle^\delta.\nom
\ene
By Theorem \ref{Propa-2} and \eq{cross}, we have
\beq
&&\Vert\langle x\rangle^\delta P_{-+}e^{-itH_0}P_{+-}\langle x\rangle^\delta\Vert\le C_{s\delta}\langle t\rangle^{-s},\nom\\
&&\Vert\langle x\rangle^\delta P_{-+}e^{-itH_0}P_{++}\langle x\rangle^\delta\Vert\le C_{s\delta}\langle t\rangle^{-s},\nom\\
&&\Vert\langle x\rangle^\delta P_{--}e^{-itH_0}P_{+-}\langle x\rangle^\delta\Vert\le C_{s\delta}\langle t\rangle^{-s}.\nom
\ene
Thus we have only to consider 
\beq
&&\langle x\rangle^\delta
P_{--}e^{-itH_0}P_{++}\langle x\rangle^\delta f(x)\label{main}\\
&&=\mbox{Os-}\int\int e^{i(x\xi-t\xi^2/2-y\xi)}\langle x\rangle^\delta p_{--}(x,\xi)p_{++}(\xi,y)\langle y\rangle^\delta f(y)dyd\wxi.\nom
\ene
Noting
\beq
|x-t\xi-y|^{-1}\le C(|x|+|t\xi|+|y|)^{-1}\label{3-est}
\ene
for $(x,\xi,y)$ satisfying $\cos(x,\xi)\le \theta-\rho/3$ and $\cos(\xi,y)\ge \theta+\rho/3$,
we integrate by parts in \eq{main} by using the relation
\beq
Qe^{i(x\xi-t\xi^2/2-y\xi)}=e^{i(x\xi-t\xi^2/2-y\xi)},\nom
\ene
where
\beq
Q=(1+|x-t\xi-y|^2)^{-1}(1+(x-t\xi-y)D_\xi).\nom
\ene
Then we obtain
\beq
\langle x\rangle^\delta P_{--}e^{-itH_0}P_{++}\langle x\rangle^\delta f(x)=\mbox{Os-}\int\int e^{i(x-y)\xi}r_t^{(k)}(x,\xi,y)f(y)dyd\wxi,\nom
\ene
where
\beq
r_t^{(k)}(x,\xi,y)=e^{-it\xi^2/2}({}^tQ)^k(\langle x\rangle^\delta p_{--}(x,\xi)p_{++}(\xi,y)\langle y\rangle^\delta)\nom
\ene
satisfies by \eq{3-est}
\beq
|\partial_x^\alpha\partial_\xi^\beta\partial_y^\gamma
r_t^{(k)}(x,\xi,y)|\le C_{\alpha\beta\gamma k}\langle t\xi\rangle^{3m_0}
\langle x\rangle^{-k/3+\delta}
\langle t\xi\rangle^{-k/3}
\langle y\rangle^{-k/3+\delta}\nom
\ene
for all $\alpha,\beta,\gamma$ with $|\alpha+\beta+\gamma|\le 3m_0$ and any integer $k\ge 0$. Therefore by Theorem \ref{L^2-bound} and \eq{energy-cut-off-2}, we obtain the theorem. 
$\Box$

\chapter{Two-Body Hamiltonian}\label{chap:6}

\section{Eigenvalues of a two-body Hamiltonian}

In this chapter we consider a perturbed Hamiltonian defined in $\HH=L^2(R^m)$ $(m=1,2,\cdots)$
\beq
H=H_0+V,\label{two-body-Hamiltonian}
\ene
where $H_0$ is the free Hamiltonian defined by \eq{free-Hamiltonian} and $V$ is a multiplication operator by a real-valued measurable function $V(x)$ that can be decomposed as a sum of two real-valued measurable functions: $V(x)=V_S(x)+V_L(x)$ satisfying the decay assumption:
\beq
&&|V_S(x)|\le C\langle x\rangle^{-1-\delta},\label{short-range}\\
&&|\partial_x^\alpha V_L(x)|\le C_\alpha \langle x\rangle^{-|\alpha|-\delta}\label{long-range}
\ene
for a constant $0<\delta<1$ and all multi-indices $\alpha$ with constants $C>0$ and $C_\alpha>0$ independent of $x\in R^m$. $H$ is a generalization of a two-body Hamiltonian that describes the two-body system in $R^3$ to a general dimension $m=1,2,\cdots$. $V_S$ and $V_L$ are called short- and long-range potentials respectively. The assumption \eq{short-range} can be weakened to allow some local singularities with respect to $x\in R^m$. E.g. 
\beq
h(R)=\Vert V_S(H_0+1)^{-1}\chi_{\{x| |x|> R\}}\Vert \in L^1((0,\infty))\label{Enss-potential}
\ene
is known sufficient (\cite{[E1]}) for some results we prove below. \eq{Enss-potential} includes the Coulomb singularities of order $1/|x|$ at $x=0$, thus together with the long-range part $V_L$, $V$ covers the Coulomb type long-range potentials. As the inclusion of singularities is of rather technical nature, we restrict the description below to the potentials satisfying \eq{short-range} and \eq{long-range}.

As is well-known (see, e.g. \cite{Schiff}, Chapter 4.), the perturbed Hamiltonian \eq{two-body-Hamiltonian} has eigenvalues in general, unlike the unperturbed Hamiltonian $H_0$. Thus we have first to specify eigenvalues and eigenspace $\HH_{p}(H)$ of $H$. Then restricting the total Hilbert space $\HH$ to the continuous spectral subspace $\HH_c(H)=\HH_{p}(H)^\perp$, we will consider the properties of scattering states $f \in \HH_c(H)$.

Before going to the investigation of eigenvalues of $H$, we see that $H$ defines a selfadjoint operator in $\HH$. In the case of the unperturbed Hamiltonian $H_0$, the selfadjointness was trivial by virtue of the expression \eq{Hamiltonian}. But in the case of the perturbed Hamiltonian, getting such an expression is a problem that will be solved later.
We recall the definition of the adjoint operator $H^*$ of $H$. Let ${\cal D}(H^*)$ be defined as the set of all $g\in\HH$ such that there exists an $f\in\HH$ satisfying
\beq
(g,Hu)=(f,u)\q \mbox{for all}\ u\in{\cal D}(H).\label{def-adjoint}
\ene
Since ${\cal D}(V)=\HH$, the domain ${\cal D}(H)$ of the Hamiltonian $H=H_0+V$ is equal to ${\cal D}(H_0)=H^2(R^m)$, which is dense in $\HH$. Thus the relation \eq{def-adjoint} determines $f$ uniquely. Then we define $H^*g=f$ for $g\in{\cal D}(H^*)$. In general, $H$ is called symmetric if $H^*$ is an extension of $H$ (in notation $H\subset H^*$) and selfadjoint when $H^*=H$. A symmetric operator $H$ is called essentially selfadjoint if the closure $H^{**}$ of $H$ is selfadjoint. In our case of $H_0$ and $V$, it is clear that $H_0$ and $V$ are selfadjoint with ${\cal D}(H_0^*)=H^2(R^m)={\cal D}(H_0)$ by \eq{Hamiltonian}, and ${\cal D}(V^*)=\HH={\cal D}(V)$ by our assumptions \eq{short-range} and \eq{long-range}. Thus ${\cal D}(H^*)={\cal D}(H_0^*)\cap {\cal D}(V^*)=H^2(R^m)={\cal D}(H)$; hence $H$ is selfadjoint.

Turning to the eigenvalues of $H$, we define $i[H,A]$ as a form sum
\beq
(i[H,A]f,g)=i(Af,Hg)-i(Hf,Ag)
\ene
for $f,g\in\SSS$,
 where as before $A=(x\cdot D_x+D_x\cdot x)/2=x\cdot D_x +m/(2i)=D_x\cdot x-m/(2i)$ is a selfadjoint operator with ${\cal D}(A)=H^1_1(R^m)$, the weighted Sobolev space, as defined in the section of Notation at the beginning. 

We first prove the non-existence of positive eigenvalues of $H$. To do so we assume that $u\in \DD(H)$ satisfies $Hu=\lambda u$ for some $\lambda>0$, and will prove that $u=0$. Then we have shown that $H$ has no positive eigenvalues. We set $P(\lambda)=E_H(\lambda)-E_H(\lambda-0)$ for the above $\lambda>0$. (See the section of Notation.) Then letting $B=(\lambda-\epsilon/\mu,\lambda+\epsilon/\mu)$ with $\epsilon>0$ sufficiently small and $\mu>1$ sufficiently large, and  using the equalities as form sums $i[H_0,A]=2H_0$ and $i[V_L,A]=-x\cdot \nabla_x V_L(x)$, we have
\beq
&&(E_H(B)i[H,A]E_H(B)u,u)\label{Mourre-2}\\
&&=i(AE_H(B)u,HE_H(B)u)-i(HE_H(B)u,AE_H(B)u)\nom\\
&&=i(AE_H(B)u,V_SE_H(B)u)-i(V_SE_H(B)u,AE_H(B)u)\nom\\
&&\q\q\q\q\q+(i[V_L,A]E_H(B)u,E_H(B)u)+(2H_0E_H(B)u,E_H(B)u)\nom\\
&&= i(x\cdot D_xE_H(B)u,V_SE_H(B)u)-i(V_SE_H(B)u,x\cdot D_xE_H(B)u)\nom\\
&&\q\q\q\q\q+m(V_SE_H(B)u,E_H(B)u)-(x\cdot \nabla_xV_LE_H(B)u,E_H(B)u)\nom\\
&&\q\q\q\q\q-(2VE_H(B)u,E_H(B)u)+(2HE_H(B)u,E_H(B)u).\nom
\ene
By \eq{short-range} and \eq{long-range}, $E_H(B)V_S(x\cdot D_x)E_H(B)$, $E_H(B)V_SE_H(B)$, $E_H(B)(x\cdot \nabla_xV_L)E_H(B)$ and $E_H(B)VE_H(B)$ are compact operators defined on $\HH$. We now assume that $\varphi(x/R)u(x)$ is not identically $0$ for any $R>0$, where $\varphi\in C^\infty(R^m)$, $0\le \varphi\le 1$, $\varphi(x)=1$ for $|x|\ge 2$ and $\varphi(x)=0$ for $|x|\le 1$. We denote by $\varphi_R$ the multiplication operator by $\varphi(x/R)$. In \eq{Mourre-2}, we replace $u$ by $u_R=\varphi_Ru/\Vert \varphi_Ru\Vert$ with $R>1$. Then since $u_R\to 0$ weakly in $\HH$ as $R\to\infty$, by taking a large $R>1$ we can bound the terms on the RHS of \eq{Mourre-2} except for the last term by an arbitrarily small constant times $\Vert u_R\Vert^2$, while the last term is bounded by $2(\lambda-\epsilon/\mu)(E_H(B)u_R,E_H(B)u_R)=2(\lambda-\epsilon/\mu)\Vert E_H(B)u_R\Vert^2$ from below. Thus we have
\beq
\liminf_{R\to\infty}(E_H(B)i[H,A]E_H(B)u_R,u_R)\ge \alpha \liminf_{R\to\infty}\Vert E_H(B)u_R\Vert^2\label{Mourre-3}
\ene
for some constant $\alpha>0$. We remark that this inequality holds uniformly in small $\epsilon>0$ and large $\mu>1$. 

On the other hand, if we set $R_\mu=\mu (\mu+iA)^{-1}$ for the same $\mu>1$, it is easy to see that $\sup_{\mu>1}\Vert R_\mu\Vert \le 1$, and s-$\lim_{\mu\to\infty}R_\mu=I$. Then we have that for any small $\epsilon>0$ there is an $M_\epsilon>1$ such that for $\mu>M_\epsilon$:
\beq
&&|(E_H(B)i[H,R_\mu]E_H(B)u_R,u_R)|\\
&&=|i(R_\mu E_H(B)u_R,(H-\la)E_H(B)u_R)-i((H-\la)E_H(B)u_R,R_\mu E_H(B)u_R)|\nom\\
&&\le 2\epsilon/\mu \Vert E_H(B)u_R\Vert^2.\nom
\ene
By a calculation, we have $-\mu[H,R_\mu]=R_\mu i[H,A]R_\mu$. Thus
we have for $\epsilon>0$ and $\mu>M_\epsilon$
\beq
|(E_H(B)R_\mu i[H,A]R_\mu E_H(B)u_R,u_R)|\le 2\epsilon \Vert E_H(B)u_R\Vert^2.
\ene
Letting $\mu\to\infty$, we obtain for an arbitrarily small $\epsilon>0$
\beq
|(P(\lambda)i[H,A]P(\lambda)u_R,u_R)|\le 2\epsilon\Vert P(\lambda)u_R\Vert^2.
\ene
This and \eq{Mourre-3} with being let $\mu\to\infty$ yield a contradiction. Therefore our assumption that $\varphi(x/R)u(x)$ is not identically $0$ for any $R>0$ is false, and we have proved that there is an $R>0$ such that $\varphi(x/R)u(x)=0$ for all $x\in R^m$. In particular we have that $u(x)=0$ for $|x|\ge 2R$. Recalling that $u$ satisfies $Hu=\lambda u$ and the well-known unique continuation theorem for the solutions of elliptic partial differential equations (see e.g., \cite{Gilberg-Trudinger}), we see that $u(x)=0$ for all $x\in R^m$. Hence $H$ has no positive eigenvalues.

Next we consider negative eigenvalues of $H$. Assume that $H u_j=\la_j u_j$ with $\la_j\le\la_0<0$ and $(u_i,u_j)_\HH=\delta_{ij}$ for $i,j=1,2,\cdots$ and that $\la_j\to\la(\le\la_0)$ as $j\to\infty$. Then
\beq
(H_0-\la_j)u_j=-Vu_j\in L^2_{\delta}\nonumber
\ene
by our assumptions \eq{short-range} and \eq{long-range}. Since $-\la_j\ge -\la_0>0$, from this relation we obtain
\beq
\sup_{j}\Vert u_j\Vert_{H^2_{\delta}}=\sup_{j}\Vert (H_0-\la_j)^{-1}Vu_j\Vert_{H^2_{\delta}}<\infty.
\ene
In particular, we have that $\{u_j\}_{j=1}^\infty$ forms a precompact subset of $\HH=L^2(R^m)$.
Therefore there is a subsequence $\{u_{j_k}\}$ of $\{u_j\}$ such that
$u_{j_k}\to u$ as $k\to\infty$ in $\HH$ for some $u\in \HH$. By $\Vert u_{j_k}\Vert_\HH=1$, we thus have $\Vert u\Vert_\HH=1$ and $\lim_{k\to\infty}(u_{j_k},u)_\HH=\Vert u\Vert_\HH^2=1$. On the other hand by our assumption $(u_i,u_j)_\HH=\delta_{ij}$, we also have $(u_{j_k},u)_\HH=\lim_{\ell\to\infty}(u_{j_k},u_{j_\ell})=0$ for any $k=1,2,\cdots$, a contradiction. Thus no negative eigenvalues accumulate to a real number other than $0$. Further taking $\la_j=\la\le\la_0<0$ in the argument, we also see that all negative eigenvalues have finite multiplicity.

Summarizing we have proved:

\begin{thm} Let the conditions \eq{short-range} and \eq{long-range} be satisfied. Then the two-body Hamiltonian $H=H_0+V$ in \eq{two-body-Hamiltonian} has no positive eigenvalues, and its negative eigenvalues are of finite multiplicity and do not accumulate other than $0$. In particular, the set of eigenvalues of $H$ is at most countable, and discrete except for the neighborhoods of $0$.
\end{thm}

\section{Wave operators}\label{wave}

We now consider the behavior of $\exp(-itH)f$ for $f\in \HH_c(H)=\HH_{p}(H)^\perp$ when time $t$ varies. As has been indicated in Theorem \ref{Enss}, $\exp(-itH)f$ for all $f\in \HH_c(H)$ scatter as $t\to\pm\infty$. In the case of the two-body Hamiltonian $H$, the theorem implies for any $f\in\HH_c(H)\cap H^2(R^m)\cap L^2_{2}(R^m)$, $R>0$, and $\phi\in C_0^\infty(R^1)$
\beq
&&\Vert F(|x|<R)\exp(-it_mH)f\Vert\to 0,\label{Enss-5.2.1}\\
&&\Vert (\phi(H)-\phi(H_0))\exp(-it_mH)f\Vert\to 0,\label{Enss-5.2.3}\\
&&\left\Vert\left(\frac{x}{t_m}-D_x\right)\exp(-it_mH)f\right\Vert\to 0\label{Enss-5.2.2}
\ene
as $m\to\pm\infty$. We note that \eq{Enss-5.2.3} implies $E_H(B)f=0$ for $f\in\HH_c(H)$ and $B\subset(-\infty,0)$, which is seen by taking $\phi\in C_0^\infty((-\infty,0))$ in \eq{Enss-5.2.3} and remembering $H_0\ge 0$. \eq{Enss-5.2.2} yields that the relative position of the two particles goes to the direction almost parallel to the relative momentum of the two particles. This is an analogue of the propagation estimates Theorems \ref{Propa-1}-\ref{Propa-3} for the free Hamiltonian. Thus we are tempted to compare the behavior of $\exp(-itH)f$ $(f\in\HH_c(H))$ with the behavior of the free propagator $\exp(-itH_0)$ as $t\to\pm\infty$. In fact we can prove the following theorem:

\begin{thm}\label{short-range-wave}
Let \eq{short-range} be satisfied and let $V_L(x)=0$ $(x\in R^m)$. Then there exist the limits
\beq
W_{\pm}g=\lim_{t\to\pm\infty}e^{itH}e^{-itH_0}g\label{short-wave}
\ene
for any $g\in \HH=L^2(R^m)$. $W_\pm$ are called the wave operators.
\end{thm}

\F
{\it Proof:} We compute 
\beq
e^{itH}e^{-itH_0}g-g =\int_0^t \frac{d}{ds}(e^{isH}e^{-isH_0}g)ds
= i\int_0^t e^{isH}V_Se^{-isH_0}gds.\label{Cook's}
\ene
Since the functions of the form $g=q(D_x)\langle x\rangle^{-1-\delta}h$ with $h\in L^2(R^m)$ and $q$ being the symbol satisfying \eq{energy-cut-off} in section \ref{propagationest} are dense in $\HH=L^2(R^m)$ and the operator $e^{itH}e^{-itH_0}$ is uniformly bounded in $t\in R^1$, we have only to show the convergence of the integral in \eq{Cook's} for $g=q(D_x)\langle x\rangle^{-1-\delta}h$, where $\delta>0$ is the constant in \eq{short-range}. By Theorem \ref{Propa-1} and \eq{short-range}, we have
\beq
\Vert V_S e^{-isH_0}q(D_x)\langle x\rangle^{-1-\delta}h\Vert\le\langle s\rangle^{-1-\delta}\Vert h\Vert\in L^1((-\infty,\infty)).
\ene
Thus the integral in \eq{Cook's} converges for such a $g$ and the proof is complete. $\Box$
\BP

By the definition \eq{short-wave} of $W_\pm$, we easily see that $W_\pm$ are isometric operators from $\HH$ into $\HH$ and the relation holds:
\beq
e^{isH}W_{\pm}=W_{\pm}e^{isH_0}\q (s\in R^1).\label{intert-1}
\ene
By \eq{resolvent-t} and its variant for $H$, we have for $\ep>0, \la\in R^1$ and $f\in \HH$
\beq
&&R_0(\la\pm i\ep)f=i\int_0^{\pm\infty}e^{is(\la\pm i\ep-H_0)}fds\\
&&R(\la\pm i\ep)f=i\int_0^{\pm\infty}e^{is(\la\pm i\ep-H)}fds.
\ene
Taking the Laplace transforms of both sides of \eq{intert-1}, we thus have
\beq
R(\la\pm i\ep)W_\pm=W_\pm R_0(\la\pm i\ep).
\ene
It follows from this that
\beq
\frac{1}{2\pi i}(R(\la+i\ep)-R(\la-i\ep))W_\pm f
=W_\pm \frac{1}{2\pi i}(R_0(\la+i\ep)-R_0(\la-i\ep)).\label{resolv-wave}
\ene
Let $\overline{E}_H(a)=\frac{1}{2}(E_H(a-0)+E_H(a))$, etc. Then
it is not difficult to see (see \cite{Yosida} p. 325 or \cite{Kato} p.359) by using \eq{res1}-\eq{res3} that for $-\infty<a<b<\infty$
\beq
&&\mbox{s-}\lim_{\ep\downarrow 0}\frac{1}{2\pi i}\int_{a}^{b}(R(\la+i\ep)-R(\la-i\ep))d\la
=\overline{E}_H(b)-\overline{E}_H(a),\\
&&\mbox{s-}\lim_{\ep\downarrow 0}\frac{1}{2\pi i}\int_{a}^{b}(R_0(\la+i\ep)-R_0(\la-i\ep))d\la
=\overline{E}_0(b)-\overline{E}_0(a).
\ene
{}From these and \eq{resolv-wave}, we now have
\beq
E_H(B)W_\pm=W_\pm E_0(B)\label{Intertwining}
\ene
for any Borel set $B\subset (-\infty,\infty)$, where $E_H(B)$ and $E_0(B)$ are spectral measures for $H$ and $H_0$ respectively. From \eq{Intertwining}, we have for any continuous function $F(\la)$,
\beq
F(H)W_\pm\supset W_\pm F(H_0).\label{Intertwining-2}
\ene
This relation is called the intertwining property of the wave operators $W_\pm$.

We define the absolutely continuous spectral subspace $\HH_{ac}(H)$ of a selfadjoint operator $H$ as the space of functions $f\in\HH$ such that
\beq
E_H(B)f=0\q \mbox{if}\q |B|=0,
\ene
where $|B|$ is the Lebesgue measure of a Borel set $B$ of $R^1$. We remark that $f\in \HH_c(H)=\HH_{p}(H)^\perp$ is equivalent to the condition
\beq
(E_H(\la)f,f)\ \mbox{is continuous with respect to}\ \la \in R^1.\label{spectral-continuous}
\ene
In fact, the eigenspace of $H$ for an eigenvalue $\la\in R^1$ is equal to $P(\la)\HH:=(E_H(\la)-E_H(\la-0))\HH$, where $E_H(\la-0)=\mbox{s-}\lim_{\mu\uparrow \la}E_H(\mu)$. Thus $\HH_{p}(H)$ is spanned by $P(\la)\HH$ with $\la\in R^1$, and the orthogonal complement $\HH_{p}(H)^\perp$ consists of those elements $f$ that satisfy \eq{spectral-continuous}.
In the case of the free Hamiltonian $H_0$, it follows from Theorem \ref{Spectral-H0} and $\HH_{p}(H_0)=\{0\}$ that 
\beq
\HH_{ac}(H_0)=\HH_c(H_0)=\HH.\label{free-completeness}
\ene
Combining this with the intertwining property \eq{Intertwining}, we see that for $f=W_\pm g$ in the range ${\cal R}(W_\pm)=W_\pm \HH$, 
\beq
E_H(B)f=E_H(B)W_\pm g=W_\pm E_0(B)g=0
\ene
for a Borel set $B$ with the Lebesgue measure $|B|=0$. Namely $W_\pm f$ $(f\in \HH)$ belongs to the absolutely continuous spectral subspace $\HH_{ac}(H)(\subset\HH_c(H))$ of $H$. Thus we have shown
\beq
{\cal R}(W_\pm)\subset \HH_{ac}(H) \subset \HH_{c}(H).
\ene
For the perturbed Hamiltonian $H$, $\HH_{p}(H)\ne \{0\}$ in general as seen above, thus \eq{free-completeness} does not necessarily hold. Nevertheless by Theorem \ref{Enss}, it is expected that 
\beq
{\cal R}(W_\pm)= \HH_{ac}(H)= \HH_c(H)\label{two-body-completeness}
\ene
holds. If this equality holds, $\HH_c(H)$ is unitarily equivalent to $\HH_c(H_0)=\HH=L^2(R^m)$ by the unitary operators
\beq
W_\pm: \HH=\HH_c(H_0)=\HH_{ac}(H_0) \longrightarrow \HH_c(H)=\HH_{ac}(H),
\ene
which intertwine $H$ and $H_0$.
The relation \eq{two-body-completeness} is called
the {\it asymptotic completeness} of the wave operators $W_\pm$. 
Assuming the asymptotic completeness holds, we define unitary operators
\beq
\FF_\pm =\FF W_\pm^*: \HH_c(H) \longrightarrow \FF \HH,
\ene
where $\FF$ is Fourier transformation as before. Then by the intertwining property \eq{Intertwining-2}, we have for $f\in \HH_c(H)\cap {\cal D}(H)$
\beq
\FF_\pm Hf(\xi)=\FF W_\pm^* Hf(\xi)=\FF H_0W_\pm^* f(\xi)=\frac{|\xi|^2}{2}\FF_\pm f(\xi).
\ene
Thus $\FF_\pm$ are considered as generalizations of Fourier transformation that diagonalize $H$ on $\HH_c(H)$ as the usual Fourier transformation $\FF$ diagonalizes $H_0$: $\FF H_0f(\xi)=(|\xi|^2/2)\FF f(\xi)$ $(f\in\HH_c(H_0)\cap {\cal D}(H_0)={\cal D}(H_0))$. Thus the asymptotic completeness gives a spectral representation of the perturbed Hamiltonian $H=H_0+V$.

\section{Asymptotic completeness}

In this section, we will be devoted in the proof of the asymptotic completeness \eq{two-body-completeness}. To do so, since by definition
\beq
\HH_{ac}(H)\subset \HH_c(H),
\ene
we have only to show that
\beq
\HH_c(H)\subset {\cal R}(W_\pm).\label{asymptotic-completeness}
\ene

The definition of wave operators $W_\pm$ in \eq{short-wave} is known working only for short-range potentials satisfying \eq{short-range} or \eq{Enss-potential}. To extend \eq{asymptotic-completeness} to include the long-range potentials satisfying \eq{long-range}, it is necessary to modify the definition \eq{short-wave}. We here choose a modification of the form
\beq
W_\pm f=\lim_{t\to\pm\infty}e^{itH}Je^{-itH_0}f,\label{modified-wave-operators}
\ene
where $J$ is an identification operator or time-independent modifier introduced in \cite{[IK]}, and will be defined as a Fourier integral operator of the form
\beq
Jf(x)&=&\mbox{Os-}\int\int e^{i(\varphi(x,\xi)-y\xi)}f(y)dyd\wxi\label{Identification}\\
&=&c_m\int e^{i\varphi(x,\xi)}{\hat f}(\xi)d\xi.\nom
\ene
Here ${\hat f}$ is the Fourier transform of $f\in \SSS(R^m)$, $c_m=(2\pi)^{-m/2}$, and the phase function $\varphi(x,\xi)$ will be constructed as a solution of an eikonal equation
\beq
\frac{1}{2}|\nabla_x\varphi(x,\xi)|^2+V_L(x)=\frac{1}{2}|\xi|^2\nom
\ene
in the forward and backward regions, in which $x\in R^m$ and $\xi\in R^m$ are almost parallel and anti-parallel to each other, respectively. With $\varphi(x,\xi)$ being well-defined, we argue as follows:
We appeal to the Cauchy criterion of convergence in proving the existence of $W_\pm$ and its asymptotic completeness: ${\cal R}(W_\pm)=\HH_c(H)$, with using Theorem \ref{Enss}. For instance, the argument for the asymptotic completeness is as follows: We consider the case $t\to +\infty$ only. The other case is treated similarly. Noting that the vectors of the form $E_H(B)g$ with $g\in\HH_c(H)$ and $B$ being a Borel subset of the interval $(0,\infty)$ are dense in $\HH_c(H)$, we evaluate the difference for $g\in \HH_c(H)$ with $g=E_H(B)g$ $(B\subset(d^2/2,b),\ 0<d,\ 0<d^2/2<b<\infty)$:
\beq
&&(e^{itH_0}J^{-1}e^{-itH}-e^{it_mH_0}J^{-1}e^{-it_mH})g\label{Cauchy-diff}\label{difference-Enss}\\
&&\q\q=e^{it_mH_0}(e^{i(t-t_m)H_0}J^{-1}e^{-i(t-t_m)H}-J^{-1})e^{-it_mH}g.\nom
\ene
Here $t_m$ is a sequence in Theorem \ref{Enss} tending to infinity as $m\to\infty$ and $J^{-1}$ is an inverse of $J$ that exists for a suitable choice of $\varphi$. By virtue of Theorem \ref{Enss}, we can insert a factor $P_+F(|x|>R_m)$ between 
$(e^{i(t-t_m)H_0}J^{-1}e^{-i(t-t_m)H}-J^{-1})$ and $e^{-it_mH}g$ on the RHS of \eq{Cauchy-diff}, where $P_+$ is a $\psi$do in \eq{ps+} satisfying \eq{energy-cut-off-2} with $\sig=d^2/2$, and $R_m$ is a suitable sequence tending to infinity as $m\to+\infty$ such that \eq{Enss-5.2.1} holds with $R>0$ replaced by $R_m>0$. For the later use, we choose the symbol $p_+(\xi,y)$ of $P_+$ such that it satisfies in the forward region
\beq
|\parti_y^\al\parti_\xi^\be p_+(\xi,y)|\le C_{\al\be}\lan y\ran^{-|\al|}\q (\cos(y,\xi)\ge \theta+\rho).\label{p+restriction}
\ene
for any multi-indices $\al,\be$, where the constant $C_{\al\be}>0$ is independent of $(\xi,y)$. Since the role of $P_+$ is to restrict the support of the applied function to the forward sector, this restriction is not an essential one.
We then estimate
\beq
&&\Vert(e^{i(t-t_m)H_0}J^{-1}e^{-i(t-t_m)H}-J^{-1})P_+F(|x|>R_m)\Vert\\
&=&\Vert e^{i(t-t_m)H_0}J^{-1}e^{-i(t-t_m)H}(J-e^{i(t-t_m)H}Je^{-i(t-t_m)H_0})J^{-1}P_+F(|x|>R_m)\Vert\nom\\
&\le & C\left\Vert\int_0^{t-t_m}\frac{d}{ds}\left(e^{isH} J e^{-isH_0}\right)J^{-1}P_+F(|x|>R_m)ds\right\Vert\nom\\
&\le &C\int_0^{t-t_m}\Vert(HJ-JH_0)e^{-isH_0}J^{-1}P_+F(|x|>R_m)\Vert ds.\nom
\ene
Here by the eikonal equation, $T=HJ-JH_0$ decays in the forward direction to the order $\langle x\rangle^{-1-\delta}$ so that we can apply Theorems \ref{Propa-2} and \ref{Propa-3} to get
\beq
&&\Vert Te^{-isH_0}J^{-1}P_+F(|x|>R_m)\Vert\label{asymptotic-c-estimate}\\
&&\le \Vert Te^{-isH_0}J^{-1}P_+\langle x\rangle^{\delta/2}\Vert \Vert\langle x\rangle^{-\delta/2}F(|x|>R_m)\Vert
\le  C\langle s\rangle^{-1-\delta/2}\langle R_m\rangle^{-\delta/2}.\nom
\ene
These yield
\beq
\Vert (e^{itH_0}J^{-1}e^{-itH}-e^{it_mH_0}J^{-1}e^{-it_mH})g\Vert \to 0
\ene
as $t>t_m\to\infty$. Thus we have a Cauchy criterion for $t>s>t_m\to\infty$:
\beq
&&\Vert (e^{itH_0}J^{-1}e^{-itH}-e^{isH_0}J^{-1}e^{-isH})g\Vert\\
&&\q\le \Vert (e^{itH_0}J^{-1}e^{-itH}-e^{it_mH_0}J^{-1}e^{-it_mH})g\Vert\nom\\
&&\q\q+\Vert (e^{isH_0}J^{-1}e^{-isH}-e^{it_mH_0}J^{-1}e^{-it_mH})g\Vert\to 0.\nom
\ene
This proves the existence of the limit
\beq
\Omega_+g=\lim_{t\to\infty}e^{itH_0}J^{-1}e^{-itH}g\label{Omega}
\ene
for $g\in \HH_c(H)$. Combining this with the existence of $W_+$, we get for $g\in\HH_c(H)$
\beq
g=W_+\Omega_+g\in {\cal R}(W_+),
\ene
which will complete the proof of the asymptotic completeness.

\BP

To construct the phase function $\varphi(x,\xi)$ with the desired properties, we need to consider the classical orbits associated with the classical Hamiltonian:
\beq
H_\rho(t,x,\xi)=\frac{1}{2}|\xi|^2+V_\rho(t,x).\label{classical-Hamiltonian}
\ene
Here $0<\rho <1$ and
\beq
V_\rho(t,x)=V_L(x)\phi(\rho x)\phi\left(\frac{\lan \log\lan t\ran\ran}{\lan t\ran}x\right),\label{Vrho}
\ene
where $\phi(x)$ is a $C^\infty(R^m)$ function satisfying
\beq
\phi(x)=
\left\{
\begin{array}{ll}
1& \q |x|\ge 2\\
0& \q |x|\le 1
\end{array}
\right.\label{cutout}
\ene
with $0\le \phi(x)\le 1$. Then $V_\rho$ satisfies
\beq
|\partial_x^\alpha V_\rho(t,x)|\le C_\alpha \rho^{\de_0}\langle t\rangle^{-\ell}\langle x\rangle^{-m}\label{potential-estimate}
\ene
for $\ell,m\ge 0$ and $0<\de_0<\de$ such that $\de_0+\ell+m<|\alpha|+\delta$.

The corresponding classical orbit $(q,p)(t,s,y,\xi)=(q(t,s,y,\xi),p(t,s,y,\xi))$ is determined by the equation
\beq
\left\{
\begin{array}{l}
q(t,s)=y+\int_s^t p(\tau,s)d\tau,\\
p(t,s)=\xi- \int_s^t \nabla_xV_\rho(\tau,q(\tau,s))d\tau.
\end{array}
\right.\label{classical-orbits}
\ene
Letting $\de_0,\de_1>0$ be fixed as $0<\de_0+\de_1<\de$, we have the following estimates for $(q,p)(t,s,y,\xi)$, which are proved by solving the equation \eq{classical-orbits} by iteration:

\begin{pro}\label{Estimates-for-qp}
 There is a constant $C_\ell>0$ $(\ell=0,1,2,\cdots)$ such that for all $(y,\xi)\in R^{2m}$, $\pm t\ge \pm s \ge 0$ and multi-index $\al$:
\beq
&&|p(s,t,y,\xi)-\xi|\le C_0\rho^\deo\lan s\ran^{-\del}.\label{pst}\\
&&|\parti_y^\al[\nabla_y q(s,t,y,\xi)-I]|\le C_{|\al|}\rho^\deo\lan s\ran^{-\del},\\
&&|\parti_y^\al[\nabla_y p(s,t,y,\xi)]|\le C_{|\al|}\rho^\deo\lan s\ran^{-1-\del}.\\
&&|\nabla_\xi q(t,s,y,\xi)-(t-s)I|\le C_0\rho^\deo\lan s\ran^{-\del}|t-s|,\label{q-2}\\
&&|\nabla_\xi p(t,s,y,\xi)-I|\le C_0\rho^\deo\lan s\ran^{-\del}.\\
&&|\nabla_y q(t,s,y,\xi)-I|\le C_0\rho^\deo\lan s\ran^{-1-\del}|t-s|,\\
&&|\nabla_y p(t,s,y,\xi)|\le C_0\rho^\deo\lan s\ran^{-1-\del}.\\
&&|\parti_\xi^\al[q(t,s,y,\xi)-y-(t-s)p(t,s,y,\xi)]|\label{qts}\\
&&\q\q\q\q\q\le C_{|\al|}\rho^\deo\min(\lan t\ran^{1-\del},|t-s|\lan s\ran^{-\del}).\nom
\ene
Further for any $\al,\be$ satisfying $|\al+\be|\ge 2$, there is a constant $C_{\al\be}>0$ such that
\beq
&&|\parti_y^\al\parti_\xi^\be q(t,s,y,\xi)|\le C_{\al\be}\rho^\deo|t-s|\lan s\ran^{-\del},\label{q-3}\\
&&|\parti_y^\al\parti_\xi^\be p(t,s,y,\xi)|\le C_{\al\be}\rho^\deo\lan s\ran^{-\del}\label{p-3}
\ene
\end{pro}

{}From this proposition, taking $\rho>0$ so small that $C_0\rho^\deo<1/2$ we obtain the following

\begin{pro}\label{Estimates-for-etay}
Take $\rho>0$ so that $C_0\rho^\deo<1/2$. Then for $\pm t\ge \pm s\ge 0$ one can construct diffeomorphisms of $R^{m}$
\beq
&&x\mapsto y(s,t,x,\xi)\\
&&\xi\mapsto \eta(t,s,x,\xi)
\ene
such that
\beq
\left\{
\begin{array}{l}
q(s,t,y(s,t,x,\xi),\xi)=x\\
p(t,s,x,\eta(t,s,x,\xi))=\xi
\end{array}
\right..\label{qp-yeta}
\ene
$y(s,t,x,\xi)$ and $\eta(t,s,x,\xi)$ are $C^\infty$ in $(x,\xi)\in R^{2m}$ and their derivatives $\parti_x^\al\parti_\xi^\be y$ and $\parti_x^\al\parti_\xi^\be\eta$ are $C^1$ in $(t,s,x,\xi)$. 
They satisfy the relation
\beq
\left\{
\begin{array}{l}
y(s,t,x,\xi)=q(t,s,x,\eta(t,s,x,\xi))\\
\eta(t,s,x,\xi)=p(s,t,y(s,t,x,\xi),\xi)
\end{array}
\right.\label{y-q-eta-p}
\ene
and the estimates
for any $\al,\be$
\beq
&&|\parti_x^\al\parti_\xi^\be[\nabla_x y(s,t,x,\xi)-I]|\le C_{\al\be}\rho^\deo\lan s\ran^{-\del},\label{y-1}\\
&&|\parti_x^\al\parti_\xi^\be[\nabla_x\eta(t,s,x,\xi)]|\le C_{\al\be}\rho^\deo\lan s\ran^{-1-\del}.\label{eta-1}\\
&&|\parti_\xi^\al[\eta(t,s,x,\xi)-\xi]|\le C_{\al}\rho^\deo\lan s\ran^{-\del}\label{eta-2}\\
&&|\parti_\xi^\al[y(s,t,x,\xi)-x-(t-s)\xi]|\label{y-2}\\
&&\q\q\q\q\q\le C_{\al}\rho^\deo\min(\lan t\ran^{1-\del},|t-s|\lan s\ran^{-\del}).\nom
\ene
Further for any $|\al+\be|\ge 2$
\beq
&&|\parti_x^\al\parti_\xi^\be \eta(t,s,x,\xi)|\le C_{\al\be}\rho^\deo\lan s\ran^{-\del},\label{eta-3}\\
&&|\parti_x^\al\parti_\xi^\be y(s,t,x,\xi)|\le C_{\al\be}\rho^\deo\lan t-s\ran\lan s\ran^{-\del}.\label{y-3}
\ene
Here the constants $C_\al, C_{\al\be}>0$ are independent of $t,s,x,\xi$
\end{pro}

The following illustration would be helpful to understand the meaning of the diffeomorphisms $y(s,t,x,\xi)$ and $\eta(t,s,x,\xi)$: Let $U(t,s)$ be the map that assigns the point $(q,p)(t,s,x,\eta)$ to the initial data $(x,\eta)$. Then
\beq
\begin{array}{ccc}
\mbox{time}\ s& \ &\mbox{time}\ t\\
\left(
\begin{array}{c}
x\\
\ \\
\eta(t,s,x,\xi)
\end{array}
\right)
&
\begin{array}{c}
U(t,s) \\
\longmapsto\\
\ 
\end{array}
&
\left(
\begin{array}{c}
y(s,t,x,\xi)\\
\ \\
\xi
\end{array}
\right)
\end{array}
\ene

We now define $\phi(t,x,\xi)$ by
\beq
\phi(t,x,\xi)=u(t,x,\eta(t,0,x,\xi)),
\ene
where
\beq
u(t,x,\eta)=x\cdot \eta+\int_0^t(H_\rho-x\cdot \nabla_xH_\rho)(\tau,q(\tau,0,x,\eta),p(\tau,0,x,\eta))d\tau.
\ene
Then it is shown by a direct calculation that $\phi(t,x,\xi)$ satisfies the Hamilton-Jacobi equation
\beq
&&\parti_t\phi(t,x,\xi)=\frac{1}{2}|\xi|^2+V_\rho(t,\nabla_\xi\phi(t,x,\xi)),\label{Hamilton-Jacobi}\\
&&\phi(0,x,\xi)=x\cdot \xi,\nom
\ene
and the relation
\beq
&&\nabla_x\phi(t,x,\xi)=\eta(t,0,x,\xi),\label{eta-phi}\\
&&\nabla_\xi\phi(t,x,\xi)=y(0,t,x,\xi).\label{y-phi}
\ene
We define for $(x,\xi)\in R^{2m}$
\beq
\phi_\pm(x,\xi)=\lim_{t\to\pm\infty}(\phi(t,x,\xi)-\phi(t,0,\xi)).\label{limit-phi}
\ene
We will show the existence of the limits below. We set for $R,d>0$ and $\sig_0\in(-1,1)$
\beq
\Gamma_\pm&=&\Gamma_\pm(R,d,\sig_0)\\
&=&\{ (x,\xi)\in R^{2m}|\ |x|\ge R,|\xi|\ge d,\pm\cos(x,\xi)\ge \pm\sigma_0\}.\nom
\ene

\begin{pro}
The limits \eq{limit-phi} exist for all $(x,\xi)\in R^{2m}$ and define $C^\infty$ functions of $(x,\xi)$. The limit $\phi_\pm(x,\xi)$ satisfies the eikonal equation: For any $d>0$ and $\sig_0\in(-1,1)$, there is a constant $R=R_d=R_{d\sig_0}>1$ such that for any $(x,\xi)\in\Gamma_\pm=\Gamma_\pm(R,d,\sig_0)$, the following relation holds:
\beq
\frac{1}{2}|\nabla_x\phi_\pm(x,\xi)|^2+V_L(x)=\frac{1}{2}|\xi|^2.
\label{eikonal-pm}
\ene
Further for any $\al,\be$ we have the estimate:
\beq
|\parti_x^\al\parti_\xi^\be(\phi_\pm(x,\xi)-x\cdot\xi)|\le C_{\al\be}|\xi|^{-1}\lan x\ran^{1-|\al|-\de},\label{phi-estimates}
\ene
where $C_{\al\be}>0$ is independent of $(x,\xi)\in \Gamma_\pm$.
\end{pro}

\F
{\it Proof:} We consider $\phi=\phi_+$ only. $\phi_-$ can be treated similarly. 
We first prove the existence of the limit \eq{limit-phi} for $t\to+\infty$. To do so, 
setting
\beq
R(t,x,\xi)=\phi(t,x,\xi)-\phi(t,0,\xi),
\ene
we show the existence of the limits 
\beq
\lim_{t\to\infty}\parti_x^\al\parti_\xi^\be R(t,x,\xi)=\lim_{t\to\infty}\int_0^t
\parti_x^\al\parti_\xi^\be \parti_t R(\tau,x,\xi)d\tau+\parti_x^\al\parti_\xi^\be(x\cdot\xi).
\ene
By Hamilton-Jacobi equation \eq{Hamilton-Jacobi},
\beq
\parti_t R(t,x,\xi)&=&\parti_t\phi(t,x,\xi)-\parti_t\phi(t,0,\xi)\label{dtR}\\
&=&V_\rho(t,\nabla_\xi\phi(t,x,\xi))-V_\rho(t,\nabla_\xi\phi(t,0,\xi))\nom\\
&=&(\nabla_\xi\phi(t,x,\xi)-\nabla_\xi\phi(t,0,\xi))\cdot
a(t,x,\xi)
\nom\\
&=&(y(0,t,x,\xi)-y(0,t,0,\xi))\cdot a((t,x,\xi)\nom\\
&=&\nabla_\xi R(t,x,\xi)\cdot a(t,x,\xi),\nom
\ene
where
\beq
&&a(t,x,\xi)=\int_0^1(\nabla_x V_\rho)
(t,\nabla_\xi\phi(t,0,\xi)+\theta\nabla_\xi R(t,x,\xi))d\theta,\label{def-a}\\
&&
\nabla_\xi R(t,x,\xi)=x\cdot\int_0^1(\nabla_xy)(0,t,\theta x,\xi)d\theta.
\ene
By \eq{y-1}, we have for any $\al,\be$
\beq
|\parti_x^\al\parti_\xi^\be\nabla_\xi R(t,x,\xi)|\le C_{\al\be}\lan x\ran.\label{R-1}
\ene
By \eq{y-2} and \eq{y-phi}, for $|\be|\ge 1$
\beq
|\parti_\xi^\be \nabla_\xi \phi(t,0,\xi)|\le C_\be |t|.\label{nablaxi-phi}
\ene
{}From this, \eq{def-a}, and \eq{R-1}, we have
\beq
|\parti_x^\al\parti_\xi^\be a(t,x,\xi)|\le C_{\al\be}\lan t\ran^{-1-\de/2}\lan x\ran^{|\al|+|\be|}.\label{a-1}
\ene
Thus by \eq{dtR}, \eq{R-1} and \eq{a-1}, there exists the limit for any $\al,\be$
\beq
\lim_{t\to\infty}\parti_x^\al\parti_\xi^\be R(t,x,\xi)=\int_0^\infty \parti_x^\al\parti_\xi^\be\left(\nabla_\xi R(t,x,\xi)\cdot a(t,x,\xi)\right)dt+\parti_x^\al\parti_\xi^\be(x\cdot \xi).
\ene
In particular, $\phi=\phi_+(x,\xi)=\lim_{t\to\infty}R(t,x,\xi)$ and $\eta(\infty,0,x,\xi)=\lim_{t\to\infty}\nabla_x\phi(t,x,\xi)$ exist and are $C^\infty$ in $(x,\xi)$.

Next we show \eq{eikonal-pm}. By the arguments above, the following limit exist:
\beq
\nabla_x\phi(x,\xi)&=&\lim_{t\to\infty}\nabla_x\phi(t,x,\xi)=\lim_{t\to\infty}\eta(t,0,x,\xi)\\
&=&\lim_{t\to\infty} p(0,t,y(0,t,x,\xi),\xi).\nom
\ene
Thus for a sufficiently large $|x|$ (i.e. for $|\rho x|\ge 2$) we have
\beq
\frac{1}{2}|\nabla_x\phi_+(x,\xi)|^2+V_L(x)=\frac{1}{2}\lim_{t\to\infty}|p(0,t,y(0,t,x,\xi),\xi)|^2+V_\rho(0,x).\label{phi-p-limit}
\ene
Set for $0\le s\le t<\infty$
\beq
f_t(s,y,\xi)=\frac{1}{2}|p(s,t,y,\xi)|^2+V_\rho(s,q(s,t,y,\xi)).
\ene
Then by \eq{classical-orbits} we have
\beq
\frac{\parti f_t}{\parti s}(s,y,\xi)&=&p(s,t,y,\xi)\cdot \parti_s p(s,t,y,\xi)\\
&&\q+(\nabla_x V_\rho)(s,q(s,t,y,\xi))\cdot\parti_s q(s,t,y,\xi)
+\frac{\parti V_\rho}{\parti t}(s,q(s,t,y,\xi))\nom\\
&=&\frac{\parti V_\rho}{\parti t}(s,q(s,t,y,\xi)).\nom
\ene
On the other hand we have from \eq{qp-yeta} and \eq{y-q-eta-p}
\beq
q(s,t,y(0,t,x,\xi),\xi)&=&q(s,t,q(t,0,x,\eta(t,0,x,\xi)),\xi)\\
&=&q(s,0,x,\eta(t,0,x,\xi)),\nom\\
p(s,t,y(0,t,x,\xi),\xi)&=&p(s,t,q(t,0,x,\eta(t,0,x,\xi)),\xi)\\
&=& p(s,0,x,\eta(t,0,x,\xi)).\nom
\ene
Now using Proposition \ref{Estimates-for-qp}, we have for $\cos(x,\xi)\ge \sig_0$
\beq
&&|q(s,t,y(0,t,x,\xi),\xi)|=|q(s,0,x,\eta(t,0,x,\xi))|\\
&&\q\ge |x+sp(s,0,x,\eta(t,0,x,\xi))|-C_0\rho^\deo\lan s\ran^{1-\del}\nom\\
&&\q=|x+sp(s,t,y(0,t,x,\xi),\xi)|-C_0\rho^\deo\lan s\ran^{1-\del}\nom\\
&&\q\ge c(|x|+s|\xi|)-C_0\rho^\deo\lan s\ran^{1-\del}-C_0\rho^\deo\lan s\ran^{1-\del},\nom
\ene
where $c>0$ is a constant independent of $s,t,x,\xi$. By $(x,\xi)\in \Gamma_+(R,d,\sig_0)$, we have $|\xi|\ge d$, and from the definition \eq{Vrho} of $V_\rho(t,x)$
\beq
\mbox{supp}\ \frac{\parti V_\rho}{\parti t}(s,x)\subset
\{x| 1\le \lan \log\lan s\ran\ran|x|/\lan s\ran\le 2\}.
\ene
Thus there is a constant $S=S_{d,\sig_0}>1$ independent of $t$ such that for any $s\in[S,t]$
\beq
\frac{\parti f_t}{\parti s}(s,y(0,t,x,\xi),\xi)=0.
\ene
For $s\in[0,S]$, taking $R=R_S>1$ large enough, we have for $|x|\ge R$ and $\cos(x,\xi)\ge \sig_0$
\beq
\frac{\parti f_t}{\parti s}(s,y(0,t,x,\xi),\xi)=0.
\ene
Therefore we have shown that for $(x,\xi)\in \Gamma_+(R,d,\sig_0)$
\beq
f_t(s,y(0,t,x,\xi),\xi)=\mbox{constant for}\ 0\le s\le t<\infty.
\ene
In particular we have
\beq
f_t(0,y(0,t,x,\xi),\xi)=f_t(t,y(0,t,x,\xi),\xi),
\ene
which means
\beq
\frac{1}{2}|p(0,t,y(0,t,x,\xi),\xi)|^2+V_\rho(0,x)=\frac{1}{2}|\xi|^2+V_\rho(t,y(0,t,x,\xi)).
\ene
Since $V_\rho(t,y)\to0$ uniformly in $y\in R^m$ when $t\to\infty$ by \eq{potential-estimate}, we have from this and \eq{phi-p-limit}
\beq
\frac{1}{2}|\nabla_x\phi_+(x,\xi)|^2+V_L(x)=\frac{1}{2}|\xi|^2\q \mbox{for}\q (x,\xi)\in \Gamma_+(R,d,\sig_0),
\ene
if $R>1$ is sufficiently large.

We finally prove the estimates \eq{phi-estimates}. We first consider the derivatives with respect to $\xi$:
\beq
\parti_\xi^\be(\phi_+(x,\xi)-x\cdot \xi)=\int_0^\infty\parti_\xi^\be \parti_t R(t,x,\xi)dt,
\ene
where as above $R(t,x,\xi)=\phi(t,x,\xi)-\phi(t,0,\xi)$. Set
\beq
\gamma(t,x,\xi)=y(0,t,x,\xi)-(x+t\xi)
\ene
for $(x,\xi)\in\Gamma_+(R,d,\sig_0)$. Then
by \eq{y-2} we have for $\theta\in[0,1]$
\beq
|\nabla_\xi\phi(t,0,\xi)+\theta\nabla_\xi R(t,x,\xi)|&=&
|y(0,t,0,\xi)+\theta(y(0,t,x,\xi)-y(0,t,0,\xi))|\label{nabphiR}\\
&=&|t\xi+\gamma(t,0,\xi)+\theta(x+\gamma(t,x,\xi)-\gamma(t,0,\xi))|\nom\\
&=&|\theta x+t\xi+(1-\theta)\gamma(t,0,\xi)+\theta\gamma(t,x,\xi)|\nom\\
&\ge& c_0(\theta|x|+t|\xi|)-c_1\rho^\deo\min(\lan t\ran^{1-\del},|t|)\nom
\ene
for some constants $c_0,c_1>0$ independent of $x,\xi,\theta$ and $t\ge0$. Thus there are constants $\rho\in(0,d)$ and $T=T_{d,\sig_0}>0$ such that for all $t\ge T$ and $(x,\xi)\in \Gamma_+(R,d,\sig_0)$
\beq
\lan \nabla_\xi\phi(t,0,\xi)+\theta\nabla_\xi R(t,x,\xi)\ran^{-1}\le C\lan \theta|x|+t|\xi|\ran^{-1}.
\ene
Therefore $a(t,x,\xi)$ defined by \eq{def-a} satisfies by \eq{R-1} and \eq{nablaxi-phi}
\beq
|\parti_\xi^\be a(t,x,\xi)|\le C_\be \int_0^1\lan \theta|x|+t|\xi|\ran^{-1-\de}d\theta.\label{est-a}
\ene
Using \eq{nabphiR}, we see that \eq{est-a} holds also for $t\in[0,T]$ if we take $\rho>0$ small enough. Therefore for all $(x,\xi)\in\Gamma_+(R,d,\sig_0)$ we have from \eq{dtR} and \eq{R-1}
\beq
|\parti_\xi^\be(\phi_+(x,\xi)-x\cdot\xi)|&\le&C_{T,\be}\lan x\ran\int_0^\infty\int_0^1\lan \theta|x|+t|\xi|\ran^{-1-\de}d\theta dt\\
&\le&C_{T,\be}\lan x\ran|\xi|^{-1}\int_0^1\lan\theta|x|\ran^{-\de}d\theta\nom\\
&\le& C_{T,\be}\lan x\ran^{1-\de}|\xi|^{-1}.\nom
\ene

We next consider
\beq
\nabla_x\phi_+(x,\xi)-\xi&=&\lim_{t\to\infty}(\nabla_x\phi(t,x,\xi)-\xi)\\
&=&\lim_{t\to\infty}(p(0,t,y(0,t,x,\xi),\xi)-\xi)\nom\\
&=&\lim_{t\to\infty}\int_0^t(\nabla_xV_\rho)\left(\tau,q(\tau,t,y(0,t,x,\xi),\xi)\right)d\tau\nom\\
&=&\lim_{t\to\infty}\int_0^t(\nabla_x V_\rho)\left(\tau,q(\tau,0,x,\eta(t,0,x,\xi))\right)d\tau\nom\\
&=&\int_0^\infty (\nabla_x V_\rho)\left(\tau,q(\tau,0,x,\eta(\infty,0,x,\xi))\right)d\tau.\nom
\ene
By \eq{pst} and \eq{qts} of Proposition \ref{Estimates-for-qp}
\beq
|q(\tau,0,x,\eta(\infty,0,x,\xi))|&\ge&
|x+\tau p(\tau,0,x,\eta(\infty,0,x,\xi))|-C_0\rho^\deo\lan \tau\ran^{1-\del}\\
&\ge&
|x+\tau p(\tau,\infty,y(0,\infty,x,\xi),\xi)|-C_0\rho^\deo\lan \tau\ran^{1-\del}\nom\\
&\ge& |x+\tau\xi|-C_0\rho^\deo\lan \tau\ran^{1-\del}-C_0\rho^\deo\lan \tau\ran^{1-\del}.\nom
\ene
Thus taking $\rho>0$ sufficiently small and $R=R_{d,\sig_0,\rho}>1$ sufficiently large, we have for $(x,\xi)\in\Gamma_+(R,d,\sig_0)$
\beq
|q(\tau,0,x,\eta(\infty,0,x,\xi))|\ge c_0(|x|+\tau|\xi|)
\ene
for some constant $c_0>0$. Therefore we obtain
\beq
|\nabla_x\phi_+(x,\xi)-\xi|\le C\int_0^\infty\lan|x|+\tau|\xi|\ran^{-1-\de}d\tau
\le C|\xi|^{-1}\lan x\ran^{-\de}.
\ene

For higher derivatives, the proof is similar. For example let us consider
\beq
\parti_\xi\parti_x\phi_+(x,\xi)-I&=&\int_0^\infty  \parti_\xi\{(\nabla_x V_\rho)\left(\tau,q(\tau,0,x,\eta(\infty,0,x,\xi))\right)\}d\tau\\
&=&\int_0^\infty  (\nabla_x\nabla_x V_\rho)\left(\tau,q(\tau,0,x,\eta(\infty,0,x,\xi))\right) \nabla_\xi q\cdot \nabla_\xi\eta d\tau,\nom
\ene
where we abbreviated $q=q(\tau,0,x,\eta(\infty,0,x,\xi))$ and $\eta=\eta(\infty,0,x,\xi)$.
The RHS is bounded by a constant times
\beq
 \int_0^\infty\lan |x|+\tau|\xi|\ran^{-2-\de}\lan\tau\ran d\tau\le c |\xi|^{-1}\lan x\ran^{-\de}
\ene
for $(x,\xi)\in\Gamma_+(R,d,\sig_0)$ by \eq{q-2} and \eq{eta-2} of Propositions \ref{Estimates-for-qp} and \ref{Estimates-for-etay}.
Other estimates are proved similarly by using \eq{q-2}, \eq{q-3}, \eq{eta-2} and \eq{eta-3}. $\Box$

\BP

Now let $-1<\sig_-<\sig_+<1$ and take two functions $\psi_\pm(\sig)\in C^\infty([-1,1])$ such that
\beq
&&0\le \psi_\pm(\sig)\le 1,\\
&&\psi_+(\sig)=
\left\{
\begin{array}{ll}
1,&\q \sig_+\le \sig\le 1\\
0,&\q -1\le \sig\le \sig_-
\end{array}
\right.,
\\
&&
\psi_-(\sig)=1-\psi_+(\sig)=
\left\{
\begin{array}{ll}
0,&\q \sig_+\le\sig\le 1\\
1,&\q -1\le\sig\le\sig_-
\end{array}
\right.,
\ene
and set
\beq
\chi_\pm(x,\xi)=\psi_\pm(\cos(x,\xi)),\q \left(\cos(x,\xi)=\frac{x\cdot \xi}{|x||\xi|}\right).
\ene
We then 
define the phase function $\varphi(x,\xi)$ by
\beq
&&\varphi(x,\xi)\label{phase-def}\\
&&=\left\{(\phi_+(x,\xi)-x\cdot\xi)\chi_+(x,\xi)+(\phi_-(x,\xi)-x\cdot \xi)\chi_-(x,\xi)\right\}\phi(2\xi/d)\phi(2x/R)+x\cdot\xi,\nom
\ene
where $\phi(x)$ is the function defined by \eq{cutout}. $\varphi(x,\xi)$ is a $C^\infty$ function of $(x,\xi)\in R^{2m}$.

Noting that $\chi_+(x,\xi)+\chi_-(x,\xi)\equiv 1$ for $x\ne0,\xi\ne0$, we have proved the following theorem.

\begin{thm}\label{phase-amplitude}
Let the notations be as above. Then for any $d>0$ and $-1<\sig_-<\sig_+<1$, there is $R=R_d=R_{d\sig_\pm}>1$ such that $R_d>1$ increases as $d>0$ decreases and the followings hold:
\MP

\F
{\rm i)} For $|\xi|\ge d$, $|x|\ge R$ and $\cos(x,\xi)\ge \sig_+$ or $\cos(x,\xi)\le\sig_-$
\beq
\frac{1}{2}|\nabla_x\varphi(x,\xi)|^2+V_L(x)=\frac{1}{2}|\xi|^2.\label{eikonal-whole}
\ene
\MP

\F
{\rm ii)} For any multi-indices $\al,\be$ there is a constant $C_{\al\be}>0$ such that
\beq
|\parti_x^\al\parti_\xi^\be(\varphi(x,\xi)-x\cdot\xi)|\le C_{\al\be}\lan x\ran^{1-\de-|\al|}\lan \xi\ran^{-1}.\label{varphi-est-1}
\ene
In particular for $|\al|\ne 0$, we have for $\de_0,\de_1\ge0$ with $\de_0+\de_1=\de$
\beq
|\parti_x^\al\parti_\xi^\be(\varphi(x,\xi)-x\cdot\xi)|\le C_{\al\be}R^{-\deo}\lan x\ran^{1-\del-|\al|}\lan \xi\ran^{-1}.\label{varphi-est-2}
\ene
\MP

\F
{\rm iii)} Set
\beq
a(x,\xi)&=&e^{-i\varphi(x,\xi)}\left(-\frac{1}{2}\Delta+V_L(x)-\frac{1}{2}|\xi|^2\right)e^{i\varphi(x,\xi)}\label{amplitude}\\
&=&\frac{1}{2}|\nabla_x\varphi(x,\xi)|^2+V_L(x)-\frac{1}{2}|\xi|^2-\frac{i}{2}\Delta_x \varphi(x,\xi).\nom
\ene
Then $a(x,\xi)$ satisfies for $|\xi|\ge d$, $|x|\ge R$ and any $\al,\be$
\beq
|\parti_x^\al\parti_\xi^\be a(x,\xi)|\le
\left\{
\begin{array}{ll}
C_{\al\be}\lan x\ran^{-1-\de-|\al|}\lan \xi\ran^{-1},&\q \cos(x,\xi)\in[-1,\sig_-]\cup[\sig_+,1]\\
C_{\al\be}\lan x\ran^{-\de-|\al|},&\q \cos(x,\xi)\in[\sig_-,\sig_+]
\end{array}
\right.\label{amplitude-T}
\ene
\end{thm}
\BP

We can now define the identification operator $J$ for $f\in\SSS(R^m)$:
\beq
Jf(x)&=&\mbox{Os-}\int\int e^{i(\varphi(x,\xi)-y\xi)}f(y)dyd\wxi\\
&=& c_m\int e^{i\varphi(x,\xi)}{\hat f}(\xi)d\xi,\nom
\ene
where $c_m=(2\pi)^{-m/2}$. We remark that this definition of $J$ depends on the choice of the constants $d>0$, $R=R_{d\sig_\pm}>1$ and $\sig_-,\sig_+$ with $-1<\sig_-<\sig_+<1$ by the definition of $\varphi(x,\xi)$ in \eq{phase-def}. But when two phase functions $\varphi_{d_1,R_{d_1}}$ and $\varphi_{d_2,R_{d_2}}$ corresponding to the constants $d_2>d_1>0$ with the same $\sig_\pm$ are given, they coincide with each other on the common region $\Gamma_\pm(R_{d_1}, d_2,\sig_\pm)$, since the limits \eq{limit-phi} exist for all $(x,\xi)\in R^{2m}$. In the following we fix a pair $(\sig_-,\sig_+)$ with $-1<\sig_-<\sig_+<1$ but vary the constants $d>0$ and $R=R_d>1$ in accordance with the context, and write $J=J_d$ when necessary to denote $J$ with the phase function satisfying \eq{eikonal-whole} for $|\xi|\ge d$. The function $a(x,\xi)$ in iii) of the theorem satisfies for $f\in\SSS$
\beq
Tf(x)&=&(HJ-JH_0)f(x)\\
&=&\int\int e^{i(\varphi(x,\xi)-y\xi)}\{a(x,\xi)+V_S(x)\}f(y)dyd\wxi\nom\\
&=& c_m \int e^{i\varphi(x,\xi)}\{a(x,\xi)+V_S(x)\}{\hat f}(\xi) d\xi.\nom
\ene
Thus Theorem \ref{phase-amplitude} tells that $T$ satisfies the properties required for our arguments of the asymptotic completeness that we have stated at the beginning of this section.

To see that the wave operators \eq{modified-wave-operators} define bounded operators, we prove that $J$ is bounded.
To this end, since $J$ is densely defined in $\HH=L^2(R^m)$, it suffices to show that the adjoint operator $J^*$ is bounded. It is defined for $f\in\SSS$ by
\beq
J^* f(x)=\mbox{Os-}\int\int e^{i(x\xi-\varphi(y,\xi))}f(y)dyd\wxi.
\ene
By the inequality (see the proof of Theorem \ref{L^2-bound})
\beq
\Vert J^*\Vert^2 \le \Vert J^{**}J^*\Vert,
\ene
we have only to prove the boundedness of the operator
\beq
J^{**}J^*f(x)
=\mbox{Os-}\int\int e^{i(\varphi(x,\xi)-\varphi(y,\xi))}f(y)dyd\wxi.\label{J**J*}
\ene
We compute
\beq
\varphi(x,\xi)-\varphi(y,\xi)&=&(x-y)\cdot\int_0^1 \nabla_x\varphi(y+\theta(x-y),\xi)d\theta\label{varphi-x-xi-y}\\
&=:&(x-y)\cdot {\widetilde \nabla}_x\varphi(x,\xi,y).\nom
\ene
By \eq{varphi-est-2}, we have
\beq
|\nabla_\xi{\widetilde \nabla}_x\varphi(x,\xi,y)-I_m|\le CR^{-\deo},\label{varphi-I}
\ene
where $I_m$ is the unit matrix of order $m$.
Thus switching to a larger $R=R_d>1$ if necessary, we can make the RHS of \eq{varphi-I} less than $1/2$. Then if we set $G_\eta(\xi)=\xi-{\widetilde \nabla}_x\varphi(x,\xi,y)+\eta$, we have
\beq
|G_\eta(\xi)-G_\eta(\xi')|&=&|(I_m-\nabla_\xi {\widetilde \nabla}_x\varphi(x,\xi,y))(\xi-\xi')|\\
&\le&\frac{1}{2}|\xi-\xi'|.\nom
\ene
 Therefore $G_\eta(\xi)$ is a contraction mapping from $R^m$ into $R^m$, and there is a unique fixed point $\xi$ of $G_\eta(\xi)$ for any $\eta\in R^m$:
\beq
G_\eta(\xi)=\xi,\q \mbox{i.e.}\q \eta={\widetilde \nabla}_x\varphi(x,\xi,y).
\ene
Thus the inverse ${\widetilde \nabla}_x\varphi^{-1}(x,\eta,y)$ of the mapping
\beq
R^m\ni\xi\mapsto \eta={\widetilde \nabla}_x\varphi(x,\xi,y)\in R^m
\ene
exists and defines a diffeomorphism of $R^m$. We make a change of variable in \eq{J**J*} by this transformation. Then denoting the Jacobian of ${\widetilde \nabla}_x\varphi^{-1}(x,\eta,y)$ by $J(x,\eta,y)=|\det(\nabla_\eta{\widetilde \nabla}_x\varphi^{-1}(x,\eta,y))|$, we obtain
\beq
J^{**}J^*f(x)=\mbox{Os-}\int\int e^{i(x-y)\eta}J(x,\eta,y)f(y)dyd\weta.\label{J**J*-eta}
\ene
Since \eq{varphi-est-1} implies the estimates
\beq
|\parti_x^\al\parti_\eta^\be\parti_y^\gamma J(x,\eta,y)|\le C_{\al\be\gamma}
\ene
for all $\al,\be,\gamma$ with constants $C_{\al\be\gamma}>0$ independent of $(x,\eta,y)$, we have from Theorem \ref{L^2-bound} that $J^{**}J^*$ is a bounded operator on $\HH=L^2(R^m)$. This proves that $J$ is extended to a bounded operator from $\HH$ into itself and $J=J^{**}$.

We next prove that $J$ has a bounded inverse. From \eq{J**J*-eta}, we have
\beq
(I-JJ^*)f(x)=\mbox{Os-}\int\int e^{i(x-y)\eta}(1-J(x,\eta,y))f(y)dyd\weta.
\ene
Let $m_0=2[m/2+1]$ as in section \ref{propagationest} and $C_0>0$ be the constant in \eq{L2bound}.
Then we have by
\eq{varphi-est-2}
\beq
\sup_{|\al+\be+\gamma|\le 3m_0}\sup_{x,\eta,y}|\parti_x^\al\parti_\eta^\be\parti_y^\gamma(1-J(x,\eta,y))|\le C_{m_0}R^{-\deo}<\frac{1}{2C_0}\label{I-JJ*}
\ene
by taking $R>1$ large enough.
Then we have from Theorem \ref{L^2-bound} that
\beq
\Vert I-JJ^* \Vert\le \frac{1}{2}.\label{1/2-bound}
\ene
Thus $JJ^*$ is invertible with the inverse
\beq
(JJ^*)^{-1}=(I-(I-JJ^*))^{-1}=\sum_{j=0}^\infty (I-JJ^*)^j,\label{P^{-1}}
\ene
whose RHS converges in operator norm by \eq{1/2-bound}. This implies that the range ${\cal R}(J)$ equals $\HH$ and $J^*$ is one-to-one.
Furthermore, \eq{I-JJ*} implies that the symbol $r(x,\eta,y)=1-J(x,\eta,y)$ of the $\psi$do $I-JJ^*$ is small so that the series of symbols
\beq
q_1(x,\eta,y)=\sum_{j=0}^\infty r_j(x,\eta,y)\label{sum-for-inverse}
\ene
converges in the Fr\'eche space $S_{0,0}$ of symbols $p(x,\eta,y)$ whose semi-norms are
\beq
|p|_\ell=\sup_{|\al+\be+\gamma|\le \ell}\sup_{x,\eta,y}|\parti_x^\al\parti_\eta^\be\parti_y^\gamma p(x,\eta,y)|<\infty\q (\ell=0,1,2,\cdots).
\ene
Here in \eq{sum-for-inverse}, $r_j(x,\eta,y)$ is the symbol of the $\psi$do $(I-JJ^*)^j$ $(j=0,1,2,\cdots)$. To see that \eq{sum-for-inverse} converges in $S_{0,0}$, we note that the inequality \eq{nu-prod} stated before Theorem \ref{L^2-bound} implies
\beq
|r_j|_\ell
\le (C_0)^j\sum_{|\ell_1+\cdots+\ell_j|\le \ell}\prod_{k=1}^j|r|_{3m_0+|\ell_k|}
\le (C_0)^j (|r|_{3m_0})^{j-\ell} (|r|_{3m_0+\ell})^\ell
\left(\sum_{|\ell_1+\cdots+\ell_j|\le \ell} 1\right),
\ene
where $r(x,\eta,y)=r_1(x,\eta,y)=(1-J)(x,\eta,y)$.
Since
\beq
\sum_{|\ell_1+\cdots+\ell_j|\le \ell} 1=
\sum_{k=0}^\ell
\left(
\begin{array}{c}
3j+k-1\\
k
\end{array}
\right)
\le C_\ell j^\ell
\ene
for some constant $C_\ell>0$ independent of $j=0,1,2,\cdots$ (recall that $\ell_i$'s are 3-dimensional multi-indices in \eq{nu-prod}),
we have
\beq
|r_j|_\ell\le C_\ell
 j^\ell (C_0|r|_{3m_0})^{j-\ell}(C_0|r|_{3m_0+\ell})^\ell.
\ene
Thus \eq{I-JJ*}:
\beq
C_0|r|_{3m_0}<\frac{1}{2},
\ene
implies
 the convergence of the series \eq{sum-for-inverse} in the symbol space $S_{0,0}$, and we have from \eq{P^{-1}} with $Q_1=q_1(X,D_x,X')$
\beq
Q_1(JJ^*)=(JJ^*)Q_1=I.\label{PJJ*=I}
\ene
We next consider for $g\in\SSS$
\beq
\FF J^*J\FF^{-1}g(\xi)=\mbox{Os-}\int\int e^{-i(\varphi(y,\xi)-\varphi(y,\eta))}g(\eta)dyd\weta.
\ene
Similarly to \eq{varphi-x-xi-y}, we write
\beq
\varphi(y,\xi)-\varphi(y,\eta)=(\xi-\eta)\cdot{\widetilde \nabla}_\xi\varphi(\xi,y,\eta),\label{varphi-xi-y-eta}
\ene
where
\beq
{\widetilde \nabla}_\xi\varphi(\xi,y,\eta)=\int_0^1\nabla_\xi\varphi(y,\eta+\theta(\xi-\eta))d\theta.
\ene
Then noting that the inequality similar to \eq{varphi-I} holds also for ${\widetilde \nabla}_\xi\varphi(\xi,y,\eta)$, we make a change of variable:
\beq
z={\widetilde \nabla}_\xi\varphi(\xi,y,\eta),
\ene
and obtain
\beq
\FF J^*J\FF^{-1}g(\xi)=\mbox{Os-}\int\int e^{-i(\xi-\eta)z}J(\xi,z,\eta)g(\eta)dzd\weta,
\ene
Here $J(\xi,z,\eta)=|\det(\nabla_y{\widetilde \nabla}_\xi\varphi^{-1}(\xi,z,\eta))|$ is a Jacobian, which belongs to $S_{0,0}$.
Arguing similarly to the case of $JJ^*$, we can now construct a $\psi$do $Q_2=q_2(X,D_x,X')$ that satisfies $q_2\in S_{0,0}$ and
\beq
Q_2(J^*J)=(J^*J)Q_2=I.\label{Q'J*J}
\ene
\eq{Q'J*J} and \eq{PJJ*=I} show that $J$ has an inverse
\beq
J^{-1}:\HH\to\HH
\ene
that is expressed as
\beq
J^{-1}=J^*Q_1=Q_2J^*=q_2(X,D_x,X')J^*.\label{J^*Q}
\ene
Since $J^*$ is bounded as we have seen, and $Q_2=q_2(X,D_x,X')$ is bounded on $\HH$ by $q_2\in S_{0,0}$ and Theorem \ref{L^2-bound}, $J^{-1}=Q_2J^*$ is also a bounded operator on $\HH$.
\BP

We are now prepared to prove the existence and asymptotic completeness of the wave operators \eq{modified-wave-operators}. Before going to the proof, we remark that the definition \eq{modified-wave-operators} should be understood as follows with taking into account the dependence of the identification operator $J=J_d$ on $d>0$. Namely for $f\in \FF^{-1}C_0^\infty(R^m)$, the support of whose Fourier transform is contained in a set $\Sigma(d):=\{\xi|\ |\xi|\ge d\}$, we define $W_\pm f$ by \eq{modified-wave-operators} with taking $J=J_d$. Noting that the phase function $\varphi(x,\xi)$ of $J_d$ is equal to
\beq
\varphi(x,\xi)=
\phi_+(x,\xi)\chi_+(x,\xi)+\phi_-(x,\xi)\chi_-(x,\xi)
\ene
in $\Sigma(d)\cap\{x|\ |x|\ge R\}$,
we have a definition of $W_\pm$ independent of $d>0$ and $R=R_d>1$ by extending this $W_\pm$ to the whole space $\HH$ by preserving the boundedness. 

\BP

Since the proof of the existence of $W_\pm$ is quite similar to and simpler than that of the asymptotic completeness, we only prove the latter. We have already stated the outline of the proof of the asymptotic completeness at the beginning of this section. There what should be noted is that we prove the existence of the limit \eq{Omega} for $g=E_H(B)g$, where the Borel set $B$ is a subset of $(d^2/2,b)$ for some $d>0$ with $0<d^2/2<b<\infty$. Then by \eq{Enss-5.2.3}, the energy restriction $E_H(B)$ is translated into the restriction $E_0(B)$ on the state $e^{-it_mH}g=E_H(B)e^{-it_mH}g$ as $m\to\infty$. Thus asymptotically as $t_m\to\infty$, we can assume that the $\xi$-support of the symbol $p_+(\xi,y)$ of $P_+$ satisfies \eq{energy-cut-off-2} with $\sig=d/2$ on the state $e^{-it_mH}g$. Then we can let $J=J_d$ in \eq{difference-Enss}, and take $P_+$ thereafter so that its symbol $p_+(\xi,y)$ satisfies \eq{p+}, \eq{p+restriction}, and \eq{energy-cut-off-2} with $\sig=d/2$. What remains to be proved is then the estimation of the following factor in \eq{asymptotic-c-estimate}:
\beq
\Vert Te^{-isH_0}J^{-1}P_+\lan x\ran^{\de/2}\Vert.
\ene
By \eq{J^*Q} we have
\beq
Te^{-isH_0}J^{-1}P_+\lan x\ran^{\de/2}=
Te^{-isH_0}Q_2J^{*}P_+\lan x\ran^{\de/2}.
\ene
By \eq{Enss-5.2.2}, the constant $\theta+\rho$ in \eq{p+} can be taken arbitrarily as far as $-1<\theta+\rho<1$. We here take $\theta+\rho=\sig_+ + \rho<1$ with $\rho>0$, where $\sig_+\in(-1,1)$ is the number specified in Theorem \ref{phase-amplitude}. We write
\beq
J^*P_+\lan x\ran^{\de/2}=J^*P_+\lan x\ran^{\de/2}JJ^{-1}.
\ene
The last factor $J^{-1}$ is bounded and can be omitted in the estimation. So we have to estimate
\beq
Te^{-isH_0}Q_2J^{*}P_+\lan x\ran^{\de/2}J.
\ene
By a calculation
\beq
\FF J^{*}P_+\lan x\ran^{\de/2}J\FF^{-1}{\hat f}(\xi)=\mbox{Os-}\int\int
e^{-i(\varphi(y,\xi)-\varphi(y,\eta))}
r_+(y,\eta){\hat f}(\eta)dyd\weta.\label{J*PJ}
\ene
Here the symbol $r_+(y,\eta)$ is given by
\beq
r_+(y,\eta)=\mbox{Os-}\int\int e^{i(y-z)\zeta}p_+(\zeta+{\widetilde \nabla}_x\varphi(y,\eta,z),z)\lan z\ran^{\de/2}dzd\wzeta\label{r+yeta}
\ene
and satisfies by \eq{p+restriction}
\beq
|r_+|_\ell^{(-1,\de/2)}:=\sup_{|\al+\be|\le \ell}\sup_{y,\eta}|\lan y\ran^{-\de/2}\lan y\ran^{|\al|}\parti_y^\al\parti_\eta^\be r_+(y,\eta)|<\infty,\q(\ell=0,1,2,\cdots).
\ene
We denote by $S_{-1,\de/2}$ the space of symbols that satisfy this estimate.
As in \eq{varphi-xi-y-eta}, we write
\beq
\varphi(y,\xi)-\varphi(y,\eta)=(\xi-\eta)\cdot{\widetilde \nabla}_\xi\varphi(\xi,y,\eta),
\ene
where
\beq
{\widetilde \nabla}_\xi\varphi(\xi,y,\eta)=\int_0^1\nabla_\xi\varphi(y,\eta+\theta(\xi-\eta))d\theta.
\ene
We then make a change of variable:
\beq
z={\widetilde \nabla}_\xi\varphi(\xi,y,\eta).
\ene
Letting $J(\xi,z,\eta)$ be the Jacobian of the inverse mapping ${\widetilde \nabla}_\xi\varphi^{-1}(\xi,z,\eta)$, we write \eq{J*PJ} as
\beq
\FF J^{*}P_+\lan x\ran^{\de/2}J\FF^{-1}{\hat f}(\xi)=\mbox{Os-}\int\int
e^{-i(\xi-\eta)z}
{\tilde r}_+(\xi,z,\eta){\hat f}(\eta)dzd\weta,\label{J*SJ-2}
\ene
where
\beq
{\tilde r}_+(\xi,z,\eta)=r_+({\widetilde \nabla}_\xi\varphi^{-1}(\xi,z,\eta),\eta)J(\xi,z,\eta).\label{tilde-r+}
\ene
By our additional assumption \eq{p+restriction} on $p_+(\xi,y)$ stated at the beginning of this section and by the estimates \eq{varphi-est-1} for $\varphi(x,\xi)$, ${\tilde r}_+(\xi,z,\eta)$ belongs to the symbol space $S_{0,\de/2}$ whose semi-norms are
\beq
|{\tilde r}_+|_\ell^{(0,\de/2)}
=\sup_{|\al+\be+\gamma|\le\ell}\sup_{\xi,z,\eta}|\lan z\ran^{-\de/2}\parti_\xi^\al\parti_z^\be\parti_\eta^\gamma {\tilde r}_+(\xi,z,\eta)|<\infty,\q (\ell=0,1,2,\cdots).
\ene
Further, by the property \eq{p+} of $p_+$, ${\tilde r}_{+L}$ defined as in Proposition \ref{pseudo-pro} from ${\tilde r}_{+}$ satisfies
\beq
|\parti_\xi^\al\parti_z^\be {\tilde r}_{+L}(\xi,z)|\le C_{\ell\al\be}\lan z\ran^{-\ell}\q 
(\cos(z,\xi)<\sig_+ +\rho/2)\label{r-1}
\ene
for any $\ell=0,1,2,\cdots$. We return to the configuration space by Fourier inversion and obtain
\beq
J^{*}P_+\lan x\ran^{\de/2}J f(x)&=&\mbox{Os-}\int\int
e^{i(x-z)\xi}
{s}(\xi,z)\lan z\ran^{\de/2}f(z)dzd\wxi\\
&=&s(D_x,X')\lan x\ran^{\de/2}f(x),\nom
\ene 
where $s(\xi,z)\in S_{0,0}$ satisfies \eq{r-1} with $\sig_++\rho/2$ replaced by $\sig_++\rho/3$ by virtue of \eq{r-1}, \eq{tilde-r+}, \eq{r+yeta}, supp $p_+\subset\{(\xi,y)|\ |\xi|\ge d/2, |y|\ge d/2\}$, and $J(\xi,z,\eta)\in S_{0,0}$. Thus the problem is reduced to the estimation of
\beq
\Vert Te^{-isH_0}Q_2s(D_x,X')\lan x\ran^{\de/2}f\Vert
\le\Vert(J^*)^{-1}\Vert\Vert J^*Te^{-isH_0}Q_2s(D_x,X')\lan x\ran^{\de/2}f\Vert
.\label{final-asymp}
\ene
Arguing as above and reducing $J^*T$ to a $\psi$do, and recalling the estimate
 \eq{amplitude-T} for the symbol of $T$ in Theorem \ref{phase-amplitude}, we can bound the RHS by a constant times $\lan s\ran^{-1-\de/2}$ by applying Theorems \ref{Propa-2} and \ref{Propa-3}, as announced in \eq{asymptotic-c-estimate}. The proof of the asymptotic completeness is complete:

\begin{thm} Let \eq{short-range} and \eq{long-range} be satisfied. Let $J$ be defined as above. Then the wave operators
\beq
W_\pm=\mbox{s-}\lim_{t\to\pm\infty}e^{itH}Je^{-itH_0}
\ene
exist and define bounded operators on $\HH$, and
the asymptotic completeness holds:
\beq
{\cal R}(W_\pm)=\HH_{ac}(H)=\HH_c(H).
\ene
Further $W_\pm$ intertwine $H$ and $H_0$: for any Borel set $B$ in $R^1$
\beq
E_H(B)W_\pm=W_\pm E_0(B).
\ene
\end{thm}

\BP

We reformulate our problem as follows.

Let an inner product $(\cdot,\cdot)_J$ in $\HH$ be defined by
\beq
(f,g)_J=(Jf,Jg)_\HH.
\ene
It is clear that the space $\HH$ becomes a Hilbert space $\HH_J$ with this inner product. Further as $J$ and $J^{-1}$ are bounded operators from $\HH$ onto $\HH$, the two inner products $(\cdot,\cdot)_\HH$ and $(\cdot,\cdot)_J$ are equivalent.

We consider an operator
\beq
H_J=J^{-1}HJ\label{H-J}
\ene
in $\HH_J$. Then it is clear that $H_J$ is a selfadjoint operator in $\HH_J$. And we have proved in the above that the limit
\beq
W_\pm^Jf=\lim_{t\to\pm \infty}e^{itH_J}e^{-itH_0}f\label{Wave-J}
\ene
exists for all $f\in \HH$, and that it is asymptotic complete
\beq
{\cal R}(W_\pm^J)=\HH_c(H)=\HH_c(H_J),\label{completeness-J}
\ene
where $\HH_c(H_J)$ is understood to be defined in the space $\HH_J$.

Those mean that we can regard that $H_J$ is a Hamiltonian obtained from $H_0$ with being perturbed by a sort of ``short-range" perturbation $V_J=H_J-H_0$ between the spaces $\HH$ and $\HH_J$:
\beq
H_J=H_0+V_J.\label{short-long}
\ene

\chapter{Many-Body Hamiltonian}\label{chap:7}

We begin with some repetition of notations given in chapter \ref{chap:2} and section \ref{section3.2} with modifications to the present situation.

\section{Preliminaries}

We consider the Schr\"odinger operator defined in $L^2(R^{\nu N})$ ($\nu\ge 1$, $N\ge 2$)
\beq
H=H_0+V,\q H_0=-\sum_{i=1}^N \frac{\hbar^2}{2m_i}\frac{\parti^2}{\parti r_i^2}.\label{tag 1.1}
\ene 
Here
\beq
V=\sum_\al V_\al(x_\al),\label{tag 1.2}
\ene
where $x_\al=r_i-r_j$, $r_i=(r_{i1},\cdots,r_{i\nu})\in R^\nu$ is the position vector of the $i$-th particle, $\frac{\parti}{\parti r_i}=\left(\frac{\parti}{\parti r_{i1}},\cdots,\frac{\parti}{\parti r_{i\nu}}\right)$, $\frac{\parti^2}{\parti r_i^2}=\sum_{j=1}^\nu \frac{\parti^2}{\parti r_{ij}^2}=\Delta_{r_i}$, $m_i>0$ is the mass of the $i$-th particle, and $\al=\{i,j\}$ is a pair with $1\le i<j\le N$. Our assumption on the decay rate of the pair potentials $V_\al(x_\al)$ is as follows.

\begin{ass}\label{ass1}
$V_\al(x)$ ($x\in R^\nu$) is split into a sum of a real-valued $C^\infty$ function $V_\al^L(x)$ and a real-valued measurable function $V_\al^S(x)$ of $x\in R^\nu$ satisfying the following conditions: There are real numbers $\ep$ and $\ep_1$ with $0<\ep,\ep_1<1$ such that for all multi-indices $\beta$
\beq
|\parti_{x}^\be V_\al^L(x)|\le C_\be\lan x\ran^{-|\be|-\ep}\label{tag 1.3}
\ene
with some constants $C_\be>0$ independent of $x\in R^\nu$, and
\beq
\lan x\ran^{1+\ep_1}V_\al^S(x)(-\Delta_x+1)^{-1} \ \mbox{is a bounded operator in}\ L^2(R^\nu).\label{tag 1.4}
\ene
Here $\Delta_x$ is a Laplacian with respect to $x$, and $\lan x\ran$ is a $C^\infty$ function of $x$ such that
$
\lan x\ran
=|x|$ for $|x|\ge1$ and
$\ge \frac{1}{2}$ for $|x|<1$.
\end{ass}

We can adopt weaker conditions on the differentiability and decay rate for higher derivatives of the long-range part $V_\al^L(x)$, but for later convenience of exposition, we adopt this form in the present paper.

The free part $H_0$ of $H$ in \eq{tag 1.1} has various forms in accordance with our choice of coordinate systems. We use the so-called Jacobi coordinates. The center of mass of our $N$-particle system is
$$
X_C=\frac{m_1r_1+\cdots+m_Nr_N}{m_1+\cdots+m_N},
$$
and the Jacobi coordinates are defined by
\beq
x_i=r_{i+1}-\frac{m_1r_1+\cdots+m_i r_i}{m_1+\cdots+m_i},\q i=1,2,\cdots,N-1.\label{tag 1.5}
\ene
Accordingly the corresponding canonically conjugate momentum operators are defined by
$$
P_C=\frac{\hbar}{i}\frac{\parti}{\parti X_C},\q
p_i=\frac{\hbar}{i}\frac{\parti}{\parti x_i}.
$$
Using these new $X_C,P_C,x_i,p_i$, we can rewrite $H_0$ in \eq{tag 1.1} as
\beq
H_0={\tilde H_0}+H_C.\label{tag 1.6}
\ene
Here
$$
{\tilde H}_0=\sum_{i=1}^{N-1} \frac{1}{2\mu_i}p_i^2=-\sum_{i=1}^{N-1} \frac{\hbar^2}{2\mu_i}\Delta_{x_i},\q H_C=\frac{1}{\sum_{j=1}^N m_j}P_C^2,
$$
where $\mu_i>0$ is the reduced mass defined by the relation:
$$
\frac{1}{\mu_i}=\frac{1}{m_{i+1}}+\frac{1}{m_1+\cdots+m_i}.
$$
The new coordinates give a decomposition $L^2(R^{\nu N})=L^2(R^\nu)\otimes L^2(R^n)$ with $n=\nu(N-1)$ and in this decomposition, $H$ is decomposed
$$
H=H_C\otimes I+I\otimes {\tilde H},\q {\tilde H}={\tilde H}_0+V.
$$
$H_C$ is a Laplacian, so we consider ${\tilde H}$ in the Hilbert space $\HH=L^2(R^n)=L^2(R^{\nu(N-1)})$. We write this ${\tilde H}$ as $H$ in the followings:
\beq
H=H_0+V=\sum_{i=1}^{N-1}\frac{1}{2\mu_i}p_i^2+\sum_{\al}V_\al(x_\al)=-\sum_{i=1}^{N-1}\frac{\hbar^2}{2\mu_i}\Delta_{x_i}+\sum_\al V_\al(x_\al).\label{tag 1.7}
\ene
This means that we consider the Hamiltonian $H$ in \eq{tag 1.1} restricted to the subspace of $R^{\nu N}$:
\beq
(m_1+\cdots+m_N)X_C=m_1r_1+\cdots+m_N r_N=0.\label{tag 1.8}
\ene
We equip this subspace with the inner product:
\beq
\lan x,y\ran=\sum_{i=1}^{N-1} \mu_i x_i\cdot y_i,\label{tag 1.9}
\ene
where $\cdot$ denotes the Euclidean scalar product. With respect to this inner product, the changes of variables between Jacobi coordinates in \eq{tag 1.5} are realized by orthogonal transformations on the space $R^n$ defined by \eq{tag 1.8}, while $\mu_i$ and $x_i$ depend on the order of the construction of the Jacobi coordinates in \eq{tag 1.5}. If we define velocity operator $v$ by
$$
v=(v_1,\cdots,v_{N-1})=(\mu_1^{-1}p_1,\cdots,\mu_{N-1}^{-1}p_{N-1}),
$$
we can write using the inner product above
\beq
H_0=\frac{1}{2}\lan v,v\ran.\label{tag 1.10}
\ene

Next we introduce clustered Jacobi coordinate.
 Let $a=\{C_1,\cdots,C_k\}$ be a disjoint decomposition of the set $\{1,2,\cdots,N\}$: 
$C_j \ne \emptyset$
 $(j=1,2,\cdots,k)$, $\cup_{j=1}^k C_j=\{1,2,\cdots,N\}$ with $C_i\cap C_j=\emptyset$ when $i\ne j$. We denote the number of elements of a set $S$ by $|S|$. Then $|a|=k$ in the present case, and we call $a$ a cluster decomposition with $|a|$ clusters $C_1,\cdots,C_{|a|}$.
A clustered Jacobi coordinate $x=(x_a,x^a)$ associated with a cluster decomposition $a=\{C_1,\cdots,C_k\}$ is obtained by first choosing a Jacobi coordinate
$$
x^{(C_\ell)}=(x_1^{(C_\ell)},\cdots,x_{|C_\ell|-1}^{(C_\ell)})\in R^{\nu(|C_\ell|-1)}\q (\ell=1,2,\cdots,k)
$$
for the $|C_\ell|$ particles in the cluster $C_\ell$ and then by choosing an intercluster Jacobi coordinate
$$
x_a=(x_1,\cdots,x_{k-1})\in R^{\nu(k-1)}
$$
for the center of mass of the $k$ clusters $C_1,\cdots,C_k$. Then $x^a=(x^{(C_1)},\cdots,x^{(C_k)})\in R^{\nu(N-k)}$ and $x=(x_a,x^a)\in R^{\nu(N-1)}=R^n$. The corresponding canonically conjugate momentum operator is
\beq
p&=&(p_a,p^a),\q p_a=(p_1,\cdots,p_{k-1}),\q p^a=(p^{(C_1)},\cdots,p^{(C_k)}),\nonumber\\
p_i&=&\frac{\hbar}{i}\frac{\parti}{\parti x_i},\q
p^{(C_\ell)}=(p_1^{(C_\ell)},\cdots,p_{|C_\ell|-1}^{(C_\ell)}),\q p_i^{(C_\ell)}=\frac{\hbar}{i}\frac{\parti}{\parti x_i^{(C_\ell)}}.\nonumber
\ene
Accordingly $\HH=L^2(R^n)$ is decomposed:
\beq
\HH&=&\HH_a\otimes \HH^a,\q \HH_a=L^2(R^{\nu(k-1)}_{x_a}),\nonumber\\ 
\HH^a&=&L^2(R^{\nu(N-k)}_{x^a})=\HH^{(C_1)}\otimes\cdots\otimes \HH^{(C_k)},\q
\HH^{(C_\ell)}=L^2(R^{\nu(|C_\ell|-1)}_{x^{(C_\ell)}}).\label{tag 1.11}
\ene
In this coordinates system, $H_0$ in \eq{tag 1.7} is decomposed:
\beq
H_0&=&T_a+H_0^a,\nom\\
T_a&=&-\sum_{\ell=1}^{k-1}\frac{\hbar^2}{2M_\ell}\Delta_{x_\ell},\label{tag 1.12}\\
H_0^a&=&\sum_{\ell=1}^k H_0^{(C_\ell)},\q H_0^{(C_\ell)}=-\sum_{i=1}^{|C_\ell|-1}\frac{\hbar^2}{2\mu_i^{(C_\ell)}}\Delta_{x_i^{(C_\ell)}},\nom
\ene
where
$\Delta_{x_\ell}$ and $\Delta_{x_i^{(C_\ell)}}$ are $\nu$-dimensional Laplacians and $M_\ell$ and $\mu_i^{(C_\ell)}$ are the reduced masses. We introduce the inner product in the space $R^n=R^{\nu(N-1)}$ as in \eq{tag 1.9}:
\beq
\lan x,y\ran&=&\lan (x_a,x^a),(y_a,y^a)\ran=\lan x_a,y_a\ran+\lan x^a,y^a\ran\nom\\
&=&\sum_{\ell=1}^{k-1}M_\ell x_\ell\cdot y_\ell + \sum_{\ell=1}^k\sum_{i=1}^{|C_\ell|-1}\mu_i^{(C_\ell)}x_i^{(C_\ell)}\cdot y_i^{(C_\ell)},\label{tag 1.13}
\ene
and velocity operator
\beq
v=(v_a,v^a)=M^{-1}p=(m_a^{-1}p_a,(\mu^a)^{-1}p^a),\label{tag 1.14}
\ene
where 
\beq
M=\left(\begin{array}{cc}
m_a & 0 \\
0 & \mu^a 
\end{array}\right)
\ene
 is the $n=\nu(N-1)$ dimensional diagonal mass matrix whose diagonals are given by
$M_1,\cdots,M_{k-1},\mu_1^{(C_1)},\cdots,\mu_{|C_k|-1}^{(C_k)}$. Then $H_0$ is written as
\beq
H_0=\frac{1}{2}\lan v,v\ran=T_a+H_0^a=\frac{1}{2}\lan v_a,v_a\ran+\frac{1}{2}\lan v^a,v^a\ran.\label{tag 1.15}
\ene

We need a notion of order in the set of cluster decompositions. A cluster decomposition $b$ is called a refinement of a cluster decomposition $a$, iff any $C_\ell\in b$ is a subset of some $D_k\in a$. When $b$ is a refinement of $a$ we denote this as $b\le a$. $b\not\leq a$ is its negation: some cluster $C_\ell\in b$ is not a subset of any $D_k\in a$. Thus for a pair $\al=\{i,j\}$, $\al\le a$ means that $\al=\{i,j\}\subset D_k$ for some $D_k\in a$, and $\al\not\leq a$ means that $\al=\{i,j\}\not\subset D_k$ for any $D_k\in a$. $b<a$ means that $b\le a$ but $b\ne a$.

We decompose the potential term $V$ in \eq{tag 1.7} as
\beq
\sum_\al V_\al(x_\al)=V_a+I_a,\label{tag 1.16}
\ene
where
\beq
V_a&=&\sum_{C_\ell\in a}V_{C_\ell},\nom\\
V_{C_\ell}&=&\sum_{\al\subset C_\ell}V_\al(x_\al),\label{tag 1.17}\\
I_a&=&\sum_{\al\not\leq a} V_\al(x_\al).\nom
\ene
By definition, $V_{C_\ell}$ depends only on the variable $x^{(C_\ell)}$ inside the cluster $C_\ell$. Similarly, $V_a$ depends only on the variable $x^a=(x^{(C_1)},\cdots,x^{(C_k)})\in R^{\nu(N-|a|)}$, while $I_a$ depends on all components of the variable $x$.

Then $H$ in \eq{tag 1.7} is decomposed:
\beq
H&=&H_a + I_a=T_a\otimes I+ I\otimes H^a + I_a,\nom\\
H_a&=&H-I_a=T_a\otimes I+ I\otimes H^a,\label{tag 1.18}\\
H^a&=&H_0^a+V_a=\sum_{C_\ell\in a} H^{(C_\ell)},\q H^{(C_\ell)}=H_0^{(C_\ell)}+V_{C_\ell},\nom
\ene
where $T_a$ is an operator in $\HH_a=L^2(R^{\nu(k-1)}_{x_a})$, $H^a$ and $H_0^a$ are operators in $\HH^a=L^2(R^{\nu(N-k)}_{x^a})$, and $H^{(C_\ell)}$ and $H_0^{(C_\ell)}$ are operators in $\HH^{(C_\ell)}=L^2(R^{\nu(|C_\ell|-1)}_{x^{(C_\ell)}})$.

We denote by $P_a$ the orthogonal projection onto the pure point spectral subspace (or eigenspace) $\HH_{pp}^a=\HH_{pp}(H^a)(\subset\HH^a)$ for $H^a$. We use the same notation $P_a$ for the obvious extention $I\otimes P_a$ to the total space $\HH$. For $|a|=N$, we set $P_a=I$. Let $M=1,2,\cdots$ and $P_a^M$ denote an $M$-dimensional partial projection of $P_a$ such that s-$\lim_{M\to\infty}P_a^M=P_a$. We define for $\ell=1,\cdots,N-1$ and an $\ell$-dimensional multi-index $M=(M_1,\cdots,M_\ell)$ ($M_j\ge 1$)
\beq
{\widehat P}_\ell^M=\left(I-\sum_{|a_\ell|=\ell}P_{a_\ell}^{M_\ell}\right)\cdots\left(I-\sum_{|a_2|=2}P_{a_2}^{M_2}\right)(I-P^{M_1}).\label{tag 1.19}
\ene
(Note that for $|a|=1$, $a=\{C\}$ with $C=\{1,2,\cdots,N\}$. Thus $P^{M_1}$ is an $M_1$-dimensional partial projection into the eigenspace of $H$.)
We further define for a $|a|$-dimensional multi-index $M_a=(M_1,\cdots,M_{|a|-1},M_{|a|})=({\widehat M}_a,M_{|a|})$
\beq
{\widetilde P}_a^{M_a}=P_a^{M_{|a|}}{\widehat P}_{|a|-1}^{{\widehat M}_a},\q
2\le |a| \le N.\label{tag 1.20}
\ene
Then it is clear that
\beq
\sum_{2\le |a|\le N}{\widetilde P}_a^{M_a}={\widehat P}_1^{M_1}=I-P^{M_1},\label{tag 1.21}
\ene
provided that the component $M_j$ of $M_a$ depends only on the number $j$ but not on $a$. In the following we use such $M_a$'s only.

Related with those notions, we denote by $\HH_c=\HH_c(H)$ the orthogonal complement $\HH_{pp}(H)^\perp$ of the eigenspace $\HH_{pp}=\HH_{pp}(H)$ for the total Hamiltonian $H$. Namely $\HH_c(H)$ is the continuous spectral subspace for $H$. We note that $\HH_c(H)=(I-P_a)\HH$ for a unique $a$ with $|a|=1$, and that for $f\in\HH$, $(I-P^{M_1})f\to (I-P_a)f\in \HH_c(H)$ as $M_1\to\infty$. We use freely the notations of functional analysis for selfadjoint operators, e.g. $E_H(\Delta)$ is the spectral measure for $H$.

To state a theorem due to Enss \cite{[En]}, we introduce an assumption:

\begin{ass}\label{ass2}
 For any cluster decomposition $a$ with $2\le |a|\le N-1$ and any integer $M=1,2,\cdots$,
\beq
\Vert |x^a|^2P_a^M\Vert<\infty.\label{tag 1.22}
\ene
\end{ass}

This assumption is concerned with the decay rate of eigenvectors of subsystem Hamiltonians. Since it is known that non-threshold eigenvectors decay exponentially (see Froese and Herbst \cite{[FH]}), this assumption is the one about threshold eigenvectors.

Let $v_a$, as above, denote the velocity operator between the clusters in $a$. It is expressed as $v_a=m_a^{-1}p_a$ for some $\nu(|a|-1)$-dimensional diagonal mass matrix $m_a$. Then we can state the theorem, which is the same as Theorem \ref{Enss}.

\begin{thm}\label{En}(\cite{[En]})
Let $N\ge 2$ and let $H$ be the Hamiltonian $H$ in \eq{tag 1.7} or \eq{tag 1.18} for an $N$-body quantum-mechanical system. Let Assumptions \ref{ass1} and \ref{ass2} be satisfied.
Let $f\in \HH$. Then there exist a sequence $t_m\to\pm\infty$ (as $m\to\pm\infty$) and a sequence $M_a^m$ of multi-indices whose components all tend to $\infty$ as $m\to\pm\infty$ such that for all cluster decompositions $a$ with $2\le |a|\le N$, for all $\varphi\in C_0^\infty(R_{x_a}^{\nu(|a|-1)})$, $R>0$, and $\alpha=\{i,j\}\not\leq a$
\beq
&&\left\Vert\frac{|x^a|^2}{t_m^2}{\widetilde P}_a^{M_a^m}e^{-it_mH/\hbar}f\right\Vert \to 0\label{tag 1.23}\\
&&\Vert F(|x_\alpha|<R){\widetilde P}_a^{M_a^m}e^{-it_mH/\hbar}f\Vert \to 0\label{tag 1.24}\\
&&\Vert (\varphi(x_a/t_m)-\varphi(v_a)){\widetilde P}_a^{M_a^m}e^{-it_mH/\hbar}f\Vert \to 0\label{tag 1.25}
\ene
as $m\to\pm\infty$. Here $F(S)$ is the characteristic function of the set defined by the condition $S$.
\end{thm}

We denote the sum of the sets of thresholds and eigenvalues of $H$ by $\TT$:
\beq
\TT=\bigcup_{1\le |a|\le N}\sig_p(H^a)={\tilde \TT}\cup\sig_p(H),\q {\tilde \TT}=\bigcup_{2\le |a|\le N}\sig_p(H^a)\label{tag 1.26}
\ene
where 
\beq
\sig_p(H^a)=\{\tau_1+\cdots+\tau_{|a|}\ |\ \tau_\ell\in\sig_p(H^{(C_\ell)})\ (C_\ell\in a)\}\label{tag 1.27}
\ene
is the set of eigenvalues of a subsystem Hamiltonian $H^a=\sum_{C_\ell\in a}H^{(C_\ell)}$. For $|a|=N$ we define $\sig_p(H^a)=\{0\}$. Similarly $\TT_a$ and ${\tilde \TT}_a$ are defined:
\beq
\TT_a=\bigcup_{b\le a}\sig_p(H^b)={\tilde \TT}_a\cup\sig_p(H^a),\q {\tilde \TT}_a=\bigcup_{b<a}\sig_p(H^b).\label{tag 1.28}
\ene
It is known (Froese and Herbst \cite{[FH]}) that these sets are subsets of $(-\infty,0]$. Further these sets form bounded, closed and countable subsets of $R^1$, and $\sig_p(H^a)$ accumulates only at ${\tilde \TT}_a$ (see Cycon {\it et al}. \cite{[C]}).

We use the notation $\Delta\subset\subset \Delta'$ for Borel sets $\Delta, \Delta'\subset R^k$ to mean that the closure $\bar{\Delta}$ of $\Delta$ is compact in $R^k$ and is a subset of the interior of $\Delta'$.
\BP

\vskip 12pt

\BP

\section{Scattering spaces}

In the following we consider the case $t\to\infty$ only. The other case $t\to-\infty$ is treated similarly. We also choose a unit system such that $\hbar=1$. We use the notation $f(t)\sim g(t)$ as $t\to\infty$ to mean that $\Vert f(t)-g(t)\Vert\to 0$ as $t\to\infty$ for $\HH$-valued functions $f(t)$ and $g(t)$ of $t>1$.

\begin{df}\label{def21} Let real numbers $r, \sig, \de$ and a cluster decomposition $b$ satisfy $0\le r\le 1$, $\sig, \de >0$ and $2\le |b|\le N$.

\F
i) Let $\Delta\subset\subset R^1-\TT$ be a closed set.
We define $S_b^{r\sig\de}(\Delta)$ for $0<r\le 1$ by
\beq
S_b^{r\sig\de}(\Delta)=\{f\in E_H(\Delta)\HH \ |\ 
 e^{-itH}f\sim \prod_{\al\not\leq b}F(|x_\al|\ge \sig t)F(|x^b|\le\de t^r)e^{-itH}f\ \mbox{as}\ t\to\infty\}.\label{tag 2.1}
\ene
For $r=0$ we define $S_b^{0\sig}(\Delta)$ by
\beq
&&S_b^{0\sig}(\Delta)=\{f\in E_H(\Delta)\HH \ |\ \nom\\
&&\lim_{R\to\infty}\limsup_{t\to\infty}\Bigl\Vert e^{-itH}f - \prod_{\al\not\leq b}F(|x_\al|\ge \sig t)F(|x^b|\le R)e^{-itH}f\Bigr\Vert=0 \}.\label{tag 2.2}
\ene
We then define the localized scattering space $S_b^r(\Delta)$ of order $r\in(0,1]$ for $H$ as the closure of
\beq
&&\bigcup_{\sig>0}\bigcap_{\de>0}S_b^{r\sig\de}(\Delta)
=\{f\in  E_H(\Delta)\HH \ |\ \exists \sig>0, \forall \de>0:\nom\\
&&\ e^{-itH}f\sim \prod_{\al\not\leq b}F(|x_\al|\ge \sig t)F(|x^b|\le\de t^r)e^{-itH}f\ \mbox{as}\ t\to\infty\}.\label{tag 2.3}
\ene
$S_b^0(\Delta)$ is defined as the closure of
\beq
&&\bigcup_{\sig>0}S_b^{0\sig}(\Delta)=\{f\in E_H(\Delta)\HH\ |\ \exists \sig>0:\nom\\
&&\lim_{R\to\infty}\limsup_{t\to\infty}\Bigl\Vert e^{-itH}f - \prod_{\al\not\leq b}F(|x_\al|\ge \sig t)F(|x^b|\le R)e^{-itH}f\Bigr\Vert=0 \}.\label{tag 2.4}
\ene

\F
ii) We define the scattering space $S_b^r$ of order $r\in [0,1]$ for $H$ as the closure of
\beq
\bigcup_{\Delta\subset\subset R^1-\TT}S_b^r(\Delta).\label{tag 2.5}
\ene
\end{df}

We note that $S_b^{r\sig\de}(\Delta)$, $S_b^{0\sig}(\Delta)$, $S_b^{r}(\Delta)$ and $S_b^{r}$ define closed subspaces of $E_H(\Delta)\HH$ and $\HH_c(H)$, respectively. 
\BP

\begin{pro}\label{pro22}
 Let $\Delta\subset\subset R^1-\TT$ and $f\in S_b^{r\sig\de}(\Delta)$ for $0< r\le 1$ or $f\in S_b^{0\sig}(\Delta)$ for $r=0$ with $\sig,\de>0$ and $2\le |b|\le N$. Then the following limit relations hold:
\MP

\F
i) Let $\al\not\leq b$. Then for $0<r\le 1$ we have when $t\to\infty$
\beq
F(|x_\al|<\sig t)F(|x^b|\le\de t^r)e^{-itH}f\to 0.\label{tag 2.6}
\ene
For $r=0$ we have
\beq
\lim_{R\to\infty}\limsup_{t\to\infty}\left\Vert F(|x_\al|< \sig t)F(|x^b|\le R)e^{-itH}f\right\Vert=0.\label{tag 2.7}
\ene
\F
ii) For $0<r\le 1$ we have when $t\to\infty$
\beq
F(|x^b|> \de t^r)e^{-itH}f\to 0.\label{tag 2.8}
\ene
For $r=0$
\beq
\lim_{R\to\infty}\limsup_{t\to\infty}\left\Vert F(|x^b|> R)e^{-itH}f\right\Vert=0.\label{tag 2.9}
\ene
\F
iii) There exists a sequence $t_m\to\infty$ as $m\to\infty$ depending on $f\in S_b^{r\sig\de}(\Delta)$ or $f\in S_b^{0\sig}(\Delta)$ such that
\beq
\left\Vert \left(\varphi\left({x_b}/{t_m}\right)-\varphi(v_b)\right)e^{-it_mH}f\right\Vert\to 0\q \mbox{as}\q m\to\infty \label{tag 2.10}
\ene
for any function $\varphi\in C_0^\infty(R^{\nu(|b|-1)}_{x_b})$.
\end{pro}
{\it Proof:}\ 
i) and ii) are clear from the definition of $S_b^{r\sig\de}(\Delta)$ or $S_b^{0\sig}(\Delta)$. We prove iii). 
Since $f\in E_H(\Delta)\HH\subset H_c(H)$, we have by \eq{tag 1.21}, Theorem \ref{Enss} and $f\in S_b^{r\sig\de}(\Delta)$ (or $f\in S_b^{0\sig}(\Delta)$)
\beq
e^{-it_mH}f\sim \sum_{d\le b}{\widetilde P}_d^{M_d^m}e^{-it_mH}f\label{tag 2.11}
\ene
along some sequence $t_m\to\infty$ depending on $f$. On each state on the right-hand side (RHS) of \eq{tag 2.11}, \eq{tag 1.25} with $a$ replaced by $d$ holds. By the restriction $d\le b$ in the sum of the RHS of \eq{tag 2.11}, we obtain \eq{tag 2.10}.\ $\Box$


\BP

The following propositions are obvious by definition.

\begin{pro}\label{pro2.3} Let $2\le |b|\le N$. If $1\ge r'\ge r> 0$, $\sig\ge \sig'>0$ and $\de'\ge \de>0$ and $\Delta\subset\subset R^1-\TT$, then $S_b^{0\sig}(\Delta)\subset S_b^{0\sig'}(\Delta)$, $S_b^{0\sig}(\Delta)\subset S_b^{r\sig\de}(\Delta)\subset S_b^{r'\sig'\de'}(\Delta)$, $S_b^0(\Delta)\subset S_b^r(\Delta)\subset S_b^{r'}(\Delta)$, $S_b^0(\Delta)\subset S_b^r(\Delta)\subset S_b^{r}$, and $S_b^0\subset S_b^r\subset S_b^{r'}$.
\end{pro}

\begin{pro}\label{pro2.4} Let $b$ and $b'$ be different cluster decompositions: $b\ne b'$. Then for any $0\le r \le 1$, $S_b^r$ and $S_{b'}^r$ are orthogonal mutually: $S_b^r\perp S_{b'}^r$.
\end{pro}

\section{A partition of unity}

To state a proposition that will play a fundamental role in our decomposition of continuous spectral subspace by $S_b^1$, we prepare some notations. Let $b$ be a cluster decomposition with $2\le|b|\le N$. For any two clusters $C_1$ and $C_2$ in $b$, we define a vector $z_{b1}$ that connects the two centers of mass of the clusters $C_1$ and $C_2$. The number of such vectors when we move over all pairs of clusters in $b$ is $k_b=\left(\begin{array}{c} |b|\\ 2\end{array}\right)$ in total. We denote these vectors by $z_{b1},z_{b2},\cdots,z_{bk_b}$.

 Let $z_{bk}$ $(1\le k\le k_b)$ connect two clusters $C_\ell$ and $C_m$ in $b$ $(\ell\ne m)$. Then for any pair $\al=\{i,j\}$ with $i\in C_\ell$ and $j\in C_m$, the vector $x_\al=x_{ij}$ is expressed like $(z_{bk},x^{(C_\ell)},x^{(C_m)})\in R^{\nu+\nu(|C_\ell|-1)+\nu(|C_m|-1)}$, where $x^{(C_\ell)}(\in R^{\nu(|C_\ell|-1)})$ and $x^{(C_m)}(\in R^{\nu(|C_m|-1)})$ are the positions of the particles $i$ and $j$ in $C_\ell$ and $C_m$, respectively. The vector expression $x_\al=(z_{bk},x^{(C_\ell)},x^{(C_m)})$ is in the space $R^{\nu+\nu(|C_\ell|-1)+\nu(|C_m|-1)}$. If we express it in the larger space $R^{\nu}_{z_{bk}}\times R^{\nu(N-|b|)}_{x^b}$, it would be $x_\al=(z_{bk},x^b)$, and $|x_\al|^2=|z_{bk}|^2+|x^b|^2$.  Thus if $|z_{bk}|^2$ is sufficiently large compared to $|x^b|^2\ge|x^{(C_\ell)}|^2+|x^{(C_m)}|^2$, e.g. if $|z_{bk}|^2> \rho>0$ and $|x^b|^2< \theta$ with $\rho\gg \theta>0$ (which means that $\rho/\theta$ is sufficiently large), then $|x_\al|^2> \rho/2$ for all $\al\not\leq b$. 

Next if $c<b$ and $|c|=|b|+1$, then just one cluster, say $C_\ell\in b$, is decomposed into two clusters $C'_\ell$ and $C''_\ell$ in $c$, and other clusters in $b$ remain the same in the finer cluster decomposition $c$. In this case, we can choose just one vector $z_{ck}$ $(1\le k\le k_c)$ that connects clusters $C'_\ell$ and $C''_\ell$ in $c$, and we can express $x^b=(z_{ck},x^c)$. The norm of this vector is written as 
\beq
|x^b|^2=|z_{ck}|^2+|x^c|^2.\label{tag 3.1}
\ene
Similarly the norm of $x=(x_b,x^b)$ is written as
\beq
|x|^2=|x_b|^2+|x^b|^2.\label{tag 3.2}
\ene
 We recall that norm is defined, as usual, from the inner product defined by \eq{tag 1.13} which changes in accordance with the cluster decomposition used in each context. E.g., in \eq{tag 3.1}, the left-hand side (LHS) is defined by using \eq{tag 1.13} for the cluster decomposition $b$ and the RHS is by using \eq{tag 1.13} for $c$.

With these preparations, we state the following lemma, which is partially a repetition of \cite{[K1]}, Lemma 2.1. We define subsets $T_b(\rho,\theta)$ and ${\tilde T}_b(\rho,\theta)$ of $R^n=R^{\nu(N-1)}$ for cluster decompositions $b$ with $2\le |b|\le N$ and real numbers $\rho,\theta$ with $1>\rho,\theta>0$:
\beq
&&T_b(\rho,\theta)=\left(\bigcap_{k=1}^{k_b}\{x\ |\ |z_{bk}|^2>\rho|x|^2\}\right)\cap\{ x\ |\ |x_b|^2>(1-\theta)|x|^2\},\label{tag 3.3}\\
&&{\tilde T}_b(\rho,\theta)=\left(\bigcap_{k=1}^{k_b}\{x\ |\ |z_{bk}|^2>\rho\}\right)\cap\{ x\ |\ |x_b|^2>1-\theta\}.\label{tag 3.4}
\ene
Subsets $S$ and $S_{\theta}$ $(\theta>0)$ of $R^n=R^{\nu(N-1)}$ are defined by
\beq
S&=&\{x\ |\ |x|^2\ge 1\},\nom\\
S_\theta&=&\{x\ |\ 1+\theta\ge |x|^2\ge 1\}.\nom
\ene

\begin{lem}\label{lemma31}
Suppose that constants $1\ge\theta_1>\rho_j>\theta_j> \rho_N>0$ satisfy $\theta_{j-1}\ge \theta_j+\rho_j$ for $j=2,3,\cdots,N-1$. 
Then the followings hold:
\BP

\F
i)
\beq
S\subset \bigcup_{2\le |b|\le N}T_b(\rho_{|b|},\theta_{|b|}).\label{tag 3.5}
\ene
\MP

\F
ii)
Let $\gamma_j>1$ $(j=1,2)$ satisfy
\beq
\gamma_1\gamma_2< r_0:=\min_{2\le j\le N-1}\{{\rho_j/\theta_j}\}.\label{tag 3.6}
\ene
If $b\not\le c$ with $|b|\ge |c|$, then
\beq
T_b(\gamma_1^{-1}\rho_{|b|},\gamma_2\theta_{|b|})\cap
T_c(\gamma_1^{-1}\rho_{|c|},\gamma_2\theta_{|c|})=\emptyset.\label{tag 3.7}
\ene
\MP

\F
iii)
For $\gamma>1$ and $2\le|b|\le N$
\beq
T_b(\rho_{|b|},\theta_{|b|})\cap S_{\theta_{N-1}}&\subset&
{\tilde T}_b(\rho_{|b|},\theta_{|b|})\cap S_{\theta_{N-1}}\nom\\
&\subset\subset& {\tilde T}_b(\gamma^{-1}\rho_{|b|},\gamma\theta_{|b|})\cap S_{\theta_{N-1}}\nom\\
&\subset& T_b({\gamma'_1}^{-1}\rho_{|b|},\gamma'_2\theta_{|b|})\cap S_{\theta_{N-1}},
\label{tag 3.8}
\ene
where
\beq
\gamma'_1=\gamma(1+\theta_{N-1}),\q \gamma'_2=(1+\gamma)(1+\theta_{N-1})^{-1}.\label{tag 3.9}
\ene
\MP

\F
iv)
If $\frac{2\gamma'_1\gamma'_2}{2-\gamma'_1} <r_0$, then for $2\le|b|\le N$
\beq
{T}_b({\gamma'_1}^{-1}\rho_{|b|},\gamma'_2\theta_{|b|})\subset \{ x\ |\ |x_\al|^2> \rho_{|b|}|x|^2/2 \ \mbox{for all}\ \al\not\leq b\}.\label{tag 3.10}
\ene
\MP

\F
v)
If $\gamma(1+\gamma)<r_0$ and $b\not\le c$ with $|b|\ge |c|$, then
\beq
T_b({\gamma'_1}^{-1}\rho_{|b|},\gamma'_2\theta_{|b|})\cap
T_c({\gamma'_1}^{-1}\rho_{|c|},\gamma'_2\theta_{|c|})=\emptyset.\label{tag 3.11}
\ene
\end{lem}
{\it Proof:} To prove \eq{tag 3.5}, suppose that $|x|^2\ge 1$ and $x$ does not belong to the set
$$
A=\bigcup_{2\le |b|\le N-1}\left[\left(\bigcap_{k=1}^{k_b}\{x\ |\ |z_{bk}|^2>\rho_{|b|}|x|^2\}\right)\cap\{ x\ |\ |x_b|^2>(1-\theta_{|b|})|x|^2\}\right].
$$
Under this assumption, we prove $|x_\al|^2>\rho_N|x|^2$ for all pairs $\al=\{i,j\}$. (Note that $z_{bk}$ for $|b|=N$ equals some $x_\al$.) Let $|b|=2$ and write $x=(z_{b1},x^b)$. Then by \eq{tag 3.1}, $1\le|x|^2=|z_{b1}|^2+|x^b|^2$. Since $x$ belongs to the complement $A^c$ of the set $A$, we have $|z_{b1}|^2\le \rho_{|b|}|x|^2$ or $|x_b|^2\le(1-\theta_{|b|})|x|^2$. If $|z_{b1}|^2\le \rho_{|b|}|x|^2$, then $|x^b|^2=|x|^2-|z_{b1}|^2\ge(1-\rho_{|b|})|x|^2\ge(\theta_1-\rho_{|b|})|x|^2\ge \theta_{|b|}|x|^2$ by $\theta_{j-1}\ge \theta_j+\rho_j$. Thus $|x_b|^2=|x|^2-|x^b|^2\le(1-\theta_{|b|})|x|^2$ for all $b$ with $|b|=2$.

Next let $|c|=3$ and assume $|x_c|^2>(1-\theta_{|c|})|x|^2$. Then by $x\in A^c$, we can choose $z_{ck}$ with $1\le k\le k_c$ such that $|z_{ck}|^2\le \rho_{|c|}|x|^2$. Let $C_\ell$ and $C_m$ be two clusters in $c$ connected by $z_{ck}$, and let $b$ be the cluster decomposition obtained by combining $C_\ell$ and $C_m$ into one cluster with retaining other clusters of $c$ in $b$. Then $|b|=2$, $x^b=(z_{ck},x^c)$, and $|x^b|^2=|z_{ck}|^2+|x^c|^2$. Thus $|x_b|^2=|x|^2-|x^b|^2=|x|^2-|z_{ck}|^2-|x^c|^2=|x_c|^2-|z_{ck}|^2>(1-\theta_{|c|}-\rho_{|c|})|x|^2\ge(1-\theta_{|b|})|x|^2$, which contradicts the result of the previous step. Thus $|x_c|^2\le(1- \theta_{|c|})|x|^2$ for all $c$ with $|c|=3$. 

Repeating this procedure, we finally arrive at $|x_d|^2\le (1-\theta_{|d|})|x|^2$, thus $|x^d|^2=|x|^2-|x_d|^2\ge \theta_{|d|}|x|^2> \rho_N|x|^2$ for all $d$ with $|d|=N-1$. Namely $|x_\al|^2>\rho_N|x|^2$ for all pairs $\al=\{i,j\}$.
The proof of \eq{tag 3.5} is complete.

We next prove \eq{tag 3.7}. By $b\not\le c$, we can take a pair $\al=\{i,j\}$ and clusters $C_\ell, C_m\in c$ such that $\al\le b$, $i\in C_\ell$, $j\in C_m$, and $\ell\ne m$. Then we can write $x_\al=(z_{ck},x^c)$ for some $1\le k\le k_c$. Thus if there is $x\in T_b(\gamma_1^{-1}\rho_{|b|},\gamma_2\theta_{|b|})\cap T_c(\gamma_1^{-1}\rho_{|c|},\gamma_2\theta_{|c|})$, then 
\beq
\gamma_2\theta_{|b|}|x|^2>|x^b|^2\ge|x_\al|^2=|z_{ck}|^2+|x^c|^2\ge |z_{ck}|^2>\gamma_1^{-1}\rho_{|c|}|x|^2.\label{tag 3.12}
\ene
 But since $|b|\ge |c|$, we have $\rho_{|c|}> \gamma_1\gamma_2\theta_{|b|}$ when $|b|=|c|$ by \eq{tag 3.6}, and $\rho_{|c|}>\gamma_1\gamma_2\theta_{|c|}\ge \gamma_1\gamma_2(\theta_{|b|}+\rho_{|b|})>\gamma_1\gamma_2\theta_{|b|}$ when $|c|<|b|$ by $\theta_{j-1}\ge \theta_j+\rho_j$, which both contradict the inequality \eq{tag 3.12}. This completes the proof of \eq{tag 3.7}.

\eq{tag 3.8} follows by a simple calculation from the inequality $|x|^2(1+\theta_{N-1})^{-1}\le 1$ that holds on $S_{\theta_{N-1}}$. \eq{tag 3.10} follows from the relation $|x_\al|^2=|z_{bk}|^2+|x^b|^2$ stated before the lemma, and \eq{tag 3.11} from $\gamma'_1\gamma'_2=\gamma(1+\gamma)$ and ii).\ $\Box$

\BP

In the followings we fix constants  $\gamma>1$ and $1\ge \theta_1>\rho_j>\theta_j>\rho_N>0$ such that
\beq
&&\theta_{j-1}\ge \theta_j+\rho_j\q(j=2,3,\cdots,N-1),\label{tag 3.13}\\
&&\max\left\{\gamma(1+\gamma),\frac{2\gamma'_1\gamma'_2}{2-\gamma'_1}\right\}<r_0=\min_{2\le j\le N-1}\{\rho_j/\theta_j\},\label{tag 3.14}
\ene
where $\gamma'_j$ $(j=1,2)$ are defined by \eq{tag 3.9}.

Let $\rho(\la)\in C^\infty(R^1)$ be such that $0\le \rho(\la)\le 1$, $\rho(\la)=1$ $(\la\le -1)$, $\rho(\la)=0$ $(\la\ge 0)$, and $\rho'(\la)\le 0$. Then we define  functions $\phi_\sigma(\la<\tau)$ and $\phi_\sigma(\la>\tau)$ of $\la\in R^1$ by
\beq
&&\phi_\sig(\la<\tau)=\rho((\la-(\tau+\sig))/\sig),\label{3.16}\\
&&\phi_\sig(\la>\tau)=1-\phi_\sig(\la<\tau-\sig)
\label{tag 3.15}
\ene
for constants $\sig>0,\tau\in R^1$.
We note that $\phi_\sig(\la<\tau)$ and $\phi_\sigma(\la>\tau)$ satisfy
\beq
&&\phi_\sig(\la<\tau)=
\left\{
\begin{array}{ll}
1 \q&(\la\le\tau)\\
0\q& (\la\ge\tau+\sig)
\end{array}\right.
\label{tag 3.16}\\
&&\phi_\sigma(\la>\tau)=\left\{
\begin{array}{ll}
0 \q &(\la\le\tau-\sig)\\
1 \q& (\la\ge\tau)
\end{array}\right.
\label{tag 3.17}\\
&&\phi'_\sig(\la<\tau)=\frac{d}{d\la}\phi_\sig(\la<\tau)\le 0,\nom\\
&&\phi'_\sig(\la>\tau)\ge 0.
\ene

We define for a cluster decomposition $b$ with $2\le |b|\le N$ 
\beq
\varphi_b(x_b)=\prod_{k=1}^{k_b}\phi_\sig(|z_{bk}|^2>\rho_{|b|})\phi_\sig(|x_b|^2>1-\theta_{|b|}),\label{tag 3.18}
\ene
where $\sig>0$ is fixed as
\beq
0<\sig<\min_{2\le j\le N-1}\{(1-\gamma^{-1})\rho_N,(1-\gamma^{-1})\rho_j,(\gamma-1)\theta_j\}.\label{tag 3.19}
\ene
Then $\varphi_b(x_b)$ satisfies for $x\in S_{\theta_{N-1}}$
\beq
\varphi_b(x_b)=\left\{
\begin{array}{l}
1\q \mbox{for}\q x\in {\tilde T}_b(\rho_{|b|},\theta_{|b|}),\\
0\q \mbox{for}\q x\not\in {\tilde T}_b(\gamma^{-1}\rho_{|b|},\gamma\theta_{|b|}).
\end{array}\right.\label{tag 3.20}
\ene
We set for $|b|=k$ $(k=2,3,\cdots,N)$
\beq
J_b(x)=\varphi_b(x_b)
\left(1-\sum_{|b_{k-1}|=k-1}\varphi_{b_{k-1}}(x_{b_{k-1}})\right)
\cdots
\left(1-\sum_{|b_{2}|=2}\varphi_{b_{2}}(x_{b_{2}})\right).\label{tag 3.21}
\ene
By v) and iii) of Lemma \ref{lemma31} and \eq{tag 3.20}, the sums on the RHS remain only in the case $b<b_j$ for $j=k-1,\cdots,2$ and $x\in S_{\theta_{N-1}}$:
\beq
J_b(x)=\varphi_b(x_b)
\left(1-\sum_{|b_{k-1}|=k-1, b<b_{k-1}}\varphi_{b_{k-1}}(x_{b_{k-1}})\right)
\cdots
\left(1-\sum_{|b_{2}|=2, b<b_2}\varphi_{b_{2}}(x_{b_{2}})\right).\label{tag 3.22}
\ene
Thus $J_b(x)$ is a function of the variable $x_b$ only:
\beq
J_b(x)=J_b(x_b)\q\mbox{when}\q x=(x_b,x^b)\in S_{\theta_{N-1}}.\label{tag 3.23}
\ene
We also note that the supports of $\varphi_{b_j}$ in each sum on the RHS of 
\eq{tag 3.21} are disjoint mutually in $S_{\theta_{N-1}}$ by iii) and v) of Lemma \ref{lemma31}.
By \eq{tag 3.5} and \eq{tag 3.8} of lemma \ref{lemma31}, and the definition \eq{tag 3.18}-\eq{tag 3.21} of $J_b(x_b)$, we therefore have
$$
\sum_{2\le|b|\le N}J_b(x_b)=1\q\mbox{on}\q S_{\theta_{N-1}}.
$$

We have constructed a partition of unity on $S_{\theta_{N-1}}$:

\begin{pro}\label{pro32} Let real numbers $1\ge\theta_1>\rho_j>\theta_j>\rho_N>0$ satisfy $\theta_{j-1}\ge \theta_j+\rho_j$ for $j=2,3,\cdots,N-1$. 
Assume that \eq{tag 3.14} holds and let $J_b(x_b)$ be defined by \eq{tag 3.18}-\eq{tag 3.22}. Then we have 
\beq
\sum_{2\le|b|\le N}J_b(x_b)=1\q\mbox{on}\q S_{\theta_{N-1}}.\label{tag 3.24}
\ene
$J_b(x_b)$ is a $C^\infty$ function of $x_b$ and satisfies $0\le J_b(x_b)\le 1$. Further on {\rm supp}\hskip2pt$J_b \cap S_{\theta_{N-1}}$ we have
\beq
|x_\al|^2>\rho_{|b|}|x|^2/2\label{tag 3.25}
\ene
for any pair $\al\not\le b$, and
\beq
\sup_{x\in R^n,2\le|b|\le N}|\nabla_{x_b} J_b(x_b)|<\infty\label{tag 3.26}
\ene
for each fixed $\sig>0$ in \eq{tag 3.18}-\eq{tag 3.19}.
\end{pro}
{\it Proof:} We have only to see \eq{tag 3.25} and \eq{tag 3.26}. But \eq{tag 3.25} is clear by \eq{tag 3.8}, \eq{tag 3.10}, \eq{tag 3.14}, \eq{tag 3.20} and \eq{tag 3.21}, and \eq{tag 3.26} follows from \eq{tag 3.15}, \eq{tag 3.18} and \eq{tag 3.22}.\ $\Box$

\vskip 12pt

\BP

\section{A decomposition of continuous spectral subspace}

The following theorem gives a decomposition of $\HH_c(H)$ by scattering spaces $S_b^1$ $(2\le|b|\le N)$. 

\begin{thm}\label{Theorem 4.1} Let Assumptions \ref{ass1} and \ref{ass2} be satisfied. Then we have
\beq
\HH_c(H)=\bigoplus_{2\le|b|\le N}S_b^1.\label{tag 4.1}
\ene
\end{thm}
{\it Proof:} Since the set
$$
\bigcup_{\Delta\subset\subset R^1-\TT}E_H(\Delta)\HH
$$
is dense in $\HH_c(H)$, and $S_b^1$ $(2\le|b|\le N)$ are closed and mutually orthogonal, it suffices to prove that any $\Phi(H)f$ with $\Phi\in C_0^\infty(R^1-\TT)$ and $f\in\HH$ can be decomposed as a sum of the elements $f_b^1$ in $S_b^1$: $\Phi(H)f=\sum_{2\le|b|\le N}f_b^1$.

We divide the proof into two steps. In the first step I), we prove existence of certain time limits. In the second step II), we prove existence of some ``boundary values" of those limits, and conclude the proof of decomposition \eq{tag 4.1}.
\BP

\F
I) Existence of some time limits:
\MP

We decompose $\Phi(H)f$ as a finite sum: $\Phi(H)f=\sum_{j_0}^{\mbox{\scriptsize{finite}}}\psi_{j_0}(H)f$, where $\psi_{j_0}\in C_0^\infty(R^1-\TT)$. In the step I), we will prove the existence of the limit
\beq
\lim_{t\to\infty}\sum_{\ell=1}^L e^{itH}
G_{b,\la_\ell}(t)^*J_b(v_b/r_\ell)G_{b,\la_\ell}(t)
e^{-itH}\psi_{j_0}(H)f,\label{tag 4.2}
\ene
under the assumption that supp\hskip2pt$\psi_{j_0}\subset \tilde\Delta\subset\subset\Delta$ for some intervals $\tilde\Delta\subset\subset\Delta\subset\subset R^1-\TT$ with $E\in \Delta$ and diam\hskip2pt$\Delta<d(E)$, where $d(E)>0$ is some small constant depending on $E\in R^1-\TT$ and diam\hskip2pt$S$ denotes the diameter of a set $S\subset R^1$.
The relevant factors in \eq{tag 4.2} will be defined in the course of the proof.
We will write $f$ for $\psi_{j_0}(H)f$ in the followings.

We take $\psi\in C_0^\infty(R^1)$ such that $\psi(\la)=1$ for $\la\in \tilde\Delta$ and supp\hskip2pt$\psi\subset{\Delta}$ for the intervals ${\tilde \Delta}\subset\subset\Delta$ above. Then $f=\psi(H)f=E_H(\Delta)f \in E_H(\Delta)\HH\subset \HH_c(H)$ and $e^{-itH}f=\psi(H)e^{-itH}f$. Thus
 we can use the decomposition \eq{tag 1.21} for the sequences $t_m$ and $M_b^m$ in Theorem \ref{En}:
\beq
e^{-it_mH}f=\psi(H)e^{-it_mH}f=\psi(H)\sum_{2\le|d|\le N}{\widetilde P}_d^{M_d^m}e^{-it_mH}f.\label{tag 4.3}
\ene
By Theorem \ref{En}-\eq{tag 1.24}
\beq
\psi(H){\widetilde P}_d^{M_d^m}e^{-it_mH}f\sim \psi(H_d){\widetilde P}_d^{M_d^m}e^{-it_mH}f\label{tag 4.4}
\ene
as $m\to\infty$. Since
\beq
{\widetilde P}_d^{M_d^m}=P_d^{M_{|d|}^m}{\widehat P}_{|d|-1}^{{\widehat M}_d^m},\q
P_d^{M_{|d|}^m}=\sum_{j=1}^{M_{|d|}^m}P_{d,E_j},\label{tag 4.5}
\ene
where $P_{d,E_j}$ is one dimensional eigenprojection for $H^d$ with eigenvalue $E_j$, the RHS of \eq{tag 4.4} equals
\beq
\sum_{j=1}^{M_{|d|}^m}\psi(T_d+E_j)P_{d,E_j}{\widehat P}_{|d|-1}^{{\widehat M}_d^m}e^{-it_mH}f.\label{tag 4.6}
\ene
By supp\hskip2pt$\psi\subset{\Delta}\subset\subset R^1-\TT$, $E_j\in \TT$, and $T_d\ge0$, we can take constants $\Lambda_d>\lambda_d>0$ independent of $j=1,2,\cdots$ such that $\Lambda_d\ge T_d\ge \la_d$ if $\psi(T_d+E_j)\ne 0$. Set $\Lambda_0=\max_{d}\Lambda_d>\la_0=\min_{d}\la_d>0$ and
\beq
\Sigma(E)=\{E-\la\ |\ \la\in\TT, E\ge \la\}.\label{tag 4.7}
\ene
 Note that we can take $\Lambda_0>\la_0>0$ so that 
\beq
\Sigma(E)\subset\subset(\la_0,\La_0) \subset (0,\infty).\label{tag 4.8}
\ene
Let $\Psi\in C_0^\infty(R^1)$ satisfy $\Psi(\la)=1$ for $\la\in [\la_0,\Lambda_0]$ and supp\hskip2pt$\Psi\subset[\la_0-\kappa,\Lambda_0+\kappa]$ for some small constant $\kappa>0$ such that the set $[\la'_0,\la_0]\cup[\Lambda_0,\Lambda'_0]$ is bounded away from $\Sigma(E)$, where $\la'_0=\la_0-2\kappa>0$ and $\Lambda'_0=\Lambda_0+2\kappa$. Then the RHS of \eq{tag 4.6} equals for any $m=1,2,\cdots$
\beq
\Psi^2(T_d)\psi(H_d){\widetilde P}_d^{M_d^m}e^{-it_mH}f
.\label{tag 4.9}
\ene
On the other hand, by Theorem \ref{En}-\eq{tag 1.23} and \eq{tag 1.25}, we have
\beq
\frac{|x^d|^2}{t_m^2}\psi(H_d){\widetilde P}_d^{M_d^m}e^{-it_mH}f\sim 0\label{tag 4.10}
\ene
and
\beq
\Psi^2(T_d)\psi(H_d){\widetilde P}_d^{M_d^m}e^{-it_mH}f\sim
\Psi^2(|x_d|^2/(2t_m^2))\psi(H_d){\widetilde P}_d^{M_d^m}e^{-it_mH}f\label{tag 4.11}
\ene
as $m\to\infty$, where to see \eq{tag 4.10} we used \eq{tag 1.23} and $i[H^d,|x^d|^2/t^2]=i[H_0^d,|x^d|^2/t^2]=2A^d/t^2$ where $A^d=(x^d\cdot p^d+p^d\cdot x^d)/2$, and to see \eq{tag 4.11} the fact that $|x_d|^2/t_m^2$ and $H_d$ commute asymptotically as $m\to\infty$ by \eq{tag 1.25}. Thus by $|x|^2=|x_d|^2+|x^d|^2$ we have
\beq
\Psi^2(T_d)\psi(H_d){\widetilde P}_d^{M_d^m}e^{-it_mH}f\sim
\Psi^2(|x|^2/(2t_m^2))\psi(H_d){\widetilde P}_d^{M_d^m}e^{-it_mH}f\label{tag 4.12}
\ene
as $m\to\infty$. From \eq{tag 4.3}-\eq{tag 4.4}, \eq{tag 4.6}, \eq{tag 4.9} and \eq{tag 4.12}, we obtain
\beq
e^{-it_mH}f\sim
\Psi^2(|x|^2/(2t_m^2))e^{-it_mH}f\label{tag 4.13}
\ene
as $m\to\infty$.

Let constants $\gamma>1$ and $1\ge \theta_1>\rho_j>\theta_j>\rho_N>0$ be fixed such that
\beq
&&\theta_{j-1}\ge \theta_j+\rho_j\q(j=2,\cdots,N-1),\label{tag 4.14}\\
&&\max\left\{\gamma(1+\gamma),\frac{2\gamma'_1\gamma'_2}{2-\gamma'_1}\right\}<r_0=\min_{2\le j\le N-1}\{\rho_j/\theta_j\}\label{tag 4.15}
\ene
for $\gamma'_j$ $(j=1,2)$ defined by \eq{tag 3.9}.
Set
\beq
\la''_0=\la'_0\theta_{N-1}>0\label{tag 4.16}
\ene
with $\la'_0=\la_0-2\kappa$ defined above.
Let $\tau_0>0$ satisfy
\beq
0<16\tau_0<\la''_0(<\la'_0<\la_0).\label{tag 4.17}
\ene
We take a finite subset $\{{\tilde \la}_\ell\}_{\ell=1}^L$ of $\TT$ such that
\beq
\TT\subset \bigcup_{\ell=1}^L({\tilde \la}_\ell-\tau_0,{\tilde\la}_\ell+\tau_0).\label{tag 4.18}
\ene
Then we can choose real numbers $\la_\ell\in R^1$, $\tau_\ell>0$ $(\ell=1,2,\cdots,L)$ and $\sig_0>0$ such that
\beq
&&\tau_\ell<\tau_0,\q \sig_0<\tau_0,\q |\la_\ell-{\tilde\la}_\ell|<\tau_0,\nom\\
&&\TT\subset\bigcup_{\ell=1}^L(\la_\ell-\tau_\ell,\la_\ell+\tau_\ell),\q (\la_\ell-\tau_\ell,\la_\ell+\tau_\ell)\subset ({\tilde \la}_\ell-\tau_0,{\tilde\la}_\ell+\tau_0),\nom\\
&&\mbox{dist}\{(\la_\ell-\tau_\ell,\la_\ell+\tau_\ell),(\la_k-\tau_k,\la_k+\tau_k)\}>4\sig_0(>0)\q\mbox{for any}\ \ell\ne k.\label{tag 4.19}
\ene
We note that for $\ell=1,\cdots,L$
\beq
\left\{\Lambda \left|\ \tau_\ell\le|\Lambda-(E-\la_\ell)|\le\tau_\ell+4\sig_0\right.\right\}
\cap \Sigma(E)=\emptyset.\label{tag 4.20}
\ene
Now let the intervals $\Delta$ and ${\tilde \Delta}$ be so small that
\beq
\mbox{diam}\hskip2pt{\tilde\Delta}<\mbox{diam}\hskip2pt{\Delta}<{\tilde \tau}_0:=\min_{1\le\ell\le L}\{\sig_0,\tau_\ell\}.\label{tag 4.21}
\ene

Returning to \eq{tag 4.6}, we have
\beq
T_d+E_j\in \mbox{supp}\hskip2pt\psi,\label{tag 4.22}
\ene
if $\psi(T_d+E_j)\ne 0$ in \eq{tag 4.6}. By supp\hskip2pt$\psi\subset{\Delta}$, diam\hskip2pt${\Delta}<{\tilde \tau}_0$, and $E\in{\Delta}$, we have from \eq{tag 4.22}
\beq
-{\tilde \tau}_0\le T_d-(E-E_j)\le{\tilde\tau}_0.\label{tag 4.23}
\ene
Thus we have asymptotically on each state in \eq{tag 4.6}
\beq
-2{\tilde\tau}_0\le\frac{|x|^2}{t_m^2}-2(E-E_j)\le2\tilde\tau_0.\label{tag 4.24}
\ene
By \eq{tag 4.19}, $E_j\in\TT$ is included in just one set $(\la_\ell-\tau_\ell,\la_\ell+\tau_\ell)$ for some $\ell=\ell(j)$ with $1\le \ell(j)\le L$. Since $|E_j-\la_\ellj|<\tau_\ellj$, we have using \eq{tag 4.21}
\beq
-2\tau_\ellj-2\sig_0\le\frac{|x|^2}{t_m^2}-2(E-\la_\ellj)\le 2\tau_\ellj+2\sig_0\label{tag 4.25}
\ene
on each state in \eq{tag 4.6}. Thus 
\beq
\sum_{\ell=1}^L \phi_{\sig_0}^2(\left||x|^2/t_m^2-2(E-\la_\ell)\right|<2\tau_\ell+2\sig_0)=1\nom
\ene
asymptotically as $m\to\infty$ on \eq{tag 4.6}.
Now by the same reasoning that led us to \eq{tag 4.13}, we see that \eq{tag 4.3} asymptotically equals as $m\to\infty$
\beq
\sum_{\ell=1}^L \sum_{2\le|d|\le N}\phi_{\sig_0}^2(\left||x|^2/t_m^2-2(E-\la_\ell)\right|<2\tau_\ell+2\sig_0)\Psi^2(|x|^2/(2t_m^2))\wtP_d^\Mdm e^{-it_mH}f.\label{tag 4.26}
\ene
Since $\phi_{\sig_0}(\left||x|^2/t_m^2-2(E-\la_\ell)\right|<2\tau_\ell+4\sig_0)=1$ on supp $\phi_{\sig_0}(\left||x|^2/t_m^2-2(E-\la_\ell)\right|<2\tau_\ell+2\sig_0)$, \eq{tag 4.26} equals
\beq
&&\sum_{\ell=1}^L \sum_{2\le|d|\le N}\phi_{\sig_0}^2(\left||x|^2/t_m^2-2(E-\la_\ell)\right|<2\tau_\ell+4\sig_0)\nom\\
&&\times\phi_{\sig_0}^2(\left||x|^2/t_m^2-2(E-\la_\ell)\right|<2\tau_\ell+2\sig_0)\Psi^2(|x|^2/(2t_m^2))\wtP_d^\Mdm e^{-it_mH}f.\label{tag 4.27}
\ene
Set
\beq
B=\lan x\ran^{-1/2}A\lan x\ran^{-1/2},\q A=\frac{1}{2}(x\cdot p+p\cdot x)=\frac{1}{2}(\lan x,v\ran+\lan v,x\ran).\label{tag 4.28}
\ene
We note by Theorem \ref{En}-\eq{tag 1.23} and \eq{tag 1.25} that on the state $\wtP_d^\Mdm e^{-it_mH}f$
\beq
B\sim \sqrt{2T_d}\sim \frac{|x|}{t_m}\label{tag 4.29}
\ene
 asymptotically as $t_m\to\infty$. Using this, we replace $\phi_{\sig_0}(\left||x|^2/t_m^2-2(E-\la_\ell)\right|<2\tau_\ell+2\sig_0)$ by $\phi_{\sig_0}(\left|B^2-2(E-\la_\ell)\right|<2\tau_\ell+2\sig_0)$ in \eq{tag 4.27}. Let $\varphi(\la)\in C^\infty_0((\sqrt{2(\la_0-2\kappa)}$, $\sqrt{2(\Lambda_0+2\kappa)}))$,
$0\le \varphi(\la)\le 1$, and $\varphi(\la)=1$ on $[\sqrt{2(\la_0-\kappa)},\sqrt{2(\La_0+\kappa)}](\supset$ supp $\Psi(\la^2/2)\cap(0,\infty))$. We insert a factor $\varphi^2(B)$ into \eq{tag 4.27} and then remove the factor $\Psi^2(|x|^2/(2t_m^2))$ using \eq{tag 4.13}:
\beq
&&\sum_{\ell=1}^L \sum_{2\le|d|\le N}\phi_{\sig_0}^2(\left||x|^2/t_m^2-2(E-\la_\ell)\right|<2\tau_\ell+4\sig_0)\nom\\
&&\times\phi_{\sig_0}^2(\left|B^2-2(E-\la_\ell)\right|<2\tau_\ell+2\sig_0)
\varphi^2(B)
\wtP_d^\Mdm e^{-it_mH}f.\label{tag 4.30}
\ene
On supp $\phi_{\sig_0}(\left||x|^2/t_m^2-2(E-\la_\ell)\right|<2\tau_\ell+4\sig_0)$ we have
\beq
0<2(E-\la_\ell)-7\tau_0\le\frac{|x|^2}{t_m^2}\le 2(E-\la_\ell)+7\tau_0.\label{tag 4.31}
\ene
Since \eq{tag 4.8}, $|\la_\ell-\tilde\la_\ell|<\tau_0$ and \eq{tag 4.17} imply
\beq
\frac{2(E-\la_\ell)+7\tau_0}{2(E-\la_\ell)-7\tau_0}-1
=\frac{14\tau_0}{2(E-\la_\ell)-7\tau_0}<\frac{14\la''_0/16}{30\la_0/16-\la''_0}
<\theta_{N-1},\label{tag 4.32}
\ene
we can apply the partition of unity in Proposition \ref{pro32} to the ring defined by \eq{tag 4.31}. Then we obtain
\beq
e^{-it_mH}f &\sim 
&\sum_{\ell=1}^L \sum_{2\le|b|\le N}\sum_{2\le|d|\le N}J_b(x_b/(r_\ell t_m))\phi_{\sig_0}^2(\left||x|^2/t_m^2-2(E-\la_\ell)\right|<2\tau_\ell+4\sig_0)\nom\\
&\times&\phi_{\sig_0}^2(\left|B^2-2(E-\la_\ell)\right|<2\tau_\ell+2\sig_0)\varphi^2(B)\wtP_d^\Mdm e^{-it_mH}f,\label{tag 4.33}
\ene
where 
\beq
r_\ell=\sqrt{2(E-\la_\ell)-7\tau_0}>0\q(\ell=1,\cdots,L).\label{tag 4.34}
\ene
By the property \eq{tag 3.25}, only the terms with $d\le b$ remain in \eq{tag 4.33}:
\beq
e^{-it_mH}f &\sim 
&\sum_{\ell=1}^L \sum_{2\le|b|\le N}\sum_{d\le b}J_b(x_b/(r_\ell t_m))\phi_{\sig_0}^2(\left||x|^2/t_m^2-2(E-\la_\ell)\right|<2\tau_\ell+4\sig_0)\nom\\
&\times&\phi_{\sig_0}^2(\left|B^2-2(E-\la_\ell)\right|<2\tau_\ell+2\sig_0)\varphi^2(B)\wtP_d^\Mdm e^{-it_mH}f.\label{tag 4.35}
\ene
Using Theorem \ref{En}-\eq{tag 1.25}, we replace $x_b/t_m$ by $v_b$, and at the same time we introduce a pseudodifferential operator into \eq{tag 4.35}:
\beq
P_b(t)=\phi_{\sig}(|x_b/t-v_b|^2<u)\label{tag 4.36}
\ene
with $u>0$ sufficiently small.
Then \eq{tag 4.35} becomes
\beq
e^{-it_mH}f &\sim 
&\sum_{\ell=1}^L \sum_{2\le|b|\le N}\sum_{d\le b}P_b^2(t_m)J_b(v_b/r_\ell)\phi_{\sig_0}^2(\left||x|^2/t_m^2-2(E-\la_\ell)\right|<2\tau_\ell+4\sig_0)\nom\\
&\times&\phi_{\sig_0}^2(\left|B^2-2(E-\la_\ell)\right|<2\tau_\ell+2\sig_0)\varphi^2(B)\wtP_d^\Mdm e^{-it_mH}f.\label{tag 4.37}
\ene
We rearrange the order of the factors on the RHS of \eq{tag 4.37} using that the factors mutually commute asymptotically as $m\to\infty$ by Theorem \ref{En}. Setting
\beq
G_{b,\la_\ell}(t)&=&P_b(t)\phi_{\sig_0}(\left||x|^2/t^2-2(E-\la_\ell)\right|<2\tau_\ell+4\sig_0)\nom\\
&\times&\phi_{\sig_0}(\left|B^2-2(E-\la_\ell)\right|<2\tau_\ell+2\sig_0)\varphi(B),\label{tag 4.38}
\ene
we obtain
\beq
e^{-it_mH}f \sim 
\sum_{\ell=1}^L \sum_{2\le|b|\le N}\sum_{d\le b}G_{b,\la_\ell}(t_m)^*
J_b(v_b/r_\ell)G_{b,\la_\ell}(t_m)\wtP_d^\Mdm e^{-it_mH}f.\label{tag 4.39}
\ene
Now by some calculus of pseudodifferential operators and Theorem \ref{En} we note that $P_b(t)J_b(v_b/r_\ell)$ yields a partition of unity ${\tilde J}_b(x_b/(r_\ell t))$ asymptotically as $m\to\infty$ whose support is close to that of $J_b(x_b/(r_\ell t))$. Then we can recover the terms with $d\not\le b$, and using \eq{tag 1.21}, we remove the sum of $\wtP_d^\Mdm$ over $2\le|d|\le N$:
\beq
e^{-it_mH}f \sim 
\sum_{\ell=1}^L \sum_{2\le|b|\le N}
G_{b,\la_\ell}(t_m)^*J_b(v_b/r_\ell)G_{b,\la_\ell}(t_m)
e^{-it_mH}f.\label{tag 4.40}
\ene
We note that on the RHS, the support with respect to $B^2/2$ of the derivative \linebreak $\phi'_{\sig_0}(\left|B^2-2(E-\la_\ell)\right|<2\tau_\ell+2\sig_0)$ is disjoint with $\Sigma(E)$ by \eq{tag 3.16} and \eq{tag 4.20}, and the support of $\varphi'(B)$ is similar by \eq{tag 4.8} and the definition of $\varphi$ above.

We prove the existence of the limit
\beq
f_{b,\ell}:=\lim_{t\to\infty}e^{itH}
G_{b,\la_\ell}(t)^*J_b(v_b/r_\ell)G_{b,\la_\ell}(t)
e^{-itH}f\label{tag 4.41}
\ene
for $\ell=1,\cdots,L$ and $b$ with $2\le|b|\le N$.

For this purpose we differentiate the function
\beq
(e^{itH}
G_{b,\la_\ell}(t)^*J_b(v_b/r_\ell)G_{b,\la_\ell}(t)
e^{-itH}f,g)\label{tag 4.42}
\ene
 with respect to $t$, where $f,g\in E_H(\Delta)\HH$. Then writing
\beq
D_t^b g(t)=i[H_b,g(t)]+\frac{dg}{dt}(t)\label{tag 4.43}
\ene
for an operator-valued function $g(t)$, we have
\beq
&&\frac{d}{dt}
(e^{itH}
G_{b,\la_\ell}(t)^*J_b(v_b/r_\ell)G_{b,\la_\ell}(t)
e^{-itH}f,g)\nom\\
&&=(e^{itH}D_t^b(\varphi(B))\phi_{\sig_0}(\left|B^2-2(E-\la_\ell)\right|<2\tau_\ell+2\sig_0)\nom\\
&&\q\times \phi_{\sig_0}(\left||x|^2/t^2-2(E-\la_\ell)\right|<2\tau_\ell+4\sig_0)P_b(t)J_b(v_b/r_\ell)G_{b,\la_\ell}(t)e^{-itH}f,g)\nom\\
&&+
(e^{itH}\varphi(B)D_t^b\left(\phi_{\sig_0}(\left|B^2-2(E-\la_\ell)\right|<2\tau_\ell+2\sig_0)\right)\nom\\
&&\q\times \phi_{\sig_0}(\left||x|^2/t^2-2(E-\la_\ell)\right|<2\tau_\ell+4\sig_0)P_b(t)J_b(v_b/r_\ell)G_{b,\la_\ell}(t)e^{-itH}f,g)\nom\\
&&+
(e^{itH}\varphi(B)\phi_{\sig_0}(\left|B^2-2(E-\la_\ell)\right|<2\tau_\ell+2\sig_0)\nom\\
&&\q\times D_t^b\left(\phi_{\sig_0}(\left||x|^2/t^2-2(E-\la_\ell)\right|<2\tau_\ell+4\sig_0)\right)P_b(t)J_b(v_b/r_\ell)G_{b,\la_\ell}(t)e^{-itH}f,g)\nom\\
&&+
(e^{itH}\varphi(B)\phi_{\sig_0}(\left|B^2-2(E-\la_\ell)\right|<2\tau_\ell+2\sig_0)\nom\\
&&\q\times \phi_{\sig_0}(\left||x|^2/t^2-2(E-\la_\ell)\right|<2\tau_\ell+4\sig_0)D_t^b\left(P_b(t)\right)J_b(v_b/r_\ell)G_{b,\la_\ell}(t)e^{-itH}f,g)\nom\\
&&+((h.c.)f,g)\nom\\
&&+
(e^{itH}i[I_b,G_{b,\la_\ell}(t)
J_b(v_b/r_\ell)G_{b,\la_\ell}(t)]e^{-itH}f,g),\label{tag 4.44}
\ene
where $(h.c.)$ denotes the adjoint of the operator in the terms preceding it.
\BP

We need the following lemmas (see \cite{[K1]}, Lemmas 4.1 and 4.2):

\BP

\begin{lem}\label{Lemma 4.2} Let Assumption \ref{ass1} be satisfied. Let $E\in R^1-\TT$. Let $F(s)\in C_0^\infty(R^1)$ satisfy $0\le F\le 1$ and the condition that the support with respect to $s^2/2$ of $F(s)$ is disjoint with $\Sigma(E)$. Then there is a constant $d(E)>0$ such that for any interval $\Delta$ around $E$ with {\rm diam} $\Delta<d(E)$, one has
\beq
\int_{-\infty}^\infty \left\Vert \frac{1}{\sqrt{\lan x\ran}}F(B)e^{-itH}E_H(\Delta)f\right\Vert^2 dt\le C\Vert f\Vert^2\label{tag 4.45}
\ene
for some constant $C>0$ independent of $f\in \HH$.
\end{lem}

\BP

\begin{lem}\label{Lemma 4.3} For the pseudodifferential operator $P_b(t)$ defined by \eq{tag 4.36} with $u>0$, there exist norm continuous bounded operators $S(t)$ and $R(t)$ such that
\beq
D_t^bP_b(t)=\frac{1}{t}S(t)+R(t)\label{tag 4.46}
\ene
and
\beq
S(t)\ge 0,\q\Vert R(t)\Vert\le C\lan t\ran^{-2}\label{tag 4.47}
\ene
for some constant $C>0$ independent of $t\in R^1$.
\end{lem}

We switch to a smaller interval $\Delta$ if necessary in the followings when we apply Lemma \ref{Lemma 4.2}.

For the first term on the RHS of \eq{tag 4.44} we have
\beq
D_t^b(\varphi(B))=\varphi'(B)i[H_b,B]+R_1\label{tag 4.48}
\ene
with
\beq
&&\Vert (H+i)^{-1}\lan x\ran^{1/2}i[H_b,B]\lan x\ran^{1/2}(H+i)^{-1}\Vert<\infty,\label{tag 4.49}\\
&&\Vert (H+i)^{-1}\lan x\ran R_1 \lan x\ran(H+i)^{-1}\Vert<\infty.\label{tag 4.50}
\ene
(See section 4 of \cite{[K1]} for a detailed argument yielding the estimates for the remainder terms $R_1$ here and $S_1(t)$, etc. below.)
By the remark after \eq{tag 4.40}, the support with respect to $B^2/2$ of $\varphi'(B)$ is disjoint with $\Sigma(E)$. Hence the condition of Lemma \ref{Lemma 4.2} is satisfied. Thus using \eq{tag 4.49}-\eq{tag 4.50} and rearranging the order of the factors in the first term on the RHS of \eq{tag 4.44} with some integrable errors, we have by Lemma \ref{Lemma 4.2}:
\beq
\mbox{the 1st term }=(e^{itH}B^{(1)}_2(t)^* B^{(1)}_1(t)e^{-itH}f,g)+(e^{itH}S_1(t)e^{-itH}f,g),\label{tag 4.51}
\ene
where $B^{(1)}_j(t)$ $(j=1,2)$ and $S_1(t)$ satisfy
\beq
&&\int_{-\infty}^\infty \Vert B^{(1)}_j(t)e^{-itH}f\Vert^2 dt\le C\Vert f\Vert^2,\label{tag 4.52}\\
&&\Vert (H+i)^{-1}S_1(t)(H+i)^{-1}\Vert\le Ct^{-2}\label{tag 4.53}
\ene
for some constant $C>0$ independent of $f\in E_H(\Delta)\HH$ and $t\in R^1$.

Similarly by another remark after \eq{tag 4.40} and Lemma \ref{Lemma 4.2}, we have a similar bound for the second term on the RHS of \eq{tag 4.44}:
\beq
\mbox{the 2nd term }=(e^{itH}B^{(2)}_2(t)^* B^{(2)}_1(t)e^{-itH}f,g)+(e^{itH}S_2(t)e^{-itH}f,g),\label{tag 4.54}
\ene
where $B^{(2)}_j(t)$ $(j=1,2)$ and $S_2(t)$ satisfy
\beq
&&\int_{-\infty}^\infty \Vert B^{(2)}_j(t)e^{-itH}f\Vert^2 dt\le C\Vert f\Vert^2,\label{tag 4.55}\\
&&\Vert (H+i)^{-1}S_2(t)(H+i)^{-1}\Vert\le Ct^{-2}\label{tag 4.56}
\ene
for some constant $C>0$ independent of $f\in E_H(\Delta)\HH$ and $t\in R^1$.

For the third term on the RHS of \eq{tag 4.44}, we have
\beq
&&\varphi(B)\phi_{\sig_0}(\left|B^2-2(E-\la_\ell)\right|<2\tau_\ell+2\sig_0)\nom\\
&&\q\times D_t^b\left(\phi_{\sig_0}(\left||x|^2/t^2-2(E-\la_\ell)\right|<2\tau_\ell+4\sig_0)\right)\nom\\
&&=\frac{2}{t}\varphi(B)\phi_{\sig_0}(\left|B^2-2(E-\la_\ell)\right|<2\tau_\ell+2\sig_0)\nom\\
&&\q\times \left(\frac{A}{t}-\frac{|x|^2}{t^2}\right)
\phi'_{\sig_0}(\left||x|^2/t^2-2(E-\la_\ell)\right|<2\tau_\ell+4\sig_0)\nom
\\
&&\q+S_3(t),\label{tag 4.57}
\ene
where $S_3(t)$ satisfies
\beq
\Vert (H+i)^{-1}S_3(t)(H+i)^{-1}\Vert\le C t^{-2},\q t>1.\label{tag 4.58}
\ene
On the support of $\phi'_{\sig_0}(\left||x|^2/t^2-2(E-\la_\ell)\right|<2\tau_\ell+4\sig_0)$, we have 
\beq
|x|/t\ge \sqrt{2(E-\la_\ell)-2\tau_\ell-5\sig_0}>0
\ene
 by \eq{tag 4.17} and \eq{tag 4.19}. Thus there is a large $T>1$ such that for $t\ge T$ we have $|x|>1$ and $\lan x\ran=|x|$, and hence
\beq
2\left(\frac{A}{t}-\frac{|x|^2}{t^2}\right)&
=&\frac{\lan x\ran}{t}\left(\frac{x}{\lan x\ran}\cdot D_x-\frac{|x|}{t}\right)+
\left(D_x\cdot\frac{x}{\lan x\ran}-\frac{|x|}{t}\right)\frac{\lan x\ran}{t}\nom
\\
&=&2\frac{\lan x\ran}{t}\left(B-\frac{|x|}{t}\right)+tS_4(t)\label{tag 4.59}
\ene
with
$\Vert S_4(t)\Vert\le Ct^{-2}$ for $t\ge T$.
By \eq{3.16}, we have
\beq
\mbox{supp}\hskip2pt\phi'_{\sig_0}(|s|<2\tau_\ell+4\sig_0)\subset I_1\cup I_2
\ene
 with
\beq
I_1=[-2\tau_\ell-5\sig_0,-2\tau_\ell-4\sig_0],\q
I_2=[2\tau_\ell+4\sig_0,2\tau_\ell+5\sig_0],\label{tag 4.60}
\ene
and 
\beq
&\phi'_{\sig_0}(|s|<2\tau_\ell+4\sig_0)\ge 0\q \mbox{for}\ s\in I_1,\label{tag 4.61}\\
&\phi'_{\sig_0}(|s|<2\tau_\ell+4\sig_0)\le 0\q \mbox{for}\ s\in I_2.\label{tag 4.62}
\ene
Consider the case $|x|^2/t^2-2(E-\la_\ell)\in I_2$. Then 
\beq
\frac{|x|^2}{t^2}\in [2(E-\la_\ell)+2\tau_\ell+4\sig_0,
2(E-\la_\ell)+2\tau_\ell+5\sig_0].\label{tag 4.63}
\ene
By the factor $\varphi(B)\phi_{\sig_0}(\left|B^2-2(E-\la_\ell)\right|<2\tau_\ell+2\sig_0)$, we have 
\beq
B^2\in [2(E-\la_\ell)-2\tau_\ell-3\sig_0,2(E-\la_\ell)+2\tau_\ell+3\sig_0]\label{tag 4.64}
\ene
 and
$B\ge \sqrt{2\la'_0}>0$. Thus
\beq
B-\frac{|x|}{t}\le 0.\label{tag 4.65}
\ene
Therefore
by \eq{tag 4.59} and \eq{tag 4.62}, \eq{tag 4.57} is positive in this case up to an integrable error. Similarly we see that \eq{tag 4.57} is positive also in the case
 $|x|^2/t^2-2(E-\la_\ell)\in I_1$. Rearranging the order of the factors in the third term on the RHS of \eq{tag 4.44} with an integrable error, we see that it has the form
\beq
\mbox{the 3rd term}=(e^{itH}A(t)^*A(t)e^{-itH}f,g)+(e^{itH}S_5(t)e^{-itH}f,g)\label{tag 4.66}
\ene
with
\beq
\Vert (H+i)^{-1}S_5(t)(H+i)^{-1}\Vert\le Ct^{-2}.\label{tag 4.67}
\ene

The fourth term on the RHS of \eq{tag 4.44} has a similar form by virtue of Lemma \ref{Lemma 4.3}.

The fifth term $((h.c.)f,f)$ is treated similarly to the terms above.

The sixth term on the RHS of \eq{tag 4.44} satisfies
\beq
|\mbox{the 6th term}|\le Ct^{-1-\min\{\ep,\ep_1\}}\Vert f\Vert \Vert g\Vert.\label{tag 4.68}
\ene
This estimate follows if we note with using \eq{tag 3.25} and some calculus of pseudodifferential operators as stated after \eq{tag 4.39} that the factor $G_{b,\la_\ell}(t)^*J_b(v_b/r_\ell)G_{b,\la_\ell}(t)$ restricts the coordinates in the region: $|x_\al|^2>\rho_{|b|}|x|^2/2$.

Summarizing we have proved that \eq{tag 4.44} is written as
\beq
&&\frac{d}{dt}
(e^{itH}
G_{b,\la_\ell}(t)^*J_b(v_b/r_\ell)G_{b,\la_\ell}(t)
e^{-itH}f,g)\nom\\
&&=(e^{itH}A(t)^*A(t)e^{-itH}f,g)+\sum_{k=1}^2(e^{itH}B^{(k)}_2(t)^*B^{(k)}_1(t)e^{-itH}f,g)\nom\\
&&\quad\quad\quad\quad +(S_6(t)f,g),\label{tag 4.69}
\ene
where with some constant $C>0$ independent of $t>T$ and $f\in \HH$
\beq
&&\int_T^\infty\Vert B^{(k)}_j(t)e^{-itH}E_H(\Delta)f\Vert^2\le C\Vert f\Vert^2,\q (j,k=1,2)\label{tag 4.70}\\
&&\Vert (H+i)^{-1}S_6(t)(H+i)^{-1}\Vert\le Ct^{-1-\min\{\ep,\ep_1\}}.\label{tag 4.71}
\ene

Integrating \eq{tag 4.69} with respect to $t$ on an interval $[T_1,T_2]\subset[T,\infty)$, we obtain
\beq
&& \left.(e^{itH}
G_{b,\la_\ell}(t)^*J_b(v_b/r_\ell)G_{b,\la_\ell}(t)
e^{-itH}f,g)\right|_{t=T_1}^{T_2}\nom\\
&&\q=
\int_{T_1}^{T_2}(A(t)e^{-itH}f,A(t)e^{-itH}g) dt\nom\\
&&\qq+
\sum_{k=1}^2\int_{T_1}^{T_2}(B^{(k)}_1(t)e^{-itH}f,B^{(k)}_2(t)e^{-itH}g)dt+\int_{T_1}^{T_2}(S_6(t)f,g)dt.\label{tag 4.72}
\ene
Hence using \eq{tag 4.70}, \eq{tag 4.71} and the uniform boundedness of $G_{b,\la_\ell}(t)$ in $t>1$, we have
\beq
\int_{T_1}^{T_2}\Vert A(t)e^{-itH}g\Vert^2 dt\le C\Vert g\Vert^2\label{tag 4.73}
\ene
for some constant $C>0$ independent of $T_2>T_1\ge T$ and $g\in E_H(\Delta)\HH$.

\eq{tag 4.73} and \eq{tag 4.72} with \eq{tag 4.70} and \eq{tag 4.71} then yield that
\beq
\left|\left.(e^{itH}
G_{b,\la_\ell}(t)^*J_b(v_b/r_\ell)G_{b,\la_\ell}(t)
e^{-itH}f,g)\right|_{t=T_1}^{T_2}\right|
\le \de(T_1)\Vert f\Vert\Vert g\Vert\label{tag 4.74}
\ene
for some $\de(T_1)>0$ with $\de(T_1)\to 0$ as $T_2>T_1\to\infty$. This means that
the limit
\beq
\tf_{b}^1=\lim_{t\to\infty}\sum_{\ell=1}^L e^{itH}
G_{b,\la_\ell}(t)^*J_b(v_b/r_\ell)G_{b,\la_\ell}(t)
e^{-itH}f\label{tag 4.75}
\ene
exists for any $f\in E_H(\Delta)\HH$ and $b$ with $2\le|b|\le N$ if $\Delta$ is an interval sufficiently small around $E\in R^1-\TT$: diam\hskip2pt$\Delta<d(E)$. Then the asymptotic decomposition \eq{tag 4.40} implies
\beq
f=\sum_{2\le|b|\le N} \tf_b^1\label{tag 4.76}
\ene
for $f=\psi(H)f=E_H(\Delta)f$.
Further by the existence of the limit \eq{tag 4.75} and $f=E_H(\Delta)f$, we see that $\tf_b^1$ satisfies
\beq
E_H(\Delta)\tf_b^1=\tf_b^1\label{tag 4.77}
\ene
in a way similar to the proof of the intertwining property of wave operators.

\MP

Now returning to the first $\Phi(H)f$, and noting that supp\hskip2pt$\Phi$ is compact in $R^1-\TT$, we take a finite number of open intervals $\Delta_{j_0}\subset\subset R^1-\TT$ such that $E_{j_0}\in\Delta_{j_0}$, diam\hskip2pt$\Delta_{j_0}<d(E_{j_0})$, and supp\hskip2pt$\Phi\subset\subset\bigcup_{j_0}^{\mbox{\scriptsize finite}}\Delta_{j_0}\subset\subset R^1-\TT$. Then we can take $\psi_{j_0}\in C_0^\infty(\Delta_{j_0})$ such that $\Phi(H)f=\sum_{j_0}^{\mbox{\scriptsize finite}}\psi_{j_0}(H)f$. Thus from \eq{tag 4.75}-\eq{tag 4.77}, we obtain the existence of the limit for $2\le|b|\le N$:
\beq
\tf_b^1=\lim_{t\to\infty}\sum_{j_0}^{\mbox{\scriptsize finite}} \sum_{\ell=1}^L e^{itH}
G_{b,\la_\ell}(t)^*J_b(v_b/r_\ell)G_{b,\la_\ell}(t)
e^{-itH}\psi_{j_0}(H)f,
\label{tag 4.78}
\ene
and the relations
\beq
\Phi(H)f=\sum_{2\le|b|\le N}\tf_b^1,\q E_H(\Delta)\tf_b^1=\tf_b^1\label{tag 4.79}
\ene
for any set $\Delta\subset\subset R^1-\TT$ with supp\hskip2pt$\Phi\subset\Delta$.

Set
\beq
\sig_j=\sqrt{\gamma^{-1}\rho_j\la'_0/2},\q
\de_j=\sqrt{\gamma\theta_j\La'_0}\q (j=2,3,\cdots,N,\q \theta_N=0).\label{tag 4.80}
\ene
Then by \eq{tag 4.78}, some calculus of pseudodifferential operators, and
\beq
\mbox{supp}\hskip2pt{\left(J_b(x_b/r_\ell)\phi_{\sig_0}(\left||x|^2-2(E-\la_\ell)\right|<2\tau_\ell+4\sig_0)\right)}\subset\subset {\tilde T}_b(\gamma^{-1}\rho_{|b|},\gamma\theta_{|b|}),\label{tag 4.81}
\ene
which follows from \eq{tag 3.19}-\eq{tag 3.21}, we see that as $t\to\infty$
\beq
e^{-itH}\tf_b^1&\sim& 
\sum_{k=1}^K \sum_{\ell=1}^L G_{b,\la_\ell}(t)^*J_b(v_b/r_\ell)G_{b,\la_\ell}(t)
 e^{-itH}E_H(\Delta_k)f\nom\\
&\sim&
\prod_{\al\not\leq b}F(|x_\al|\ge \sig_{|b|} t)F(|x^b|\le \de_{|b|} t)\nom\\
&&\qq\qq\times \sum_{k=1}^K\sum_{\ell=1}^LG_{b,\la_\ell}(t)^*J_b(v_b/r_\ell)G_{b,\la_\ell}(t)
 e^{-itH}E_H(\Delta_k)f\nom\\
&\sim& \prod_{\al\not\leq b}F(|x_\al|\ge \sig_{|b|} t)F(|x^b|\le \de_{|b|} t) e^{-itH}\tf_b^1.\label{tag 4.82}
\ene
\eq{tag 4.79} and \eq{tag 4.82} imply
\beq
\tf_b^1\in S_b^{1\sig_{|b|}\de_{|b|}}(\Delta)\q(2\le|b|\le N).\label{tag 4.83}
\ene

\BP


\F
II) A refinement:
\BP

As in \eq{tag 4.18}-\eq{tag 4.19}, we take a finite subset $\{{\tilde\la}^b_\ell\}_{\ell=1}^{L_b}$ of $\TT_b$ for a constant $\tau_0^b>0$ with $\tau_0^b<\tau_0$ such that
\beq
\TT_b\subset\bigcup_{\ell=1}^{L_b}({\tilde\la}^b_\ell-\tau_0^b,{\tilde\la}^b_\ell+\tau^b_0),\label{tag 4.84}
\ene
and choose real numbers $\la_\ell^b\in R^1$, $\tau_\ell^b>0$ $(\ell=1,\cdots,L_b)$ and $\sig_0^b>0$ such that
\beq
&&\tau_\ell^b<\tau_0^b,\q \sig_0^b<\tau_0^b,\q |\la_\ell^b-{\tilde\la}^b_\ell|<\tau_0^b,\nom\\
&&\TT_b\subset\bigcup_{\ell=1}^{L_b}(\la^b_\ell-\tau^b_\ell,\la^b_\ell+\tau^b_\ell),\q (\la^b_\ell-\tau^b_\ell,\la^b_\ell+\tau^b_\ell)\subset ({\tilde \la}^b_\ell-\tau_0^b,{\tilde\la}^b_\ell+\tau_0^b),\nom\\
&&\mbox{dist}\{(\la^b_\ell-\tau^b_\ell,\la^b_\ell+\tau^b_\ell),(\la^b_k-\tau^b_k,\la_k^b+\tau_k^b)\}>4\sig_0^b(>0)\q\mbox{for any}\ \ell\ne k.\label{tag 4.85}
\ene 
We set $\TT^{F}_b=\{\la^b_\ell\}_{\ell=1}^{L_b}$ and
\beq
{\tilde\tau}_0^b=\min_{1\le\ell\le L_b}\{\sig_0^b,\tau_\ell^b\}.\label{tag 4.86}
\ene
Then, we take $\psi_1(\la)\in C_0^\infty(R^1)$ such that
\beq
&&0\le\psi_1\le 1,\label{tag 4.87}\\
&&\psi_1(\la)=\left\{
\begin{array}{ll}
1&\q  \mbox{for any}\ \la\ \mbox{with}\ |\la-\la^b_\ell|\le {\tilde\tau}_0^b/2\ \mbox{for some}\ \la^b_\ell\in\TT^F_b\\
0&\q \mbox{for any}\ \la\ \mbox{with}\ |\la-\la^b_\ell|\ge {\tilde\tau}_0^b\ \mbox{for all}\ \la^b_\ell\in\TT^F_b
\end{array}\right.
\label{tag 4.88}
\ene
and
we divide \eq{tag 4.78} as follows:
\beq
\tf_b^1=h_b+g_b,\label{tag 4.89}
\ene
where
\beq
h_b&=&\lim_{t\to\infty}\sum_{j_0}^{\mbox{\scriptsize finite}} \sum_{\ell=1}^L e^{itH}
G_{b,\la_\ell}(t)^*\psi_1(H^b)J_b(v_b/r_\ell)G_{b,\la_\ell}(t)
e^{-itH}\psi_{j_0}(H)f,\label{tag 4.90}\\
g_b&=&\lim_{t\to\infty}\sum_{j_0}^{\mbox{\scriptsize finite}} \sum_{\ell=1}^L e^{itH}
G_{b,\la_\ell}(t)^*(I-\psi_1)(H^b)J_b(v_b/r_\ell)G_{b,\la_\ell}(t)
e^{-itH}\psi_{j_0}(H)f.
\label{tag 4.91}
\ene
The proof of the existence of these limits is similar to that of ${\tilde f}_b^1$ in \eq{tag 4.78}, since the change in the present case is the appearance of the commutator $[H,\psi_1(H^b)]=[I_b,\psi(H^b)]$ whose treatment is quite the same as  that of the commutators including $I_b$ in \eq{tag 4.68}.
We introduce the decomposition \eq{tag 1.21} into $h_b$ and $g_b$ on the left of $e^{-itH}\psi_{j_0}(H)f$ as in \eq{tag 4.3}.
Then by the factor $J_b(v_b/r_\ell)G_{b,\la_\ell}(t)$, we see that only the terms with $d\le b$ in the sum in \eq{tag 4.3} remain asymptotically as $t=t_m\to\infty$ by the arguments similar to step I). On each summand $P_{d,E_j}\whP_{|d|-1}^{{\widehat M}_{d}^m}$ in these terms (see \eq{tag 4.5}), $H^b$ asymptotically equals $H^b_d=T_d^b+H^d=T_d^b+E_j\sim|x^b_d|^2/(2t_m^2)+E_j\sim|x^b|^2/(2t_m^2)+E_j$, where for $d\le b$, $H^b_d=T^b_d+H^d=H_d-T_b$, $T^b_d=T_d-T_b$ and $x^b=(x^b_d,x^d)$ is a clustered Jacobi coordinate inside the coordinate $x^b$. 
 Thus we have
\beq
&&\psi_1(H^b)J_b(v_b/r_\ell)G_{b,\la_\ell}(t)P_{d,E_j}\whP_{|d|-1}^{{\widehat M}_{d}^m}e^{-it_mH}\psi_{j_0}(H)f\nom\\
&&\sim\psi_1(|x^b|^2/(2t_m^2)+E_j)J_b(v_b/r_\ell)G_{b,\la_\ell}(t)P_{d,E_j}\whP_{|d|-1}^{{\widehat M}_{d}^m}e^{-it_mH}\psi_{j_0}(H)f\label{tag 4.92}
\ene
as $m\to\infty$. If $\psi_1(|x^b|^2/(2t_m^2)+E_j)\ne 0$, then for some $\ell=1,\cdots,L_b$
\beq
\left|\frac{|x^b|^2}{2t_m^2}-(\la_\ell^b-E_j)\right|\le {\tilde\tau}_0^b.\label{tag 4.93}
\ene
If $\ell$ is a (unique) $\ellj$ such that $E_j\in(\la_\ellj^b-\tau_\ellj^b,\la_\ellj^b+\tau_\ellj^b)$, we have
\beq
\frac{|x^b|^2}{2t_m^2}\le\tau_\ellj^b+{\tilde\tau}_0^b<\tau_0^b+{\tilde\tau}_0^b.\label{tag 4.94}
\ene
Thus setting $\deL=\sqrt{2(\tau_0^b+{\tilde\tau}_0^b)}$, we have
\beq
|x^b|\le\deL t_m.\label{tag 4.95}
\ene
If $\ell\ne\ellj$, we have by \eq{tag 4.93}
\beq
0\le \la_\ell^b-E_j+{\tilde\tau}_0^b,\nom
\ene
from which and \eq{tag 4.85}-\eq{tag 4.86} follows
\beq
\la_\ell^b-E_j\ge 4\sig_0^b.\nom
\ene
Thus from \eq{tag 4.93}
\beq
\frac{|x^b|^2}{2t_m^2}\ge 4\sig_0^b-{\tilde\tau}_0^b\ge 3\sig_0^b\ge 3{\tilde\tau}_0^b.\label{tag 4.96}
\ene
Setting $\sigL=\sqrt{6{\tilde\tau}_0^b}$ we then have for $\ell\ne \ellj$
\beq
|x^b|\ge \sigL t_m.\label{tag 4.97}
\ene
Therefore $h_b$ can be decomposed as
\beq
h_b=f_{b}^\deL+g_{b1}^\sigL,\label{tag 4.98}
\ene
where
\beq
&&f_{b}^\deL=\lim_{t\to\infty}\sum_{j_0}^{\mbox{\scriptsize finite}} \sum_{\ell=1}^L e^{itH}
G_{b,\la_\ell}(t)^*F(|x^b|\le\deL t)\psi_1(H^b)J_b(v_b/r_\ell)G_{b,\la_\ell}(t)
e^{-itH}\psi_{j_0}(H)f,\nom\\
\label{tag 4.99}\\
&&g_{b1}^\sigL=\lim_{t\to\infty}\sum_{j_0}^{\mbox{\scriptsize finite}} \sum_{\ell=1}^L e^{itH}
G_{b,\la_\ell}(t)^*F(|x^b|\ge \sigL t)\psi_1(H^b)J_b(v_b/r_\ell)G_{b,\la_\ell}(t)
e^{-itH}\psi_{j_0}(H)f,\nom\\
\label{tag 4.100}
\ene
The existence of the limit \eq{tag 4.99} is proved similarly to that of \eq{tag 4.90} by rewriting the factor $F(|x^b|\le\deL t)$ as a smooth one and absorbing it into $J_b(v_b/r_\ell)G_{b,\la_\ell}(t)$ with changing the constants in it suitably. The existence of \eq{tag 4.100} then follows from this, \eq{tag 4.78} and \eq{tag 4.98}.

For $g_b$, similarly to $g_{b1}^\sigL$ we obtain
\beq
&&g_b=\lim_{t\to\infty}\sum_{j_0}^{\mbox{\scriptsize finite}} \sum_{\ell=1}^L e^{itH}
G_{b,\la_\ell}(t)^*F(|x^b|\ge\sigL t)(I-\psi_1)(H^b)\label{tag 4.101}\\
&&\times J_b(v_b/r_\ell)G_{b,\la_\ell}(t)
e^{-itH}\psi_{j_0}(H)f.\nom
\ene
Setting
\beq
g_b^\sigL&=&g_{b1}^\sigL+g_b\label{tag 4.102}\\
&=&\lim_{t\to\infty}\sum_{j_0}^{\mbox{\scriptsize finite}} \sum_{\ell=1}^L e^{itH}
G_{b,\la_\ell}(t)^*F(|x^b|\ge\sigL t)J_b(v_b/r_\ell)G_{b,\la_\ell}(t)
e^{-itH}\psi_{j_0}(H)f,\nom
\ene
we obtain a decomposition of ${\tilde f}_b^1$:
\beq
{\tilde f}_b^1=f_b^\deL+g_b^\sigL,\label{tag 4.103}
\ene
where $f_b^\deL$ and $g_b^\sigL$ satisfy
\beq
e^{-itH}f_b^\deL&\sim& 
\prod_{\al\not\leq b}F(|x_\al|\ge \sig_{|b|} t)F(|x^b|\le \deL t)e^{-itH}f_b^\deL,\label{tag 4.104}\\
e^{-itH}g_b^\sigL&\sim& 
\prod_{\al\not\leq b}F(|x_\al|\ge \sig_{|b|} t)F(|x^b|\le \de_{|b|} t)F(|x^b|\ge\sigL t)e^{-itH}g_b^\sigL.\label{tag 4.105}
\ene
We can prove the existence of the limits
\beq
f_b^1=\lim_{\deL\downarrow0}f_b^\deL,\q
g_b^1=\lim_{\sigL\downarrow0}g_b^\sigL\label{tag 4.106}
\ene
in the same way as in Enss \cite{[E3]}, Lemma 4.8, because we can take $\psi_1$ in \eq{tag 4.87}-\eq{tag 4.88} monotonically decreasing when ${\tilde\tau}_0^b\downarrow0$ and the factors $F(|x^b|\le \deL t)$ and $F(|x^b|\ge\sigL t)$ can be treated similarly to $\psi_1$ by regarding $x^b/t$ as a single variable. Further we have as in \eq{tag 4.77}
\beq
E_H(\Delta)f_b^1=f_b^1,\q E_H(\Delta)g_b^1=g_b^1,\label{tag 4.107}
\ene
which, \eq{tag 4.104} and \eq{tag 4.106} imply
\beq
f_b^1 \in S_b^1.\label{tag 4.108}
\ene

Thus we have a decomposition:
\beq
{\tilde f}_b^1=f_b^1+g_b^1,\q f_b^1\in S_b^1.\label{tag 4.109}
\ene
$g_b^\sigL$ can be decomposed further by using the partition of unity of the ring $\sigL\le |x^b|/t\le \de_{|b|}$ with regarding $x^b$ as a total variable $x$ in Proposition \ref{pro32}. Arguing similarly to steps I) and II), we can prove that $g_b^1$ can be decomposed as a sum of the elements $f_d^1$ of $S_d^1$ with $d< b$.
Combining this with \eq{tag 4.76}, \eq{tag 4.108} and \eq{tag 4.109}, we obtain \eq{tag 4.1}.
\ $\Box$

\BP

We remark that Theorem \ref{Theorem 4.1} implies the asymptotic completeness when the long-range part $V_\al^L$ vanishes for all pairs $\al$, because in this case we see straightforwardly that $S_b^1=\RR(W_b^\pm)$, where $W_b^\pm$ are the short-range wave operators defined by
\beq
W_b^\pm=\mbox{s-}\lim_{t\to\pm\infty}e^{itH}e^{-itH_b}P_b.\label{tag 4.110}
\ene
For the case when long-range part does not vanish, 
we have the following 

\begin{thm}\label{Theorem 4.2} Let Assumptions \ref{ass1} and \ref{ass2} be satisfied. Then 
\MP

\F
i) For $2(2+\ep)^{-1}<r\le1$
\beq
S_b^r=S_b^1.\label{tag 4.111}
\ene
\MP

\F
ii) If $\ep>2(2+\ep)^{-1}$, i.e. when $\ep>\sqrt{3}-1$, we have for all $r$ with $0\le r\le 1$
\beq
S_b^r=S_b^1.\label{tag 4.112}
\ene
\MP

\F
iii) If $\ep>1/2$ and $V^L_\al(x_\al)\ge 0$ for all pairs $\al$, then we have for all $r$ with $0\le r\le 1$
\beq
S_b^r=S_b^1.\label{tag 4.113}
\ene
\end{thm}
{\it Proof:} i) and ii) follow from Proposition 5.8 of \cite{[De]} and Proposition \ref{pro2.3} above. \eq{tag 4.112} for $r=0$ follows from the proof of Proposition 5.8 of \cite{[De]}. iii) follows from Theorem 1.1 and Proposition 4.3 of \cite{[Ki($N$)]} and \eq{tag 4.116} below: Note that $\RR(\Omega_b^\psi)$ in Theorem 1.1-(1.31) of \cite{[Ki($N$)]} constitutes a dense subset of $S_b^1$, when $\psi$ varies in $C_0^\infty(R^1-\TT)$.
\ $\Box$

{}From Theorem \ref{Theorem 4.2}-ii), iii) and Theorem \ref{Theorem 4.1} follows

\begin{thm}\label{Theorem 4.3} Let Assumptions \ref{ass1} and \ref{ass2} be satisfied with
 $\ep>2(2+\ep)^{-1}$ or with $\ep>1/2$ and $V^L_\al(x_\al)\ge 0$ for all pairs $\al$. Then we have for all $r$ with $0\le r\le 1$
\beq
\bigoplus_{2\le|b|\le N} S_b^r =\HH_c(H).\label{tag 4.114}
\ene
\end{thm}

In the next section, we will construct modified wave operators: 
\beq
W_b^\pm=\mbox{s-}\lim_{t\to\pm\infty}e^{itH}J_be^{-itH_b}P_b\label{tag 4.115}
\ene
with $J_b$ being an extension of $J$ of \cite{[IK]} to the $N$-body case.
We will then prove
\beq
\RR(W_b^\pm)=S_b^0,\label{tag 4.116}
\ene
which and Theorems \ref{Theorem 4.2} and \ref{Theorem 4.3} imply

\begin{thm}\label{Theorem 4.4} Let Assumptions \ref{ass1} and \ref{ass2} be satisfied with
 $\ep>2(2+\ep)^{-1}$ or with $\ep>1/2$ and $V^L_\al(x_\al)\ge 0$ for all pairs $\al$. Then we have for all $r$ with $0\le r\le 1$
\beq
\RR(W_b^\pm)=S_b^r,\label{tag 4.117}
\ene
and
\beq
\bigoplus_{2\le|b|\le N}\RR(W_b^\pm)=\HH_c(H).\label{tag 4.118}
\ene
\end{thm}

One might expect that \eq{tag 4.114} and \eq{tag 4.118} are always true, but it is denied:

\begin{thm}\label{Theorem 4.5} Let Assumptions \ref{ass1} and \ref{ass2} be satisfied and let $N\ge 3$. 
Then the followings hold:
\MP

\F
i) Let $2\le|b|\le N$ and
let $E_b(r)$ be the orthogonal projection onto $S_b^r$ $(0\le r\le 1)$. Then $E_b(r_1)\le E_b(r_2)$ for $0\le r_1\le r_2\le 1$, and the discontinuous points of $E_b(r)$ with respect to $r\in[0,1]$ in the strong operator topology are at most countable.
\MP

\F
ii) Let $0<\ep<1/2$ in Assumption \ref{ass1}. Then there are long-range pair potentials $V_\al(x_\al)$ such that for some cluster decomposition $b$ with $2\le|b|\le N$, $E_b(r)$ is discontinuous at $r=r_0$, where $\ep<r_0:=(\ep+1)/3<1/2$. In particular, there are real numbers $r_1$ and $r_2$ with $0\le r_1<r_0<r_2\le 1$ such that 
\beq
S_b^{r_1}\mbox{ is a proper subset of } S_b^{r_2}.\label{tag 4.119}
\ene
\end{thm}
{\it Proof:} i) By Proposition \ref{pro2.3}, $S_b^r$ $(0\le r\le 1)$ is a family of closed subspaces of a separable Hilbert space $\HH$ that increases when $r\in[0,1]$ increases. Thus the corresponding orthogonal projection $E_b(r)$ $(0\le r\le 1)$ onto $S_b^r$ increases as $r$ increases, and hence has at most a countable number of discontinuous points with respect to $r\in[0,1]$ in the strong operator topology.

ii) holds by Theorem 4.3 of \cite{[Yaf]}, Theorem \ref{Theorem 4.1} and Proposition \ref{pro2.3}, for $b$, $|b|=N-1$, with a suitable choice of pair potentials that satisfy Assumption \ref{ass1}.  In fact, the sum of the ranges $\RR(W_n)$ of Yafaev's wave operators $W_n$ $(n=1,2,\cdots)$ in Theorem 4.3 of \cite{[Yaf]} constitutes a subspace of $(E_b(r_0+0)-E_b(r_0-0))\HH$ for $b$ with $|b|= N-1$ by his construction of $W_n$, which means that $E_b(r)$ is discontinuous at $r=r_0$. Here $E_b(r_0\pm 0)=\mbox{s-}\lim_{r\to r_0\pm 0}E_b(r)$. \ $\Box$

\vskip 12pt

\BP

\section{A characterization of the ranges of wave operators}\label{characterization}

The purpose in this section is to prove relation \eq{tag 4.116} for general long-range pair potentials $V_\alpha(x_\alpha)$ under Assumptions \ref{ass1} and \ref{ass2}. The inclusion
\beq
\RR(W_b^\pm)\subset S_b^0\label{tag 5.1}
\ene
is a trivial relation for any form of definition of the wave operators $W_b^\pm$. Thus our main concern is to prove the reverse inclusion
\beq
S_b^0\subset \RR(W_b^\pm).\label{tag 5.2}
\ene
The proof of this inclusion is essentially the same for any definition of wave operators and is not difficult in the light of Enss method \cite{[E1]}. As announced, we here consider the wave operators of the form
\beq
W_b^\pm=\mbox{s-}\lim_{t\to\pm\infty}e^{itH}J_be^{-itH_b}P_b,\label{tag 5.3}
\ene
where $J_b$ is an extension of the identification operator or stationary modifier introduced in \cite{[IK]} for two-body long-range case. The first task in this section is to construct $J_b$. Our relation \eq{tag 5.2} then follows from the definition of the scattering spaces $S_b^0$ and properties of $J_b$ by Enss method.

To make the descriptions simple we hereafter consider the case $V_\alpha^S=0$ for all pairs $\alpha$. The recovery of the short-range potentials in the following arguments is easy.

Let a $C^\infty$ function $\chi_0(x)$ of $x\in R^\nu$ satisfy
\beq
\chi_0(x)=\left\{\begin{array}{ll}
1 & (|x|\ge 2)\\
0 & (|x|\le 1).
\end{array}\right. \label{tag 5.4}
\ene
To define $J_b$ we introduce time-dependent potentials $I_{b\rho}(x_b,t)$ for $\rho\in(0,1)$:
\beq
I_{b\rho}(x_b,t)=I_b(x_b,0)\prod_{k=1}^{k_b}\chi_0(\rho z_{bk})\chi_0(\langle \log\langle t\rangle\rangle z_{bk}/\langle t\rangle).\label{tag 5.5}
\ene
Then $I_{b\rho}(x_b,t)$ satisfies
\beq
|\partial_{x_b}^\beta I_{b\rho}(x_b,t)|\le C_\beta \rho^{\epsilon_0}\langle t\rangle^{-\ell}\label{tag 5.6}
\ene
for any $\ell\ge 0$ and $0<\epsilon_0<\epsilon$ with $\epsilon_0+\ell<|\beta|
+\epsilon$, where $C_\beta>0$ is a constant independent of $t,x_b$ and $\rho$.

Then we can apply almost the same arguments as in section 2 of \cite{[IK]} to get a solution $\varphi_b(x_b,\xi_b)$ of the eikonal equation:
\beq
\frac{1}{2}|\nabla_{x_b}\varphi_b(x_b,\xi_b)|^2+I_b(x_b,0)=\frac{1}{2}|\xi_b|^2
\label{tag 5.7}
\ene
in some conic region in phase space. More exactly we have the following theorems. Let
\beq
\cos(z_{bk},\zeta_{bk}):=\frac{z_{bk}\cdot \zeta_{bk}}{|z_{bk}|_e|\zeta_{bk}|_e},\nom
\ene
where $|z_{bk}|_e= (z_{bk}\cdot z_{bk})^{1/2}$ is the Euclidean norm. We then set for $R_0,d>0$ and $\theta\in(0,1)$
$$
\Gamma_{\pm}(R_0,d,\theta)=\{(x_b,\xi_b)\ |\ |z_{kb}|\ge R_0,|\zeta_{bk}|\ge d, \pm\cos(z_{bk},\zeta_{bk})\ge \theta\ (k=1,\cdots,k_b)\},
$$
where $\zeta_{bk}$ is the variable conjugate to $z_{bk}$.

\begin{thm}\label{Theorem 5.1} Let Assumption \ref{ass1} be satisfied with $V_\alpha^S=0$ for all pairs $\alpha$. Then there exists a $C^\infty$ function $\phi_b^\pm(x_b,\xi_b)$ that satisfies the following properties: For any $0<\theta,d<1$, there exists a constant $R_0> 1$ such that for any $(x_b,\xi_b)\in\Gamma_{\pm}(R_0,d,\theta)$
\beq
\frac{1}{2}|\nabla_{x_b}\phi_b^\pm(x_b,\xi_b)|^2+I_b(x_b,0)=\frac{1}{2}|\xi_b|^2
\label{tag 5.8}
\ene
and
\beq
|\partial_{x_b}^\alpha\partial_{\xi_b}^\beta(\phi_b^\pm(x_b,\xi_b)- x_b\cdot\xi_b)|\le
\left\{
\begin{array}{ll}
 C_{\alpha\beta}\left(\max_{1\le k\le k_b}\langle z_{bk}\rangle\right)^{1-\epsilon},& \alpha=0\\
 C_{\alpha\beta}\left(\min_{1\le k\le k_b}\langle z_{bk}\rangle\right)^{1-\epsilon-|\alpha|},& \alpha\ne 0,
\end{array}\right.
\label{tag 5.9}
\ene
where $C_{\alpha\beta}>0$ is a constant independent of $(x_b,\xi_b)\in\Gamma_\pm(R_0,d,\theta)$.
\end{thm}

{}From this we can derive the following theorem in quite the same way as that for Theorem 2.5 of \cite{[IK]}. Let $0<\theta<1$ and let $\psi_\pm(\tau)\in C^\infty([-1,1])$ satisfy
\beq
&0&\le \psi_\pm(\tau)\le 1,\nom\\
&&\psi_+(\tau)=\left\{
\begin{array}{ll}
 1 &\mbox{for}\ \theta\le \tau\le 1,\nom\\
0&\mbox{for}\ -1\le\tau\le\theta/2,
\end{array}\right.\\
&&\psi_-(\tau)=\left\{
\begin{array}{ll}
 0 &\mbox{for}\ -\theta/2\le\tau\le 1,\nom\\
1&\mbox{for}\ -1\le \tau\le-\theta.
\end{array}\right.
\ene
We set
\beq
\chi_\pm(x_b,\xi_b)=\prod_{k=1}^{k_b}\psi_\pm(\cos(z_{bk},\zeta_{bk}))\nom
\ene
and define $\varphi_b(x_b,\xi_b)=\varphi_{b,\theta,d,R_0}(x_b,\xi_b)$ by
\beq
\varphi_b(x_b,\xi_b)&=&\{(\phi_b^+(x_b,\xi_b)-x_b\cdot\xi_b)\chi_+(x_b,\xi_b)+
(\phi_b^-(x_b,\xi_b)-x_b\cdot\xi_b)\chi_-(x_b,\xi_b)\}\nom\\
&\times&\prod_{k=1}^{k_b}\chi_0(2\zeta_{bk}/d)\chi_0(2z_{bk}/R_0)+x_b\cdot\xi_b
\ene
for
$d,R_0>0$. Note that $\varphi_{b,\theta,d,R_0}(x_b,\xi_b)=\varphi_{b,\theta,d',R_0'}(x_b,\xi_b)$ when $|z_{bk}|\ge\max(R_0,R_0')$, $|\zeta_{bk}|\ge\max(d,d')$ for all $k$. We then have

\begin{thm}\label{Theorem 5.2} Let Assumption \ref{ass1} be satisfied with $V_\alpha^S=0$ for all pairs $\alpha$. Let $0<\theta<1$ and $d>0$. Then there exists a constant $R_0>1$ such that the $C^\infty$ function $\varphi_b(x_b,\xi_b)$ defined above satisfies the following properties.
\SP

\F
i) For $(x_b,\xi_b)\in \Gamma_+(R_0,d,\theta)\cup\Gamma_-(R_0,d,\theta)$, $\varphi_b$ is a solution of
\beq
\frac{1}{2}|\nabla_{x_b}\varphi_b(x_b,\xi_b)|^2+I_b(x_b,0)=\frac{1}{2}|\xi_b|^2.
\label{tag 5.10}
\ene

\F
ii) For any $(x_b,\xi_b)\in R^{2\nu(|b|-1)}$ and multi-indices $\alpha,\beta$, $\varphi_b$ satisfies
\beq
|\partial_{x_b}^\alpha\partial_{\xi_b}^\beta(\varphi_b(x_b,\xi_b)-x_b\cdot\xi_b)|
\le
\left\{\begin{array}{ll}
 C_{\alpha\beta}\left(\max\langle z_{bk}\rangle\right)^{1-\epsilon}, &\alpha=0\\
 C_{\alpha\beta}\left(\min\langle z_{bk}\rangle\right)^{1-\epsilon-|\alpha|}, &\alpha\ne0.
\end{array}\right.\label{tag 5.11}
\ene
In particular, if $\alpha\ne 0$,
\beq
|\partial_{x_b}^\alpha\partial_{\xi_b}^\beta(\varphi_b(x_b,\xi_b)-x_b\cdot\xi_b
)|\le C_{\alpha\beta}R_0^{-\epsilon_0}
\left(\min\langle z_{bk}\rangle\right)^{1-\epsilon_1-|\alpha|}\label{tag 5.12}
\ene
for any $\epsilon_0,\epsilon_1\ge 0$ with $\epsilon_0+\epsilon_1=\epsilon$. Further
\beq
\varphi_b(x_b,\xi_b)=x_b\cdot\xi_b\q \mbox{when } |z_{bk}|\le R_0/2\  \mbox{or}\ |\zeta_{bk}|\le d/2\ \mbox{for some}\ k.\label{tag 5.13}
\ene

\F
iii) Let
\beq
a_b(x_b,\xi_b)=e^{-i\varphi_b(x_b,\xi_b)}\left(T_b+I_b(x_b,0)-\frac{1}{2}|\xi_b|^2\right) e^{i\varphi_b(x_b,\xi_b)}.\label{tag 5.14}
\ene
Then
\beq
a_b(x_b,\xi_b)=\frac{1}{2}|\nabla_{x_b}\varphi_b(x_b,\xi_b)|^2+I_b(x_b,0)-\frac{1}{2}|\xi_b|^2+i(T_b\varphi_b)(x_b,\xi_b)\label{tag 5.15}
\ene
and
\beq
|\partial_{x_b}^\alpha\partial_{\xi_b}^\beta a_b(x_b,\xi_b)|
\le\left\{
\begin{array}{ll}
C_{\alpha\beta}\left(\min\langle z_{bk}\rangle\right)^{-1-\epsilon-|\alpha|},& (x_b,\xi_b)\in\Gamma_+(R_0,d,\theta)\cup\Gamma_-(R_0,d,\theta)\\
C_{\alpha\beta}\left(\min\langle z_{bk}\rangle\right)^{-\epsilon-|\alpha|}\langle\xi_b\rangle,&
\mbox{otherwise}.
\end{array}\right.\label{tag 5.16}
\ene
\end{thm}

We now define $J_b=J_{b,\theta,d,R_0}$ by
\beq
J_bf(x_b)=(2\pi)^{-\nu(|b|-1)}\int_{R^{\nu(|b|-1)}}\int_{R^{\nu(|b|-1)}}e^{i(\varphi_b(x_b,\xi_b)-y_b\cdot\xi_b)}f(y_b)dy_bd\xi_b\label{tag 5.17}
\ene
for $f\in \HH_b=L^2(R^{\nu(|b|-1)})$ as an oscillatory integral (see e.g. \cite{[KK]}).
Wave operators $W_b^\pm$ are now defined by
\beq
W_b^\pm=\mbox{s-}\lim_{t\to\pm\infty}e^{itH}J_be^{-itH_b}P_b.\label{tag 5.18}
\ene
We note that this definition depends on $\theta,d,R_0$, but applying stationary phase method to $e^{-itT_b}$ in $e^{-itH_b}=e^{-itT_b}\otimes e^{-itH^b}$ on the RHS we see that the dependence disappears in the limit $t\to\pm\infty$ by the remark made just before Theorem \ref{Theorem 5.2}. Further the asymptotic behavior seen by the stationary phase method tells that the inclusion \eq{tag 5.1} holds:
\beq
\RR(W_b^\pm)\subset S_b^0,\label{tag 5.19}
\ene
if the limits \eq{tag 5.18} exist. The existence of \eq{tag 5.18} follows from Theorem \ref{Theorem 5.2}-iii), the asymptotic behavior of $e^{-itT_b}$ and Assumptions \ref{ass1}-\ref{ass2} by noting the relations
\beq
&&(HJ_b-J_bH_b)e^{-itH_b}P_bf(x)=((T_b+I_b(x_b,x^b))J_b-J_bT_b)e^{-itH_b}P_bf(x),\nom\\
&&((T_b+I_b(x_b,0))J_b-J_bT_b)g(x_b)\nom\\
&&\hskip20pt =(2\pi)^{-\nu(|b|-1)}\int_{R^{\nu(|b|-1)}}\int_{R^{\nu(|b|-1)}}e^{i(\varphi_b(x_b,\xi_b)-y_b\cdot\xi_b)}a_b(x_b,\xi_b)g(y_b)dy_bd\xi_b
\label{tag 5.20}
\ene
and the fact that $\mbox{s-}\lim_{M\to\infty}P_b^M=P_b$. Thus to prove the reverse inclusion \eq{tag 5.2}
\beq
S_b^0\subset\RR(W_b^\pm),\label{tag 5.21}
\ene
it suffices to prove that
\beq
f\in S_b^0\ominus \RR(W_b^\pm)\label{tag 5.22}
\ene
implies
\beq
f=0.\label{tag 5.23}
\ene
To see this we consider the case $t\to+\infty$ and the quantity
\beq
(I-e^{isH}J_be^{-isH_b}J_b^{-1})e^{-itH}f=(J_b-e^{isH}J_be^{-isH_b})J_b^{-1}e^{-itH}f\label{tag 5.24}
\ene
for $f\in S_b^0(\Delta)$, $\Delta\subset\subset R^1-\TT$, and $t,s\ge 0$ and use Enss method. Here the existence of $J_b^{-1}$ follows from Theorem 3.3 of \cite{[K4]} by taking $R_0>0$ in \eq{tag 5.12} large enough (with a slight adaptation to the present case for phases and symbols satisfying the estimates in Theorem \ref{Theorem 5.2}). \eq{tag 5.24} equals
\beq
-i\int_0^s e^{iuH}(HJ_b-J_bH_b)e^{-iuH_b}J_b^{-1}du\ e^{-itH}f.\label{tag 5.25}
\ene
By Definition \ref{def21}-i)-\eq{tag 2.4} of $S_b^0(\Delta)$, we approximate $f$ by $h \in S_b^{0\sigma}(\Delta)$ for some small $\sigma>0$ with an arbitrarily small error $\delta>0$ so that $\Vert f-h\Vert<\delta$. Then we have for any sufficiently large $R>0$
\beq
\limsup_{t\to\infty}\left\Vert e^{-itH}h-\prod_{\alpha\not\le b}F(|x_\alpha|\ge \sigma t)F(|x^b|\le R)e^{-itH}h\right\Vert<\delta.\label{tag 5.26}
\ene
Proposition \ref{pro22}-iii) and $h \in S_b^{0\sigma}(\Delta)$ yield that for some sequence $t_m\to\infty$ (as $m\to\infty$)
\beq
\Vert(\varphi(x_b/t_m)-\varphi(v_b))e^{-it_mH}h\Vert\to0 \q \mbox{as}\q m\to\infty\nom
\ene
for any $\varphi\in C_0^\infty(R^{\nu(|b|-1)})$. Replacing $t$ and $f$ in \eq{tag 5.25} by $t_m$ and $h$, we can therefore insert or remove the factor
\beq
\Phi=\prod_{k=1}^{k_b}Q_k\tilde F(|p_{bk}|\ge \sigma')\tilde F(|p_b|\le S)\tilde F(|z_{bk}|/t\ge \sigma')\tilde F(|x^b|\le R) \label{tag 5.27}
\ene
to or from the left of $e^{-it_mH}h$ in \eq{tag 5.25} anytime with an error $\delta>0$. Here $p_{bk}=\frac{1}{i}\frac{\partial}{\partial z_{bk}}$, $\sigma'>0$ is a small number with $\sigma'<\sigma$, $\tilde F(|p_b|\le S)$ comes from $E_H(\Delta)$ in $h=E_H(\Delta)h$, and $\tilde F(\tau\le S)$ is a smooth characteristic function of the set $\{\tau\in R^1|\ \tau\le S\}$ with a slope independent of $S$, and $Q_k$ is a pseudodifferential operator
\beq
Q_kg(x_b)=(2\pi)^{-\nu(|b|-1)}\int_{R^{\nu(|b|-1)}}\int_{R^{\nu(|b|-1)}}
e^{(x_b\cdot \xi_b-y_b\cdot\xi_b)}
q_k(z_{bk},\zeta_{bk})g(y_b)dy_bd\xi_b\label{tag 5.28}
\ene
with symbol $q_k(z_{bk},\zeta_{bk})$ satisfying
\beq
\left.\begin{array}{ll}
&|\partial_{z_{bk}}^\beta\partial_{\zeta_{bk}}^\gamma q_k(z_{bk},\zeta_{bk})|
\le C_{\beta\gamma}\langle z_{bk}\rangle^{-|\beta|}
\langle \zeta_{bk}\rangle^{-|\gamma|},\\
&q_k(z_{bk},\zeta_{bk})=0\q \mbox{for}\q \cos(z_{bk},\zeta_{bk})\le \theta\ \mbox{or} \ |z_{bk}|\le R_0.
\end{array}\right.
\label{tag 5.29}
\ene
The order of products in \eq{tag 5.25} of factors in \eq{tag 5.27} and $J_b^{-1}$ may be arbitrary because these factors are mutually commutative asymptotically as $t\to\infty$ by virtue of \eq{tag 5.26}.
We note that $d>0$ in the definition of $J_b=J_{b,\theta,d,R_0}$ can be taken smaller than $\sigma' >0$ beforehand since $W_b^+$ is independent of $d>0$ as mentioned. Thus we can assume the following in addition to \eq{tag 5.29}:
\beq
q(z_{bk},\zeta_{bk})=0\q \mbox{for}\q |\zeta_{bk}|\le d.\label{tag 5.30}
\ene

We now insert the decomposition \eq{tag 1.21} to the left of $e^{-it_mH}h$ in \eq{tag 5.25} with noting $(I-P^{M_1^m})f=f$ by $f\in \HH_c(H)$. Then by $\Vert(I-P^{M_1^m})h-h\Vert<2\delta$ and by inserting the factor \eq{tag 5.27} to the left of $e^{-it_mH}h$ after the insertion of \eq{tag 1.21}, we have
\beq
\limsup_{m\to\infty}\Vert(I-P_b^{M_{|b|}^m})\Phi e^{-it_mH}h\Vert<3\delta.\label{tag 5.31}
\ene
By the factor $\tilde F(|x^b|\le R)$ in \eq{tag 5.27} and $E_H(\Delta)$ in $h=E_H(\Delta)h$, $P_b^{M_{|b|}^m}$ in \eq{tag 5.31} converges to $P_b$ as $m\to\infty$ in operator norm in the expression \eq{tag 5.31}.
It thus suffices to consider the quantity
\beq
\int_0^s e^{iuH}((T_b+I_b(x_b,x^b))J_b-J_bT_b)e^{-iuH_b}J_b^{-1}du\ P_b^{M_{|b|}^{m_0}}\Phi e^{-it_mH}h\label{tag 5.32}
\ene
for some large but fixed $m_0$ with an error $\delta>0$.
Since $P_b^{M_{|b|}^{m_0}}=\sum_{j=1}^{M_{|b|}^{m_0}} P_{b,E_j}$ $(0\le M_{|b|}^{m_0} <\infty)$ with $P_{b,E_j}$ being one dimensional eigenprojection of $H^b$ corresponding to eigenvalue $E_j$, \eq{tag 5.32} is reduced to considering
\beq
\int_0^s e^{-iuE_j} e^{iuH}((T_b+I_b(x_b,x^b))J_b-J_bT_b)P_{b,E_j}e^{-iuT_b}J_b^{-1}\Phi du\  e^{-it_mH}h.\label{tag 5.33}
\ene
By Assumptions \ref{ass1}-\ref{ass2}, the factor $P_{b,E_j}$ bounds the variable $x^b$ and yields a short-range error of order $O(\left(\min\langle z_{bk}\rangle\right)^{-1-\ep})$ on the left of $e^{-iuT_b}$ when we replace $I_b(x_b,x^b)$ by $I_b(x_b,0)$, and we have that \eq{tag 5.33} equals
\beq
\int_0^s e^{-iuE_j} e^{iuH}P_{b,E_j}O(\langle x^b\rangle)((T_b+I_b(x_b,0))J_b-J_bT_b&+&O(\left(\min\langle z_{bk}\rangle\right)^{-1-\ep}))\label{tag 5.34}\\
&\times& e^{-iuT_b}J_b^{-1}\Phi du\  e^{-it_mH}h,\nom
\ene
where $O(\langle x^b\rangle)$ is an operator such that $\langle x^b\rangle^{-1}O(\langle x^b\rangle)$ is bounded.
Using \eq{tag 5.20} and the estimate \eq{tag 5.16} in Theorem \ref{Theorem 5.2}-iii) and applying the propagation estimates in Lemma 3.3-ii) of \cite{[IK]} (again with a slight adaptation to the present case), we now get the estimate:
\beq
\Vert ((T_b+I_b(x_b,0))J_b-J_bT_b+O(\left(\min\langle z_{bk}\rangle\right)^{-1-\ep}))
e^{-iuT_b}J_b^{-1}\Phi\left(\min\langle z_{bk}\rangle\right)^{\ep/2}\Vert \le C\langle u\rangle^{-1-\ep/2}\nom\\
\label{tag 5.35}
\ene
for some constant $C>0$ independent of $u\ge 0$. On the other hand \eq{tag 5.26} yields
 that
\beq
\Vert\left(\min\langle z_{bk}\rangle\right)^{-\ep/2}e^{-it_mH}h\Vert\nom
\ene
is asymptotically less than $2\delta$ as $m\to\infty$.
This and \eq{tag 5.35} prove that the norm of \eq{tag 5.32} is asymptotically less than a constant times $\delta$ as $m\to\infty$.

Returning to \eq{tag 5.24} we have proved that
\beq
&&\limsup_{m\to\infty}\sup_{s\ge 0}\Vert (I-e^{isH}J_b e^{-isH_b}J_b^{-1})e^{-it_mH} f\Vert\nom\\
&&\q \approx_\delta \limsup_{m\to\infty}\sup_{s\ge 0}\Vert (I-e^{isH}J_b e^{-isH_b}J_b^{-1})P_b^{M_{|b|}^m}e^{-it_mH} f\Vert\le C \delta,
\label{tag 5.36}
\ene
where $a\approx_\delta b$ means that $|a-b|\le C\delta$ for some constant $C>0$.
Since wave operator $W_b^+=\mbox{s-}\lim_{s\to\infty}e^{isH}J_be^{-isH_b}P_b$ exists, \eq{tag 5.36} yields
\beq
\limsup_{m\to\infty}\Vert (I-W_b^+J_b^{-1})P_b^{M_{|b|}^m}e^{-it_mH}f\Vert \le C\delta.\label{tag 5.37}
\ene
By the arguments above deriving \eq{tag 5.31} we can remove $P_b^{M_{|b|}^m}$ and get
\beq
\limsup_{m\to\infty}\Vert (I-W_b^+ J_b^{-1})e^{-it_mH}f\Vert \le C\delta.\label{tag 5.38}
\ene
Since we assumed \eq{tag 5.22}, $f$ is orthogonal to $\RR(W_b^+)$. Thus taking the inner product of the vector inside the norm in \eq{tag 5.38} with $e^{-it_mH}f$, we have
\beq
\Vert f\Vert^2=\lim_{m\to\infty}|(e^{-it_mH}f, e^{-it_mH}f)|
=\lim_{m\to\infty}|(e^{-it_mH}f,(I-W_b^+ J_b^{-1})e^{-it_mH}f)|
\le C\delta\Vert f\Vert.\nom
\ene
As $\delta>0$ is arbitrary, this gives $f=0$, proving \eq{tag 5.23}. The proof of \eq{tag 4.116} is complete.

\BP

\newpage



\noindent
{\bf Exercise}
\BP

\noindent
{\bf 1}. With $W_\pm$ being defined by \eq{modified-wave-operators} we consider
$$
W_\pm W_\pm^*f
$$
for $f$ belonging to a suitable subspace $\DD$ of $\HH_c(H)$. Show that the operator
$$
JJ^*-I
$$
defines a compact operator on $\DD$, and prove the asymptotic completeness of $W_\pm$ without utilizing the existence of the inverse $J^{-1}$ of $J$.

We remark that the existence of the inverse is required in section \ref{characterization} as the inverse $J_b^{-1}$ of $J_b$.

\BP

\noindent
{\bf 2}. Let $\chi(x)$ $(x\in R^1)$ be a real-valued $C^\infty$ function with compact support such that $\chi(0)=1$, and let $g=g(x)$ be a complex-valued $C^\infty$ function with compact support defined on $R^1$.
Set for $t>0$
$$
f(t,x)=\lim_{\epsilon\downarrow 0}\frac{1}{\sqrt{2\pi it}}\int_{-\infty}^{\infty} e^{i\frac{(x-y)^2}{2t}}g(y)\chi(\epsilon y)dy,
$$
where $\arg i=\displaystyle\frac{\pi}{2}$. Show the following.

\BP

\F
i) For $t>0$
$$
\lim_{\epsilon\downarrow 0}\frac{1}{\sqrt{2\pi it}}\int_{-\infty}^{\infty} e^{i\frac{(x-y)^2}{2t}} \chi(\epsilon y)dy =1.
$$

\MP

\F
ii) For $t>0$ the following holds uniformly in $x\in R^1$
$$
|f(t,x)-g(x)|\le C\sqrt{t},
$$
where $C>0$ is a constant depending only on $g$.
\BP

\F
{\bf 3}. Let $f$ be a bounded, uniformly continuous function from $R^1$ to $\C$. Let $L^1$ be the totality of the Lesbegue integrable functions on $R^1$. For $\varphi\in L^1$, define
$$
(\varphi* f)(t)=\int_{-\infty}^\infty\varphi(t-r)f(r)dr.
$$
Prove for a given $A\in \C$ that the following two conditions 1) and 2) are  mutually equivalent.
\begin{namelist}{888}
\item[1)] The limit
$$
\lim_{t\to\infty}f(t)
$$
exists and equals $A$.
\item[2)] For any $\varphi\in L^1$, the limit
$$
\lim_{t\to\infty}(\varphi* f)(t)
$$
exists and equals
$$
A\int_{-\infty}^\infty\varphi(r)dr.
$$
\end{namelist}

\noindent
{\bf 4}. Let ${\cal H}$ be a Hilbert space and let $H_1,H_2$ be selfadjoint operators defined in ${\cal H}$ that satisfy the following relation for some bounded operators $A_1,A_2$ defined on ${\cal H}$
$$
H_2=H_1+A_2^*A_1.
$$
Assume that there exist constants $C_j>0$ $(j=1,2)$ such that
$$
\int_0^\infty\Vert A_je^{-itH_j}f\Vert^2dt\le C_j\Vert f\Vert^2\quad(\forall f\in {\cal H}, \ j=1,2)
$$
hold. Then show that for any $f\in{\cal H}$, the limit
$$
Wf=\lim_{t\to\infty}e^{itH_2}e^{-itH_1}f
$$
exists in ${\cal H}$ and defines a unitary operator on ${\cal H}$.\footnote[5]{This is a simplified version of the result of T. Kato, {\it Wave operators and similarity for some non-selfadjoint operators}, Math. Annalen 162 (1966), 258-279.}

\part{Observation}\label{observation-part}

\chapter{Principle of General Relativity}\label{chap:8}

\F
We now see how we can combine relativity and quantum mechanics in our formulation.
\BP

\BP

\F
We note that the center of mass of a local system 
$(H_{n\ell},\HH_{n\ell})$ is always at the origin of the space
 coordinate system $x_{(H_{n\ell},\HH_{n\ell})}\in R^3$ for the
 local system by the requirement: $\sum_{j\in F^\ell_{n+1}} m_j X_j=0$ in
 Axiom \ref{axiom2}, and that the space coordinate system describes
 just the relative motions inside a local system
 by our formulation. The center of mass of a local system,
 therefore, cannot be identified from the local system
 itself, except the fact that it is at the origin of the coordinates.

Moreover, as any two local systems $(H_{n\ell},\HH_{n\ell})$ and $(H_{mk},\HH_{mk})$ are independent mutually in the sense that QM inside $(H_{n\ell},\HH_{n\ell})$ does not affect the QM of another system $(H_{mk},\HH_{mk})$, we see
 that the time coordinates $t_{(H_{n\ell},\HH_{n\ell})}$
 and $t_{(H_{mk},\HH_{mk})}$, and the space coordinates 
$x_{(H_{n\ell},\HH_{n\ell})}\in R^3$ and 
$x_{(H_{mk},\HH_{mk})}\in R^3$ of these two local systems are
 independent mutually. Thus the space-time coordinates 
$(t_{(H_{n\ell},\HH_{n\ell})},x_{(H_{n\ell},\HH_{n\ell})})$ and 
$(t_{(H_{mk},\HH_{mk})},$ $x_{(H_{mk},\HH_{mk})})$ are independent
 between two different local systems $(H_{n\ell},\HH_{n\ell})$ and 
$(H_{mk},\HH_{mk})$. In particular, insofar as the systems are
 considered as quantum-mechanical ones, there is no relation between 
their centers of mass. In other words, the center of mass of any 
local system cannot be identified by other local systems
 quantum-mechanically.

Summing these two considerations, we conclude: 
\begin{namelist}{888}
\item[(1)] The center of mass of a local system $(H_{n\ell},\HH_{n\ell})$
 cannot be identified {\it quantum-mechani-cally} by any local system 
$(H_{mk},\HH_{mk})$ including the case 
$(H_{mk},\HH_{mk})=(H_{n\ell},\HH_{n\ell})$.

\item[(2)] There is no {\it quantum-mechanical} relation between any two
 local coordinates 
$(t_{(H_{n\ell},\HH_{n\ell})},$ $x_{(H_{n\ell},\HH_{n\ell})})$ and 
$(t_{(H_{mk},\HH_{mk})},x_{(H_{mk},\HH_{mk})})$ of two different
 local systems $(H_{n\ell},\HH_{n\ell})$ and $(H_{mk},\HH_{mk})$.
\end{namelist}
\MP

\F
Utilizing these properties of the centers of mass and the
 coordinates of local systems, we may make any postulates concerning
\begin{namelist}{888}
\item[(1)] the motions of the {\it centers of mass} of various local
 systems,
\end{namelist}
 and
\begin{namelist}{888}
\item[(2)] the relation between two local coordinates of any
 two local systems.
\end{namelist}
\MP

\F
In particular, we may impose {\it classical}
 postulates on them as far as the
 postulates are consistent in themselves. 
\MP

Thus we assume an arbitrary but fixed transformation:
\beq
y_2=f_{21}(y_1)\label{eqTr}
\ene
between the coordinate systems $y_j=(y_j^\mu)_{\mu=0}^3=(y_j^0,y_j^1,y_j^2,y_j^3)=(ct_j,x_j)$ for $j=1,2$,
 where $c$ is the speed of light in vacuum and $(t_j,x_j)$ is
 the space-time coordinates of the local system
 $L_j=(H_{n_j\ell_j},\HH_{n_j\ell_j})$. 
We regard these coordinates $y_j=(ct_j,x_j)$ as {\it classical}
coordinates, when we consider 
the motions of centers of mass and the relations of
 coordinates of various local systems.
We can now postulate the general principle of relativity on
 the physics of the centers of mass:

\MP

\hyphenation{centers}

\begin{axm}\label{axiom4} 
The laws of physics which control the relative motions of the
 centers of mass of local systems are covariant under the change of
 the reference frame coordinates from
 $(ct_{(H_{mk},\HH_{mk})},$ $x_{(H_{mk},\HH_{mk})})$ to 
$(ct_{(H_{n\ell},\HH_{n\ell})},x_{(H_{n\ell},\HH_{n\ell})})$
 for any
 pair $(H_{mk},\HH_{mk})$ and $(H_{n\ell},\HH_{n\ell})$ of
 local systems.
\end{axm}

\MP

We note that this axiom is
 consistent with the Euclidean metric adopted for the 
quantum-mechanical coordinates inside a local system, 
because Axiom \ref{axiom4} is concerned with classical motions of the centers
 of mass {\it outside} local systems, and we are dealing here with
 a different aspect of nature from the quantum-mechanical one 
{\it inside} a local system.

Axiom \ref{axiom4} implies the invariance of the distance under the change of
 coordinates between two local systems. Thus the metric tensor
 $g_{\mu\nu}(ct,x)$ which appears here satisfies the transformation rule:
\beq
g^1_{\mu\nu}(y_1)=g^2_{\alpha\beta}(f_{21}(y_1))
\frac{\partial f^\alpha_{21}}{\partial y_1^\mu}(y_1)
\frac{\partial f_{21}^\beta}{\partial y_1^\nu}(y_1),\label{eq5}
\ene
where $y_1=(ct_1,x_1)$; $y_2=f_{21}(y_1)$ is the transformation \eq{eqTr}
 in the above from $y_1=(ct_1,x_1)$ to $y_2=(ct_2,x_2)$; and 
$g_{\mu\nu}^j(y_j)$ is the metric tensor expressed in the classical
 coordinates $y_j=(ct_j,x_j)$ for $j=1,2$.

The second postulate is the principle of equivalence, which asserts
 that the classical coordinate system 
$(ct_{(H_{n\ell},\HH_{n\ell})},x_{(H_{n\ell},\HH_{n\ell})})$ is a
 local Lorentz system of coordinates, insofar as it is concerned
 with the classical behavior of the center of mass of the local
 system $(H_{n\ell},\HH_{n\ell})$:

\MP

\begin{axm}\label{axiom5} 
The metric or the gravitational tensor $g_{\mu\nu}$ for the center
 of mass of a local system $(H_{n\ell},\HH_{n\ell})$ in the
 coordinates 
$(ct_{(H_{n\ell},\HH_{n\ell})},x_{(H_{n\ell},\HH_{n\ell})})$ of
 itself are equal to $\eta_{\mu\nu}$, where $\eta_{\mu\nu}=0$ for 
$\mu\ne\nu$, $=1$ for $\mu=\nu=1,2,3$, and $=-1$ for $\mu=\nu=0$.
\end{axm}

\MP

Since, at the center of mass, the classical space
 coordinates $x=0$, Axiom \ref{axiom5} together with the transformation rule
 \eq{eq5} in the above yields
\beq
g^1_{\mu\nu}(f^{-1}_{21}(ct_2,0))=\eta_{\alpha\beta}
\frac{\partial f^\alpha_{21}}{\partial y_1^\mu}(f^{-1}_{21}(ct_2,0))
\frac{\partial f_{21}^\beta}{\partial y_1^\nu}(f^{-1}_{21}(ct_2,0)).
\label{eq6-1}
\ene
Also by the same reason, the relativistic
 proper time 
$d\tau=\sqrt{-g_{\mu\nu}(ct,0)dy^\mu dy^\nu}$ \linebreak
$=\sqrt{-\eta_{\mu\nu}dy^\mu dy^\nu}$ at the origin of a local system
 is equal to $c$ times the quantum-mechanical proper time $dt$
 of the system.

\BP

By the fact that the classical Axioms \ref{axiom4} and \ref{axiom5} of physics
 are imposed on the centers of mass which are uncontrollable 
 quantum-mechanically, and on the relation between the coordinates of
 different, therefore quantum-mechanically non-related local systems,
 the consistency of classical relativistic Axioms \ref{axiom4} and \ref{axiom5}
 with quantum-mechanical Axioms \ref{axiom2} and \ref{axiom3} is clear:

\BP

\begin{thm} Axioms \ref{axiom2}, \ref{axiom3}, \ref{axiom4}, and \ref{axiom5} are consistent.
\end{thm}


For the sake of completeness we state a formal proof of this theorem.
\BP

\F
{\it Proof:} 
The local coordinate system 
$(t_{(H_{n\ell},\HH_{n\ell})}, x_{(H_{n\ell},\HH_{n\ell})})$ 
 is determined only within each local system $(H_{n\ell},\HH_{n\ell})$ by Definition \ref{Time},
 through the quantum-mechanical {\it internal} motions of the system.
This coordinate system is independent of the local coordinate system
 $(t_{(H_{mk},\HH_{mk})},x_{(H_{mk},\HH_{mk})})$  of any other local system
 $(H_{mk},\HH_{mk})$.
 This is due to the mutual  independence of the $L^2$ representations (given by Axiom \ref{axiom2}) of
 the base Hilbert spaces $\HH_{n\ell}$ and $\HH_{mk}$.

The relativity axioms, Axioms \ref{axiom4} and \ref{axiom5}, are concerned merely with the 
{\it centers} of mass of local systems $(H_{mk},\HH_{mk})$, {\it observed} 
by an observer system $(H_{n\ell},\HH_{n\ell})$ with coordinate system
 $(t_{(H_{n\ell},\HH_{n\ell})}, x_{(H_{n\ell},\HH_{n\ell})})$.
 This {\it observer's} coordinate system $(t_{(H_{n\ell},\HH_{n\ell})},
 x_{(H_{n\ell},\HH_{n\ell})})$ is {\it independent of} the coordinate system
 $(t_{(H_{mk},\HH_{mk})}, x_{(H_{mk},\HH_{mk})})$ of the {\it observed}
 system $(H_{mk},\HH_{mk})$, as stated in the previous paragraph.
Because of this independence, the system $(H_{mk},\HH_{mk})$ can follow
 quantum mechanics (Axioms \ref{axiom2} and \ref{axiom3}) {\it inside} the system 
{\it with respect to} its own coordinate system $(t_{(H_{mk},\HH_{mk})},
 x_{(H_{mk},\HH_{mk})})$, {\it as well as} its {\it center} of mass
 can follow     general relativity (Axiom \ref{axiom4}) or any other given
 postulates {\it with respect to} the observer's coordinate system
 $(t_{(H_{n\ell},\HH_{n\ell})}, x_{(H_{n\ell},\HH_{n\ell})})$.
 This is the case, even if the coordinate system
 $(t_{(H_{n\ell},\HH_{n\ell})},  x_{(H_{n\ell},\HH_{n\ell})})$ of the
 observer coincides with the coordinate system $(t_{(H_{mk},\HH_{mk})},
 x_{(H_{mk},\HH_{mk})})$ of the observed system itself, because the motion
 of the {\it center} of mass and the internal {\it relative} motion of a local
 system are mutually independent.
Therefore, the local Lorentz postulate (Axiom \ref{axiom5}) for the {\it center} 
of mass of the system $(H_{n\ell},\HH_{n\ell})$ also does not contradict the
 Euclidean postulates in Axioms \ref{axiom2} and \ref{axiom3} of the internal space-time of that system.

In this sense, Axioms \ref{axiom4} and \ref{axiom5} are chosen so that the relativity theory
 holds between the {\it observed} motions of {\it centers} of mass of local 
systems, and {\it have nothing to do with} the {\it internal} motion of each
 local system, which obeys Axioms \ref{axiom2} and \ref{axiom3}.
Thus Axioms \ref{axiom4} and \ref{axiom5} are consistent with Axioms \ref{axiom2} and \ref{axiom3}.   $\Box$

\chapter{Observation}\label{chap:9}

\section{Preliminaries}

\F
Thus far, we did not mention any thoughts about the physics which
 is actually observed. We have just given two aspects of nature which
 are mutually independent. We will introduce a procedure which yields what
 we observe when we see nature. This procedure will not be
 contradictory with the two aspects of nature which we have discussed,
 as the procedure is concerned solely with ``{\it how nature looks, at
 the observer}," i.e. it is solely concerned with
 ``{\it at the place of the observer, how nature looks},"
 with some abuse of the word ``place."
 The validity of the procedure should be judged merely
 through the comparison between the observation and the prediction
 given by our procedure.

We remark that our approach to observation differs from the traditional approach like canonical quantization of gravity (see e.g., \cite{I}) or from quantum gravity by Ashtekar et al. \cite{AS}. They try to ``quantize" gravity with regarding the two aspects: quantum physics and gravity as lying on the same level. We do not regard gravity as an actual force or something similar, but we regard it as something `fictitious' as assumed in axiom \ref{axiom5}. Nevertheless as we will see below, we can explain some of the relativistic quantum mechanical phenomena.
\BP

 We note that, in observation, we can observe only a finite number of disjoint
 systems, say $L_1,\cdots,L_k$ with $k\ge1$ a finite integer. We
 cannot grasp an infinite number of systems at a time. Further each
 system $L_j$ must have only a finite number of elements by the same
 reason. Thus these systems $L_1,\cdots,L_k$ may be identified with
 local systems in the sense of the part \ref{LocalSystem} and the later chapter \ref{StationaryUniverse}.

Local systems are quantum-mechanical systems, and their coordinates
 are confined to their insides insofar as we appeal to Axioms \ref{axiom2}--\ref{axiom3}.
 However we postulated Axioms \ref{axiom4} and \ref{axiom5} on the classical aspects of
 those coordinates, which make the local coordinates of a local
 system a classical reference frame for the centers of mass of other
 local systems. This leaves us the room to define observation as the
 {\it classical} observation of the centers of mass of local systems
 $L_1,\cdots,L_k$. We call this an observation of $L=(L_1,\cdots,L_k)$
 inquiring into sub-systems $L_1,\cdots,L_k$, where $L$ is a local
 system consisting of the particles which belong to one of the local
 systems $L_1,\cdots,L_k$.

When we observe the sub-local systems $L_1,\cdots,L_k$ of $L$,
 we observe the relations or motions among these sub-systems.
 Internally the local system $L$ behaves following the Hamiltonian
 $H_L$ associated to the local system $L$. However the actual
 observation differs from what the pure quantum-mechanical
 calculation gives for the system $L$. For example,
 when an electron is scattered by a nucleus with relative
 velocity close to that of light, the observation is different
 from the pure quantum-mechanical prediction.

%
\MP

The quantum-mechanical process inside the local system $L$ is
 described by the evolution
$$
\exp(-it_LH_L)f,
$$
where $f$ is the initial state of the system and $t_L$ is
 the local time of the system $L$. The Hamiltonian $H_L$
 is decomposed as
 follows in virtue of the local Hamiltonians $H_1,\cdots,H_k$,
 which correspond to the sub-local systems $L_1,\cdots,L_k$:
$$
H_L=H^b+T+I,\quad H^b=H_1+\cdots+H_k.
$$
Here $b=(C_1,\cdots,C_k)$ is the cluster decomposition corresponding
 to the decomposition $L=(L_1,\cdots,L_k)$ of $L$; $H^b=H_1+\cdots+H_k$
 is the sum of the internal energies $H_j$ inside $L_j$, and is an operator
 defined
 in the internal state space $\HH^b=\HH^b_1\otimes\cdots\otimes\HH^b_k$;
 $T=T_b$ denotes the intercluster free energy among the clusters
 $C_1,\cdots,C_k$ defined in the external state space $\HH_b$;
 and $I=I_b=I_b(x)=I_b(x_b,x^b)$ is the sum of the intercluster
 interactions between various two different clusters in the cluster
 decomposition $b$ (cf. section \ref{section3.2}).

The main concern in this process would be the case that the 
clusters $C_1,\cdots,C_k$ form asymptotically bound states as
 $t_L\to\infty$, since other cases are hard to be observed along
 the process when the observer's concern is, as is usually the case in the observation of scattering process, upon the final state of the {\it bound} sub-systems $L_1,\cdots,L_k$.

The evolution $\exp(-it_LH_L)f$ then behaves asymptotically as
 $t_L\to\infty$ as follows for some bound states $g_1,\cdots,g_k$
 ($g_j\in\HH^b_j$) of local Hamiltonians $H_1,\cdots,H_k$ and for
 some $g_0$ belonging to the external state space $\HH_b$:
\beq
\exp(-it_LH_L)f\sim \exp(-it_L h_b)g_0
\otimes\exp(-it_LH_1)g_1\otimes\cdots\otimes \exp(-it_LH_k)g_k,
\quad k\ge 1,
\label{eq6}
\ene
where $h_b=T_b+I_b(x_b,0)$. It is easy to see that
 $g=g_0\otimes g_1\otimes\cdots\otimes g_k$ is given by
$$
g=g_0\otimes g_1\otimes\cdots\otimes g_k=\Omega_b^{+\ast} f
=P_b \Omega_b^{+\ast} f,
$$
provided that the decomposition of the evolution $\exp(-it_LH_L)f$
 is of the simple form as in \eq{eq6}. Here $\Omega_b^{+\ast}$ is the
 adjoint operator of a canonical wave operator (\cite{[De]})
 corresponding to the
 cluster decomposition $b$:
$$
\Omega_b^+=s{\mbox{-}}\lim_{t\to\infty}\exp(itH_L)\cdot
\exp(-ith_b)\otimes\exp(-itH_1)\otimes\cdots\otimes\exp(-itH_k)P_b,
$$
where $P_b$ is the eigenprojection onto the eigenspace of the
 Hamiltonian $H^b=H_1+\cdots+H_k$. The process \eq{eq6} just describes
 the quantum-mechanical process inside the local system $L$, and
 does not specify any meaning related with observation up to the
 present stage.

To see what we observe in actual observations, let us reflect
what we observe in scattering process. We note that the
 observation of scattering processes is concerned with
 their initial and final stages.
 At the final stage of observation of scattering processes,
 the quantities observed are firstly the points hit by the scattered
 particles on a screen. If the circumstances
 are properly set up, one can further indicate the momentum
 of the scattered particles at the final stage to the extent
 that the uncertainty principle allows. Consider, for example,
 a scattering process of an electron by a nucleus. Given the
 magnitude of initial momentum of an electron relative to the
 nucleus, one can infer the magnitude of momentum of the electron
 at the final stage to be equal to the initial one by the law
 of conservation of energy, since the electron is far away from the nucleus at the initial and final stages, and hence the potential
 energy between them can be neglected compared to the
 relative kinetic energy. The direction of momentum at the final
 stage can also be specified, up to the error due to the
 uncertainty principle, by setting a sequence of slits toward
 the desired direction at each point on the screen. Then the
 observer detects only the electrons scattered to that direction.
 The magnitude of momentum at initial stage can be selected in
 advance by applying a uniform magnetic field to the electrons,
 perpendicularly to their momenta, so that they circulate around
 circles with the radius proportional to the magnitude of momentum,
 and by setting a sequence of slits that select the desired stream from those
 electrons. The selection of magnitude of initial momentum
 makes the direction of momentum ambiguous due to the
 uncertainty principle, since the sequence of slits lets the
 position of electrons accurate to some extent.
 To sum up, the sequences of slits at the initial and final stages
 necessarily require to take into account the uncertainty
 principle so that some ambiguity remains in the observation.
 
However, in the actual observation of a {\it single} particle,
 we {\it have to decide} at which point on the screen the 
particle hits and which momentum the particle has, using
 the prepared apparatus like the sequence of slits located at each 
point on the screen. Even if we impose an interval for the 
observed values, we {\it have to assume} that the edges of 
the interval are sharply designated. These are the assumption 
which we always impose on what is called ``observation." That is to say, 
we idealize the situation in any observation or in any measurement
 of a single particle so that the observed values for each particle
 are sharp for both of the configuration and momentum. In this 
sense, the values observed actually for each particle must be 
classical. We have then necessary and sufficient conditions
 to make predictions about the differential cross section, as we 
will see in section \ref{O1}.

 Summarizing, we observe just the classical quantities for each 
particle at the final stage of all observations. In other words, 
 even if we cannot know the values actually,
we have to {\it presuppose} that the values observed for each 
particle have sharp values.
 We can apply to this fact the remark stated in the 
fourth paragraph of this section about the possibility of defining 
observation as that of the {\it classical} centers of mass of 
local systems, and may assume that the actually observed values 
follow the classical Axioms \ref{axiom4} and \ref{axiom5}. Those sharp values actually 
observed for each particle will give, when summed over the large 
number of particles, the probabilistic nature of quantum physical 
phenomena of scattering processes.

Theoretically, the quantum-mechanical, probabilistic
 nature of scattering processes is described by differential
 cross section, defined as the square of the absolute value
 of the scattering amplitude gotten from scattering operators
 $S_{bd}=W_b^{+\ast} W_d^- $, where $W_b^\pm$ are usual wave
 operators. Given the magnitude of the initial momentum of the
 incoming particle and the scattering angle, the differential
 cross section gives a prediction about the probability at which
 point and to which direction on the screen each particle hits
 on the average. However, as we have remarked, the idealized 
point on the screen hit by each particle and the scattering angle 
given as an idealized difference between the directions of the 
initial and final momenta of each particle have sharp values, and 
the observation at the final stage is {\it classical}. We are 
then required to correct these classical observations by 
taking into account the classical relativistic effects to those 
classical quantities.
\BP

\section{The first step}\label{O1}

As the first step of the relativistic modification of the
 scattering process, we consider the scattering amplitude 
$\SSS(E,\theta)$, where $E$ denotes the energy level of the scattering
 process and $\theta$ is a parameter describing the direction of the
 scattered particles. Following our remark made in the previous
 paragraph, we make the following postulate on the scattering
 amplitude observed in actual experiment:
\BP

\begin{axm}\label{axiom61prime}
 When one observes the final stage of scattering
 phenomena, the total energy $E$ of the scattering process should be
 regarded as a classical quantity and is replaced by a relativistic
 quantity, which obeys the relativistic change of coordinates from
 the scattering system to the observer's system. 
\end{axm}
\BP

Since it is not known much about $\SSS(E,\theta)$ in the many body
 case, we consider an example of the two body case. Consider a
 scattering phenomenon of an electron by a Coulomb potential 
$Ze^2/r$, where $Z$ is a real number, $r=|x|$, and $x$ is the
 position vector of the electron relative to the scatterer. We assume
 that the mass of the scatterer is large enough compared to that of
 the electron and that $|Z|/137\ll 1$. Then quantum mechanics gives
 the differential cross section in a Born approximation:
$$
\frac{d\sigma}{d\Omega}=|{\cal S}(E,\theta)|^2 
\approx \frac{Z^2e^4}{16E^2\sin^4(\theta/2)},
$$
where $\theta$ is the scattering angle and $E$ is the total
 energy of the system of the electron and the scatterer.
 We assume that the observer is stationary with respect to
 the center of mass of this system of an electron and the
 scatterer. Then, since the electron is far away from the
 scatterer after the scattering and the mass of the scatterer
 is much larger than that of the electron, we may suppose that
 the energy $E$ in the formula in the above can be replaced by the
 {\it classical} kinetic energy of the electron by Axiom \ref{axiom61prime}.
 Then, assuming that the speed $v$ of the electron relative to
 the observer is small compared to the speed $c$ of light 
in vacuum and denoting the rest mass of the electron by $m$,
 we have by Axiom \ref{axiom61prime} that $E$ is observed to have the 
following relativistic value:
$$
E'=c\sqrt{p^2+m^2c^2}-mc^2
=\frac{mc^2}{\sqrt{1-(v/c)^2}}-mc^2\approx 
\frac{mv^2}{2\sqrt{1-(v/c)^2}},
$$
where $p=mv/\sqrt{1-(v/c)^2}$ is the relativistic momentum of
 the electron. Thus the differential cross section should be
 observed approximately equal to
\beq
\frac{d\sigma}{d\Omega}
\approx \frac{Z^2e^4}{4m^2v^4\sin^4(\theta/2)}(1-(v/c)^2).\label{eq7}
\ene
This coincides with the usual relativistic prediction obtained
 from the Klein-Gordon equation by a Born approximation. See 
\cite{[Kitada-ToL]}, p.297, for a case which involves the spin of the electron.

Before proceeding to the inclusion of gravity in the general $k$ cluster
 case, we review this two body case. We note that the two body case
 corresponds to the case $k=2$, where $L_1$ and $L_2$ consist of
 single particle, therefore the corresponding Hamiltonians $H_1$
 and $H_2$ are zero operators on $\HH^0={\mbox{\bf C}}=$ the complex
 numbers. The scattering amplitude ${\cal S}(E,\theta)$ in this 
case is an integral kernel of the scattering matrix 
${\widehat S}={\cal F}S{\cal F}^{-1}$, where $S=W^{+\ast}W^-$ is
 a scattering operator; 
$W^\pm=s$-$\lim_{t\to\pm\infty}\exp(itH_L)\exp(-itT)$ are
 wave operators ($T$ is negative Laplacian for short-range
 potentials under an appropriate unit system, while it has
 to be modified when long-range potentials are included);
 and $\cal F$ is Fourier transformation so that 
${\cal F}T{\cal F}^{-1}$ is a multiplication operator by 
$|\xi|^2$ in the momentum representation $L^2(R^3_\xi)$.
 By definition, $S$ commutes with $T$. This makes ${\widehat S}$
 decomposable with respect to $|\xi|^2={\cal F}T{\cal F}^{-1}$. Namely,
 for {\it a.e.} $E>0$, there is a unitary operator ${\cal S}(E)$
 on $L^2(S^2)$, $S^2$ being two dimensional sphere with radius
 one, such that for {\it a.e.} $E>0$ and $\omega\in S^2$
$$
({\widehat S}h)(\sqrt{E}\omega)
=\left({\cal S}(E)h(\sqrt{E}\cdot)\right)(\omega),
\quad h\in L^2(R^3_\xi)
=L^2((0,\infty),L^2(S^2_\omega),|\xi|^2d|\xi|).
$$
Thus ${\widehat S}$ can be written as ${\widehat S}=\{{\cal S}(E)\}_{E>0}$.
 It is known \cite{[I-Ki]} that ${\cal S}(E)$ can be expressed as 
$$
({\cal S}(E)\varphi)(\theta)=\varphi(\theta)
-2\pi i \sqrt{E}\int_{S^2}
{\cal S}(E,\theta,\omega)\varphi(\omega)d\omega
$$
for $\varphi\in L^2(S^2)$. The integral kernel 
${\cal S}(E,\theta,\omega)$ with $\omega$ being 
the direction of initial wave, is the scattering amplitude
 ${\cal S}(E,\theta)$ stated in the above and
 $|{\cal S}(E,\theta,\omega)|^2$ is called differential
 cross section. These are the most important quantities in physics
 in the sense that they are the {\it only} quantities which can be
 observed in actual physical observation. 

The energy level $E$ in the previous example thus corresponds
 to the energy shell $T=E$, and the replacement of $E$ by $E'$
 in the above means that $T$ is replaced by a {\it classical
 relativistic} quantity $E'=c\sqrt{p^2+m^2c^2}-mc^2$. We have then
 seen that the calculation in the above gives a correct relativistic
 result, which explains the actual observation.

Axiom \ref{axiom61prime} is concerned with the observation of the final stage
 of scattering phenomena. To include the gravity into our
 consideration, we extend Axiom \ref{axiom61prime} to the intermediate process
 of quantum-mechanical evolution. The intermediate process cannot
 be an object of any {\it actual} observation, because the
 intermediate observation would change the process itself,
 consequently the result observed at the final stage would be
 altered. Our next Axiom \ref{axiom61} is an extension of Axiom \ref{axiom61prime} from
 the {\it actual} observation to the {\it ideal} observation in
 the sense that Axiom \ref{axiom61} is concerned with such invisible
 intermediate processes and modifies the {\it ideal} intermediate
 classical quantities by relativistic change of coordinates. The
 spirit of the treatment developed below is to trace the
 quantum-mechanical paths by ideal observations so that the
 quantities will be transformed into classical quantities at
 each step, but the quantum-mechanical paths will {\it not} be altered
 owing to the {\it ideality} of the observations. The classical
 Hamiltonian obtained at the last step will be ``requantized" to
 recapture the quantum-mechanical nature of the process, therefore
 the ideality of the intermediate observations will be realized
 in the final expression of the propagator of the observed system.


\BP

\section{The second step}\label{O2}

With these remarks in mind, we return to the general
 $k$ cluster case, and consider a way to include gravity
 in our framework.

In the scattering process into $k\ge1$ clusters, what we observe
 are the centers of mass of those $k$ clusters $C_1,\cdots,C_k$,
 and of the combined system $L=(L_1,\cdots,L_k)$. In the example
 of the two body case of section \ref{O1}, only the combined
 system $L=(L_1,L_2)$ appears due to $H_1=H_2=0$, therefore the
 replacement of $T$ by $E'$ is concerned with the free energy
 between two clusters $C_1$ and $C_2$ of the combined system 
$L=(L_1,L_2)$. 

Following this treatment of $T$ in the section \ref{O1}, we
 replace $T=T_b$ in the exponent of 
$\exp(-it_Lh_b)=\exp(-it_L(T_b+I_b(x_b,0)))$ on the right hand
 side of the asymptotic relation \eq{eq6} by the relativistic kinetic
 energy $T'_b$ among the clusters $C_1,\cdots,C_k$ around the
 center of mass of $L=(L_1,\cdots,L_k)$, defined by
\beq
T'_b=\sum_{j=1}^k\left(c\sqrt{p_j^2+m_j^2c^2}-m_jc^2\right).
\label{eq8}
\ene
Here $m_j>0$ is the rest mass of the cluster $C_j$, which involves
 all the internal energies like the kinetic energies inside $C_j$
 and the rest masses of the particles inside $C_j$, and $p_j$ is
 the relativistic momentum of the center of mass of $C_j$ inside
 $L$ around the center of mass of $L$. For simplicity, we assume
 that the center of mass of $L$ is stationary relative to the
 observer. Then we can set in the exponent of 
$\exp(-it_L(T'_b+I_b(x_b,0)))$
\beq
t_L=t_O,\label{eq10}
\ene
where $t_O$ is the observer's time.

 For the factors $\exp(-it_LH_j)$ on the right hand side of \eq{eq6},
 the object of the {\it ideal} observation is the centers of
 mass of the $k$ number of clusters $C_1,\cdots,C_k$. These are
 the ones which now require the relativistic treatment. Since we
 identify the clusters $C_1,\cdots,C_k$ as their centers of mass
 moving in a classical fashion, $t_L$ in the exponent of
 $\exp(-it_LH_j)$ should be replaced by $c^{-1}$ times
 the classical relativistic proper time at the origin of the local
 system $L_j$, which is equal to the quantum-mechanical local time
 $t_j$ of the sub-local system $L_j$. By the same reason and by the
 fact that $H_j$ is the internal energy of the cluster $C_j$
 relative to its center of mass, it would be justified to
 replace the Hamiltonian $H_j$ in the exponent of $\exp(-it_jH_j)$
 by the classical relativistic energy {\it inside} the cluster
 $C_j$ around its center of mass
\beq
H'_j=m_jc^2, \label{eq9}
\ene
where $m_j>0$ is the same as in the above.

Summing up, we arrive at the following postulate, which has the
 same spirit as in Axiom \ref{axiom61prime} and includes Axiom \ref{axiom61prime} as a special
 case concerned with actual observation:
\BP

\begin{axm}\label{axiom61}
 In either actual or ideal observation, the
 space-time coordinates $(ct_L,x_L)$ and the four momentum
 $p=(p^\mu)=(E_L/c,p_L)$ of the observed system $L$ should be
 replaced by classical relativistic quantities, which are transformed
 into the classical quantities $(ct_O,x_O)$ and $p=(E_O/c,p_O)$ in
 the observer's system $L_O$ according to the relativistic change
 of coordinates specified in Axioms \ref{axiom4} and \ref{axiom5}. Here $t_L$ is the local
 time of the system $L$ and $x_L$ is the internal space coordinates
 inside the system $L$; and $E_L$ is the internal energy of the system
 $L$ and $p_L$ is the momentum of the center of mass of the system $L$.
\end{axm}
\BP

In the case of the present scattering process into $k$ clusters,
 the system $L$ in this axiom is each of the local systems $L_j$
 $(j=1,2,\cdots,k)$ and $L$.

We continue to consider the $k$ centers of mass of the clusters
 $C_1,\cdots,C_k$. At the final stage of the scattering process,
 the velocities of the centers of mass of the clusters
 $C_1,\cdots,C_k$ would be steady, say $v_1,\cdots,v_k$, 
relative to the observer's system. Thus, according to Axiom \ref{axiom61}, 
the local times $t_j$ $(j=1,2,\cdots,k)$ in the exponent of 
$\exp(-it_jH'_j)$, which are equal to $c^{-1}$ times the
 relativistic proper times at the origins $x_j=0$ of the
 local systems $L_j$, are expressed in the observer's time
 coordinate $t_O$ by
\beq
t_j=t_O\sqrt{1-(v_j/c)^2}\approx t_O\left(1-v_j^2/(2c^2)\right),
\quad j=1,2,\cdots,k,\label{eq11}
\ene
where we have assumed $|v_j/c|\ll 1$ and used Axioms \ref{axiom4} and \ref{axiom5} 
to deduce the Lorentz transformation:
$$
t_j=\frac{t_O-(v_j/c^2)x_O}{\sqrt{1-(v_j/c)^2}},\quad 
x_j=\frac{x_O-v_jt_O}{\sqrt{1-(v_j/c)^2}}.
$$
(For simplicity, we wrote the Lorentz transformation for the
 case of 2-dimensional space-time.)

Inserting \eq{eq8}, \eq{eq10}, \eq{eq9} and \eq{eq11} into the right-hand side of
 \eq{eq6}, we obtain a classical approximation of the evolution:
\beq
\exp\left(-it_O[(T'_b+I_b(x_b,0)+H'_1+\cdots+H'_k)
-(m_1v_1^2/2+\cdots+m_kv_k^2/2)]\right)\label{eq12}
\ene
under the assumption that  $|v_j/c|\ll 1$ for all $j=1,2,\cdots,k$.

 What we want to clarify is the final stage of the scattering
 process. Thus as we have mentioned, we may assume that
 all clusters $C_1,\cdots,C_k$ are far
 away from any of the other clusters and moving almost in steady
 velocities $v_1,\cdots,v_k$ relative to the observer. We denote by
 $r_{ij}$ the distance between two centers of mass of the clusters
 $C_i$ and $C_j$ for $1\le i<j\le k$. Then, according to our spirit
 that we are observing the behavior of the centers of mass of the
 clusters $C_1,\cdots,C_k$ in {\it classical} fashion following Axioms
 \ref{axiom4} and \ref{axiom5}, the clusters $C_1,\cdots,C_k$ can be regarded to have
 gravitation among them. This gravitation can be calculated if we
 assume Einstein's field equation, $|v_j/c|\ll 1$, and certain
 conditions that the gravitation is weak (see \cite{[M]}, section 17.4),
 in addition to our Axioms \ref{axiom4} and \ref{axiom5}. As an approximation of the 
 first order, we obtain the gravitational potential of Newtonian
 type for, e.g., the pair of the clusters 
$C_1$ and $U_1=\bigcup_{i=2}^kC_i$:
$$
-G\sum_{i=2}^km_1m_i/r_{1i},
$$
where $G$ is Newton's gravitational constant.

Considering the $k$ body classical problem for the $k$ clusters
 $C_1,\cdots,C_k$ moving in the sum of these gravitational fields,
 we see that the sum of the kinetic energies of $C_1,\cdots,C_k$
 and the gravitational potentials among them is constant by the
 classical law of conservation of energy:
$$
m_1v_1^2/2+\cdots+m_kv_k^2/2-G\sum_{1\le i<j\le k}m_im_j/r_{ij}
={\mbox{constant}}.
$$
Assuming that $v_j\to v_{j\infty}$ as time tends to infinity, we
 have constant $=m_1v_{1\infty}^2/2+\cdots+m_kv_{k\infty}^2/2$.
 Inserting this relation into \eq{eq12} in the above, we obtain the
 following as a classical approximation of the evolution \eq{eq6}:
\beq
\exp\left(-it_O\left[T'_b+I_b(x_b,0)
+\sum_{j=1}^k(m_jc^2-m_jv_{j\infty}^2/2)
-G\sum_{1\le i<j\le k}m_im_j/r_{ij}\right]\right).
\label{eq13}
\ene
What we do at this stage are {\it ideal} observations, and these
 observations should not give any sharp classical values. Thus
 we have to consider \eq{eq13} as a {\it quantum-mechanical evolution}
 and we have to recapture the quantum-mechanical feature of the
 process. To do so we replace $p_j$ in $T'_b$ in \eq{eq13} by a
 quantum-mechanical momentum $D_j$, where $D_j$ is a differential
 operator $-i\frac{\partial}{\partial x_j}
=-i\left(\frac{\partial}{\partial x_{j1}},
\frac{\partial}{\partial x_{j2}},
\frac{\partial}{\partial x_{j3}}\right)$ with respect 
to the 3-dimensional coordinates $x_j$ of the center of mass
 of the cluster $C_j$. Thus  the actual process should be
 described by \eq{eq13} with $T'_b$ replaced by a quantum-mechanical
 Hamiltonian
$$
\tT_b=\sum_{j=1}^k\left(c\sqrt{D_j^2+m_j^2c^2}-m_jc^2\right).
$$
This procedure may be called ``requantization," and is summarized
 as the following axiom concerning the ideal observation.
\BP

\begin{axm}\label{axiom62}
 In the expression describing the classical process
 at the time of the {\it ideal} observation, the intercluster momentum
 $p_j=(p_{j1},p_{j2},p_{j3})$ should be replaced by a quantum-mechanical
 momentum $D_j=-i\left(\frac{\partial}{\partial x_{j1}},
\frac{\partial}{\partial x_{j2}},\frac{\partial}{\partial x_{j3}}\right)$.
 Then this gives the evolution describing the intermediate 
{\it quantum-mechanical} process.
\end{axm}
\BP

We thus arrive at an approximation for a quantum-mechanical
 Hamiltonian including gravitational effect up to a constant term,
 which depends on the system $L$ and its decomposition into
 $L_1,\cdots,L_k$, but not affecting the quantum-mechanical
 evolution, therefore can be eliminated:
\SP

$$
\tH_L=\tT_b+I_b(x_b,0)-G\sum_{1\le i<j\le k}m_im_j/r_{ij}\qqq\qqq\qqq\qqq
$$
\vskip-8pt
\beq
=\sum_{j=1}^k\left(c\sqrt{D_j^2+m_j^2c^2}-m_jc^2\right)+I_b(x_b,0)
-G\sum_{1\le i<j\le k}m_im_j/r_{ij}.
\label{eq14}
\ene
\SP

\F
 We remark that the gravitational terms here come from the substitution
 of local times $t_j$ to the time $t_L$ in the factors $\exp(-it_L H_j)$
 on the right-hand side of \eq{eq6}. This form of Hamiltonian in \eq{eq14} is
 actually used in \cite{[Li]} with $I_b=0$ to explain the stability and
 instability of cold stars of large mass, showing the effectiveness
 of the Hamiltonian.

 Summarizing these arguments from \eq{eq6} to \eq{eq14}, we have obtained
 the following {\it interpretation} of the observation of the
 quantum-mechanical evolution: To get our prediction for the
 observation of local systems $L_1,\cdots,L_k$, the
 quantum-mechanical evolution of the combined local system
 $L=(L_1,\cdots,L_k)$
$$
\exp(-it_LH_L)f
$$
should be replaced by the following evolution, in the
 approximation of the first order under the assumption that
 $|v_j/c|\ll 1$ $(j=1,2,\cdots,k)$ and the gravitation is weak,
\beq
(\exp(-it_O\tH_L)\otimes 
\underbrace{I\otimes\cdots\otimes I}_{\mbox{\scriptsize $k$ factors}})P_b
\Omega_b^{+\ast}f,
\label{eq15}
\ene
provided that the original evolution $\exp(-it_LH_L)f$ decomposes
 into $k$ number of clusters $C_1,\cdots,C_k$ as $t_L\to\infty$ in
 the sense of \eq{eq6}. Here $b$ is the cluster decomposition
 $b=(C_1,\cdots,C_k)$ that corresponds to the decomposition
 $L=(L_1,\cdots,L_k)$ of $L$; $t_O$ is the observer's time; and
\beq
\tH_L={\widetilde T}_b+I_b(x_b,0)-G\sum_{1\le i<j\le k}m_im_j/r_{ij}
 \label{eq16}
\ene
is the relativistic Hamiltonian inside $L$ given by \eq{eq14}, which
 describes the motion of the centers of mass of the clusters
 $C_1,\cdots,C_k$. 

We remark that \eq{eq15} may produce a bound state combining 
$C_1,\cdots,C_k$ as $t_O\to\infty$ therefore for all $t_O$, 
due to the gravitational potentials in the exponent. Note that
 this is not prohibited by our assumption that $\exp(-it_LH_L)f$
 has to decompose into $k$ clusters $C_1,\cdots,C_k$, because the
 assumption is concerned with the original Hamiltonian $H_L$ but
 not with the resultant Hamiltonian $\tH_L$.
\BP

\hyphenation{dif-fer-en-tial}
\hyphenation{small}
\hyphenation{since}
\hyphenation{stated}

 Extending our primitive assumption Axiom \ref{axiom61prime}, which was valid for
 an example stated in section \ref{O1}, we have arrived at a
 relativistic Hamiltonian $\tH_L$, which would describe approximately
 the intermediate process, under the assumption
 that the gravitation is weak and the velocities of the particles
 are small compared to $c$, by using the Lorentz transformation.
 We note that, since we started our argument from the asymptotic
 relation \eq{eq6}, which is concerned with the final stage of
 scattering processes, we could assume that the velocities
 of particles are almost steady relative to the observer
 in the correspondent
 classical expressions of the processes, therefore we could appeal to
 the Lorentz transformations when performing the change of coordinates
 in the relevant arguments. 

 The final values of scattering amplitude should
 be calculated by using the Hamiltonian $\tH_L$. Then they would
 explain actual observations. This is our prediction for the observation
 of relativistic quantum-mechanical phenomena including the effects by
 gravity and quantum-mechanical forces.

 In the example discussed in section \ref{O1}, this approach gives
 the same result as \eq{eq7} in the approximation of the first order,
 showing the consistency of our spirit (see \cite{[Ki2]}).


\hyphenation{physics}

\BP

\newpage



\noindent
{\bf Exercise}
\BP

Let us consider the light clock which Einstein used in his definition of an ideal clock in special theory of relativity. A light clock consists of two mirrors stood parallel to each other with light running mirrored to each other continuously. The time is then measured as the number of counts that the light hits the mirrors. This clock is placed stationary to an inertial frame of reference, and the time of the frame is defined by the number of the light-hits of this clock. Insofar as the light is considered as a classical wave and the frame is an inertial one which has no acceleration, this clock can measure the time of the frame. One feature of this clock is that the time is defined by utilizing the distance between the two mirrors and the velocity of light in vacuum which is assumed as an absolute constant in special theory of relativity. Thus time is measured only after the distance between the mirrors and the velocity of light are given, and it is not that time measures the motion or velocity of light wave. 

This light clock occupies a certain volume in space, and if an acceleration exists, the mirrors in the clock should be distorted according to general theory of relativity. In this case the number of counts of light between two mirrors cannot be regarded as giving the time of a certain definite frame of reference. To any point in the clock, different metrical tensor may be associated, and one cannot determine the time of which point the clock measures. Here appears a problem of the size of the actual clock which cannot be infinitesimally small. In this sense, the operational definition of the clock in general theory of relativity has a problem. It seems that this problem may be avoided by an interpretation that the general theory of relativity is an approximation of reality, and the theory gives a sufficiently good approximation as experiments and astronomical observations show. This problem, however, will be a cause of difficulty when one tries to quantize the field equation of general theory of relativity, for the expected quantized theory should be covariant under the diffeomorphism which transforms a point of a space-time manifold to a point of the manifold, and no point can accommodate any clocks with actual sizes.

To have sound foundations to resolve these problems, we therefore have to find, firstly, a definition of clock and time which should be given through length (or positions) and velocity (or motion) to accord with the spirit of Einstein's light clock, and secondly, our notion of time should have a certain ``good" residence just as the inertial frame of reference in special theory of relativity accommodates the light clock. 

As a residence, we prepare a Euclidean quantum space, and within that space we have defined a quantum-mechanical clock which measures the common parameter of quantum-mechanical motions of particles in a (local) system consisting of a finite number of particles. Since clocks thus defined are proper to each local system, and local systems are mutually independent as concerns the relation among the coordinates of these systems, we can impose relativistic change of coordinates among them. And the change of coordinates gives a relation among those local systems which would yield relativistic quantum-mechanical Hamiltonians, explaining the actual observations.

These are an outline of what we did in this book. The problem between quantum mechanics and relativistic theories has been noted \cite{B} already in 1932 by John von Neumann in the footnote on page 6 of the English translation \cite{N} of his book ``Die Mathematische Grundlagen der Quantenmechanik," Springer-Verlag, Berlin:
\begin{quotation}
\F
-- in all attempts to develop a general and relativistic theory of
electromagnetism, in spite of noteworthy partial successes, the theory (of
Quantum Mechanics) seems to lead to great difficulties, which apparently
cannot be overcome without the introduction of wholly new ideas.
\end{quotation}

We below will consider the problem in the case of quantum mechanics and special theory of relativity, and will see a solution in the case of one particle in 3 dimensional space $R^3$ first. After then we will see a sketchy explanation of general $N$-particle case.
\BP

\F
{\bf I. One particle case with mass $m$}
\BP

\F
We consider this one particle in the universe expressed as a vector bundle $X\times R^6$ with base space $X$ being the curved Riemannian space where relativistic CM (classical mechanics) holds and with a fibre $R^6$ (a phase space associated to each point $x\in X$) with an Euclidean structure on which QM (quantum mechanics) holds. Then inside the QM system of the one particle (which we call a ``local system" of the particle), the particle follows the Schr\"odinger equation when the local time $t$ of the system is given as in definition \ref{Time}, and in the classical space outside the local system the particle is observed as a classical particle moving with velocity $v$ relative to the observer in the observer's time.
\MP

\F
The motion inside a local system corresponds to the usually
 conjectured ``invisible" {\it zitterbewegung} of particles like
 virtual photons associated around elementary particles. This
 motion exists even when the particle is at ``zero-point"
 energy, e.g. Not only that but the spin, the vortices observed
 at very low temperatures in superfluidity etc., would be
 interpreted \cite{Natarajan} as this internal motion inside
 a local system.
\MP

\F
Instead of imposing connections among those fibres
 as is usually expected for the terminology 
``vector bundle," we postulate a relation between
 the internal and external worlds by actualizing the
 ``invisible" internal motion as velocity $u$ inside a
 local system in the following two axioms, which is
 assumed to hold between the two worlds on the
 occasion of observation:
\MP

\F
A1. Let $u$ be the QM velocity of the particle inside the local system. Then $u$ and
$v$ satisfy
$$
|u|^2+|v|^2=c^2,
$$
where $c$ is the velocity of light in vacuum.
\MP

\F
A2. The magnitude of momentum inside the local system is observed constant
independent of the velocity $v$ relative to the observer:
$$
m^2|u|^2=(m_0)^2c^2,
$$
where $m_0$ is the rest mass of the particle and $m$ is the observed mass of the
particle moving with velocity $v$ relative to the observer.
\BP

\F
These axioms are an extension of Einstein's principle \cite{[Ein]} of the constancy of the velocity of light in vacuum to a principle of the constancy of the velocity of everything when the velocities in the internal quantum mechanical space and the external classical relativistic space are summed. See Natarajan \cite{Na} for the natural motivation for these axioms corresponding to postulates IV and V in \cite{Na}, whereas Natarajan considers both of internal and external worlds are classical. We consider the internal world quantum mechanical. Thus the above axioms need a justification in order for these to have a consistent meaning, which discussion is given below. This will then yield a unification of quantum mechanics and special theory of relativity.

\BP

\F
To see the consequences of those axioms, we first consider the internal motion inside the QM local system of the particle.
\MP

\F
Let $H$ be the Hamiltonian of the particle inside the QM space:
$$
H = -(\hbar^2)/(2m)(\Delta_x)
= \frac{1}{2m}\left(\frac{\hbar}{i}\right)^2\left(\frac{\partial}{\partial x}\right)^2 = \frac{1}{2m}P^2,
$$
where
$$
\frac{\partial}{\partial x}=\left(\frac{\partial}{\partial x_1},\frac{\partial}{\partial x_2},\frac{\partial}{\partial x_3}\right),
\quad
\left(\frac{\partial}{\partial x}\right)^2=\left(\frac{\partial}{\partial x_1}\right)^2+\left(\frac{\partial}{\partial x_2}\right)^2+\left(\frac{\partial}{\partial x_3}\right)^2,
\quad
P=\frac{\hbar}{i}\frac{\partial}{\partial x}
$$
are partial differentiation with respect to 3 dimensional space coordinate
$x=(x_1,x_2,x_3)$ and $\hbar=h/(2\pi)$ with $h$ being Planck constant. $P$ is identified with the QM momentum of the particle.
\MP

\F
We define a clock of the local system as in definition \ref{Time}. A remark should be added. There we defined a local clock for the particle number $N\ge 2$. The one particle case we are now considering can be regarded in this context a system of 2 particles with the ocnfiguration $x$ above being the relative 3-dimensional coordinates inside the 2 body system. With this understanding, the local time of the system is defined by the $t$ on the exponent of the local clock $\exp(-itH/\hbar)$.

\BP

\F
If $t$ is given as such, we can describe the QM motion of the particle by a solution
$$
\phi(t) = \exp(-itH/\hbar)\phi(0).
$$
of the Schr\"odinger equation
$$
\frac{\hbar}{i}\frac{d}{dt}\phi(t) + H \phi(t) = 0.
$$

\MP

\F
If we insert the QM velocity $P/m$ into $u$ in postulates A1 and A2, they become
\MP

\F
A1. $|P/m|^2+|v|^2=c^2$.
\MP

\F
A2. $|P|^2=(m_0)^2c^2$.
\BP

\F
But $P$ is a differential operator and the rest are numerical quantities, so these
conditions are meaningless. To make these two postulates meaningful, we move to a
momentum space by spectral representation or by Fourier transformation as
follows.
\MP

\F
In our one particle case, we actually considering a two-body system so that there is an interaction term between the two partilces. But for the sake of simplicity we consider the system as free from interaction. Then the solution
$\phi(t)=\exp(-itH/\hbar)\phi(0)$ is given by using Fourier transformation ${\cal F}$:
$$
{\cal F}f(p)=(2\pi \hbar)^{-3/2}\int_{R^3} \exp(-ip\cdot x/\hbar)f(x)dx,
$$
where $p \in R^3$, $p\cdot x=p_1x_1+p_2x_2+p_3x_3$,
as follows:
$$
\phi(t)=\phi(t,x)={\cal F}^{-1}\exp(-itp^2/(2m\hbar)){\cal F}\phi(0).
$$
Also the Hamiltonian $H$ is given by
$$
H=\frac{1}{2m}{\cal F}^{-1}p^2{\cal F}.
$$
${\cal F}$ is a unitary operator from ${\cal H}=L^2(R^3)$ onto itself. Here $L^2(R^3)$ is the Hilbert
space of Lebesgue measurable complex-valued functions $f(x)$ on $R^3$ that satisfies
$$
\int_{R^3}|f(x)|^2dx < \infty,
$$
and has inner product and norm:
$$
(f,g)=\int_{R^3}f(x)\overline{g(x)} dx,\quad \Vert f\Vert=\sqrt{(f,f)},
$$
where
$\overline{g(x)}$ is the complex conjugate to a complex number $g(x)$.
\MP

\F
From this we construct a spectral representation of $H$ as follows:
\MP

\F
Let ${\cal F}(\lambda)$ $(\lambda>0)$ be a map from a subspace (exactly speaking, $L^2_{s}(R^3)$ with
$s>1/2$, see chapter \ref{chap:5}) of $L^2(R^3)$ into
$L^2(S^2)$ ($S^2$ is the unit sphere in $R^3$) defined by
$$
{\cal F}(\lambda)f(\omega)=(2\lambda)^{1/4}({\cal F}f)(\sqrt{2\lambda}\omega),
$$
where $\omega$ is in $S^2$, i.e. $\omega \in R^3$ and $|\omega|=1$.
Then by the above equation we have
$$
{\cal F}(\lambda)Hf(\omega)=(\lambda/m){\cal F}(\lambda)f(\omega).
$$
Thus $H$ is identified with a multiplication by $\lambda/m$ when transformed by
${\cal F}(\lambda)$ into a spectral representation space
$$
\widehat{{\cal H}} = L^2((0,\infty), L^2(S^2), d\lambda),
$$
where $\{{\cal F}(\lambda)\}_{\lambda>0}$ is extended to a unitary operator from ${\cal H}$ onto $\widehat{\cal H}$ in the following sense
$$
\int_0^\infty\Vert {\cal F}(\lambda)f\Vert_{L^2(S^2)}^2 d\lambda =\Vert f\Vert_{L^2(R^3)}^2.
$$
(For more details, see chapter \ref{chap:5}, section \ref{spectral}.)
\MP

\F
Summing up we can regard $H$ as a multiplication operator $\lambda/m$ by moving to a
momentum space representation.
\MP

\F
Originally, $H$ is
$$
H = \frac{1}{2m}P^2,
$$
and $P$ is a 3 dimensional QM momentum. Thus formally we have a correspondence
$$
\lambda \leftrightarrow P^2/2.
$$
Thus if we denote the QM velocity of the particle inside the local system by $u$,
it is
$$
u=P/m
$$
and satisfies a relation
$$
u^2=P^2/m^2 \leftrightarrow 2\lambda/m^2.
$$
Therefore our postulates A1 and A2 above are restated as follows:
\MP

\F
A1. $2\lambda/m^2 + |v|^2=c^2$.
\MP

\F
A2. $2\lambda=(m_0)^2c^2$.
\MP

\F
These are now meaningful, as the relevant quantities are all numeric. These
axioms correspond to a requirement that we think all things in the local system of the one particle, on an energy shell
$H=\lambda/m=(m_0)^2c^2/(2m)$ of the Hamiltonian $H$, whenever considering the observation of the particle.
\MP

\F
Show that the follwoing relation holds.
\BP

\F
Proposition 1. $m=m_0/\sqrt{1-(v/c)^2} (\ge m_0)$.
\BP

\F
We define:
\MP

\F
Definition. The period $p(v)$ of a local system moving with velocity $v$ relative
to the observer is defined by the relation:
$$
p(v)\lambda/(\hbar m)=2\pi.
$$
Thus
$$
p(v) = hm/\lambda = 2hm/[(m_0)^2c^2].
$$
This gives a period of the local system with the clock on the energy shell
$H=\lambda/m$:
$$
exp(-itH/\hbar)=exp(-it\lambda/(\hbar m)).
$$
In particular, when $v=0$, the period $p(0)$ takes the minimum value:
$$
p(0)=hm_0/\lambda=2hm_0/[(m_0)^2c^2]=2h/(m_0 c^2).
$$
This we call the least period of time (LPT) of
 the local system. This gives a minimum cycle or period
 proper to the local system. 

\MP

\F
The general $p(v)$ is related to this by virtue of the Proposition 1 as follows:
\BP

\F
Proposition 2.
$$
p(v)=hm/\lambda=p(0)/\sqrt{1-(v/c)^2} (\ge p(0)).
$$
This means that the time $p(v)$ that the clock of a local system,
 moving with velocity $v$ relative to the observer, rounds $1$ cycle
 when it is seen from the observer, is longer than the time $p(0)$
 that the observer's clock rounds $1$ cycle, and the ratio is given
 by $p(v)/p(0)=1/\sqrt{1-(v/c)^2} (\ge 1)$. Thus time, measured
 by our QM clock, of a local system moving with velocity $v$
 relative to the observer becomes slow with the rate
 $\sqrt{1-(v/c)^2}$, which is exactly the same as the rate
 that the special theory of relativity gives. This yields that
 the QM clock obeys the same transformation rule as that for
 classical relativistic clocks like light clock discussed
 at the beginning of this exercise, and shows that quantum mechanical clock
 is equivalent to the relativistic classical clock.
\BP

These mean also that the space-time measured by using
 QM clock defined as the QM evolution of a local
 system follows the classical relativistic change of coordinates of
 space-time. Thus giving a consistent unification of QM and special
 relativistic CM. 
\BP

As for the validness of the name LPT, we see how it gives the Planck time:
$$
t_P=\sqrt{Gh/c^5} = 1.35125\times 10^{-43}{\rm\ s},
$$
where $G$ is the gravitational constant. In fact, given Planck mass:
$$
m_0=m_P = \sqrt{hc/G} = 5.45604\times 10^{-5} {\rm\ g},
$$
our LPT yields
$$
p(0) = 2h/(\sqrt{hc/G}c^2)
= 2\sqrt{hG/c^5} = 2t_P,
$$
which is 2 times Planck time.
\BP

\vskip12pt


\F
{\bf II. $N$-particles case with masses $m_j$ $(j=1,2,\cdots,N)$}
\BP

\F
We consider the case after the local system $L$ of $N$ particles are scattered sufficiently. Then the
system's solution asymptotically behaves as follows as the system's time $t=t_L$ goes to $\infty$ 
(see \eq{eq6}):
$$
\exp(-it_LH_L/\hbar)f
\sim \exp(-it_L h_b/\hbar)g_0
\otimes\exp(-it_LH_1/\hbar)g_1\otimes\cdots\otimes \exp(-it_LH_k/\hbar)g_k,
$$
where $h_b=T_b+I_b(x_b,0)$ and $k \ge 1$.
\MP

\F
Now for getting the above result also in this case it suffices to note
that the Hamiltonians $H_\ell$ $(\ell=1,2,...,k)$ of each scattered cluster are treated just
as in the case I) above but with using a general theorem on the spectral
representation of self-adjoint operators $H_\ell$. $H_\ell$ are not necessarily free
Hamiltonians and we cannot use the Fourier transformation, but we can use
spectral representation theorem so that each $H_\ell$ is expressed, unitarily equivalently, as $\lambda_\ell/M_\ell$ in
some appropriate Hilbert space, where $M_\ell$ is the mass of the $\ell$-th cluster.
\MP

\F
Then it is done quite analogously to I) to derive the propositions 1 and 2 above
in the present case. Of course these relations depend on $\ell$, and show that the dependence of mass and time of each cluster on the relative velocity to the observer is exactly the same as the special theory of relativity gives as in the case I) above.
\BP

One of the important consequences of these arguments is that the quantum clock is equal to the classical relativistic clock, which has remained unexplained as one of the greatest mysteries in modern physics in spite of the observed fact that they coincide with high precision.

\part{Conclusions}\label{Conclusions}

\chapter{Inconsistency of Mathematics?}\label{inconsistency-chap}

To begin with stating our conclusive thought of this book, we consider that almost all of what we think would be able to be translated into mathematical words insofar as we consider about physical universe as we see in ordinary physical works after Galileo, Descartes, Newton, till the present age. Thus to consider our description of the universe, it would be inevitable to think about metamathematics and set theory which construct the basis of the modern mathematics. In the next chapter, we will try to describe the universe as a contradictory aspect of our language, or exactly speaking, of our mathematical language, in that it could include all sentences as at least meaningful ones so is regarded contradictory as a whole of those meaningful sentences. To be prepared for that purpose, in this chapter we will try to see if mathematics or set theory which is thought to be a basis of modern mathematics is consistent or not. As is well-known, in 1903, Russell's paradoxical set produced immense discussions about the foundation of mathematics, so follows Hilbert's formalism of mathematics, but this direction was negatively answered by G\"odel \cite{[G]}. We will see in this chapter how this theorem of G\"odel seems to give a problem that looks telling that mathematics itself is inconsistent.

\BP

We consider a formal set theory $S$, where we can develop a number theory. As no generality is lost, in the following we consider a number theory that can be regarded as a subsystem of $S$, and will call it $S^{(0)}$.

\begin{df}\label{Definition 1} 1) We assume that a G\"odel numbering of the system $S^{(0)}$ is given, and denote a formula with the G\"odel number $n$ by $A_n$.

\F
2) $\mbox{\AAA}^{(0)}(a,b)$ is a predicate meaning that ``$a$ is the G\"odel number of a formula $A$ with just one free variable (which we denote by $A(a)$), and $b$ is the G\"odel number of a proof of the formula $A(\aaa)$ in $S^{(0)}$," and $\mbox{\BBB}^{(0)}(a,c)$ is a predicate meaning that ``$a$ is the G\"odel number of a formula $A(a)$, and $c$ is the G\"odel number of a proof of the formula $\neg A(\aaa)$ in $S^{(0)}$." Here $\aaa$ denotes the formal natural number corresponding to an intuitive natural number $a$ of the meta level.
\end{df}

\begin{df}\label{Definition 2} Let $\mbox{\PP}(x_1,\cdots.x_n)$ be an intuitive-theoretic predicate. We say that\linebreak $\mbox{\PP}(x_1,\cdots,x_n)$ is {\it numeralwise expressible} in the formal system $S^{(0)}$, if there is a formula $P(x_1,\cdots,x_n)$ with no free variables other than the distinct variables $x_1,\cdots,x_n$ such that, for each particular $n$-tuple of natural numbers $x_1,\cdots,x_n$, the following holds:
\MP

\F
i) if $\mbox{\PP}(x_1,\cdots,x_n)$ is true, then $\vdash P(\xx_1,\cdots,\xx_n).$
\MP

\F
and
\MP

\F
ii) if $\mbox{\PP}(x_1,\cdots,x_n)$ is false, then $\vdash \neg P(\xx_1,\cdots,\xx_n).$
\end{df}

\F
Here ``true" means ``provable on the meta level," and $\vdash P$ means that a formula $P$ is provable in the formal system, e.g., $S^{(0)}$.

\begin{lem}\label{Lemma 1} There is a G\"odel numbering of the formal objects of the system $S^{(0)}$ such that the predicates $\mbox{\AAA}^{(0)}(a,b)$ and $\mbox{\BBB}^{(0)}(a,c)$ defined above are primitive recursive and hence numeralwise expressible in $S^{(0)}$ with the associated formulas $A^{(0)}(a,b)$ and $B^{(0)}(a,c)$. (See \cite{K}.)
\end{lem}

\begin{df}\label{Definition 3} Let $q^{(0)}$ be the G\"odel number of a formula: 
$$
\forall b [\neg A^{(0)}(a,b)\vee \exists c(c\le b \hskip3pt\&\hskip2pt B^{(0)}(a,c))].
$$
Namely
\beq
A_{q^{(0)}}(a)=\forall b [\neg A^{(0)}(a,b)\vee \exists c(c\le b \hskip3pt\&\hskip2pt B^{(0)}(a,c))].\nonumber
\ene
In particular
\beq
A_{q^{(0)}}(\qqqq^{(0)})=\forall b [\neg A^{(0)}(\qqqq^{(0)},b)\vee \exists c(c\le b \hskip3pt\&\hskip2pt B^{(0)}(\qqqq^{(0)},c))].\nonumber
\ene
\end{df}

\BP

Assume that $S^{(0)}$ is consistent.
\BP

Suppose that
\beq
\vdash A_{q^{(0)}}(\qqqq^{(0)})\mbox{ in } S^{(0)},\nonumber
\ene
and let $k^{(0)}$ be the G\"odel number of the proof of $A_{q^{(0)}}(\qqqq^{(0)})$. Then by the numeralwise expressibility of $\AAA^{(0)}(a,b)$
\beq
\vdash A^{(0)}(\qqqq^{(0)},\kk^{(0)}).\label{Rosser1}
\ene
Under our hypothesis of consistency,
$$
\vdash A_{q^{(0)}}(\qqqq^{(0)})\mbox{ in } S^{(0)}
$$
implies
$$
\mbox{not } \vdash \neg A_{q^{(0)}}(\qqqq^{(0)})\mbox{ in } S^{(0)}.
$$
Hence, for any integer $\ell$, $\BBB^{(0)}(q^{(0)},\ell)$ is false. In particular, $\BBB^{(0)}(q^{(0)},0)$, $\cdots,$ $\BBB^{(0)}(q^{(0)},k^{(0)})$ are false. By virtue of the numeralwise expressibility of $\BBB^{(0)}(a,c)$, from these follows that
$$
\vdash \neg B^{(0)}(\qqqq^{(0)},0),\vdash \neg B^{(0)}(\qqqq^{(0)},1),\cdots, \vdash \neg B^{(0)}(\qqqq^{(0)},\kk^{(0)}).
$$
Hence
$$
\vdash \forall c (c\le \kk^{(0)} \supset \neg B^{(0)}(\qqqq^{(0)},c)).
$$
This together with $\vdash A^{(0)}(\qqqq^{(0)},\kk^{(0)})$ in \eq{Rosser1} gives
$$
\vdash \exists b [A^{(0)}(\qqqq^{(0)},b)\hskip3pt\&\hskip2pt\forall c(c\le b\supset\neg B^{(0)}(\qqqq^{(0)},c))].
$$
This is equivalent to
$$
\vdash \neg A_{q^{(0)}}(\qqqq^{(0)})\mbox{ in } S^{(0)}.
$$
A contradiction with our consistency hypothesis of $S^{(0)}$. Thus
$$
\mbox{not } \vdash A_{q^{(0)}}(\qqqq^{(0)})\mbox{ in } S^{(0)}.
$$

Reversely, suppose that
\beq
\vdash \neg A_{q^{(0)}}(\qqqq^{(0)})\mbox{ in } S^{(0)}.\nonumber
\ene
Then there is a G\"odel number $k^{(0)}$ of the proof of $\neg A_{q^{(0)}}(\qqqq^{(0)})$ in $S^{(0)}$, and we have
$$
\mbox{\BBB}^{(0)}(q^{(0)},k^{(0)}) \mbox{ is true.}
$$
Thus 
$$
\vdash B^{(0)}(\qqqq^{(0)},\kk^{(0)}),
$$
from which follows
\beq
\vdash \forall b[b\ge \kk^{(0)}\supset \exists c(c\le b\hskip3pt\&\hskip2pt B^{(0)}(\qqqq^{(0)},c))].\label{(1)}
\ene
As $\neg A_{q^{(0)}}(\qqqq^{(0)})$ is provable in $S^{(0)}$, there is no proof of $A_{q^{(0)}}(\qqqq^{(0)})$ in $S^{(0)}$ by our consistency assumption of $S^{(0)}$. Therefore
\beq
\vdash \neg A^{(0)}(\qqqq^{(0)},0), \vdash \neg A^{(1)}(\qqqq^{(1)},1),\cdots,\vdash \neg A^{(0)}(\qqqq^{(0)},\kk^{(0)}-1)\nonumber
\ene
hold. Thus
\beq
\vdash \forall b[b<\kk^{(0)}\supset \neg A^{(0)}(\qqqq^{(0)},b)]\label{(2)}.
\ene
Combining \eq{(1)} and \eq{(2)}, we obtain
$$
\vdash \forall b[\neg A^{(0)}(\qqqq^{(0)},b)\vee \exists c(c\le b\hskip3pt\&\hskip2pt B^{(0)}(\qqqq^{(0)},c))],
$$
which is
$$
\vdash A_{q^{(0)}}(\qqqq^{(0)}).
$$
A contradiction with our consistency assumption of $S^{(0)}$. Thus we have
\beq
\mbox{not }\vdash \neg A_{q^{(0)}}(\qqqq^{(0)})\mbox{ in } S^{(0)}.\nonumber
\ene

We have reproduced Rosser's form of G\"odel incompleteness theorem.

\begin{lem}\label{Lemma 2} Assume $S^{(0)}$ is consistent. Then neither $A_{q^{(0)}}(\qqqq^{(0)})$ nor $\neg A_{q^{(0)}}(\qqqq^{(0)})$ is provable in $S^{(0)}$.
\end{lem}

Thus, we can add either one of $A_{q^{(0)}}(\qqqq^{(0)})$ or $\neg A_{q^{(0)}}(\qqqq^{(0)})$, which we will denote $A_{(0)}$ hereafter, as a new axiom of $S^{(0)}$ without introducing any contradiction. Namely, let $S^{(1)}$ be an extension of the formal system $S^{(0)}$ with an additional axiom $A_{(0)}$. Then by Lemma \ref{Lemma 2}
\beq
S^{(1)} \mbox{ is consistent.}\label{consis}
\ene
We now extend definitions \ref{Definition 1} and \ref{Definition 3} to the extended system $S^{(1)}$ as follows with noting that the numeralwise expressibility of the predicates $\mbox{\AAA}^{(1)}(a,b)$ and $\mbox{\BBB}^{(1)}(a,c)$ defined below can be extended to the new system $S^{(1)}$ with {\it the same} G\"odel numbering {\it as} the one given in Lemma \ref{Lemma 1} for $S^{(0)}$.
\vskip12pt
\F
1) $\mbox{\AAA}^{(1)}(a,b)$ is a predicate meaning that ``$a$ is the G\"odel number of a formula $A(a)$, and $b$ is the G\"odel number of a proof of the formula $A(\aaa)$ in $S^{(1)}$," and $\mbox{\BBB}^{(1)}(a,c)$ is a predicate meaning that ``$a$ is the G\"odel number of a formula $A(a)$, and $c$ is the G\"odel number of a proof of the formula $\neg A(\aaa)$ in $S^{(1)}$."
\MP

\F
2) Let $q^{(1)}$ be the G\"odel number of a formula: 
$$
\forall b [\neg A^{(1)}(a,b)\vee \exists c(c\le b \hskip3pt\&\hskip2pt B^{(1)}(a,c))].
$$
\BP

\F
By the extended numeralwise expressibility, we have in the way similar to Lemma \ref{Lemma 2} by using the consistency \eq{consis} of $S^{(1)}$
\beq
\mbox{not } \vdash A_{q^{(1)}}(\qqqq^{(1)}) \quad\mbox {  and  } \quad\mbox{not } \vdash \neg A_{q^{(1)}}(\qqqq^{(1)}) \quad \mbox{ in } S^{(1)}.\label{pos1}\nonumber
\ene

Continuing the similar procedure, we get for any natural number $n(\ge 0)$ that
\beq
S^{(n)} \mbox{ is consistent,}\label{consis2}
\ene
\beq
\mbox{not } \vdash A_{q^{(n)}}(\qqqq^{(n)}) \quad\mbox {  and  }\quad \mbox{not } \vdash \neg A_{q^{(n)}}(\qqqq^{(n)}) \quad \mbox{ in } S^{(n)}.
\label{posn}
\ene

We now let $S^{(\omega)}$ the extended system of $S^{(0)}$ that includes all of the formulas $A_{(n)}(=A_{q^{(n)}}(\qqqq^{(n)})$ or $\neg A_{q^{(n)}}(\qqqq^{(n)}))$ as its axioms. By \eq{consis2} $S^{(\omega)}$ is consistent. We note that the formula $A_{(n)}$ is recursively defined if we have already constructed the system $S^{(n)}$. Moreover, if we let ${\tilde q}(n)$ be the G\"odel number of the formula $A_{(n)}$, we have ${\tilde q}(i)\ne{\tilde q}(j)$ for all natural numbers $i<j$ as $A_{(j)}$ is not provable in $S^{(i+1)}$ for $i<j$. Thus $\sup_{i\le n}{\tilde q}(i)$ goes to infinity as $n$ tends to infinity. Further we note that ${\tilde q}(n)$ is a recursive function of $n$. Then given a formula $A_r$ with G\"odel number $r$, restricting our attention to the formulas $A_{(n)}$ with $\tilde{q}(n)\le r$, we can determine in $S^{(\omega)}$ recursively if that given formula $A_r$ is an axiom of the form $A_{(n)}$ or not. Thus the addition of all $A_{(n)}$ retains the recursive definition of the following predicates $\mbox{\AAA}^{(\omega)}(a,b)$ and $\mbox{\BBB}^{(\omega)}(a,c)$ defined in the same way as above.
\BP

\F
$\mbox{\AAA}^{(\omega)}(a,b)$ is a predicate meaning that ``$a$ is the G\"odel number of a formula $A(a)$, and $b$ is the G\"odel number of a proof of the formula $A(\aaa)$ in $S^{(\omega)}$," and $\mbox{\BBB}^{(\omega)}(a,c)$ is a predicate meaning that ``$a$ is the G\"odel number of a formula $A(a)$, and $c$ is the G\"odel number of a proof of the formula $\neg A(\aaa)$ in $S^{(\omega)}$."
\BP

\F
Then we see that the predicates $\mbox{\AAA}^{(\omega)}(a,b)$ and $\mbox{\BBB}^{(\omega)}(a,c)$ are numeralwise expressible in $S^{(\omega)}$ and the G\"odel number $q^{(\omega)}$ of the formula: 
$$
\forall b [\neg A^{(\omega)}(a,b)\vee \exists c(c\le b \hskip3pt\&\hskip2pt B^{(\omega)}(a,c))],
$$
denoted by $A_{q^{(\omega)}}(a)$, is well-defined.

As before, we continue the similar procedure, transfinite inductively. In this process, from the nature of our extension procedure, at each step $\alpha$ where we construct the $\alpha$-th consistent system $S^{(\alpha)}$ from the preceding systems $S^{(\gamma)}$ with $\gamma<\alpha$, the predicates $\AAA^{(\alpha)}(a,b)$ and $\BBB^{(\alpha)}(a,c)$ must be recursively defined from the preceding predicates $\AAA^{(\gamma)}(a,b)$ and $\BBB^{(\gamma)}(a,c)$ $(\gamma<\alpha)$ so as for $S^{(\alpha)}$ to be further extended with retaining consistency. For this to hold, it is necessary and sufficient that the ordinal $\alpha$ is a recursive ordinal. 

Is there any ordinal that is not recursive? A function $F(x)$ is called recursive if it has the form:
$$
F(x)=G(x,F|x),
$$
where $F|x$ is a restriction of $F$ to a domain $x$, and $G$ is a given function. Consider the formula for an ordinal $x$:
$$
x=\{y | y\in x\}.
$$
This meets the above requirement of the recursive definition of the ordinal $x$, although this is tautological and may not be considered a definition usually. But if we see the structure that it contains the domain $x$ and by using that domain only it defines $x$ itself, it is not so unreasonable to think that there is no nonrecursive ordinal.

There is, however, a possibility (\cite{M}) that the condition whether or not a nonrecursive ordinal exists in ZFC is independent of the axioms of ZFC. In that case we have two alternatives.
\MP

\F
Case i) There is no nonrecursive ordinal, and hence all ordinals are recursive.
\MP

\F
In this case, the extension of the system $S^{(\alpha)}$ above is always possible. Thus we can extend $S^{(\alpha)}$ indefinitely forever. However, in this process, we cannot reach the step where the number of added axioms is the cardinality $\aleph_1$ of the first uncountable ordinal, as the number of added axioms is at most countable by the nature of formal system. Thus there must be a least countable ordinal $\beta$ such that the already constructed consistent system $S^{(\beta)}$ is not extendable with retaining consistency. This contradicts the unlimited extendibility stated above, and we have a contradiction. Insofar as we assume that every ordinal is recursive, the only possibility remaining is to conclude that set theory is inconsistent.
\MP

\F
Case ii) There is a nonrecursive ordinal, thus there is a least nonrecursive ordinal $\omega_1$ usually called Church-Kleene ordinal (\cite{F}, \cite{T}).
\MP

\F
In this case the above extension of $S^{(\alpha)}$ is possible if and only if $\alpha<\omega_1$. We note that $\omega_1$ is a limit ordinal. For if it is a successor of an ordinal $\delta$, then $\delta<\omega_1$ is recursive, hence so is $\omega_1=\delta+1$, a contradiction with the nonrecursiveness of $\omega_1$. Therefore we can construct, in the same way as that for $S^{(\omega)}$, a consistent system $S^{(\omega_1)}$, which cannot be extended further with retaining consistency by the nonrecursiveness of $\omega_1$.

On the other hand, as we have seen in the discussion of case i), there must be a least countable ordinal $\beta$ such that the already constructed consistent system $S^{(\beta)}$ is not extendable with retaining consistency. Since $\beta$ is the least ordinal such that $S^{(\beta)}$ is not extendable, for any $\alpha<\beta$ the system $S^{(\alpha)}$ is consistently extendable. Whence by the reasoning above about the recursiveness of $\alpha$ with which $S^{(\alpha)}$ is consistently extendable, $\alpha$ is recursive and we have $\alpha<\omega_1$ if $\alpha<\beta$. Thus
\beq
\beta\le \omega_1.\label{beta}
\ene
Reversely, when $\alpha<\omega_1$, $\alpha$ is a recursive ordinal. Thus by the same reasoning as above about the recursiveness of $\alpha$, $S^{(\alpha)}$ is consistently extendable. Therefore $\alpha<\beta$ if  $\alpha<\omega_1$. This and \eq{beta} give
$$
\beta=\omega_1.
$$

\BP

Summarizing, we have proved

\begin{thm}\label{inconsistency} Assume that $S^{(0)}$ is consistent. Suppose that the condition whether or not there is a nonrecursive ordinal is independent of the axioms of ZFC. Then there are the following two alternatives:
\MP

\F
i) There is no nonrecursive ordinal, and hence all ordinals are recursive.
\begin{quotation}
\F
In this case set theory is inconsistent.
\end{quotation}

\F
ii) There is a nonrecursive ordinal, thus there is a least countable nonrecursive ordinal $\beta$.
\begin{quotation}
\F
In this case the corresponding system $S^{(\beta)}$ is consistent and cannot be extended further with retaining consistency.
\end{quotation}
\end{thm}

We remark that this is a metamathematical theorem. 

Thus the inconsistency in i) of this theorem does not give any proof in ZFC of the existence of nonrecursive ordinal. To know whether a nonrecursive ordinal exists or not, we need a proof in ZFC or if such a statement is independent of the axioms of ZFC, we need to add an axiom that determines which the case is. In the latter case, the above theorem shows a direction in which the extended ZFC can be consistent if the original ZFC is consistent.

Further, as the above theorem is a metamathematical theorem, even if there is no nonrecursive ordinal, the case i) of the theorem does not yield that set theory is inconsistent in the sense that we can find a concrete inconsistent proposition like Russell's paradox inside the set theory. Rather it would be said that we may not find such an inconsistent proposition insofar as we work inside the set theory ZFC. Thus this theorem should not be interpreted as stating any concrete inconsistency of set theory.

\chapter{Stationary Universe}\label{StationaryUniverse}

By nature what is called the universe must be a closed universe, within which
 is all. We will characterize it by a certain
 quantum-mechanical condition.

We consider a metatheory of a formal set theory $S$. We name this metatheory $M_S$, indicating that it is a Meta-theory of $S$ as well as a Meta-Scientific theory as Ronald Swan \cite{[Swan]} refers to. The following arguments are all made in $M_S$.
\BP

We define in $M_S$
$$
\phi =\mbox{ the class of all well-formed formulae of }S.
$$
This $\phi$ is a countable set in the context of $M_S$.
\MP

\F
We identify $\phi$ with the set of truth values (in complex numbers $\C$) of well-formed formulae (wff's) in $\phi$. In this identification, we define a map $T$ from $\phi$ to $\phi$ by
$$
T(\wedge(q))=\mbox{the truth value of a well-formed formula }[\wedge(q)\mbox{ and not }\wedge(q)]
$$
for $q\subset\phi$, with $\wedge(q)$ denoting the conjunction of $q$.
\MP

\F
We note that every subset $q$ of $\phi$ becomes false by adding some well-formed formula $f$ of $\phi$. Hence, the conjunction of $q'=q\cup \{f\}$ is false and satisfies
\MP

$$
T(\wedge(q'))=\wedge(q').
$$

\F
In this sense, $\phi$ is a fixed point of the map $T$.
\MP

\F
Moreover, we have the followings.

\begin{enumerate}
\item
In the sense that any subset $q$ of $\phi$ is false if some well-formed formula is added to $q$, $\phi$ is {\bf inconsistent}.

\item
As $\phi$ is the class of all possible well-formed formulae, $\phi$ is {\bf absolute}.

\item
As $\phi$ is the totality of well-formed formulae, $\phi$ includes the well-formed formula whose meaning is that ``$\phi$ is the class of all well-formed formulae in $S$" in some G\"odel type correspondence between $S$ and $M_S$. In this sense $\phi$ includes (the definition of) $\phi$ itself. Thus $\phi$ is {\bf self-referential} and {\bf self-creative}, and is {\bf self-identical}, just as in M. C. Escher's lithograph in 1948, entitled ``pencil drawing."

\end{enumerate}

\F
The item 3 implies that $\phi$ is a non-well founded set (see \cite{Wegner}).
\MP

\F
The class $\phi$ is the first world, the Universe, which is completely chaotic. In other words, $\phi$ is ``{\bf absolute inconsistent self-identity}" in the sense of Kitarou Nishida \cite{[Nishida]}, whose meaning was later clarified by Ronald Swan \cite{[Swan]} in the form stated above. In this clarification, $\phi$ can be thought ``absolute nothingness" in Hegel's sense.
\MP

\F
The Universe $\phi$ is contradictory, and hence its truth value is constantly oscillating between the two extremal values or poles, truth and false, or $+1$ and $-1$, or more generally, inside a unit sphere of $\C$. Namely, the class $\phi$ as a set of wff's of the set theory $S$ is countable, but the values which the elements of $\phi$ take vary on a unit sphere. In other words, the Universe $\phi$ is a stationary oscillation, when we see its meaning.
\MP

\F
Oscillation is expressed by exponential functions: $\exp(ix\cdot p)$, where $x=(x_1,\cdots,x_d), p=(p_1,\cdots,p_d) \in R^d$ and $x\cdot p = \sum_{i=1}^d x_i p_i$, where $d$ is a positive integer suggesting the dimension of space.
\MP

\F
This $\exp(ix\cdot p)$ is an eigenfunction of the negative Laplacian $-\Delta$:
$$
-\Delta=-\sum_{i=1}^d \frac{\partial^2}{\partial x_i^2}.
$$
Namely
$$
-\Delta \exp(ix\cdot p) = p^2 \exp(ix\cdot p).
$$
\MP

\F
This is generalized to some extent. I.e. if a perturbation $V=V(x)$ satisfies that
$$
H=-\Delta+V(x)\ \mbox{is a self-adjoint operator on}\ \HH=L^2(R^d),
$$
then
\MP

\F
\begin{center}
$\phi$ is expressed as an eigenfunction of $H$.
\end{center}
\MP

\F
Considering the absolute nature of the Universe $\phi$, we will be led to think that the Hamiltonian $H$ of $\phi$ is a Hamiltonian of infinite degree of freedom on a Hilbert space:
$$
\UU=\{\phi\}=\bigoplus_{n=0}^\infty \left(\bigoplus_{\ell=0}^\infty
\HH^n  \right) \q (\HH^n=\underbrace{\HH\otimes\cdots\otimes
\HH}_{\mbox{\scriptsize $n$ factors}}).
$$
$\UU$ is called a Hilbert space of possible universes. An element 
$\phi$ of $\UU$ is called a universe and is of the form of an
 infinite matrix $(\phi_{n\ell})$ with components $\phi_{n\ell} 
\in \HH^n$. $\phi=0$ means $\phi_{n\ell}=0$ for all $n, \ell$.

Let $\OO=\{ S\}$ be the totality of the selfadjoint operators $S$
 in $\UU$ of the form $S\phi=(S_{n\ell}\phi_{n\ell})$ for
$\phi=(\phi_{n\ell})\in{\cal D}(S)\subset\UU$, where each component
 $S_{n\ell}$ is a selfadjoint operator in $\HH^n$. 

We now postulate that the Universe $\phi$ is an eigenfunction of the total Hamiltonian $H=H_{\mbox{\scriptsize{\it total}}}$.
\BP

\F
{\bf Axiom 1}.
There is a selfadjoint operator $H_{\mbox{\scriptsize{\it total}}}=(H_{n\ell})\in\OO$ in 
$\UU$ such that for some $\phi\in \UU-\{0\}$ and $\la\in R^1$
\beq
H_{\mbox{\scriptsize{\it total}}}\phi\approx\la \phi \label{eq1}
\ene
in the following sense: Let $F_n$ be a finite subset of 
${\N}=\{1,2,\cdots\}$ with $\sharp(F_n)(=$ the number of elements in
 $F_n)=n$ and let $\{ F_n^\ell\}_{\ell=0}^\infty$ be the totality of
 such $F_n$ (note: the set $\{ F_n^\ell\}_{\ell=0}^\infty$ is countable). Then the formula \eq{eq1} in the above means that there are
 integral sequences $\{n_k\}_{k=1}^\infty$ and 
$\{ \ell_k\}_{k=1}^\infty$ and a real sequence 
$\{\la_{n_k \ell_k}\}_{k=1}^\infty$
such that 
$F_{n_k}^{\ell_k}\subset F_{n_{k+1}}^{\ell_{k+1}}$; 
$\bigcup_{k=1}^\infty F_{n_k}^{\ell_k} = \N$;
\beq
H_{n_k\ell_k}\phi_{n_k\ell_k}=\la_{n_k\ell_k}\phi_{n_k\ell_k},
\q \phi_{n_k\ell_k}\ne0,\q k=1,2,3,\cdots; \label{eq2}
\ene
and
$$
\la_{n_k\ell_k}\to\la\q {\mbox{as}}\q k\to\infty.
$$ 
\BP

\F
Here we should repeat a remark in Axiom \ref{axiom3}: the subscript $\ell$ in $H_{n\ell}$ distinguishes different local systems with the same number $N=n+1$ of particles.
\BP

 $H=H_{\mbox{\scriptsize{\it total}}}$ is an infinite matrix $(H_{n\ell})$ of selfadjoint operators
 $H_{n\ell}$ in $\HH^n$. Axiom 1 asserts that this matrix converges
 in the sense of \eq{eq1} on our universe $\phi$. We remark that our
 universe $\phi$ is not determined uniquely by this condition.

 The universe as a state $\phi$ is a whole, within which is all.
 As such a whole, the state $\phi$ can follow the two ways: The one
 is that $\phi$ develops along a global time $T$ in the grand
 universe $\UU$ under a propagation $\exp(-iTH_{\mbox{\scriptsize{\it total}}})$, and another is
 that $\phi$ is a bound state of $H_{\mbox{\scriptsize{\it total}}}$. If there were such a global
 time $T$ as in the first case, all phenomena had to develop along
 that global time $T$, and the locality of time would be lost. We
 could then {\it not} construct a notion of local times compatible
 with general theory of relativity. The only one possibility is
 therefore to adopt the stationary universe $\phi$ of Axiom 1.

In every finite part of $\phi$, a local existence in $\phi$ is expressed by a superposition of exponential functions
$$
\psi(x)=(2\pi)^{-d(N-1)/2}\int_{R^{d(N-1)}}\exp(ix\cdot p) g(p)dp
$$
for some natural number $N=n+1\ge 2$ with $n$ corresponding to the superscript $n$ in $\HH^n$ of the definition of $\UU$ above. The function $g(p)$ is called Fourier transform of $\psi(x)$ and satisfies
$$
g(p)=\FF\psi(p):=(2\pi)^{-d(N-1)/2}\int_{R^{d(N-1)}}\exp(-ip\cdot y)\psi(y)dy.
$$
A finite subset of wff's in $\phi$ corresponds to a partial Hamiltonian $H$ of $H_{\mbox{\scriptsize{\it total}}}$ of finite degree of freedom, as the content/freedom that is given by a finite number of wff's in $\phi$ corresponds to a finite degree, $n=N-1$, of freedom of a partial wave function $\psi(x)$ of the total wave function $\phi$. If such a partial Hamiltonian $H$ of $H_{\mbox{\scriptsize{\it total}}}$ satisfies some conditions, we can get a similar expansion of a local existence $\psi(x)$ by using generalized eigenfunctions of $H$. This is known as a spectral representation of $H$ in a general setting, but we here are speaking of a more specific expression called generalized Fourier transform or generalized eigenfunction expansion associated with Hamiltonian $H$ (originated by Teruo Ikebe \cite{[Ikebe]}).
\MP

We call $p$ momentum conjugate to $x$. More precisely we define momentum operator $P=(P_1,\cdots,P_d)$ conjugate to configuration operator $X=(X_1,\cdots,X_d)$ $(X_j=\mbox{multiplication operator by configuration }x_j)$ by
$$
P_j=\FF^{-1}p_j\FF=\frac{1}{i}\frac{\partial}{\partial x_j}\qq(j=1,\cdots,d).
$$
Then $P$ and $X$ satisfy
$$
[P_j,X_\ell]=P_jX_\ell-X_\ell P_j =\delta_{j\ell}\frac{1}{i}.
$$
This shows that what we are dealing with is quantum mechanics. So to accord with actual observation, we modify the definition of $P$
$$
P_j=\frac{\hbar}{i}\frac{\partial}{\partial x_j},
$$
where $\hbar=h/(2\pi)$, and $h$ is Planck constant. Accordingly, the Fourier and inverse Fourier transformations are modified
\beq
&&\FF\psi(p)=g(p)=(2\pi\hbar)^{-d(N-1)/2}\int_{R^{d(N-1)}}\exp(-ip\cdot y/\hbar)\psi(y)dy,\nonumber\\
&&\FF^{-1}g(x)=(2\pi\hbar)^{-d(N-1)/2}\int_{R^{d(N-1)}}\exp(ix\cdot p/\hbar)g(p)dp.\nonumber
\ene

To sum our arguments up to here, we have constructed quantum mechanics as a semantics of the class $\phi$ of all well-formed formulae of a formal set theory $S$. Quantum mechanics is, in this context, given as an interpretation of set theory.
\BP

We continue to complete our semantics of the Universe $\phi$.
\BP

A local existence is of finite nature, and it is so local that it cannot know the existence of the infinite Universe, and is self-centered. In other words, a local coordinates system starts from its own origin, and it is the self-centered origin of the local system. All things are measured with respect to this local origin.

Therefore we have our second and third principles.
\BP

\F
{\bf Axiom 2}. (a simplified version of Axiom \ref{axiom2}) A local system is of finite nature, having its own origin of position $X$ and momentum $P$, independent of others' origins and others' inside worlds.
\BP

\F
{\bf Axiom 3}. (a simplified version of Axiom \ref{axiom3}) The nature of locality is expressed by a local Hamiltonian
$$
H=-\frac{1}{2}\Delta + V
$$
up to some perturbation $V$, that does not violate the oscillatory nature of local existence. Here $\Delta=\sum_{j=1}^{N-1}\frac{\hbar^2}{\mu_j}\sum_{k=1}^d\frac{\partial^2}{\partial x_{jk}^2}$, the number $N$ corresponds to the number of quantum particles of the local system, and $\mu_j$ is the reduced masses of the particles of the local system.
\BP

A local existence (or local system) is oscillating as a sum or integral of generalized eigenfunctions of $H$. In this sense, the locality or local system is a {\it{stationary oscillating system}}.

A local oscillation may be an eigenfunction of the local Hamiltonian $H$. However, by the very nature that locality is a self-centered existence of finite nature, it is shown that it cannot be an eigenstate of $H$, or more precisely speaking, there is at least one Universe wave function $\phi$ every part of which is not an eigenfunction of the local system Hamiltonian $H$. (See chapter \ref{LocalMotion}. See also \cite{[Kitada-localtime]}, \cite{[Kitada-localsystem]}, \cite{[Kitada-ToL]}, \cite{[Ki-Fl2]}, \cite{[Ki-Fl]}.)

To express this oscillation explicitly in some ``outer coordinate," we force the locality or local system to oscillate along an ``afterward-introduced" real-valued parameter $t$. The oscillation is then expressed by using the Hamiltonian $H$
$$
\exp(-2\pi itH/h).
$$
This operator is known in quantum mechanics as the evolution operator of the local system. We will call it the local clock of the system, and we will call $t$ the local time of the system.

Using our self-centered coordinates of our local system in Axiom 2, that is, letting $x$ be position coordinates and $v=m^{-1}P$ be velocity coordinates inside the local system ($m$ being some diagonal mass matrix), we can prove, by virtue of the fact that a local oscillation $\psi(x)$ is not an eigenfunction of $H$, that
$$
\left(\frac{x}{t}-v\right)\exp(-itH/\hbar)\psi(x) \to 0
$$
as $t$ tends to $\pm \infty$ along some sequence in some spectral decomposition of $\exp(-itH/\hbar)\psi$ (Theorem \ref{Enss} or see \cite{[Kitada-ToL]}). This means that the word ``local clock" is appropriate for the operator $\exp(-itH/\hbar)$ and so is ``local time" for the parameter $t$. Therefore we also have seen that ``time" exists locally and only locally, exactly by the fact that locality is a self-centered existence of finite nature. This fact corresponds to Ronald Swan's statements in page 27 of \cite{[Swan]} ``localization must be completely, or unconditionally, circumstantial" and ``localization is not self-creative."

\MP

Let $P_H$ denote the orthogonal projection onto the space of
bound states for a selfadjoint operator $H$. We call the set of all states
 orthogonal to the space of bound states a scattering space, and its element as a scattering state. Let $\phi=(\phi_{n\ell})$ with 
$\phi_{n\ell}=\phi_{n\ell} (x_1,\cdots,x_n)\in L^2(R^{dn})$ 
be the universe in Axiom 1, and let $\{n_k\}$ and $\{\ell_k\}$ be
 the sequences specified there. Let $x^{(n,\ell)}$ denote the
 relative coordinates of $n+1$ particles in $F_{n+1}^\ell$.
\MP


\F
{\bf Definition 1.}
\begin{namelist}{888}
\item[(1)]
We define $\HH_{n\ell}$ as the sub-Hilbert space of $\HH^n$
 generated by the functions $\phi_{n_k\ell_k}$ $(x^{(n,\ell)},y)$ of
 $x^{(n,\ell)}\in R^{dn}$ with regarding $y\in R^{d(n_k-n)}$ as a
 parameter, where $k$ moves over a set $\{k\ |\ n_k\ge n, 
F_{n+1}^\ell\subset F_{n_k+1}^{\ell_k}, k\in \N\}$.
\item[(2)]
$\HH_{n\ell}$ is called a {\it local universe} of  $\phi$.
\item[(3)]
$\HH_{n\ell}$ is said to be non-trivial if 
$(I-P_{H_{n\ell}})\HH_{n\ell}\ne\{0\}$.
\end{namelist}

The total universe $\phi$ is a single element in  $\UU$. The local
 universe $\HH_{n\ell}$ can be richer and may have elements more
 than one. This is because we consider the subsystems of the
universe consisting of a finite number of particles. These
 subsystems receive the influence from the other particles of
 infinite number outside the subsystems, and may vary to constitute
 a non-trivial subspace $\HH_{n\ell}$. We will consider this point in chapter \ref{LocalMotion}.

\hyphenation{Coulomb}

We can now define local system.

\BP

\F
{\bf Definition 2.}
\begin{namelist}{888}
\item[(1)]
The restriction of $H_{\mbox{\scriptsize{\it total}}}$ to $\HH_{n\ell}$ is also denoted by the same
 notation  $H_{n\ell}$ as the $(n,\ell)$-th component of $H_{\mbox{\scriptsize{\it total}}}$.
\item[(2)]
We call the pair
$(H_{n\ell},\HH_{n\ell})$  a local system.
\item[(3)]
The unitary group
$e^{-itH_{n\ell}}$ $(t\in R^1)$ on $\HH_{n\ell}$ is called the 
{\it local} or {\it proper clock} of the local system $(H_{n\ell},\HH_{n\ell})$, if
 $\HH_{n\ell}$ is non-trivial: 
$(I-P_{H_{n\ell}})\HH_{n\ell}\ne \{0\}$.
(Note that the clock is defined only for $N=n+1\ge 2$, since 
$H_{0\ell}=0$ and $P_{H_{0\ell}}=I$.)
\item[(4)]
The universe $\phi$ is called {\it rich} if $\HH_{n\ell}$ equals
$\HH^n=L^2(R^{dn})$ for all $n\ge 1$, $\ell\ge0$. For a rich
 universe $\phi$, $H_{n\ell}$ equals the $(n,\ell)$-th component of
 $H_{\mbox{\scriptsize{\it total}}}$.
\end{namelist}


\F
{\bf Definition 3.}
\begin{namelist}{888}
\item[(1)]
The parameter $t$ in the exponent of the local clock 
$e^{-itH_{n\ell}}=e^{-itH_{(N-1)\ell}}$ of a local system 
$(H_{n\ell},\HH_{n\ell})$ is called the (quantum-mechanical) 
{\it proper time} or {\it local time} of the local system 
$(H_{n\ell}, \HH_{n\ell})$, if $(I-P_{H_{n\ell}})\HH_{n\ell}\ne
 \{0\}$. 
\item[(2)]
This time $t$ is denoted by $t_{(H_{n\ell},\HH_{n\ell})}$ indicating
 the local system under consideration.
\end{namelist}

This definition is a one reverse to the usual definition of the
 motion or dynamics of the $N$-body quantum systems, where the time
 $t$ is given {\it a priori} and then the motion of the particles is
 defined by $e^{-itH_{(N-1)\ell}}f$ for a given initial state $f$ of
 the system.

{\it Time} is thus defined only for local systems
$(H_{n\ell},\HH_{n\ell})$  and is determined by the associated
 local clock $e^{-itH_{n\ell}}$. Therefore there are infinitely
 many number of times $t=t_{(H_{n\ell},\HH_{n\ell})}$ each of which
 is proper to the local system $(H_{n\ell},\HH_{n\ell})$. In this
 sense time is a local notion. There is no time for the total
 universe $\phi$ in Axiom 1, which is a bound state of the total
Hamiltonian $H_{\mbox{\scriptsize{\it total}}}$ in the sense of the condition \eq{eq1} of 
Axiom 1.

\BP

Once given the local time, the local system obeys Schr\"odinger equation
$$
\left(\frac{\hbar}{i}\frac{d}{dt}+H\right)\exp(-itH/\hbar)\psi(x) =0.
$$

\BP

All up to now can be expressed on a Euclidean space $R^d$. We need not worry about any curvature as we consider ourselves with respect to our own coordinates.

But when we look at the outside world, our view will be distorted due to the finiteness of our ability. As equivalent existences as localities, we are all subject to one and the same law of distortion.

\MP

Among local systems, we thus pose a law of democracy.

\MP

\F
{\bf Axiom 4}. (a simplified version of Axiom \ref{axiom4}) General Principle of Relativity. Physical worlds or laws are the same for all local observers.
\BP

As a locality, we cannot distinguish between the actual force and the fictitious force, as far as the force is caused by the distortions that our confrontations to the outside world produce.

We thus have the fifth axiom.
\BP

\F
{\bf Axiom 5}. (a simplified version of Axiom \ref{axiom5}) Principle of Equivalence. For any gravitational force, we can choose a coordinate system (as a function of time $t$) where the effect of gravitation vanishes.
\BP

Axioms 4 and 5 are concerned with the distortion of our view when we meet the outside, while Axioms 1--3 are about the inside world which is independently conceived as its own. The oscillatory nature of local systems in Axiom 3 is a consequence of the locality of the system and the stationary nature of the Universe, so that the oscillation is due to the intrinsic ``internal" cause, while the distortion of our view to the outside is due to observational ``external" cause.

Those two aspects, the internal and the external aspects, are independent mutually, because the internal coordinate system of a local system is a relative one inside the local system and does not have any relation with the external coordinates. Therefore, when we are inside, we are free from the distortion, while when we are meeting the outside, we are in a state that we forget the inside and see a curved world. Thus Axioms 1--5 are consistent.

\BP

Quantum mechanics is introduced as a semantic interpretation of a formal set theory, and general relativity is set as a democracy principle among finite, local systems. The origin of local time is in this finitude of local existence, and it gives the general relativistic proper time of each system.

Set theory is a purely inward thought. Physics obtained as semantics of the set theory is a look at it from the outside. The obtained QM itself is a description of the inside world that breeds set theory. The self-reference prevails everywhere and at every stage.

\chapter{Existence of Local Motion}\label{LocalMotion}

We are in a position to see how the stationary nature of the universe and the existence of local motion and hence local time are compatibly incorporated into our formulation.

\BP

\section{G\"odel's theorem}\label{Goedel}

Our starting point is the incompleteness theorem proved by
 G\"odel \cite{[G]}. It states that any consistent
 formal theory that can
 describe number theory includes an infinite number of undecidable
 propositions (see chapter \ref{inconsistency-chap}). The physical world includes at least
 natural numbers, and it is described by a system of words, which
 can be translated into a formal physics theory. The theory of
 physics, if consistent, 
 therefore includes an undecidable proposition, i.e. a proposition
 whose correctness cannot be known by human beings until one finds
 a phenomenon or observation that supports the proposition or
 denies
 the proposition. Such propositions exist infinitely according to
 G\"odel's theorem. Thus human beings, or any other finite entity,
 will never be able to reach a ``final" theory that can express
 the totality of the phenomena in the universe.

Thus we have to assume that any human observer sees a part
 or subsystem $L$ of the universe and never gets the total
 Hamiltonian $H_{\mbox{\scriptsize{\it total}}}$ in \eq{eq1} by his observation. Here the
 total Hamiltonian $H_{\mbox{\scriptsize{\it total}}}$ is an {\it ideal} Hamiltonian
that might be gotten by ``God." In other words, a consequence
 from G\"odel's theorem is that the Hamiltonian that an
 observer assumes with his observable universe
 is a part $H_L$ of $H_{\mbox{\scriptsize{\it total}}}$. Stating explicitly, 
the consequence from G\"odel's theorem is the
 following proposition
\beq
H_{\mbox{\scriptsize{\it total}}}=H_L+I+H_E,\q H_E\ne 0,\label{G2}
\ene
where $H_E$ is an unknown Hamiltonian describing
 the system $E$ exterior to the realm of the observer,
 whose existence, i.e. $H_E\ne 0$, is assured by G\"odel's
 theorem. This unknown system $E$ includes
all that is unknown to the observer. 
E.g., it might contain particles which
 exist near us but have not been discovered yet,
 or are unobservable for some reason at the time of
 observation.
The term $I$ is an unknown interaction between
 the observed system $L$ and the unknown system $E$. 
Since the exterior system $E$ is assured to exist
by G\"odel's theorem, the interaction $I$ does not vanish:
 In fact assume $I$ vanishes. Then the observed system $L$
 and the exterior system $E$ do not interact, which is
 the same as that the exterior system $E$ does not exist
 for the observer. 
On the other hand,
assigning the so-called G\"odel number to each
 proposition in number theory, G\"odel constructs 
undecidable
propositions in number theory by a diagonal argument, 
which shows that any consistent formal
 theory has a region exterior to the knowable world
(see \cite{[G]}).
Thus the observer 
 must be able to construct a proposition by G\"odel's
 procedure that proves
 $E$ exists, which means $I\ne 0$.
By the same reason, $I$ is not
 a constant operator:
\beq
I \ne \mbox{constant operator}.\label{G3}
\ene
For suppose it is a constant operator. Then
 the systems $L$ and $E$ do not change no matter how far or
 how near they are located because the interaction
 between $L$ and $E$ is a constant operator.
 This is the same situation as that the interaction does not
 exist, thus reduces to the case $I=0$ above.

We now arrive at the following observation:
For an observer, the observable universe is a part $L$ 
of the total universe and it looks as though it follows the
 Hamiltonian $H_L$, not following the total Hamiltonian $H_{\mbox{\scriptsize{\it total}}}$.
 And the state of the system $L$ is described by a part
 $\phi(\cdot,y)$ of the state $\phi$ of the total universe,
 where $y$ is an unknown coordinate of system $L$ inside
 the total universe, and $\cdot$ is the variable controllable
 by the observer, which we will denote by $x$.
\BP


\section{Local time exists}\label{Exists}

In the following argument, we assume an exact relation:
\beq
H_{\mbox{\scriptsize{\it total}}}\phi=0 \label{eq1-prime}
\ene
instead of \eq{eq1}, for simplicity.

Assume now, as is usually expected under condition \eq{eq1-prime},
 that there is no
 local time of $L$, i.e. that the state $\phi(x,y)$
 is an eigenstate of the local Hamiltonian $H_L$ for
 some $y=y_0$ and a real number $\mu$:
\beq
H_L\phi(x,y_0)=\mu\phi(x,y_0).\label{G4}
\ene
Then from \eq{G2}, \eq{eq1-prime} and \eq{G4} follows that
\beq
&&0=H_{\mbox{\scriptsize{\it total}}}\phi(x,y_0)
=H_L\phi(x,y_0)+I(x,y_0)\phi(x,y_0)+H_E\phi(x,y_0)\nonumber\\
&&\ \hskip5pt=(\mu+I(x,y_0))\phi(x,y_0)+H_E\phi(x,y_0).\label{G5}
\ene
Here $x$ varies over the possible positions of the particles
 inside
 $L$. On the other hand, since $H_E$ is the Hamiltonian
 describing the system $E$ exterior to $L$, it does not
 affect the variable $x$ and acts only on the variable $y$.
 Thus $H_E\phi(x,y_0)$ varies as a bare function $\phi(x,y_0)$
 insofar as the variable $x$ is concerned.
Equation \eq{G5} is now written: For all $x$
\beq
H_E\phi(x,y_0)=-(\mu+I(x,y_0))\phi(x,y_0).\label{G6}
\ene
As we have seen in \eq{G3}, the interaction $I$ 
is not a constant operator and varies when $x$
 varies\footnote[6]{Note that G\"odel's theorem
 applies to any fixed $y=y_0$ in \eq{G3}. Namely,
 for any position $y_0$ of the system $L$ in the
 universe, the observer must be able to know
 that the exterior system $E$ exists because
 G\"odel's theorem is a universal statement
 valid throughout the universe.
 Hence $I(x,y_0)$ is not a constant operator
 with respect to
 $x$ for any fixed $y_0$.},
 whereas the action
 of $H_E$ on $\phi$ does not.
 Thus there is a nonempty set of points $x_0$
 where $H_E\phi(x_0,y_0)$ and $-(\mu+I(x_0,y_0))\phi(x_0,y_0)$
 are different, and \eq{G6} does not hold at such points
 $x_0$. If $I$ is assumed to be continuous in the variables
 $x$ and $y$, these points $x_0$ constitutes a set of
 positive measure. This then implies that our assumption
 \eq{G4} is wrong. Thus a subsystem $L$ of the universe cannot
 be a bound state with respect to the observer's Hamiltonian
 $H_L$. This means that the system $L$ is observed as
 a non-stationary system, therefore there must be observed
 a motion inside the system $L$. This proves that the
 ``time" of the local system $L$ {\it exists for the
 observer} as a measure of motion, whereas the total
 universe is stationary and does not have ``time."

\section{A refined argument}\label{refined}

To show the argument in section \ref{Exists} more explicitly,
we consider a simple case of
$$
H_{\mbox{\scriptsize{\it total}}}=\frac{1}{2}\sum_{k=1}^N
h^{ab}(X_k)p_{ka} p_{kb}+V(X).
$$
Here $N$ $(1\le N\le \infty)$ is the number of particles
 in the universe, $h^{ab}$ is a three-metric, 
$X_k\in R^d$ is the position of the $k$-th particle, 
$p_{ka}$ is a functional derivative corresponding to
 momenta of the $k$-th particle, and
$V(X)$ is a potential. The configuration
 $X=(X_1,X_2,\cdots,X_N)$ of total particles is decomposed
 as $X=(x,y)$ accordingly to if the $k$-th particle
 is inside $L$ or not, i.e. if the $k$-th particle is
 in $L$, $X_k$ is a component of $x$ and if not it is
 that of $y$. $H_{\mbox{\scriptsize{\it total}}}$ is decomposed as follows:
$$
H_{\mbox{\scriptsize{\it total}}}=H_L+I+H_E.
$$
Here $H_L$ is the Hamiltonian of a subsystem $L$ that
 acts only on $x$, $H_E$ is the Hamiltonian describing the
 exterior $E$ of $L$ that acts only on $y$, and
$I=I(x,y)$ is the interaction between the systems $L$ and $E$.
 Note that $H_L$ and $H_E$ commute.

\BP

\begin{thm} Let $P$ denote the eigenprojection
onto the space of all bound states of $H_{\mbox{\scriptsize{\it total}}}$.
Let $P_L$ be the eigenprojection for $H_L$. Then we have
\begin{equation}
(1-P_L)P \ne 0,\label{G7}
\end{equation}
unless the interaction $I=I(x,y)$ is a constant with
 respect to $x$ for any $y$.
\end{thm}
\MP

\noindent
{\it Proof:} 
Assume that \eq{G7} is incorrect. Then we have
$$
P_LP=P.
$$
Taking the adjoint operators on the both sides, we then have
$$
PP_L=P.
$$
Thus $[P_L,P] = P_LP - PP_L = 0$.
 But in generic this does not hold because
$$
[H_L,H_{\mbox{\scriptsize{\it total}}}] = [H_L, H_L+I+H_E] = [H_L,I]\ne 0,
$$
unless $I(x,y)$ is equal to a constant with respect to $x$.
 Q.E.D.

\BP

\noindent
{\bf Remark.} In the context of chapter \ref{StationaryUniverse},
 the theorem implies the following: 
$$
(1-P_L)P \UU \ne \{ 0 \},
$$
where $\UU$ is a Hilbert space consisting of all
 possible states $\phi$ of the total universe.
 This relation implies that there is a vector $\phi\ne 0$
 in $\UU$ which satisfies $H_{\mbox{\scriptsize{\it total}}}\phi=\lambda \phi$ for
 a real number $\lambda$ while $H_L \Phi \ne \mu \Phi$
 for any real number $\mu$, where $\Phi=\phi(\cdot,y)$
 is a state vector of the subsystem $L$ with an appropriate
 choice of the position $y$ of the subsystem. Thus the space
generated by $\phi(\cdot,y)$'s when $y$ varies is non-trivial in the
sense of Definition 1 in chapter \ref{StationaryUniverse},
 which proves
for the universe $\phi$ 
 that any local system $L$ is non-trivial, and hence
proves the existence of local time for any local system
of the universe $\phi$. Thus we have at least one 
stationary universe $\phi$ where
every local system has its local time.
\BP

\newpage



\noindent
{\bf Exercise}
\BP

\noindent
We consider the systems $S^{(\alpha)}$ defined in chapter \ref{inconsistency-chap}.

Let ${\tilde q}(\alpha)$ denote the G\"odel number of Rosser formula or its negation $A_{(\alpha)}$ ($=A_{q^{(\alpha)}}(\qqqq^{(\alpha)})$ or $\neg A_{q^{(\alpha)}}(\qqqq^{(\alpha)})$), if the Rosser formula $A_{q^{(\alpha)}}(\qqqq^{(\alpha)})$ is well-defined.

By ``recursive ordinals" we mean those defined by Rogers \cite{R}. Then that $\alpha$ is a recursive ordinal means that $\alpha<\omega_1^{CK}$, where $\omega_1^{CK}$ is the so-called Church-Kleene ordinal (\cite{F}, \cite{T}).

\BP

\F
{\bf Lemma}.
The number ${\tilde q}(\alpha)$ is recursively defined for countable recursive ordinals $\alpha<\omega_1^{CK}$. Here `recursively defined' means that ${\tilde q}(\alpha)$ is defined inductively starting from $0$.
\MP

\F
{\bf Remark}. The original meaning of `recursive' is `inductive.' The meaning of the word `recursive' in the following is the one that matches the spirit of Kleene \cite{K} (especially, the spirit of the inductive construction of metamathematical predicates described in section 51 of \cite{K}).
\MP

\F
{\it Proof}. The well-definedness of ${\tilde q}(0)$ is assured by Rosser-G\"odel theorem as explained in chapter \ref{inconsistency-chap}.

We make an induction hypothesis that for each $\delta<\alpha$, the G\"odel number ${\tilde q}(\gamma)$ of the formula $A_{(\gamma)}$ ($=A_{q^{(\gamma)}}(\qqqq^{(\gamma)})$ or $\neg A_{q^{(\gamma)}}(\qqqq^{(\gamma)})$) with $\gamma\le\delta$ is recursively defined for $\gamma \le \delta$.

We want to prove that the G\"odel number ${\tilde q}(\gamma)$ is recursively well-defined for $\gamma \le \alpha$.

i) When $\alpha = \delta+1$, by induction hypothesis we can determine recursively whether or not a given formula $A_r$ with G\"odel number $r$ is equal to one of the axiom formulas
$A_{(\gamma)}$ ($\gamma\le\delta$) of $S^{(\alpha)}$. 
In fact, we have only to see, for a finite number
of $\gamma$'s with ${\tilde q}(\gamma)\le r$ and $\gamma\le\delta$, if we have
$A_{(\gamma)}=A_r$ or not. By induction hypothesis that ${\tilde q}(\gamma)$ is recursively well-defined for $\gamma \le \delta$, this is then decided
recursively.

Thus G\"odel predicate $\mbox{\AAA}^{(\alpha)}(a,b)$ and Rosser predicate $\mbox{\BBB}^{(\alpha)}(a,c)$ with
superscript $\alpha$ are recursively defined, and hence are numeralwise
expressible in $S^{(\alpha)}$. Then the Rosser formula $A_{q^{(\alpha)}}(\qqqq^{(\alpha)})$ is well-defined, and the G\"odel number ${\tilde q}(\alpha)$
of Rosser formula or its negation $A_{(\alpha)}$ ($=A_{q^{(\alpha)}}(\qqqq^{(\alpha)})$ or $\neg A_{q^{(\alpha)}}(\qqqq^{(\alpha)})$) is defined recursively.
Thus ${\tilde q}(\gamma)$ is recursively well-defined for $\gamma \le \alpha$.

ii) If $\alpha$ is a countable {\it recursive} limit ordinal, then there is an increasing
sequence of recursive ordinals $\alpha_n<\alpha$ such that 
\beq
\alpha= \bigcup_{n=0}^\infty \alpha_n.\label{al}
\ene
In the system $S^{(\alpha)}$, the totality of the added axioms $A_{(\gamma)}$ $(\gamma<\alpha)$
is the sum of the added axioms $A_{(\gamma)}$ $(\gamma<\alpha_n)$ of $S^{(\alpha_n)}$. By induction hypothesis, ${\tilde q}(\gamma)$ is recursively defined for $\gamma < \alpha_n$. Thus in each $S^{(\alpha_n)}$ we can determine recursively whether or not a given formula $A_r$ is an axiom of $S^{(\alpha_n)}$ by seeing, for a finite number of $\gamma$'s with ${\tilde q}(\gamma) \le r$ and $\gamma<\alpha_n$, if $A_{(\gamma)}=A_r$ or not.

This is extended to $S^{(\alpha)}$. To see this, 
we have only to see the $\gamma$'s with ${\tilde q}(\gamma) \le r$ and
$\gamma<\alpha$, and determine for those finite number of $\gamma$'s if
$A_{(\gamma)}=A_r$ or not. By \eq{al}, 
$$
{\tilde q}(\gamma) \le r\ \mbox{ and }\ \gamma<\alpha
\Leftrightarrow
\exists n \ \mbox{ such that }
{\tilde q}(\gamma) \le r\ \mbox{ and }\ \gamma<\alpha_n.
$$
Then by induction on $n$ with using the result in the above paragraph for $S^{(\alpha_n)}$ and noting that the bound $r$ on ${\tilde q}(\gamma)$ is uniform in $n$, we can show that the condition whether or not ${\tilde q}(\gamma) \le r$ and
$\gamma<\alpha$ is recursively determined. Whence the question whether or not a given formula $A_r$ is one of the axioms $A_{(\gamma)}$ of $S^{(\alpha)}$ with ${\tilde q}(\gamma) \le r$ and $\gamma<\alpha$ is determined recursively.
Thus G\"odel predicate $\mbox{\AAA}^{(\alpha)}(a,b)$ and Rosser predicate $\mbox{\BBB}^{(\alpha)}(a,c)$ with superscript
$\alpha$ are recursively defined, and hence are numeralwise expressible in
$S^{(\alpha)}$. Therefore the Rosser formula $A_{q^{(\alpha)}}(\qqqq^{(\alpha)})$ is well-defined, and the G\"odel number ${\tilde q}(\alpha)$ of Rosser formula
or its negation $A_{(\alpha)}$ ($=A_{q^{(\alpha)}}(\qqqq^{(\alpha)})$ or $\neg A_{q^{(\alpha)}}(\qqqq^{(\alpha)})$) is defined recursively.
Thus ${\tilde q}(\gamma)$ is recursively well-defined for $\gamma \le \alpha$.
This completes the proof of the lemma.

\BP


\F

Assume now that $\alpha$ is a countable limit ordinal such that there is an increasing sequence of recursive ordinals $\alpha_n<\alpha$ with
\beq
\alpha= \bigcup_{n=0}^\infty \alpha_n.\label{al2}
\ene
An actual example of such an $\alpha$ is the Church-Kleene ordinal $\omega_1^{CK}$.

In the system $S^{(\alpha)}$, the totality of the added axioms $A_{(\gamma)}$ $(\gamma<\alpha)$
is the sum of the added axioms $A_{(\gamma)}$ $(\gamma<\alpha_n)$ of $S^{(\alpha_n)}$. By the lemma, ${\tilde q}(\gamma)$ is recursively defined for $\gamma < \alpha_n$. Thus in each $S^{(\alpha_n)}$ we can determine recursively whether or not a given formula $A_r$ is an axiom of $S^{(\alpha_n)}$ by seeing, for a finite number of $\gamma$'s with ${\tilde q}(\gamma) \le r$ and $\gamma<\alpha_n$, if $A_{(\gamma)}=A_r$ or not.

This is extended to $S^{(\alpha)}$. To see this, 
we have only to see the $\gamma$'s with ${\tilde q}(\gamma) \le r$ and
$\gamma<\alpha$, and determine for those finite number of $\gamma$'s if
$A_{(\gamma)}=A_r$ or not. By \eq{al2}, 
$$
{\tilde q}(\gamma) \le r\ \mbox{ and }\ \gamma<\alpha
\Leftrightarrow
\exists n \ \mbox{ such that }
{\tilde q}(\gamma) \le r\ \mbox{ and }\ \gamma<\alpha_n.
$$
Then by induction on $n$ with using the above result for $S^{(\alpha_n)}$ in the preceding paragraph and noting that the bound $r$ on ${\tilde q}(\gamma)$ is uniform in $n$, we can show that the condition whether or not ${\tilde q}(\gamma) \le r$ and
$\gamma<\alpha$ is recursively determined. Then within those finite number of $\gamma$'s with ${\tilde q}(\gamma)\le r$ and $\gamma<\alpha$, we can decide recursively if for some $\gamma<\alpha$ with ${\tilde q}(\gamma)\le r$, we have $A_r=A_{(\gamma)}$ or not. Therefore we can determine recursively whether or not a given formula $A_r$ is an axiom of $S^{(\alpha)}$.

Therefore G\"odel predicate $\mbox{\AAA}^{(\alpha)}(a,b)$ and Rosser predicate $\mbox{\BBB}^{(\alpha)}(a,c)$ are recursively defined, and hence are numeralwise expressible in
$S^{(\alpha)}$. Then the G\"odel number $q^{(\alpha)}$ of the formula
$$
\forall b [\neg A^{(\alpha)}(a,b)\vee \exists c(c\le b \hskip3pt\&\hskip2pt B^{(\alpha)}(a,c))]
$$
is well-defined, and hence Rosser formula $A_{q^{(\alpha)}}(\qqqq^{(\alpha)})$ is well-defined and Rosser-G\"odel theorem applies to the system $S^{(\alpha)}$. Therefore we can extend $S^{(\alpha)}$ consistently by adding one of Rosser formula or its negation $A_{(\alpha)}$ ($=A_{q^{(\alpha)}}(\qqqq^{(\alpha)})$ or $\neg A_{q^{(\alpha)}}(\qqqq^{(\alpha)})$) to the axioms of $S^{(\alpha)}$ and get a consistent system $S^{(\alpha+1)}$.

\BP

In particular if we assume a least nonrecursive ordinal $\omega_1^{CK}$ exists and take $\alpha=\omega_1^{CK}$, we get a consistent system $S^{(\omega_1^{CK}+1)}$. This contradicts the case ii) of theorem \ref{inconsistency} in chapter \ref{inconsistency-chap}. We leave the following problem to the reader.

\BP

\F
{\bf Question}. The least nonrecursive ordinal, the so-called Church-Kleene ordinal $\omega_1^{CK}$ has been assumed to give a bound on recursive construction of formal systems (see \cite{F}, \cite{S}, \cite{T}). However the above argument seems to question if $\omega_1^{CK}$ really exists in usual set theoretic sense. How should we think?

\BP

\end{document}